\documentclass[a4paper,11pt]{article}
\usepackage[T1]{fontenc}
\usepackage[utf8]{inputenc}
\usepackage{lmodern}
\usepackage{jheppub}
\usepackage{mathtools}
\usepackage{microtype}
\usepackage{booktabs,array,multirow,caption}
\usepackage{enumitem}
\usepackage{needspace}
\definecolor{muted}{gray}{0.32}
\usepackage[table]{xcolor}
\usepackage{microtype}
\allowdisplaybreaks[1]
\setlist[itemize]{leftmargin=1.8em,itemsep=2pt,topsep=4pt}
\makeatletter
\renewcommand\paragraph{\@startsection{paragraph}{4}{\z@}%
  {-2.75ex \@plus -1ex \@minus -.2ex}%
  {1.1ex \@plus .2ex}%
  {\normalfont\normalsize\bfseries}}
\renewcommand*\l@paragraph{\@dottedtocline{4}{7.0em}{4.6em}}
\makeatother
\hypersetup{pdftitle={Machine learning for negative-weight reduction in MC@NLO},
            pdfsubject={JHEP-style typesetting of manuscript v004},
            bookmarksnumbered=true}

\def\beq{\begin{equation}}
\def\eeq{\end{equation}}
\def\beqn{\begin{eqnarray}}
\def\eeqn{\end{eqnarray}}
\def\abs#1{\left|#1\right|}
\newcommand\as{\alpha_{\scriptscriptstyle S}}
\newcommand\conf{{\cal K}}
\newcommand\clH{{\mathbb H}}
\newcommand\clS{{\mathbb S}}
\newcommand\meas{\chi}
\newcommand\phsp{d\phi}
\newcommand\phspnpo{\phsp_{n+1}}
\newcommand\Phsp{\Phi}
\newcommand\Phspnpo{\Phsp^{(n+1)}}
\newcommand\ep{\epsilon}
\newcommand\ord{{\cal O}}
\newcommand{\Ct}{\$_{\tau}}
\newcommand{\Ca}{\$_{A}}
\newcommand{\Tnonlin}{\Delta T_{\mathrm{train}}^{\mathrm{nonlin}}}
\newcommand{\dlin}{\Delta\tau_{\mathrm{train}}^{\mathrm{lin}}}
\newcommand{\epsml}{\epsilon}
\newcommand{\fom}{\operatorname{fom}_{\tau}}
\newcommand{\msl}{\texttt{min\_samples\_leaf}}

\newcommand\pt{p_{\scriptscriptstyle T}}

\title{On the reduction of negative-weight events in MC@NLO
with Machine Learning techniques}

\author[a,b]{L. Beccatini,}
\author[c]{V. Fossa,}
\author[d]{S. Frixione,}
\author[e]{A. Valassi,}
\author[f]{M. Zaro}

\affiliation[a]{Centre for Cosmology, Particle Physics and Phenomenology (CP3),
Universit\'e Catholique de Louvain, Chemin du Cyclotron, Louvain-la-Neuve,
B-1348, Belgium}
\affiliation[b]{Dipartimento di Fisica e Astronomia, Universit\`a di
Bologna, and INFN, Sezione di Bologna, via Irnerio 46, I-40126 Bologna, Italy}
\affiliation[c]{Faculty of Aerospace Engineering,
Delft University of Technology,
Kluyverweg 1,\\ 2629 HS Delft, The Netherlands}
\affiliation[d]{INFN, Sezione di Genova, Via Dodecaneso 33, I-16146, 
Genoa, Italy}
\affiliation[e]{Experimental Physics Department, CERN, 1211 Geneva, Switzerland}
\affiliation[f]{TIFLab, Dipartimento di Fisica, Universit\`a degli
Studi di Milano and INFN, Sezione di Milano, Via Celoria 16,
I-20133 Milan, Italy}

\emailAdd{luca.beccatini@unibo.it}
\emailAdd{veronica.fossa@cern.ch}
\emailAdd{Stefano.Frixione@cern.ch}
\emailAdd{andrea.valassi@cern.ch}
\emailAdd{marco.zaro@mi.infn.it}

\abstract{We demonstrate that Machine Learning techniques can be
successfully employed to re-define $\clS$-event MC@NLO short-distance
cross sections in a manner which in most cases leads to a reduction
of their fractions of negative weights, with a modest increase of
computational costs. We work with both
Decision Trees and Normalizing Flows, and consider several processes,
in both $e^+e^-$ and $pp$ collisions, that vastly differ in their
kinematical and dynamical complexity. In those situations where the
ML-induced reduction of negative weights is not significant we are
able to identify the primary reason(s) for this behaviour, thus paving
the way for it to be addressed in future studies. This work
is still to be considered as a proof of concept, given that the
proposed methodology is not implemented in a public code; yet,
it is based upon a fully realistic simulation environment, in which
no simplifying assumptions are made.
}


\preprint{
\begin{flushright}
TIF--UNIMI--2025--31\\
\today
\end{flushright}
}

\begin{document}
\maketitle

\section{Introduction}\label{sec:introduction}
The spectacular performances of the LHC and of the
experiments which observe the results of the collisions that
happen there have turned high-energy hadronic physics into a
field of high-precision measurements, to an extent that was difficult
to anticipate at the beginning of its operations. Most of the
dominant uncertainties affecting data are of theoretical origin,
and thus the necessity of increasing the accuracy of predictions
is accepted as a matter of fact by the theory community. Part of
that effort concerns tools which are employed by the experiments
themselves, namely the Monte Carlo (MC) programs that can simulate
hard collisions and their accompanying phenomena in minute details.

While the majority of the number of events simulated by the LHC
collaborations are still based on an underlying Leading Order (LO)
description, such a dominance is not overwhelming as it used to be,
since Next-to-Leading Order (NLO) accurate simulations are nowadays
considered as standard, and occupy a significant part of the
computing resources allocated by experiments, in keeping
with the sought increase of precision of the predictions.

Unfortunately, NLO simulations have a number of issues which do not
affect their LO counterparts. Namely, they are slower and more complicated
to run, they can reach smaller hard-event-level final-state multiplicities,
and feature events with negative weights, which effectively reduce
the statistical power of their predictions.

This is the reason why the reduction of negative weights in NLO
simulations has been a constant topic of discussion for the past
twenty-five years, after the invention of MC@NLO~\cite{Frixione:2002ik};
it is both a conceptually interesting problem, and a very practical
issue as well, since such weights cost real money. At the level of
formalisms, the most developed (in terms of both corollary theoretical
ideas and applications) is Powheg~\cite{Nason:2004rx,arXiv:0709.2092,
Alioli:2010xd}; other proposals~\cite{Jadach:2015mza,arXiv:2111.03553,
arXiv:2409.16417,arXiv:2504.05377,arXiv:2511.16605,arXiv:2605.22108}
are more recent and studied in a much less systematic manner.
A different viewpoint is that which informs the works aimed at
solving the problem starting from an existing NLO-accurate
method (typically, MC@NLO), as is done e.g.~in refs.~\cite{Frederix:2020trv,
arXiv:2110.15211,arXiv:2310.04160,Whitehead:CHEP2024,arXiv:2502.08052,
arXiv:2502.08053,arXiv:2602.18124,arXiv:2609.23500}, or by means of
a-posteriori manipulations of both-sign event files~\cite{arXiv:2005.09375,
arXiv:2007.11586,arXiv:2109.07851,arXiv:2303.15246,arXiv:2411.11651,
arXiv:2505.03724,arXiv:2510.16217,arXiv:2605.13237,arXiv:2607.08723,
arXiv:2609.30357,arXiv:1912.08824,arXiv:2011.13445}.

The present work belongs to the former category of the two we have
just mentioned; more specifically, the idea is that of modifying the
MC@NLO short-distance $\clS$-event cross section in a local way,
so that the number of negative weights is reduced without affecting
physical predictions. While our work can be seen as a generalization
of the Born-spreading technique of ref.~\cite{arXiv:2310.04160},
its defining characteristics lies in the fact that the modifications
mentioned before are entirely determined by means of Machine Learning (ML)
techniques; in particular, we employ both Decision Trees (DT) and
Normalizing Flows (NF). The main reason for selecting such approaches lies in
their suitability for efficiently evaluating integrals of the learned
function over the subspace of radiation variables, as is required by
the procedure we are going to define. We stress again that the method deals
only with $\clS$ events, and that $\clH$ events are not relevant in this
context. Having said that, it is interesting to observe that several of the
papers cited above focus on $\clH$ events instead. We shall in fact explicitly
consider some MC@NLO-$\Delta$~\cite{Frederix:2020trv} simulations,
which by construction reduce the number of negative $\clH$ events,
``moving'' some of them into the $\clS$-event set, where they can be
handled with the procedures we present in this work, thereby showing
a synergy between the two approaches.

This paper is organised as follows: we start in sect.~\ref{sec:ml}
with an elementary introduction to the basic ML concepts we shall use
in this paper. In sect.~\ref{sec:mcnlo} we set up the notation, and
write the MC@NLO and MC@NLO-$\Delta$ cross sections accordingly.
We introduce the ML-driven method we propose in this work in
sect.~\ref{subsec:learning-functions}, with DT- and NF-specific
strategies discussed in sects.~\ref{sssec:bdt-construction}
and~\ref{sssec:nf-construction}, respectively. In sect.~\ref{sec:fom}
we define a metric which we use to assess whether our ML-based procedure
is viable, in that it provides one with benefits superior to its overall
cost. We then document in sect.~\ref{sec:res} how the procedure performs
in the context of actual $e^+e^-$- and $pp$-collisions processes.
The DT- and NF-induced results are discussed separately, in
sects.~\ref{par:dtimpl:practical} and~\ref{subsec:results-dt} for the
former, and in sects.~\ref{subsec:nfimpl:practical} and~\ref{subsec:results-nf}
for the latter. We point out that we identify general issues outside of
scope of this work, that are independent of which ML approach one
adopts, and which we discuss within the DT sections. This by no means
implies that we consider DTs superior to NFs for the problem at hand.
Our overall conclusions and outlook are presented in sect.~\ref{sec:concl};
these are mostly non-technical, since DT- and NF-specific summaries
are given in sect.~\ref{sec:DTconcl} and~\ref{sec:NFconcl}, respectively.
Additional material on DTs and NFs can be found in the appendices.

\section{Introduction to machine learning}\label{sec:ml}
Machine Learning (ML) is a collection of computational methods that construct
models from data, rather than relying exclusively on explicitly programmed
rules. The central idea is to approximate an unknown function or probability
distribution using a parametrized model whose parameters (called {\em trainable})
are determined through an optimization procedure.

Suppose we have a dataset
\beq
\label{eq:ml:display-1}
  \mathcal{D}=\{(x_i,y_i)\}_{i=1}^{N},
\eeq
where $x_i$ denotes a set of input variables and $y_i$ a corresponding
target. We introduce a parametrized function $f_\theta(x)$, where $\theta$
represents the model trainable parameters. The objective is to determine $\theta$
so that $f_\theta(x_i)$ approximates $y_i$. This is typically achieved by
minimizing a \emph{loss function}:
\beq
\label{eq:ml:display-2}
  \theta^*=\underset{\theta}{\operatorname{argmin}}\,
    \mathcal{L}(\theta).
\eeq
For example, in regression\footnote{A note for the non-initiated: in
machine learning, regression is the task of learning a function (the regressor)
that predicts a numerical quantity from a set of input variables,
using examples with known target values to minimise a suitable
prediction error and generalise to previously unseen inputs; unlike
classification, its output is a numerical estimate rather than a label
that identifies an object which belongs to a discrete set.} one may
choose the mean squared error,
\beq
\label{eq:ml:display-3}
  \mathcal{L}(\theta)=\frac{1}{N}\sum_{i=1}^N
    \left[f_\theta(x_i)-y_i\right]^2.
\eeq
A crucial requirement is that the resulting model perform well on previously
unseen data, rather than merely reproducing its training sample. This
property is called \emph{generalization}.

In addition to the trainable parameters, a model is characterized by
the so-called {\em hyperparameters}. These specify the model structure,
and/or how training is performed. 

There are several broad classes of ML methods. In \emph{supervised learning},
the training data contain known targets; typical tasks are regression and
classification. In \emph{unsupervised learning}, the algorithm identifies
patterns or structures in data without prescribed targets. \emph{Generative
modelling} aims to learn a probability distribution and generate new samples
consistent with it. These categories overlap: in particular, generative models
can be trained using supervised or unsupervised objectives.

Common ML algorithms include decision trees, which partition the input
space into regions; Boosted Decision Trees, which train successive
trees adaptively to improve the combined prediction; neural networks,
which construct flexible functions through compositions of
parametrized transformations; and Normalizing Flows, which learn
probability densities through invertible transformations.

For the negative-weight reduction problem in MC@NLO simulations, ML will
be used to construct a correction to an event-weight function.
Schematically, denoting by $w(\rho,x)$ that function, with $\rho$ and
$x$ two independent (sets of) variables, the correction will amount to the
replacement:
\beq
\label{eq:ml:display-4}
  w(\rho,x)\longrightarrow w'(\rho,x)=w(\rho,x)+f_\theta(\rho,x),
\eeq
where the learned function $f_\theta(\rho,x)$ is constrained so that
\beq
\label{eq:ml:display-5}
  \int dx\,f_\theta(\rho,x)=0\qquad\text{for every fixed }\rho.
\eeq
The objective is to reduce the fraction of negative-weight events without
modifying the physical cross section and relevant differential distributions.
Integrals analogous to that in eq.~(\ref{eq:ml:display-5}) need to be computed
on the fly (i.e.~very efficiently) in our procedure, which is what has
been alluded to in the introduction when discussing the motivations
beyond our choices of DT and NF as ML tools. For example, while the
present regression task could be also handled by unconstrained Neural Networks,
the computation of an integral with high efficiency would not.

\subsection{Decision Trees}\label{subsec:bdt}

Decision Trees~\cite{breiman1984cart,quinlan1986induction,quinlan1993c45}
approximate a function
by recursively partitioning the space of its arguments. We use trees for
regression: the prediction is a real-valued approximation to the chosen
target, and not a classification of events according to the sign of
their weights.

A decision tree divides the space of input variables into
progressively smaller regions by asking a sequence of simple
questions. For example, if the input variables describe the kinematics
of a particle collision, a tree might first separate events according
to whether their energy exceeds a given threshold, and then further
subdivide them according to their angles or transverse momenta. Each
final region in the input-variable space, called a {\em leaf}, is
assigned a numerical prediction. In this way, a decision tree
approximates a potentially complicated function by a collection of
simpler, locally constant functions.

For learning inputs $z$, a constant-leaf regression tree has the
representation~\cite{SklearnTrees}
\beq
 T(z)=\sum_{\ell=1}^{L}a_\ell\,\mathbf{1}_{R_\ell}(z),
 \label{eq:app:single-tree}
\eeq
where $R_\ell$ is the region associated with leaf $\ell$, $a_\ell$ is its
prediction, and $L$ is the number of leaves; the quantity
$\mathbf{1}_{R_\ell}(z)$ is equal to one if $z\in R_\ell$, and
zero otherwise. With axis-aligned splits,
the regions $R_\ell$ are hyperrectangles in the learning variables;
their geometry is discussed in appendix~\ref{app:bdt-granularity}.

In the general ML notation, a tree defines a parametrized function
$f_\theta(z)$, where $\theta$ specifies its split variables,
thresholds and leaf predictions. The tree is constructed by a sequence
of greedy split optimizations\footnote{``Greedy split optimization''
means choosing, at each node, the split that gives the largest
immediate improvement in the fitting criterion, without looking ahead
to how that choice will affect subsequent splits. The tree is therefore
built through a sequence of locally optimal decisions, rather than by
optimizing its entire structure simultaneously.\label{ft:greed}}, not
by a gradient-descent training loop. For squared-error regression,
such as the one predominantly used in the DT applications in this
work, the prediction in each leaf is the mean of the training targets
reaching that leaf, using the supplied sample weights when
present~\cite{SklearnTrees}. After training, the tree is held fixed
during integration and event generation.

An important consideration is the balance between accuracy and complexity.
Trees with very fine subdivisions can describe the training data extremely
accurately, but may also reproduce statistical fluctuations that are not
representative of the underlying distribution. This phenomenon, known as
\emph{overfitting}, can be controlled by limiting tree complexity, adjusting
the minimum number of events assigned to each leaf, and evaluating the model
on independent data.

In the context of complex Monte Carlo simulations, DTs constitute an
attractive choice owing to their ability to learn intricate
dependences on several kinematic variables, without requiring an
explicit analytical model of the corrections. Their effectiveness
ultimately depends on whether the information contained in the
training variables is sufficient to identify the structures
which one seeks to reproduce.

Training determines the split variables, their thresholds, and the
predictions assigned to leaves. The principal settings chosen in
advance (i.e.~the model hyperparameters)
include~\cite{SklearnTreeRegressor}\footnote{The extended names in
round brackets are the variables used in ref.~\cite{SklearnTreeRegressor}.}:
\begin{itemize}
\setlength{\itemsep}{1pt}\setlength{\topsep}{3pt}
\item \textbf{Minimum samples per leaf} ($n_{\min}$,
\texttt{min\_samples\_leaf}):
  the smallest number of training events allowed in a final region;
  this is the principal granularity control varied in our DT study.
\item \textbf{Maximum tree depth} ($d_{\max}$, \texttt{max\_depth}):
  the maximum number of successive splits along a root-to-leaf path.
\item \textbf{Maximum number of leaves} (\texttt{max\_leaf\_nodes}):
  a bound on the total number of terminal regions.
\item \textbf{Minimum samples for a split} (\texttt{min\_samples\_split}):
  the minimum node occupancy required before a split is considered.
\item \textbf{Split criterion and pruning controls}: the criterion used to
  select subdivisions and any controls that remove unnecessary branches.
\end{itemize}
The numerical settings we have used are given in
section~\ref{subsec:results-dt}. 

{\em Extensions to tree ensembles.\label{sssec:tree-ensembles}}
Decision trees can also be combined into ensembles. In Boosted Decision
Trees (BDTs), successive trees are trained using information from the
preceding predictions. For gradient boosting with squared-error
regression, each additional tree is fitted to the residuals of the
current ensemble~\cite{SklearnEnsembles}:
\beqn
 F_m(z)&=&F_{m-1}(z)+\eta_{\mathrm{BDT}}T_m(z),
 \nonumber\\
 r_i^{(m)}&=&y_i-F_{m-1}(z_i).
 \label{eq:ml:gradient-boosting}
\eeqn
Here $F_0$ is an initial prediction, $T_m$ is the tree fitted at stage
$m$, and $\eta_{\mathrm{BDT}}$ is the learning rate. This is a specific
gradient-boosting example, rather than a definition of every boosting
algorithm. The sequential dependence concerns training: the final
prediction is an additive combination of trees evaluated on the same
input, not a composition of their outputs.

The structural hyperparameters retain their meaning for each
constituent tree, while the number of boosting stages, learning rate
and possible subsampling introduce additional controls. Random forests
instead average predictions from randomized trees~\cite{SklearnEnsembles};
these ensembles are possible extensions of the present DT implementation.
Their further use should be assessed together with their training,
evaluation and integration costs through the figure of merit
of section~\ref{sec:fom}. The radiation integral of an additive ensemble
is discussed in appendix~\ref{app:tree-ensembles}.

\subsection{Normalizing Flows}\label{subsec:nf}

Normalizing Flows~\cite{Tabak:2013cnz,
rippel2013highdimensionalprobabilityestimationdeep,dinh2014nice,
rezende2015variational,Papamakarios2021} are a class of ML models designed to
describe probability distributions. While many conventional ML algorithms aim
to predict a numerical quantity from a set of input variables, Normalizing
Flows focus on learning how probability is distributed over a potentially
complicated multidimensional space. This makes them particularly useful in
problems involving event generation, statistical sampling, and Monte Carlo
integration.

The central idea is to start from a simple, known probability
distribution, and transform it into a more complicated one through a
sequence of reversible mathematical operations. For example, one may
start from points uniformly distributed inside a multidimensional unit
cube. A Normalizing Flow progressively transforms these points so that
they become more concentrated in some regions and less concentrated in
others, according to a probability distribution learned from the data.
In general, Normalizing Flows are characterized by their flexibility in
describing complicated distributions, their ability to generate new
samples efficiently, and the normalization properties of their
learned densities.

Each transformation is constructed so that it can be inverted and its effect
on probability densities can be computed. This is an essential feature: it
allows the model not only to generate points according to the learned
distribution, but also to evaluate the corresponding probability density.
Furthermore, since the starting distribution is normalized and the
transformations correctly account for changes in volume, the resulting
probability distribution is automatically normalized.

In keeping with the general ML framework, the transformations depend on a set
of parameters $\theta$, whose values are determined through training. The
resulting model is a parametrized probability density, which we may denote
by $f_\theta(z)$. A loss function $\mathcal{L}(\theta)$ specifies the desired
properties of this density, and training determines the parameters that
optimize it. Depending on the application, the objective may be to reproduce
a given distribution or to construct one that improves the efficiency of a
numerical procedure.

A particularly useful extension is the \emph{conditional Normalizing Flow},
in which the learned probability distribution depends on additional input
variables. In our case, for example, we shall construct a distribution
of radiation variables whose shape depends on the underlying Born kinematics.
Thus, the same model can describe different radiation patterns for different
Born configurations.

Training determines the weights within the internal neural networks. Typical
hyperparameter settings include:
\begin{itemize}
\setlength{\itemsep}{0pt}\setlength{\topsep}{2pt}
\item \textbf{Number and type of transformations} ($N_{\mathrm{trans}}$, $\mathcal{T}$):
  how many invertible steps are used (e.g.\ affine or spline-based).
\item \textbf{Spline bins} (for spline-based flows): how flexibly each step
  can adapt its shape.
\item \textbf{Network size} ($L$, $n_{\mathrm{units}}$): the number of internal
  layers and units per layer.
\item \textbf{Training settings}: learning rate ($\eta_{\mathrm{NF}}$), batch
  size ($n_{\mathrm{batch}}$), and number of epochs ($N_{\mathrm{epoch}}$).
\end{itemize}
The parameters which appear in the last item are those that are
easiest to handle on an individual-run basis, and therefore we
elaborate further on them.

The learning rate $\eta_{\mathrm{NF}}$ controls the scale of each update
of the trainable parameters of the flows and their auxiliary networks.
For ordinary gradient descent, for example, the update\footnote{This
simple equation exemplifies how a hyperparameter ($\eta_{\mathrm{NF}}$),
which is typically fixed, controls a property of the trainable parameters
($\theta$), which change on the fly.} is
\mbox{$\theta\longrightarrow\theta-\eta_{\mathrm{NF}}
\nabla_\theta \mathcal{L}(\theta)$};
other optimizers modify the update direction while retaining a
learning-rate scale. Too large a learning rate can destabilise training,
whereas too small a value can make progress unnecessarily slow.

The batch size $n_{\mathrm{batch}}$ specifies how many configurations are
processed together to evaluate the loss and its gradients for one parameter
update. When this number is smaller than the training sample, these groups
are called \emph{mini-batches}. The parameters are updated after each batch,
rather than only after the complete sample has been processed. The actual
size of the last batch may be smaller than $n_{\mathrm{batch}}$.

For a fixed training sample, an \emph{epoch} is one complete pass through
that sample. The number of epochs $N_{\mathrm{epoch}}$ therefore counts
repeated passes over the same configurations, typically with their order
reshuffled between passes; it does not count independent new samples.
With one update per batch and no discarded configurations, the number of
updates is
\beq
 N_{\mathrm{update}}=N_{\mathrm{epoch}}
 \left\lceil\frac{N_{\mathrm{train}}}{n_{\mathrm{batch}}}\right\rceil.
 \label{eq:nftrain:update-count}
\eeq
Thus changing the batch size also changes the number of updates at a fixed
number of epochs.

\section{MC@NLO definitions and learning functions}\label{sec:mcnlo}
In this section, we first collect the basic MC@NLO and MC@NLO-$\Delta$
formulae, and then discuss how these can be employed in the context of
an ML-driven negative-weight reduction procedure.  These formulae are
understood for a fixed real-emission subprocess and FKS pair, with the
physical sums over these labels implicit. Also, it will soon become
apparent that the applied ML methodology is identical for MC@NLO
and MC@NLO-$\Delta$; thus, the latter will be generally understood.

\subsection{Phase-space variables}\label{subsec:phase-space}
Henceforth, we consider a process which, at the Born level, features
$n$ final-state particles. By
\beq
  \meas_n\,,\qquad \meas_r\,,
  \label{eq:mc:source-17}
\eeq
we collectively denote, respectively, the $3n-4$ phase-space variables at the
Born level, and the three variables relevant to the radiation of an
extra parton at the real-emission level - the latter may be equivalently called
radiation or FKS variables. The $(n+1)$-body phase space is written
as follows:
\beqn
\phspnpo&=&\Phspnpo(\meas_{n+1})\,d\meas_{n+1}\equiv
\Phspnpo(\meas_n,\meas_r)\,d\meas_n d\meas_r,
\label{eq:phase-space}
\\
\meas_{n+1}&=&\meas_n\cup\meas_r,
\label{eq:phase-split}
\\
\meas_r&=&\{\xi,y,\varphi\}.
\label{eq:radiation-vars}
\eeqn
In the FKS parametrisation (which uses the incoming-parton c.m.~frame
as a reference), $\xi$ is the rescaled energy of the FKS
parton, $y$ the cosine of the angle between that parton and its
sister, and $\varphi$ an azimuthal angle. The soft and collinear
limits are $\xi\to0$ and $y\to1$, respectively. Furthermore
\beq
0\le\xi\le\xi_{\rm max}(\meas_n)\,,\;\;\;\;\;\;\;\;
-1\le y\le 1\,,\;\;\;\;\;\;\;\;
0\le\varphi\le 2\pi\,.
\eeq
It is then convenient to introduce the rescaled variables
\beq
\hat{\xi}=\frac{\xi}{\xi_{\rm max}(\meas_n)}\,,\;\;\;\;\;\;\;\;
\hat{y}=\frac{1}{2}\big(1-y\big)\,,\;\;\;\;\;\;\;\;
\hat{\varphi}=\frac{\varphi}{2\pi}\,,
\label{FKSvresc}
\eeq
so that the soft and collinear limits are $\hat{\xi}\to 0$ and
$\hat{y}\to 0$, respectively, and importance sampling towards those
limits can be achieved by means of a power law $\hat{\xi}^k$, 
$\hat{y}^k$, with $k>1$.

This parametrisation is valid in each FKS sector; we remind the
reader that FKS sectors (and thereby the FKS variables relevant
to each of them) are identified by a pair of integer indices
(see footnote~\ref{ft:sectors} for their physical interpretation);
for the sake of a simpler notation, these will be systematically omitted
in what follows. The two additional Bjorken-$x$ integration variables will
also be understood.

Event configurations with real-emission ($\clH$ events) and Born
kinematics ($\clS$ events) will be denoted by $\conf^{(\clH)}$ and
$\conf^{(\clS)}$, respectively. They are such that:
\beq
\label{eq:mc:display-2}
 \conf^{(\clS)}=\conf^{(\clS)}(\meas_n),
 \qquad
 \conf^{(\clH)}=\conf^{(\clH)}(\meas_n,\meas_r).
\eeq
We shall often denote a fully differential cross section times its
relevant infinitesimal measure by a shorthand notation such as
$d\sigma$. In a more explicit manner, this implies that a function
$\varsigma$ exists, so that
\beq
d\sigma=\varsigma(\meas_n,\meas_r)d\meas_n d\meas_r=
\frac{\varsigma(\meas_n,\meas_r)}{\Phspnpo(\meas_n,\meas_r)}\,\phspnpo.
\label{eq:sigdiff1}
\eeq
The function $\varsigma$ is then the cross section fully differential
in $(\meas_n,\meas_r)$, or can be used to construct a Born-level
differential quantity:
\beq
\varsigma(\meas_n,\meas_r)=\frac{d\sigma}{d\meas_n d\meas_r}\,,\;\;\;\;\;\;\;
\int d\meas_r\varsigma(\meas_n,\meas_r)=
\int d\meas_r\frac{d\sigma}{d\meas_n d\meas_r}\,,
\label{eq:sigdiff2}
\eeq
respectively.
These identities will typically be used in the following to deal with
the $\clS$-event cross section and its contributions.

If $\Theta(\meas_n,\meas_r)$ denotes the integration domain of our
phase-space variables, it is non restrictive to assume that\footnote{Since
we ignore Bjorken $x$'s; these would imply an additional
$[0,1]^2$ integration domain.}
\beq
\Theta(\meas_n,\meas_r)\subseteq[0,1]^m\otimes [0,1]^3\,,
\;\;\;\;\;\;\;\;m=3n-4\,.
\label{rescvar1}
\eeq
Furthermore, in the context of MC@NLO implementations 
based on FKS subtraction, the phase-space parametrization can
be set up in such a way that the above relationship becomes
more stringent, and reads as follows:
\beq
\Theta(\meas_n,\meas_r)=[0,1]^m\otimes\Theta(\meas_n)\,,\;\;\;\;\;\;\;\;
\Theta(\meas_n)\subseteq[0,1]^3\,.
\label{eq:fdomain}
\eeq
In other words, for any given Born-level configuration $\meas_n$,
the FKS radiation variables must belong to the subspace $\Theta(\meas_n)$
for the physical kinematical configuration to exist. This implies that,
if for a generic set $s$ relevant to variables $v$ we define the
{\em indicator}
\beqn
&&\mathbf{1}_{s}(v)=1\;\;\;\;{\rm if}~v\in s\,,
\label{Onein}
\\
&&\mathbf{1}_{s}(v)=0\;\;\;\;{\rm if}~v\notin s\,,
\label{Oneout}
\eeqn
then for any function such as that of eq.~(\ref{eq:sigdiff2})
the following identity holds true:
\beq
\varsigma(\meas_n,\meas_r)=
\varsigma(\meas_n,\meas_r)\mathbf{1}_{\Theta(\meas_n)}(\meas_r)
\label{eq:seqsone}
\eeq
so that
\beq
\int_{\Theta(\meas_n,\meas_r)}d\meas_n d\meas_r\varsigma(\meas_n,\meas_r)=
\int_{[0,1]^{m+3}}d\meas_n d\meas_r
\varsigma(\meas_n,\meas_r)\mathbf{1}_{\Theta(\meas_n)}(\meas_r)\,.
\label{eq:seqsoneint}
\eeq
Since strictly speaking $\varsigma(\meas_n,\meas_r)$ is still
undefined for \mbox{$\meas_r\in [0,1]^3\setminus\Theta(\meas_n)$},
one extends it by zero outside its physical domain,
with the same symbol $\varsigma$ denoting this extension. In numerical
implementations, the off-domain branch is assigned zero without
evaluating the physical expression. Formally, bearing eqs.~(\ref{Onein})
and~(\ref{Oneout}) in mind, one defines the zero-extension on a given
function $F(v)$ that exists in a domain $s$ thus:
\beqn
\mathbf{1}_s(v)F(v)
\;:=\;
\begin{cases}
F(v),&v\in s,\\
0,&v\notin s.
\end{cases}
\label{eq:exttozero}
\eeqn
We stress that an expression $\mathbf 1_s(v)F(v)$ denotes $F(v)$ on
$s$, extended by zero outside $s$, without requiring the
evaluation of $F$ there. This convention applies to the complete
expression multiplied by the indicator and does not modify other
occurrences of $F$.

\subsection{Fixed-order cross section}\label{subsec:fixed-order}
We now introduce symbolically fixed-order short-distance cross sections.
We denote the Born and virtual contributions by $d\sigma^{(B)}$
and $d\sigma^{(V)}$, respectively; the latter is IR- and UV-finite,
defined according to FKS conventions. The remaining fixed-order
contributions are organised according to the labels $S^\prime,C,SC$,
with the latter two standard in MC@NLO implementations, and 
$S^\prime$ defined below. With this, the fixed-order NLO cross
section differential in a generic kinematic configuration $\conf$ is:
\beqn
\begin{aligned}
\frac{d\sigma^{(\mathrm{NLO})}}{d\conf}
={}&\delta\!\left(\conf-\conf^{(\clH)}\right)
\,d\sigma^{(\mathrm{NLO},E)}
\\
&+\delta\!\left(\conf-\conf^{(\clS)}\right)
\left[d\sigma^{(B)}+d\sigma^{(V)}
+\sum_{\alpha=S^\prime,C,SC}d\sigma^{(\mathrm{NLO},\alpha)}\right].
\end{aligned}
\label{eq:fixed-order}
\eeqn
Here $d\sigma^{(\mathrm{NLO},E)}$ is the real-emission contribution. The
quantity $d\sigma^{(\mathrm{NLO},S)}$ is the one conventionally employed in
the MC@NLO formulation, as in ref.~\cite{Frederix:2020trv}, and includes the
Born and virtual contributions. We define the label $S^\prime$ by separating
the latter contributions to $d\sigma^{(\mathrm{NLO},S)}$, namely:
\beq
d\sigma^{(\mathrm{NLO},S)}
=d\sigma^{(B)}+d\sigma^{(V)}
+d\sigma^{(\mathrm{NLO},S^\prime)}.
\label{eq:split}
\eeq
It follows that
\beq
\sum_{\alpha=S,C,SC}d\sigma^{(\mathrm{NLO},\alpha)}
=d\sigma^{(B)}+d\sigma^{(V)}
+\sum_{\alpha=S^\prime,C,SC}d\sigma^{(\mathrm{NLO},\alpha)}.
\label{eq:sum-split}
\eeq
Therefore, the sum over $S^\prime,C,SC$ collects the remaining IR- and
UV-finite contributions of soft, collinear and soft-collinear origin,
including the corresponding subtraction terms. 

\subsection{Generating functionals and matched short-distance cross
sections\label{subsec:functionals}\label{subsec:short-distance}}

For standard MC@NLO, the generating functional is
\beq
\mathcal F_{\mathrm{MC@NLO}}
=\mathcal F_{\mathrm{MC}}\!\left(\conf^{(\clH)}\right)
\,d\sigma^{(\clH)}
+\mathcal F_{\mathrm{MC}}\!\left(\conf^{(\clS)}\right)
\,d\sigma^{(\clS)},
\label{eq:mcnlo}
\eeq
while for MC@NLO-$\Delta$, it is
\beq
\mathcal F_{\mathrm{MC@NLO}\text{-}\Delta}
=\mathcal F_{\mathrm{MC}}\!\left(\conf^{(\clH)}\right)
\,d\sigma^{(\Delta,\clH)}
+\mathcal F_{\mathrm{MC}}\!\left(\conf^{(\clS)}\right)
\,d\sigma^{(\Delta,\clS)}.
\label{eq:mcnlo-delta}
\eeq
In both cases, the same MC counterterms ($d\sigma^{(\mathrm{MC})}$)
appear. In addition to those, the MC@NLO short-distance cross sections
read thus:
\beq
d\sigma^{(\clH)}
=d\sigma^{(\mathrm{NLO},E)}-d\sigma^{(\mathrm{MC})},
\label{eq:H}
\eeq
\beq
d\sigma^{(\clS)}
=d\sigma^{(B)}+d\sigma^{(V)}+d\sigma^{(\mathrm{MC})}
+\sum_{\alpha=S^\prime,C,SC}d\sigma^{(\mathrm{NLO},\alpha)}.
\label{eq:S}
\eeq
while MC@NLO-$\Delta$ ones are:
\beq
d\sigma^{(\Delta,\clH)}
=\left[d\sigma^{(\mathrm{NLO},E)}-d\sigma^{(\mathrm{MC})}\right]\Delta,
\label{eq:H-delta}
\eeq
\beqn
\begin{aligned}
d\sigma^{(\Delta,\clS)}
={}&d\sigma^{(B)}+d\sigma^{(V)}
+\sum_{\alpha=S^\prime,C,SC}d\sigma^{(\mathrm{NLO},\alpha)}
\\
&+d\sigma^{(\mathrm{MC})}\Delta
+d\sigma^{(\mathrm{NLO},E)}(1-\Delta).
\end{aligned}
\label{eq:S-delta}
\eeqn
Fuller details on eqs.~(\ref{eq:H-delta}) and~(\ref{eq:S-delta})
can be found in ref.~\cite{Frederix:2020trv}. Here, it suffices to
say that $\Delta$ is a perturbative ($\Delta=1+\ord(\as)$) function
with support in the $(n+1)$-body phase space, which obeys the following
properties:
\beq
0\leq\Delta\leq 1,\qquad
\Delta\longrightarrow0\ \text{in soft/collinear limits},
\qquad
\Delta\longrightarrow1\ \text{in hard regions}.
\eeq
Standard MC@NLO follows from MC@NLO-$\Delta$ by setting $\Delta=1$
in the latter.

By $\mathcal F_{\mathrm{MC}}(\conf)$ we have denoted the
generating functional of the parton shower starting from
configuration $\conf$. By means of the formal replacement:
\beq
{\mathcal F}_{\mathrm{MC}}\!\left(\conf^{(\clS/\clH)}\right)
\;\longrightarrow\;
\delta\!\left(\conf-\conf^{(\clS/\clH)}\right)
\eeq
in the equations above we can obtain the hard-event-level differential
MC@NLO and MC@NLO-$\Delta$ cross sections, i.e~basically the
quantities stored in LHE files:
\beq
\frac{d\sigma^{(\mathrm{MC@NLO})}_{\rm LHE}}{d\conf}
=\delta\!\left(\conf-\conf^{(\clH)}\right)
\,d\sigma^{(\clH)}+\delta\!\left(\conf-\conf^{(\clS)}\right)d\sigma^{(\clS)},
\label{eq:MCNLOLHE}
\eeq
and analogously for MC@NLO-$\Delta$. It follows that, by introducing
a function
\beq
  \Omega(\meas_n,\meas_r)
  \label{eq:mc:source-19}
\eeq
that defines an ``observable'' (e.g.~a transverse momentum, a rapidity,
the invariant mass of a pair), then the LHE-level MC@NLO cross section
differential in that observable at a value $\Omega_0$ is given by
\beqn
  \frac{d\sigma^{(\mathrm{MC@NLO})}_{\rm LHE}}{d\Omega}(\Omega_0)
  &=&\int d\meas_n\,d\meas_r\,\frac{d\sigma^{(\clH)}}{d\meas_n\,d\meas_r}
  \delta\left(\Omega_0-\Omega(\meas_n,\meas_r)\right)
  \nonumber\\
  &+&\int d\meas_n\,d\meas_r\,\frac{d\sigma^{(\clS)}}{d\meas_n\,d\meas_r}
  \delta\left(\Omega_0-\Omega(\meas_n,0)\right)\,.
  \label{eq:MCNLOLHEOm}
\eeqn
These hard-event kinematical distributions in general do not coincide
with their NLO fixed-order counterparts; this follows immediately from
eqs.~(\ref{eq:MCNLOLHEOm}) and~(\ref{eq:fixed-order}), which lead to:
\beq
\frac{d\sigma^{(\mathrm{MC@NLO})}_{\rm LHE}}{d\Omega}(\Omega_0)-
\frac{d\sigma^{(\mathrm{NLO})}}{d\Omega}(\Omega_0)=
\int d\sigma^{(\mathrm{MC})}\Big[
\delta\left(\Omega_0-\Omega(\meas_n,0)\right)-
\delta\left(\Omega_0-\Omega(\meas_n,\meas_r)\right)\Big].
\eeq
The r.h.s.~of this equation indeed does not vanish in general as a
differential distribution. Its integral over all possible $\Omega_0$
values is, however, equal to zero, which is a consequence of the exact
rate-conservation property of the MC@NLO matching.

The rightmost term on the r.h.s.~of eq.~(\ref{eq:MCNLOLHEOm}) renders it
manifest that while $\clS$ hard events have an $n$-body kinematics (here
denoted by the second argument of the function $\Omega$ being set
equal to zero), their corresponding short-distance cross
sections have support in the $(n+1)$-body phase space. As is
explained in detail in ref.~\cite{Frederix:2020trv}, this is
a major source of negative-weight $\clS$ events, the unweighting
that defines them being carried out in an $(n+1)$-body dimensional
phase-space, where the three FKS variables are redundant. This
observation leads one to using an equivalent ``folded'' representation
of the MC@NLO cross section, namely:
\beqn
  \frac{d\sigma^{(\mathrm{MC@NLO})}_{\rm LHE,fold}}{d\Omega}(\Omega_0)
  &=&\int d\meas_n\,d\meas_r\,\frac{d\sigma^{(\clH)}}{d\meas_n\,d\meas_r}
  \delta\left(\Omega_0-\Omega(\meas_n,\meas_r)\right)
  \nonumber\\
  &+&\int d\meas_n\delta\left(\Omega_0-\Omega(\meas_n,0)\right)
  \int\,d\meas_r\,\frac{d\sigma^{(\clS)}}{d\meas_n\,d\meas_r}
  \,.
  \label{eq:MCNLOLHEfold}
\eeqn
In other words, folding means integrating the $\clS$ short-distance cross
sections over the FKS variables prior to unweighting. This maximally
reduces the number of negative-weights $\clS$ events, but it is
computationally very expensive\footnote{\label{ft:fold}In practice,
folding can never be implemented as is implied by eq.~(\ref{eq:MCNLOLHEfold}),
since this would require an infinite amount of CPU time. Therefore,
one resorts to an approximated computation of the inner integral
by means of appropriately weighted Riemann sums over
$n_\xi\times n_y\times n_\phi$ FKS-variable values, which are determined
by the integrator. This is what is called an $(n_\xi,n_y,n_\phi)$ folding.}.
The basic idea of this paper is that of a modification of the $\clS$-event
short-distance cross section determined by ML techniques, which helps
reduce the number of negative weights without being computationally costly.

Equation~(\ref{eq:MCNLOLHEOm}) understands the integration domain
introduced in sect.~\ref{subsec:phase-space}. Therefore, in keeping
with eqs.~(\ref{eq:seqsone}) and~(\ref{eq:seqsoneint}), one can
rewrite it as follows:
\beqn
  \frac{d\sigma^{(\mathrm{MC@NLO})}_{\rm LHE}}{d\Omega}(\Omega_0)
  &=&\int_{[0,1]^{m+3}} d\meas_n\,d\meas_r\,
  \mathbf{1}_{\Theta(\meas_n)}(\meas_r)
  \frac{d\sigma^{(\clH)}}{d\meas_n\,d\meas_r}
  \delta\left(\Omega_0-\Omega(\meas_n,\meas_r)\right)
  \nonumber\\
  &+&\int_{[0,1]^{m+3}} d\meas_n\,d\meas_r\,
  \mathbf{1}_{\Theta(\meas_n)}(\meas_r)
  \frac{d\sigma^{(\clS)}}{d\meas_n\,d\meas_r}
  \delta\left(\Omega_0-\Omega(\meas_n,0)\right)\,.
  \label{eq:MCNLOLHEOm2}
\eeqn

\subsection{Learning functions\label{subsec:learning-functions}
\label{subsec:dsigmaK}}
We now proceed to manipulate the $\clS$-event short-distance
cross sections with the aim of reducing their induced negative weights.
The $\clH$-event cross sections will not be relevant in this procedure,
and will be thus ignored. Likewise, we shall understand that all
operations that will follow will be at the level of hard-event
configurations, i.e.~the LHE files.

It will turn out to be convenient to introduce a quantity $d\sigma_K$,
obtained from the $\clS$-event cross sections by explicitly separating
Born and virtual contributions. Thus, for standard MC@NLO we have
\beqn
\begin{aligned}
d\sigma_K
&\equiv d\sigma^{(\clS)}-d\sigma^{(B)}-d\sigma^{(V)}
\\
&=d\sigma^{(\mathrm{MC})}
+\sum_{\alpha=S^\prime,C,SC}d\sigma^{(\mathrm{NLO},\alpha)},
\end{aligned}
\label{eq:dsigmaK}
\eeqn
while for MC@NLO-$\Delta$
\beqn
\begin{aligned}
d\sigma_K^{(\Delta)}
&\equiv d\sigma^{(\Delta,\clS)}-d\sigma^{(B)}-d\sigma^{(V)}
\\
&=d\sigma^{(\mathrm{MC})}\Delta
+\sum_{\alpha=S^\prime,C,SC}d\sigma^{(\mathrm{NLO},\alpha)}
+d\sigma^{(\mathrm{NLO},E)}(1-\Delta),
\end{aligned}
\label{eq:dsigmaK-delta}
\eeqn
whereby
\beq
\label{eq:mc:display-1}
 d\sigma_K^{(\Delta)}=d\sigma_K+(1-\Delta)d\sigma^{(\clH)}.
\eeq
We denote the differential cross section stemming from $d\sigma_K$
by $K$ so that, in terms of the standard phase-space variables
(in keeping with the conventions for generic differential quantities,
eqs.~(\ref{eq:sigdiff1})--(\ref{eq:sigdiff2})):
\beqn
\begin{aligned}
K(\meas_n,\meas_r)
&\equiv\frac{d\sigma_K}{d\meas_n\,d\meas_r},
\\
d\sigma_K&=K(\meas_n,\meas_r)\,d\meas_n\,d\meas_r,
\qquad
\sigma_K=\int d\meas_n\,d\meas_r\,K(\meas_n,\meas_r).
\end{aligned}
\label{eq:K-density}
\eeqn
Note that any phase-space jacobian is therefore included in
$K(\meas_n,\meas_r)$. By means of analogous formulae, we define
$K^{(\Delta)}(\meas_n,\meas_r)$ starting from $d\sigma_K^{(\Delta)}$.

In the context of an ML-learning approach, which we generically denote
by $M$ so that, in our practical applications
\beq
M\in\{\mathrm{DT},\mathrm{NF}\},
\eeq
the $\clS$-event cross sections will be modified by formally adding
a contribution to the relevant $d\sigma_K$:
\beqn
d\sigma_K&\longrightarrow&
d\sigma_{K,M}=d\sigma_K+d\delta\sigma_{K,M},
\label{eq:dsigKM}
\\
d\sigma_K^{(\Delta)}&\longrightarrow&
d\sigma_{K,M}^{(\Delta)}=d\sigma_K^{(\Delta)}+d\delta\sigma_{K,M}^{(\Delta)}.
\label{eq:dsigKDM}
\eeqn
These new contributions are constructed through ML; they are locally
non-zero, but their integrals over the FKS variables, at given Born
configurations, are equal to zero:
\beq
\int_{{\rm rad}}d\delta\sigma_{K,M}=
\int_{{\rm rad}}d\delta\sigma_{K,M}^{(\Delta)}=0\,,
\eeq
which gives a specific implementation meaning to the generic constraint
of eq.~(\ref{eq:ml:display-5}). We shall soon see that this is a sufficient
condition to ensure that the ML-modified MC@NLO cross sections, i.e.
\beqn
\begin{aligned}
d\sigma^{(\clS)}_M
&=d\sigma^{(B)}+d\sigma^{(V)}+d\sigma_{K,M},
\\
d\sigma^{(\Delta,\clS)}_M
&=d\sigma^{(B)}+d\sigma^{(V)}+d\sigma_{K,M}^{(\Delta)},
\end{aligned}
\label{eq:ml:reconstructed-S}
\eeqn
are unbiased by ML; in other words, they give a statistically equivalent
set of hard $\clS$ events as the original formulation, which aims at resulting
in a smaller number of negative weights. Equation~(\ref{eq:ml:reconstructed-S})
explicitly implies that the distinction between the DT and NF constructions
(or any other ML technique) concerns, and is limited to, the added
zero-integral correction, and not a different treatment of the Born
and virtual contributions. The added terms stem from a loss function
which may be evaluated by using the full $\clS$-event weight, and not
only its $d\sigma_K$ contribution; we stress that this does not
require absorbing the Born and virtual terms into the definition
of $d\sigma_K$.

The zero-integral condition can be easily related to the normalization
of the learned function. NFs provide one with such a normalization through
their construction. For the DT method, the normalization of the correction
is instead enforced by explicitly computing the radiation integral of the
learned regressor. In both cases, consistent measures and integration
domains fulfill the required zero-integral constraint {\em independently of
the quality of the fit}, provided that the relevant integrals are evaluated
exactly. Numerically approximated normalizations require a corresponding accuracy
check.

\subsubsection{Decision Trees}\label{sssec:bdt-construction}
We start by considering the construction used with DT regressors.
However, we emphasize that our argument does not depend on
the learned function being represented by a single tree: it applies to any
integrable approximant. Decision trees specify the implementation
considered here; extensions to tree ensembles are discussed briefly
in section~\ref{subsec:bdt} and appendix~\ref{app:tree-ensembles}.

We remind the reader that, according to eq.~(\ref{eq:fdomain}),
$\Theta(\meas_n)$ denotes the complete physical radiation domain at
fixed Born configuration, identified by $\meas_n$. We also recall
that, when such a domain does not coincide with the full radiation
phase-space, any function defined in its complement w.r.t.~to the
latter will be set equal to zero there. Such a condition is enforced
by means of the quantity
\beq
\mathbf{1}_{\Theta(\meas_n)}(\meas_r),
\eeq
defined as is shown in eqs.~(\ref{Onein}) and~(\ref{Oneout}).

Now, let $\widetilde{K}(\meas_n,\meas_r)$ denote a learned approximant
of $K$. The notation understands that $(\meas_n,\meas_r)$ are the
independent variables in which $\widetilde{K}$ is learned. While this
is too restrictive, and therefore will be generalised soon, it is
sufficient to sketch the basic steps of the strategy we aim to follow.
This is based on the following identity:
\beqn
  \sigma_K&=&\int d\meas_n\,d\meas_r\left[K(\meas_n,\meas_r)-
  \mathbf{1}_{\Theta(\meas_n)}(\meas_r)\widetilde K(\meas_n,\meas_r)\right]
\nonumber\\*
  &+&\int d\meas_n\,d\meas_r\,\mathbf{1}_{\Theta(\meas_n)}(\meas_r)
  \widetilde K(\meas_n,\meas_r)\,.
  \label{eq:mc:source-22}
\eeqn
Note that the multiplication of $\widetilde{K}$ by
$\mathbf{1}_{\Theta(\meas_n)}(\meas_r)$ is not strictly necessary --
we would have an identity also without it. However, this imposes the
condition that the contribution of $\widetilde{K}$ vanishes where
$K(\meas_n,\meas_r)$ also vanishes (since when
\mbox{$\meas_r\notin\Theta(\meas_n)$} we are outside of the physical
domain, where we simply do not know how to define the latter
function). Equation~(\ref{eq:mc:source-22}) can, in keeping with
eq.~(\ref{eq:dsigKM}), be rewritten as follows:
\beqn
\begin{aligned}
d\sigma_{K,\mathrm{DT}}={}&d\sigma_K
+\mathbf 1_{\Theta(\meas_n)}(\meas_r)
\left[-\widetilde K(\meas_n,\meas_r)+\frac{I_K(\meas_n)}{V(\meas_n)}\right]
\,d\meas_n\,d\meas_r,
\end{aligned}
\label{eq:bdt:dsigmaK}
\eeqn
where we have introduced\footnote{Whenever the domain used to
normalize $I_K$ has zero volume, $-\tilde{K}+I_K/V$ is defined to
vanish, and the ratio of $I_K$ to the volume is not evaluated.}:
\beqn
I_K(\meas_n)&=&
\int d\meas_r\,\mathbf{1}_{\Theta(\meas_n)}(\meas_r)
\widetilde K(\meas_n,\meas_r)\equiv
\int_{\Theta(\meas_n)}d\meas_r\,\widetilde K(\meas_n,\meas_r),
\label{eq:bdt:normalization1}
\\
V(\meas_n)&=&
\int d\meas_r\,\mathbf{1}_{\Theta(\meas_n)}(\meas_r)
\equiv\int_{\Theta(\meas_n)}d\meas_r,
\label{eq:bdt:normalization2}
\eeqn
so that, by construction:
\beq
\int d\meas_r\,
\mathbf 1_{\Theta(\meas_n)}(\meas_r)
\left[-\widetilde
K(\meas_n,\meas_r)+\frac{I_K(\meas_n)}{V(\meas_n)}\right]=0,
\label{eq:Ktildenorm}
\eeq
for any given $\meas_n$. We point out that we have {\em chosen}
to redistribute the added-back integral $I_K$ uniformly over the
physical radiation domain with respect to the measure $d\chi_r$;
while natural and simplest, this choice is not mandatory.
Equations~(\ref{eq:bdt:dsigmaK}) and~(\ref{eq:Ktildenorm}) then imply:
\beq
\int_{\Theta(\meas_n)}d\sigma_{K,\mathrm{DT}}=
\int_{\Theta(\meas_n)}d\sigma_K\,,
\label{eq:rateNB}
\eeq
i.e.~the added ML-driven contribution does not bias $\clS$-event
rates.

In order to check that this property holds also for differential
distributions, we manipulate further eq.~(\ref{eq:bdt:dsigmaK}) by
means of an explicit reference to an observable $\Omega$. Since
from eq.~(\ref{eq:MCNLOLHEOm}) the $\clS$-event contribution to
an MC@NLO hard-event file is:
\beq
  \frac{d\sigma^{(\mathrm{MC@NLO},\clS)}_{{\rm LHE},\mathrm{DT}}}
  {d\Omega}(\Omega_0)
  =\int d\meas_n\,d\meas_r\,
  \frac{d\sigma^{(\clS)}_{\mathrm{DT}}}{d\meas_n\,d\meas_r}
  \delta\left(\Omega_0-\Omega(\meas_n,0)\right)\,,
  \label{eq:MCNLOLHEOmS}
\eeq
with the short-distance cross section in the integrand on the r.h.s.~being
given in eq.~(\ref{eq:ml:reconstructed-S}). There, the contribution relevant
to the DT procedure which is non-trivial is the r.h.s.~of
eq.~(\ref{eq:bdt:dsigmaK}), which we then need to rewrite as follows:
\beqn
  \frac{d\sigma_{K,\mathrm{DT}}}{d\Omega}(\Omega_0)
  &=&\int d\meas_n\,d\meas_r\,K(\meas_n,\meas_r)
  \delta\left(\Omega_0-\Omega(\meas_n,0)\right)
  \label{eq:mc:source-29}
  \\
  &+&\int d\meas_n\,d\meas_r
\mathbf 1_{\Theta(\meas_n)}(\meas_r)
\left[-\widetilde K(\meas_n,\meas_r)+
  \frac{I_K(\meas_n)}{V(\meas_n)}\right]
  \delta\left(\Omega_0-\Omega(\meas_n,0)\right)
\nonumber\\*
  &=&\int d\meas_n\,d\meas_r\,K(\meas_n,\meas_r)
  \delta\left(\Omega_0-\Omega(\meas_n,0)\right)
  \label{eq:mc:source-291}\\
  &+&\int d\meas_n\,\delta\left(\Omega_0-\Omega(\meas_n,0)\right)
\int d\meas_r
\mathbf 1_{\Theta(\meas_n)}(\meas_r)
\left[-\widetilde K(\meas_n,\meas_r)+
  \frac{I_K(\meas_n)}{V(\meas_n)}\right]
  \nonumber
\nonumber\\*
  &=&\frac{d\sigma_K}{d\Omega}(\Omega_0),
\eeqn
thanks to the vanishing of the rightmost term on the r.h.s.~of
eq.~(\ref{eq:mc:source-291}), which follows from eq.~(\ref{eq:Ktildenorm}).
If the relationship
\beq
\frac{d\sigma_{K,\mathrm{DT}}}{d\Omega}(\Omega_0)=
\frac{d\sigma_K}{d\Omega}(\Omega_0)
\label{eq:Knobias}
\eeq
can be established for any possible observable $\Omega$ constructed
with $\clS$-event four-momenta, then one can conclude that the DT
procedure advocated here does not bias the original MC@NLO (or
MC@NLO-$\Delta$) cross section. From the derivation above, we see
that a sufficient condition for this statement to be true is
eq.~(\ref{eq:Ktildenorm}).

The procedure outlined so far implies learning $\widetilde{K}$ in
terms of $\meas_n$ and $\meas_r$; as has been anticipated, this is
correct, but too restrictive. Therefore, let us consider alternative
sets of Born-level and FKS variables, respectively, which we denote
by $\bar\meas_n$ and $\bar\meas_r$; we assume a mapping in which the
Born variables do not depend on the FKS variables, so that the
jacobian is block triangular. We employ them as follows: we start
by defining the counterpart of $\Theta(\meas_n)$ in the space of the
barred variables $\bar\meas_r$; in order to avoid abuse of notation
and confusion, we shall include a ``phys'' index, thus:
\beq
\Theta_{\rm phys}(\bar\meas_n):=\Big\{\bar\meas_r\in[0,1]^3\,\Big|\,
\exists\meas_n(\bar\meas_n)\,\&\,
\exists\meas_r(\bar\meas_n,\bar\meas_r)\,\&\,
\meas_r(\bar\meas_n,\bar\meas_r)\in\Theta(\meas_n(\bar\meas_n))
\Big\}\,.
\eeq
We shall also employ the notation:
\beq
\mathbf{1}_{\rm phys}(\bar\meas_n,\bar\meas_r)=
\mathbf{1}_{\Theta_{\rm phys}(\bar\meas_n)}(\bar\meas_r)\,,
\;\;\;\;\;\;\;\;
\forall\,\,\bar\meas_n,\,\bar\meas_r\,.
\label{eq:eqeq1}
\eeq
We then rewrite $\sigma_K$ in eq.~(\ref{eq:K-density}) identically as
follows :
\beq
  \sigma_K=\int d\bar\meas_n\,d\bar\meas_r\,
  \mathbf{1}_{\rm phys}(\bar\meas_n,\bar\meas_r)
  \abs{\frac{\partial(\meas_n,\meas_r)}
  {\partial(\bar\meas_n,\bar\meas_r)}}
  K(\meas_n(\bar\meas_n),\meas_r(\bar\meas_n,\bar\meas_r)).
  \label{eq:mc:source-31}
\eeq
Next, we generalise eq.\ \eqref{eq:mc:source-22} as follows:
\beqn
  \sigma_K&=&\int d\bar\meas_n\,d\bar\meas_r\Bigg[
  \mathbf{1}_{\rm phys}(\bar\meas_n,\bar\meas_r)
  \abs{\frac{\partial(\meas_n,\meas_r)}
  {\partial(\bar\meas_n,\bar\meas_r)}}
  K(\meas_n(\bar\meas_n),\meas_r(\bar\meas_n,\bar\meas_r))
  \nonumber\\
  &&\phantom{aaaaaaa}
  -\mathbf{1}_{\tilde{\Theta}(\bar\meas_n)}(\bar\meas_r)
  \widetilde K(\bar\meas_n,\bar\meas_r)\Bigg]
  +\int d\bar\meas_n\,d\bar\meas_r\,
\mathbf{1}_{\tilde{\Theta}(\bar\meas_n)}(\bar\meas_r)
\widetilde K(\bar\meas_n,\bar\meas_r)\,,
  \label{eq:mc:source-32}
\eeqn
where $\tilde{\Theta}(\bar\meas_n)\subseteq [0,1]^3$ can be
freely chosen as part of the regressor definition (we call it
the {\em candidate domain}); this implies that, in general
\beq
\Theta_{\rm phys}(\bar\meas_n)\ne
\tilde{\Theta}(\bar\meas_n)\,,
\eeq
i.e.~the supports of the functions $K$ and $\tilde{K}$ in the space
spanned by the \mbox{$(\bar\meas_n,\bar\meas_r)$} need not coincide
(this is at variance with the previous construction in terms of the
\mbox{$(\meas_n,\meas_r)$} variables). 
Note the minor abuse of notation: the functions $\widetilde{K}$ of
eqs.~(\ref{eq:mc:source-22}) and~(\ref{eq:mc:source-32}) are not the
same function; they can be told apart by considering their arguments,
i.e.~the variables in terms of which any $\widetilde{K}$ is learnt.

By proceeding as before we arrive at:
\beqn
\begin{aligned}
d\sigma_{K,\mathrm{DT}}={}&d\sigma_K
+\mathbf{1}_{\tilde{\Theta}(\bar\meas_n)}(\bar\meas_r)
\left[-\widetilde K(\bar\meas_n,\bar\meas_r)+
\frac{I_K(\bar\meas_n)}{\bar{V}(\bar\meas_n)}\right]
\,d\bar\meas_n\,d\bar\meas_r,
\end{aligned}
\label{eq:bdt:dsigmaKbar}
\eeqn
where now
\beqn
d\sigma_K&=&d\bar\meas_n\,d\bar\meas_r\,
\mathbf{1}_{\rm phys}(\bar\meas_n,\bar\meas_r)
  \abs{\frac{\partial(\meas_n,\meas_r)}
  {\partial(\bar\meas_n,\bar\meas_r)}}
  K(\meas_n(\bar\meas_n),\meas_r(\bar\meas_n,\bar\meas_r)),
\\
I_K(\bar\meas_n)&=&
\int d\bar\meas_r\,
\mathbf{1}_{\tilde{\Theta}(\bar\meas_n)}(\bar\meas_r)
\widetilde K(\bar\meas_n,\bar\meas_r)\equiv
\int_{\tilde{\Theta}(\bar\meas_n)}
d\bar\meas_r\,\widetilde K(\bar\meas_n,\bar\meas_r),
\label{eq:bdt:normalizationbar1}
\\
\bar{V}(\bar\meas_n)&=&
\int d\bar\meas_r\,
\mathbf{1}_{\tilde{\Theta}(\bar\meas_n)}(\bar\meas_r)
\equiv
\int_{\tilde{\Theta}(\bar\meas_n)}d\bar\meas_r.
\label{eq:bdt:normalizationbar2}
\eeqn
The analogue of eq.~(\ref{eq:Ktildenorm}), then, holds:
\beq
\int d\bar\meas_r\,
\mathbf{1}_{\tilde{\Theta}(\bar\meas_n)}(\bar\meas_r)
\left[-\widetilde
K(\bar\meas_n,\bar\meas_r)+
\frac{I_K(\bar\meas_n)}{\bar{V}(\bar\meas_n)}\right]=0,
\label{eq:Ktildenormbar}
\eeq
for any given $\bar\meas_n$. Let us then consider the analogue
of eq.~(\ref{eq:rateNB}):
\beqn
&&\int_{[0,1]^{m}\otimes\Theta_{\rm phys}(\bar\meas_n)}
d\sigma_{K,\mathrm{DT}}=
\int_{[0,1]^{m+3}}
\mathbf{1}_{\rm phys}(\bar\meas_n,\bar\meas_r)
d\sigma_{K,\mathrm{DT}}
\label{eq:rateNB20}
\\&&\phantom{aaaaa}
=\int_{[0,1]^{m}\otimes\Theta(\meas_n)}d\sigma_K
\label{eq:rateNB2}
\\&&\phantom{aaaaa}
+\int_{[0,1]^{m+3}} d\bar\meas_n d\bar\meas_r\,
\mathbf{1}_{\rm phys}(\bar\meas_n,\bar\meas_r)
\mathbf{1}_{\tilde{\Theta}(\bar\meas_n)}(\bar\meas_r)
\left[-\widetilde
K(\bar\meas_n,\bar\meas_r)+
\frac{I_K(\bar\meas_n)}{\bar{V}(\bar\meas_n)}\right].
\nonumber
\eeqn
A sufficient condition for the absence of bias on total rates
emerges from our ability to exploit eq.~(\ref{eq:Ktildenormbar}).
In turn, this happens if
\beq
\mathbf{1}_{\rm phys}(\bar\meas_n,\bar\meas_r)
\mathbf{1}_{\tilde{\Theta}(\bar\meas_n)}(\bar\meas_r)=
\mathbf{1}_{\tilde{\Theta}(\bar\meas_n)}(\bar\meas_r)
\;\;\;\;\Longleftrightarrow\;\;\;\;
\tilde{\Theta}(\bar\meas_n)\,\subseteq\,
\Theta_{\rm phys}(\bar\meas_n).
\label{eq:Thetanobias}
\eeq
In general, we may write
\beqn
\tilde{\Theta}(\bar\meas_n)&=&
\left(
\tilde{\Theta}(\bar\meas_n)\bigcap\Theta_{\rm phys}(\bar\meas_n)
\right)
\bigcup
\left(
\tilde{\Theta}(\bar\meas_n)\bigcap\Theta_{\rm phys}^c(\bar\meas_n)
\right),
\\
\Theta_{\rm phys}^c(\bar\meas_n)&=&
[0,1]^3\setminus \Theta_{\rm phys}(\bar\meas_n),
\eeqn
so that
\beq
\mathbf{1}_{\rm phys}(\bar\meas_n,\bar\meas_r)
\mathbf{1}_{\tilde{\Theta}(\bar\meas_n)}(\bar\meas_r)=
\mathbf{1}_{\tilde{\Theta}(\bar\meas_n)}(\bar\meas_r)-
\mathbf{1}_{\tilde{\Theta}(\bar\meas_n)\bigcap\Theta_{\rm phys}^c(\bar\meas_n)}
(\bar\meas_r).
\label{eq:Iident}
\eeq
By replacing this identity on the rightmost side of eq.~(\ref{eq:rateNB2}),
and by exploiting eq.~(\ref{eq:Ktildenormbar}), we obtain
\beqn
&&\int_{[0,1]^{m}\otimes\Theta_{\rm phys}(\bar\meas_n)}
d\sigma_{K,\mathrm{DT}}=
\int_{[0,1]^{m}\otimes\Theta(\meas_n)}d\sigma_K
\label{eq:rateNB3}
\\&&\phantom{aaaaa}
-\int_{[0,1]^{m+3}} d\bar\meas_n d\bar\meas_r\,
\mathbf{1}_{\tilde{\Theta}(\bar\meas_n)
\bigcap\Theta_{\rm phys}^c(\bar\meas_n)}(\bar\meas_r)
\left[-\widetilde
K(\bar\meas_n,\bar\meas_r)+
\frac{I_K(\bar\meas_n)}{\bar{V}(\bar\meas_n)}\right].
\nonumber
\eeqn
The rightmost term on the r.h.s.~of this equation is generally
different from zero, and thus allows one to quantify the bias
on the total rate.

At the differential level w.r.t.~the observable $\Omega$ considered before,
one needs to generalize what had been done in eq.~(\ref{eq:mc:source-29}),
bearing in mind that the integrand must include the factor
\mbox{$\mathbf{1}_{\Theta(\meas_n)}(\meas_r)$}, as is shown in
eq.~(\ref{eq:MCNLOLHEOm2}), in spite of the fact that the function
that defines the observable has a trivial dependence upon $\meas_r$
in the case of $\clS$ events. Equation~(\ref{eq:mc:source-29})
becomes:
\beq
  \frac{d\sigma_{K,\mathrm{DT}}}{d\Omega}(\Omega_0)
=
\int_{[0,1]^{m+3}}
\mathbf{1}_{\rm phys}(\bar\meas_n,\bar\meas_r)
d\sigma_{K,\mathrm{DT}}
\delta\left(\Omega_0-\Omega(\meas_n(\bar\meas_n),0)\right).
\label{eq:dsigKDT}
\eeq
We can then proceed as it has been done for the total rates,
and arrive at the following expression:
\beqn
\frac{d\sigma_{K,\mathrm{DT}}}{d\Omega}(\Omega_0)&=&
\frac{d\sigma_K}{d\Omega}(\Omega_0)
\nonumber
\\&-&
\int_{[0,1]^{m}} d\bar\meas_n 
\delta\left(\Omega_0-\Omega(\meas_n(\bar\meas_n),0)\right)
\nonumber
\\&&\phantom{aa}\times
\int_{[0,1]^3} d\bar\meas_r\,
\mathbf{1}_{\tilde{\Theta}(\bar\meas_n)
\bigcap\Theta_{\rm phys}^c(\bar\meas_n)}(\bar\meas_r)
\left[-\widetilde
K(\bar\meas_n,\bar\meas_r)+
\frac{I_K(\bar\meas_n)}{\bar{V}(\bar\meas_n)}\right].
\label{eq:DTbiasdiff}
\eeqn
The rightmost term on the r.h.s.~of eq.~(\ref{eq:DTbiasdiff}) in
general does not vanish, and therefore constitutes the bias
stemming from ML on the differential distribution in $\Omega$.
It is apparent that its integral over all physical values of
$\Omega_0$ coincides with the total-rate result of eq.~(\ref{eq:rateNB3}).

Note that it would have been justified to employ \mbox{$[0,1]^{m+3}$}
as an integration range on the l.h.s.~of eq.~(\ref{eq:rateNB20}).
This would have changed the integrand of the rightmost term on the
r.h.s.~of eq.~(\ref{eq:rateNB2}) (while the leftmost one would
not have changed, $d\sigma_K$ being only defined in the physical
range), with \mbox{$\mathbf{1}_{\rm phys}(\bar\meas_n,\bar\meas_r)$}
removed from there, thereby leading to the absence of any bias.
While formally this is correct, it does not take into account the
fact that the total rate must also be obtained as a full integral
over any observable $\Omega$. But, when employing an observable, the
physical range must always be included in the definition of the
integration range, as is shown in eq.~(\ref{eq:dsigKDT}), thus
contradicting the statement made at the beginning of this paragraph.
Therefore, that statement must be considered void in the context of
sensible physical quantities.

For a DT in the learning variables $z=(\bar\meas_n,\bar\meas_r)$,
the leaf representation in eq.~(\ref{eq:app:single-tree}) makes the
radiation integral particularly transparent:
\beq
 I_K(\bar\meas_n)=\sum_{\ell=1}^{L}a_\ell
 \int_{\tilde{\Theta}(\bar\meas_n)}d\bar\meas_r\,
 \mathbf 1_{R_\ell}(\bar\meas_n,\bar\meas_r),
 \label{eq:dt:conditional-integral}
\eeq
whence the potential effectiveness of a DT approach to computing
these integrals. At fixed Born information, each term is the leaf
prediction ($a_\ell$) multiplied by the volume of its allowed
radiation slice
($\tilde{\Theta}(\bar\meas_n)\bigcap R_\ell^\prime(\bar\meas_n)$,
with $R_\ell^\prime(\bar\meas_n)=
\{\bar\chi_r:(\bar\meas_n,\bar\meas_r)\in R_\ell\}$).
For rectangular radiation
domains, or suitable unions of such domains, these volumes are
straightforward to evaluate. Any other radiation-domain mask can make
the intersections more complicated: a piecewise-constant regressor does
not by itself remove this geometrical problem. The domain restrictions
discussed below must therefore be respected also in the leaf integrals.

A further generalisation of the results above stems from considering
a given function $f(\bar\meas_n)>0$ -- as the notation suggests, such a
function has support in the Born phase space, and is non-null everywhere.
With this, eq.~(\ref{eq:mc:source-32}) is generalised as follows:
\beqn
  \sigma_K&=&\int d\bar\meas_n\,d\bar\meas_r\Bigg[
  \mathbf{1}_{\rm phys}(\bar\meas_n,\bar\meas_r)
  \abs{\frac{\partial(\meas_n,\meas_r)}
  {\partial(\bar\meas_n,\bar\meas_r)}}
  K(\meas_n(\bar\meas_n),\meas_r(\bar\meas_n,\bar\meas_r))
  \label{eq:mc:source-44}
\\
  &&\phantom{a}
  -\mathbf{1}_{\tilde{\Theta}(\bar\meas_n)}(\bar\meas_r)f(\bar\meas_n)
  \widetilde K(\bar\meas_n,\bar\meas_r)\Bigg]
  +\int d\bar\meas_n\,d\bar\meas_r\,
\mathbf{1}_{\tilde{\Theta}(\bar\meas_n)}(\bar\meas_r)f(\bar\meas_n)
\widetilde K(\bar\meas_n,\bar\meas_r)\,.
\nonumber
\eeqn
Roughly speaking, the function $f$ may be used in order to ``flatten''
the $\widetilde K$ learned starting from eq.~(\ref{eq:mc:source-32}),
thus implying a possibly easier learning procedure. We are now led
to the analogue of eq.~(\ref{eq:bdt:dsigmaKbar}), namely:
\beqn
\begin{aligned}
d\sigma_{K,\mathrm{DT}}={}&d\sigma_K
+\mathbf{1}_{\tilde{\Theta}(\bar\meas_n)}(\bar\meas_r)
f(\bar\meas_n)\left[-\widetilde K(\bar\meas_n,\bar\meas_r)+
\frac{I_K(\bar\meas_n)}{\bar{V}(\bar\meas_n)}\right]
\,d\bar\meas_n\,d\bar\meas_r,
\end{aligned}
\label{eq:bdt:dsigmaKbar2}
\eeqn
with eqs.~(\ref{eq:bdt:normalizationbar1}) and~(\ref{eq:bdt:normalizationbar2})
unchanged. Owing to the independence of $f$ of the radiation variables, the
added regressor minus its integral-over-volume still integrates to
zero on $\tilde{\Theta}$.

The prescription given above employs the domain $\tilde{\Theta}(\bar\meas_n)$
at each Born configuration $\bar\meas_n$, as in eq.~(\ref{eq:bdt:dsigmaKbar}),
and it includes every radiation configuration for which the regressor
and its integral can be evaluated. While user-defined, the domain
itself is allowed to have a Born-dependent shape, since ideally
\mbox{$\Theta_{\rm phys}(\bar\meas_n)=\tilde{\Theta}(\bar\meas_n)$}
-- this would guarantee the absence of biases and at the same time
would be numerically stable, forcing the regressor to have the same
support as the cross section. Unfortunately, this may prove computationally
expensive, unless a pre-tabulated computation of the domain
and its induced volume $\bar{V}$ can be achieved\footnote{Any tabulation
or numerical approximation of the normalization must preserve the required
radiation average to controlled accuracy; the corresponding uncertainty
must be included in the validation of Born-level distributions.}.
A cheaper option is to first choose $\tilde{\Theta}(\bar\meas_n)$ with
some simplifying assumptions, and then to actually adopt the
smaller domain defined thus:
\beq
\mathcal{L}(\bar\meas_n)\subseteq\tilde{\Theta}(\bar\meas_n)
\bigcap\Theta_{\mathrm{phys}}(\bar\meas_n).
\eeq
In other words, $\mathcal{L}(\bar\meas_n)$ would replace
$\tilde{\Theta}(\bar\meas_n)$ in the formulae given above.
By construction, the ML-modified cross section so defined
would be free of biases, but the support of the regressor would
be smaller than that of the $K$ function.

A particularly simple option of this kind is a \emph{single fixed}
domain $\mathcal L$ valid for all Born configurations, such as
\beq
\mathcal{L}\subseteq\bigcap_{\bar\meas_n}
\left[\tilde{\Theta}(\bar\meas_n)
\bigcap\Theta_{\mathrm{phys}}(\bar\meas_n)
\right].
\label{eq:Lconstant}
\eeq
However, in a fully general application a nonempty domain of this kind
\emph{need not exist}: the intersection may be empty (or have zero
volume). Even when it exists, it may be unnecessarily small. Thus, a
fixed restricted domain is only a special-purpose simplification, and
its reliability as a general solution must be assessed (as shall do
later heuristically); the complete
Born-dependent physical domain remains the preferred formulation. It
may require a more involved computation of $I_K(\bar\meas_n)$ and
$\bar V(\bar\meas_n)$, but their common support and normalisation are
necessary to guarantee the preservation of Born-level observables
with maximal numerical efficiency.

\subsubsection{Normalizing Flows}\label{sssec:nf-construction}
The NF construction can be discussed by using the same variables
and domains introduced for DTs. We work with the learning variables
$(\bar\meas_n,\bar\meas_r)$, and distinguish the candidate radiation domain
$\tilde\Theta(\bar\meas_n)$, on which the NF-induced extra cross section
contribution is defined and normalised, from the physical radiation domain
$\Theta_{\rm phys}(\bar\meas_n)$. The latter is determined by the
existence and validity of the maps to $(\meas_n,\meas_r)$, as in
sect.~\ref{sssec:bdt-construction}. Products containing indicators
are understood according to the zero-extension convention of
eq.~(\ref{eq:exttozero}). As before, the MC@NLO $\clS$-event remainder is
$d\sigma_K$; the Born and virtual contributions are unchanged.

Let $F_1$ and $F_2$ be two conditional NF densities with respect to the
same measure $d\bar\meas_r$, with
\beq
 F_i(\bar\meas_r\mid\bar\meas_n)\geq0,\qquad
 \int_{\tilde\Theta(\bar\meas_n)}d\bar\meas_r\,
 F_i(\bar\meas_r\mid\bar\meas_n)=1,\qquad i=1,2.
 \label{eq:nf:normalization-tilde}
\eeq
That is, the conditioning inputs are $\bar\meas_n$, and not the
physical momenta. Equation~(\ref{eq:nf:normalization-tilde})
is a normalisation on $\tilde\Theta$, not a statement about its
intersection with $\Theta_{\rm phys}$; note that, analogously to
the case of DTs, the densities $F_i$ are extended by zero outside
$\tilde\Theta$. The normalizing flows will be multiplied by a
common prefactor defined thus:
\beq
 C\bigl(\meas_n(\bar\meas_n),\bar\meas_n\bigr)
 =\exp\!\left[\operatorname{MLP}(\bar\meas_n)\right]\,
   \frac{d\sigma^{(B)}(\meas_n(\bar\meas_n))}{d\bar\meas_n}\,.
 \label{eq:nf:Born-amplitude}
\eeq
Here, MLP stands for MultiLayerPerceptron, which is learned
simultaneously with the NFs, and whose inputs are the same
Born variables which condition the NFs. 

With these definitions, the NF-modified cross section is the
direct counterpart of the DT-modified one, eq.~(\ref{eq:bdt:dsigmaKbar}):
\beq
 \begin{aligned}
 d\sigma_{K,\mathrm{NF}}={}&d\sigma_K
 +d\bar\meas_n\,d\bar\meas_r\,
 \mathbf1_{\rm phys}(\bar\meas_n,\bar\meas_r)
 \mathbf1_{\tilde\Theta(\bar\meas_n)}(\bar\meas_r)
 \\
 &\hspace{1em}\times
 C\bigl(\meas_n(\bar\meas_n),\bar\meas_n\bigr)
 \Big[F_1(\bar\meas_r\mid\bar\meas_n)
       -F_2(\bar\meas_r\mid\bar\meas_n)\Big].
 \end{aligned}
 \label{eq:nf:two-flow}
\eeq
Both indicators are part of the definition, stemming respectively from
the integration range in eq.~(\ref{eq:nf:normalization-tilde}), and from the
fact that the short-distance cross section will have to be integrated over
the physical range (see the comment which follows eq.~(\ref{eq:DTbiasdiff})).
This is in keeping with the original $d\sigma_K$ having physical
support, as in the DT construction.
At an unphysical configuration, the leftmost indicator sets the
NF-modified additional cross section equal to zero, without evaluating
the Born cross section or $C$. This does not imply setting the auxiliary
$F_i$ equal to zero at the same barred coordinates.

We remind the reader that, according to eq.~(\ref{eq:dsigmaK}),
the full $\clS$-event NF-modified short-distance cross section is
\beq
 d\sigma^{(\clS)}_{\mathrm{NF}}
 =d\sigma^{(B)}+d\sigma^{(V)}+d\sigma_{K,\mathrm{NF}},
 \label{eq:nf:S-reconstruction}
\eeq
with the Born and virtual terms unchanged: using a Born factor in
$C$ does not replace or redistribute the original $d\sigma^{(B)}$;
it fixes a scale for the contribution added to $d\sigma_K$. The same
formulae apply to MC@NLO-$\Delta$ (see eq.~(\ref{eq:dsigmaK-delta})),
after replacing the original remainder by $d\sigma_K^{(\Delta)}$ and
learning the corresponding correction.

The relation to the DT construction is algebraic. By keeping $C$ outside
the learned shape, as a Born-only scaling factor on physical support,
one may identify the dimensionless auxiliary approximant and its
candidate-domain integral as
\beq
 \begin{aligned}
 \widetilde K_{\mathrm{NF}}(\bar\meas_n,\bar\meas_r)
 &=-F_1(\bar\meas_r\mid\bar\meas_n),
 \\
 I_{K,\mathrm{NF}}(\bar\meas_n)
 &=\int_{\tilde\Theta(\bar\meas_n)}d\bar\meas_r\,
       \widetilde K_{\mathrm{NF}}(\bar\meas_n,\bar\meas_r)=-1.
 \end{aligned}
 \label{eq:nf:bdt-identification}
\eeq
Rather than distributing this integral uniformly, we now distribute it
according to $F_2$. This gives
\beq
 -\widetilde K_{\mathrm{NF}}+I_{K,\mathrm{NF}}F_2=F_1-F_2.
 \label{eq:nf:subtract-add-identity}
\eeq
The density arguments are suppressed in this equation for simplicity.
The integral over the candidate domain of this quantity is equal to zero;
this is true by definition on the r.h.s.~(owing to the normalization
conditions of the NFs, eq.~(\ref{eq:nf:normalization-tilde})), and
by construction on the l.h.s.. The multiplication of the two sides
of eq.~(\ref{eq:nf:subtract-add-identity}) by $C$ and by the
two indicators in eq.~(\ref{eq:nf:two-flow}) gives the ``physical
correction'', as previously was the case in the DT-based approach.
Conversely, we point out that these formulae do not imply that
the NFs are trained by direct regression of $K$.

With this, the analogue of eq.~(\ref{eq:rateNB2}) is:
\beqn
&&\int_{[0,1]^{m}\otimes\Theta_{\rm phys}(\bar\meas_n)}
d\sigma_{K,\mathrm{NF}}=
\int_{[0,1]^{m}\otimes\Theta(\meas_n)}d\sigma_K
\label{eq:rateNF2}
\\*&&\phantom{aaaaa}
+\int_{[0,1]^{m+3}} d\bar\meas_n d\bar\meas_r\,
\mathbf{1}_{\rm phys}(\bar\meas_n,\bar\meas_r)
\mathbf{1}_{\tilde{\Theta}(\bar\meas_n)}(\bar\meas_r)
\nonumber
\\*&&\phantom{aaaaaaaaaa}\phantom{\Big[}\times
 C\bigl(\meas_n(\bar\meas_n),\bar\meas_n\bigr)
 \bigl[F_1(\bar\meas_r\mid\bar\meas_n)
       -F_2(\bar\meas_r\mid\bar\meas_n)\bigr].
\nonumber
\eeqn
By using again eq.~(\ref{eq:Iident}) we obtain the analogue
of eq.~(\ref{eq:rateNB3}):
\beqn
&&\int_{[0,1]^{m}\otimes\Theta_{\rm phys}(\bar\meas_n)}
d\sigma_{K,\mathrm{NF}}=
\int_{[0,1]^{m}\otimes\Theta(\meas_n)}d\sigma_K
\label{eq:rateNF3}
\\&&\phantom{aaa}
-\int_{[0,1]^{m+3}} d\bar\meas_n d\bar\meas_r\,
\mathbf{1}_{\tilde{\Theta}(\bar\meas_n)
\bigcap\Theta_{\rm phys}^c(\bar\meas_n)}(\bar\meas_r)
 C\bigl(\meas_n(\bar\meas_n),\bar\meas_n\bigr)
 \bigl[F_1(\bar\meas_r\mid\bar\meas_n)
       -F_2(\bar\meas_r\mid\bar\meas_n)\bigr].
\nonumber
\eeqn
with the rightmost term on the r.h.s.~of this equation the
bias on the total rate.

The counterpart of this result at the differential level can be
obtained by starting from the analogue of eq.~(\ref{eq:dsigKDT}):
\beq
  \frac{d\sigma_{K,\mathrm{NF}}}{d\Omega}(\Omega_0)
=
\int_{[0,1]^{m+3}}
\mathbf{1}_{\rm phys}(\bar\meas_n,\bar\meas_r)
d\sigma_{K,\mathrm{NF}}
\delta\left(\Omega_0-\Omega(\meas_n(\bar\meas_n),0)\right).
\label{eq:dsigKNF}
\eeq
With the very same steps as before, we thus arrive at the
analogue of eq.~(\ref{eq:DTbiasdiff}):
\beqn
&&\frac{d\sigma_{K,\mathrm{NF}}}{d\Omega}(\Omega_0)=
\frac{d\sigma_K}{d\Omega}(\Omega_0)
\label{eq:NFbiasdiff}
\\&&\phantom{aaa}
-\int_{[0,1]^{m}} d\bar\meas_n 
\delta\left(\Omega_0-\Omega(\meas_n(\bar\meas_n),0)\right)
\nonumber
\\&&\phantom{aaaa}\times
\int_{[0,1]^3} d\bar\meas_r\,
\mathbf{1}_{\tilde{\Theta}(\bar\meas_n)
\bigcap\Theta_{\rm phys}^c(\bar\meas_n)}(\bar\meas_r)
 C\bigl(\meas_n(\bar\meas_n),\bar\meas_n\bigr)
 \bigl[F_1(\bar\meas_r\mid\bar\meas_n)
       -F_2(\bar\meas_r\mid\bar\meas_n)\bigr],
\nonumber
\eeqn
which again allows one to quantify the bias on the observable $\Omega$.
More explicitly, if one defines:
\beq
 q_i(\bar\meas_n)=
\int_{[0,1]^3} d\bar\meas_r\,
\mathbf{1}_{\tilde{\Theta}(\bar\meas_n)
\bigcap\Theta_{\rm phys}^c(\bar\meas_n)}(\bar\meas_r)
 F_i(\bar\meas_r\mid\bar\meas_n),\qquad i=1,2.
 \label{eq:nf:excluded-probabilities}
\eeq
then:
\beqn
&&\frac{d\sigma_{K,\mathrm{NF}}}{d\Omega}(\Omega_0)=
\frac{d\sigma_K}{d\Omega}(\Omega_0)
\label{eq:NFbiasdiffint}
\\&&\phantom{aaa}
-\int_{[0,1]^{m}} d\bar\meas_n 
\delta\left(\Omega_0-\Omega(\meas_n(\bar\meas_n),0)\right)
 C\bigl(\meas_n(\bar\meas_n),\bar\meas_n\bigr)
 \bigl[q_1(\bar\meas_n)-q_2(\bar\meas_n)\bigr].
\nonumber
\eeqn
Thus, the absence of biases requires
\mbox{$q_1(\bar\meas_n)=q_2(\bar\meas_n)$} for all $\bar\meas_n$ which
contribute to the observable computed at $\Omega_0$. Clearly, this
holds identically when $\tilde\Theta\subseteq\Theta_{\rm phys}$
(which is the same sufficient condition encountered in case of DTs),
and also when the two densities assign equal, possibly nonzero,
probability in the integration range that defines $q_i$ in
eq.~(\ref{eq:nf:excluded-probabilities}). Still, the training
does not impose $q_1=q_2$ as an identity, but attempts to minimize
both of these quantities in that integration range.

In order to sketch the strategy used to achieve that goal,
we start by reminding the reader that, in the NF approach,
the loss function is evaluated on the complete modified $\clS$
contribution, and not on $d\sigma_K$ alone\footnote{This is at
variance with what is done in the DT approach. This is not a
fundamental difference -- DTs could be trained on the complete
cross section, and NFs on $d\sigma_K$.}. In the barred coordinate
measure its density is
\beq
 \begin{aligned}
 \bar\varsigma^{(\clS)}_{\mathrm{NF}}(\bar\meas_n,\bar\meas_r)
 &\equiv\frac{d\sigma^{(\clS)}_{\mathrm{NF}}}
                   {d\bar\meas_n\,d\bar\meas_r}
 \\
 &=\mathbf1_{\rm phys}(\bar\meas_n,\bar\meas_r)
 \Bigg\{
 \frac{d\sigma^{(B)}}{d\bar\meas_n\,d\bar\meas_r}
 +\frac{d\sigma^{(V)}}{d\bar\meas_n\,d\bar\meas_r}
 +\frac{d\sigma_K}{d\bar\meas_n\,d\bar\meas_r}
 \\
 &\hspace{3em}
 +\mathbf1_{\tilde\Theta(\bar\meas_n)}(\bar\meas_r)
 C\bigl(\meas_n(\bar\meas_n),\bar\meas_n\bigr)
  \bigl[F_1(\bar\meas_r\mid\bar\meas_n)
       -F_2(\bar\meas_r\mid\bar\meas_n)\bigr] \Bigg\}.
 \end{aligned}
 \label{eq:nf:training-density}
\eeq
The individual contributions are expressed in the barred measure and
therefore already contain the jacobians of the physical coordinate
maps. By definition, the quantity in eq.~(\ref{eq:nf:training-density})
is equal to zero on unphysical points.

The complete loss is defined as follows:
\beq
 L_{\rm tot}=L_{\rm neg}
      +\lambda_{\rm unph}L_{\rm unph}
      +\lambda_{\rm reg}L_{\rm reg}.
 \label{eq:nf:loss-total}
\eeq
Here, the auxiliary coefficients $\lambda_{\rm unph}$ and $\lambda_{\rm reg}$
are determined by fixing the ratios of the gradient norms
of their respective losses over the gradient norm stemming from
$L_{\rm neg}$ to be equal to some constants $l_{\rm unph}$ and
$l_{\rm reg}$, which we empirically fix to be in the range
$[0.03,0.3]$, with the actual value chosen according to the
process type (most often it is set equal to $0.1$). More precisely:
\beq
\lambda_j=
\frac{\abs{\nabla_\theta L_{\rm neg}}}{\abs{\nabla_\theta L_j}}\,l_j\,,
\;\;\;\;\;\;\;\;
j\in\{\mathrm{unph},\mathrm{reg}\},
\eeq
with the coefficients held fixed during the corresponding parameter update, and 
with $\theta$ being the full set of trainable parameters relevant to the loss
functions. The individual
losses $L_{\rm neg}$, $L_{\rm unph}$, and $L_{\rm reg}$ are constructed in
order to reduce the number of negative weights, to minimize the NFs
in the unphysical range, and to penalise large values in the NFs,
respectively. More in detail, their definitions are as follows:
\beq
 L_{\rm neg}
 =\frac{1}{M_\varsigma}\left(
 \bigl\langle|\bar\varsigma^{(\clS)}_{\mathrm{NF}}|\bigr\rangle^{\,2}
 -\bigl\langle\bar\varsigma^{(\clS)}_{\mathrm{NF}}\bigr\rangle^{\,2}
 \right),
 \label{eq:nf:loss-negative}
\eeq
with $M_\varsigma$ the maximum\footnote{Since in the actual
implementation the sample is divided in batches, $M_\varsigma$ is
in fact the average of the maxima of the weights in absolute value
within the various batches -- see eq.~(\ref{avgNormepch}).} of the
weights in absolute value.
By restricting the average to unphysical points, we next define:
\beq
 L_{\rm unph}
 =\sum_{i=1}^{2}
   \bigl\langle\log(F_i+\epsilon)\bigr\rangle_{{\rm unph}}\,,
 \label{eq:nf:loss-unphysical}
\eeq
with $\epsilon$ an extremely small technical parameter whose role is that
of avoiding divergences. The penalty induced by this loss encourages
suppression of both densities on the complement of the physical range,
but its value is not an upper bound on either $q_i$, nor can it impose
equality of their integrals. Furthermore, sampling small density values on
large-measure regions can still miss spikes in the NFs with small-measure
supports. Therefore, the regularisation term in eq.~(\ref{eq:nf:loss-total})
is engineered to penalise large densities:
\beq
 L_{\rm reg}
 =\sum_{i=1}^{2}\langle F_i^2\rangle.
 \label{eq:nf:loss-regularization}
\eeq
In particular, this loss function controls the concentrations that can
generate large corrections and statistical fluctuations; it does not
supply one with a normalisation constraint.

\section{Figure of merit}\label{sec:fom}
Naively, one may think that any mechanism which reduces the number
of negative weights is worth applying. However, this crude argument
neglects the fact that such a reduction may have non-negligible
computing costs, which could offset any gains achieved thanks to
having a smaller number of negative-weight events.

We shall define a figure of merit (f.o.m.~henceforth) as a single number
associated with any given ML result, which gives a robust estimate of
whether an ML-improved MC@NLO simulation is ultimately economically convenient
w.r.t.~its MC@NLO-only counterpart, by accounting for the additional
costs incurred by all of the ML operations, versus the savings
made possible by the reduction of negative weights. If the savings
will be larger than the costs, then the f.o.m.~will be positive, and
negative otherwise. More generally, the larger the f.o.m.~the larger
the savings; hence, later we shall use the values of the f.o.m.~obtained
with different ML hyperparameters to empirically establish an ``ideal''
choice for them.

For unweighted events with signs $\pm1$, let $f$ be the fraction of
negative weights. The relative number of events needed to achieve the
same statistical precision as an ideal positive-weight sample
is~\cite{Frederix:2020trv}\footnote{We
emphasise that this is a purely statistical argument, which does not
take into account the fact that the positive-only sample may be of a lower
perturbative accuracy w.r.t.~the mixed-sign-event one. The relative
theoretical precision of a pair of simulations will not be discussed
here, but it must be clear that it constitutes an important aspect
for reaching a final decision on which approach is best pursued.
In the present context, the positive-weight sample will only serve
as a normalization reference.}
\beq
  c(f)=\frac{1}{(1-2f)^2}\,.
  \label{eq:fom:source-1}
\eeq
Let $N$ be the number of positive-weight events required for,
say, a physics analysis. A statistically-equivalent sample with
negative-weight fraction $f$ thus requires $Nc(f)$ events. We denote
by $\tau_+$ the time per hard event in the positive-weight reference
simulation, up to and including unweighting. Throughout this
work, generation times include the time spent writing the event
information required by our implementation. Consequently, the timing
ratios and figures of merit reported below characterize the present
implementation, including its dumping overhead. We define the effective
costs associated with hard-event generation and with all of the
subsequent operations (e.g.~showering, full or fast detector simulation,
analysis) as follows
\beq
  \Ct\equiv n_\tau p_\tau\,,\qquad
         \Ca\equiv n_A p_A\frac{\tau_A}{\tau_+}\,,
  \label{eq:fom:source-2}
\eeq
respectively. Here, $n_i$ ($i=\tau, A$) is the number of allocated
CPU cores (or, more generally, compute units), $p_i$ is the corresponding
cost per core-unit time, and $\tau_A$ is the downstream (showering,
full-simulation, etc) time per event. Thus $\Ct$ is the hard-generation
cost per unit time, whereas $\Ca$ is an effective downstream cost
coefficient expressed in units of the reference generation time $\tau_+$.
With these definitions, the downstream cost per event is simply
$\Ca\tau_+=n_A p_A\tau_A$, and the ratio of effective costs is therefore
\beq
  \frac{\Ca}{\Ct}=\frac{n_A p_A}{n_\tau p_\tau}\frac{\tau_A}{\tau_+}\,.
  \label{eq:fom:source-3}
\eeq
If the hard-generation and downstream phases run on identical
computing configurations, $n_A=n_\tau$ and $p_A=p_\tau$, and
therefore
\beq
  \frac{\Ca}{\Ct}=\frac{\tau_A}{\tau_+}\,.
  \label{eq:fom:source-4}
\eeq
The reference hard-generation cost (i.e., that associated with a
positive-definite simulation) is
\beq
  C_+=\Ct N\tau_+\,.
  \label{eq:fom:source-5}
\eeq
For the original MC@NLO simulation, let $f_0$ be the negative-weight
fraction and $\Delta\tau_\pm^{(0)}$ the additional\footnote{Additional
must be interpreted in a broad sense, since $\Delta\tau_\pm^{(0)}$ may
also be negative -- for example when the reference simulation is
obtained by means of an expensive procedure which removes all of
the original negative weights.} per-event generation time relative to
$\tau_+$. Its hard-event cost is
\beq
  C_0^{\mathrm{hard}}=\Ct N c(f_0)\bigl(\tau_+ +\Delta\tau_\pm^{(0)}\bigr).
  \label{eq:fom:source-6}
\eeq
The extra cost relative to the ideal positive-weight simulation is therefore:
\beq
  \Delta C_0^{\mathrm{hard}}=C_0^{\mathrm{hard}}-C_+
  =\Ct N\tau_+\bigl[c(f_0)(1+\delta_0)-1\bigr],
  \label{eq:fom:source-7}
\eeq
where we have introduced the dimensionless overhead
\beq
  \delta_0=\frac{\Delta\tau_\pm^{(0)}}{\tau_+}.
  \label{eq:fom:delta0}
\eeq
Note that if the positive-only reference sample is of the same
perturbative accuracy as the mixed-sign one, we expect
$\abs{\Delta\tau_\pm^{(0)}}\ll\tau_+$; indeed, in the simplified
estimates below, we shall neglect differences in the baseline
per-event generation times, while retaining the explicitly identified
ML-specific overhead. However, the general formula does not require this
approximation.

The pre-ML MC@NLO simulation whose cost has been discussed so far
generates the events employed in the ML training.
Denote by $N_{\mathrm{train}}$ the number of events upon which such
a training is based. By employing again $\tau_+$ as a reference time,
we write the total training time as follows:
\beq
  T_{\mathrm{train}}=N_{\mathrm{train}}\Bigl(
       \tau_+ +\Delta\tau_\pm^{(0)}+\dlin\Bigr)
      +\Tnonlin(N_{\mathrm{train}}).
  \label{eq:fom:source-8}
\eeq
Here $\dlin$ is the \emph{additional} time per training event from
operations whose total cost scales linearly with $N_{\mathrm{train}}$;
the first two terms in parentheses are the cost of generating that
event. The term $\Tnonlin(N_{\mathrm{train}})$ is a total time,
encompassing fixed setup costs and any other, not necessarily linear,
dependence on the training-set size; this quantity thus cannot be
assumed to be proportional to $N_{\mathrm{train}}$. 

The separation of the total training time into a linear and a
non-linear component as a function of the number of events is
in practice difficult to determine -- for example, such a separation
may depend on several hyperparameters. As a matter of fact, we expect
the non-linear contribution to be negligible w.r.t.~the linear one
unless $N_{\mathrm{train}}$ is small; therefore,
in the simplified estimates below we shall neglect it.
This is an additional approximation, to be checked against the measured
timings. Alternatively, the measured total training time can be used
directly through eq.~(\ref{eq:totalT}), without separating its linear and
nonlinear components. However, doing so renders it complicated the
extrapolation of results to large values of $N_{\mathrm{train}}$.

Let $f_1$ be the negative-weight fraction of the post-ML MC@NLO
simulation. Denote the analogue of the pre-ML parameter
$\Delta\tau_\pm^{(0)}$ by $\Delta\tau_\pm^{(1)}$ (these two
quantities are not necessarily equal to one another, since the
respective underlying cross sections are different), and by
$\Delta\tau_{\mathrm{extra}}$ an additional ML-specific time
per \emph{produced event} (e.g. inference and integration of a
learned contribution). With this, the post-ML hard-event cost,
including training, is:
\beq
  C_1^{\mathrm{hard}}=\Ct\Bigl\{Nc(f_1)
     \bigl(\tau_++\Delta\tau_\pm^{(1)}+\Delta\tau_{\mathrm{extra}}\bigr)+
     T_{\mathrm{train}}\Bigr\}.
  \label{eq:fom:source-9}
\eeq
After defining the dimensionless ratios
\beq
  \delta_1=\frac{\Delta\tau_\pm^{(1)}}{\tau_+},\qquad
  \epsml=\frac{\Delta\tau_{\mathrm{extra}}}{\tau_+},\qquad
  \eta_{\mathrm{lin}}=\frac{\dlin}{\tau_+},
  \label{eq:fom:dimensionless-overheads}
\eeq
one obtains for the extra cost of the post-ML simulation, in addition
to that of the ideal positive-weight reference, the following result:
\beqn
  \Delta C_1^{\mathrm{hard}}&=&
  C_1^{\mathrm{hard}}-C_+
  \\
   &=&\Ct N\tau_+\Biggl[c(f_1)(1+\delta_1+\epsml)-1
   \nonumber
   \\&&\phantom{\Ct N\tau_+\Biggl[}
   +\frac{N_{\mathrm{train}}}{N}
          (1+\delta_0+\eta_{\mathrm{lin}})
          +\frac{\Tnonlin(N_{\mathrm{train}})}{N\tau_+}\Biggr].
  \label{eq:fom:source-10}
\eeqn
The difference between the pre- and post-ML hard-event costs in
MC@NLO simulations is then equal to:
\beqn
  G_{\mathrm{hard}}&=&
  C_0^{\mathrm{hard}}-C_1^{\mathrm{hard}}
  \\
   &=&\Ct N\tau_+\Biggl[c(f_0)(1+\delta_0)
                -c(f_1)(1+\delta_1+\epsml)
\nonumber\\&&\phantom{\Ct N\tau_+\Biggl[}
   -\frac{N_{\mathrm{train}}}{N}
       (1+\delta_0+\eta_{\mathrm{lin}})
       -\frac{\Tnonlin(N_{\mathrm{train}})}{N\tau_+}\Biggr].
  \label{eq:fom:source-11}
\eeqn
A positive $G_{\mathrm{hard}}$ implies a net hard-event
generation saving.

After training, the mass-production phase can begin; this can be
split conceptually in a hard-event generation and unweighting step,
and in a subsequent step that features at least a shower phase.
The unitary costs for these two steps have been introduced above,
and denoted by $\Ct$ and $\Ca$ respectively. The actual costs
for the shower and any downstream steps in the original and
ML-improved MC@NLO simulations are\footnote{We assume that the
average downstream processing cost per produced event is unchanged
by the ML modification.}
\beq
\$_A N\tau_+ c(f_0)\,,\;\;\;\;\;\;\;\;
\$_A N\tau_+ c(f_1)\,,
\eeq
respectively, whence the difference between the pre- and post-ML
downstream costs is 
\beq
G_{\mathrm{downs}}=\$_A N\tau_+ \Big[c(f_0)-c(f_1)\Big]\,.
\eeq
By combining this result with that of eq.~(\ref{eq:fom:source-11})
one obtains the total (possible) financial gain of the whole chain
\beq
G=G_{\mathrm{hard}}+G_{\mathrm{downs}}\,.
\eeq
By normalizing this quantity by means of the total cost of the
ideal positive-only simulation one obtains the sought figure of merit:
\beq
 \fom :=\frac{G}{(\Ct+\Ca)N\tau_+}\,.
\eeq
Straightforward algebra, after having introduced the following parameters
in keeping with eq.~(\ref{eq:fom:source-8}):
\beqn
\label{eq:fom:display-1}
   r_\tau&=&
   \frac{\Ct}{\Ct+\Ca}
     =\frac{n_\tau p_\tau}{n_\tau p_\tau+n_A p_A\tau_A/\tau_+},
     \\
   \alpha&=&r_\tau\frac{N_{\mathrm{train}}}{N},
   \\
   \beta&=&r_\tau\frac{\Tnonlin(N_{\mathrm{train}})}{N\tau_+}.
\eeqn
leads one to the explicit result:
\beq
 \fom = c(f_0)-c(f_1)
+r_\tau\!\Big[c(f_0)\delta_0-c(f_1)(\delta_1+\epsilon)\Big]
-\alpha(1+\delta_0+\eta_{\mathrm{lin}})-\beta.
 \label{eq:fom:def2}
\eeq
Note that by construction
\beq
\alpha(1+\delta_0+\eta_{\mathrm{lin}})+\beta=
r_\tau\frac{T_{\mathrm{train}}}{N\tau_+},
\label{eq:totalT}
\eeq
which may be used in the case of highly non-linear scaling of the
training time.

Equation~(\ref{eq:fom:def2}) gives the f.o.m.~which we sought to define,
which accounts for all sources of additional costs and savings
w.r.t.~a standard MC@NLO simulation. Having said that, we have
already argued that a sensible result can also be obtained
after considering that f.o.m.~in the limit:
\beq
\delta_0\,\longrightarrow\,0\,,\;\;\;\;\;\;
\delta_1\,\longrightarrow\,0\,,\;\;\;\;\;\;
\beta\,\longrightarrow\,0\,,\;\;\;\;\;\;
n_\tau p_\tau=n_A p_A\,,
\eeq
where the rightmost equation implies equal aggregate cost rates
(e.g.~with identical computing resources) in the generation and
downstream phases.
In this scenario, eq.~(\ref{eq:fom:def2}) reads as follows:
\beqn
 \fom\simeq
 c(f_0)-c(f_1)
  -\frac{\tau_+}{\tau_+ +\tau_A}\left[
  c(f_1)\frac{\Delta\tau_{\mathrm{extra}}}{\tau_+}
+\left(1+\frac{\dlin}{\tau_+}\right)\,
\frac{N_{\mathrm{train}}}{N}\right],
 \label{eq:fom:def3}
\eeqn
where, for ease of understanding, we have used the explicit
definitions of the parameters which appear in eq.~(\ref{eq:fom:def2}).

The simplified f.o.m.~of eq.~(\ref{eq:fom:def3}) is what will be used
in sect.~\ref{sec:res} when assessing the possible savings stemming
from ML-modified MC@NLO simulations. It should be clear that, among
the quantities which appear on the r.h.s.~of eq.~(\ref{eq:fom:def3}),
$\tau_A$ and $N$ play a special role since, at variance with the other
variables there (which are measured during the short-distance simulations),
they are determined by factors not associated with such simulations.
In particular, $N$ (the positive-weight-equivalent event
count required by the analysis) is chosen
by the user depending on needs of his/her physics analysis;
while $\tau_A$ (the downstream time per event) depends on the
kind of post-generation operations one must carry out. These operations
are vastly different in theoretical and experimental applications,
since only in the latter one goes through a (full) detector simulation.
For these reasons, we shall consider the four different scenarios
obtained by combining the following settings:
\beqn
&&\frac{\tau_+}{\tau_+ +\tau_A}=0.9\,,\;\;\;\;\;\;\;\;
\frac{\tau_+}{\tau_+ +\tau_A}=0.15\,,
\label{eq:rtausimp}
\\\phantom{\Big]}
&&N=10N_{\mathrm{train}}\,,\;\;\;\;\;\;\;\;
N=100N_{\mathrm{train}}\,,
\label{eq:NvsNtr}
\eeqn
where the leftmost (rightmost) value in eq.~(\ref{eq:rtausimp}) is
assumed to be representative of an illustrative theoretical (experimental)
setup. Conversely, the two values in eq.~(\ref{eq:NvsNtr}) are
thought to constitute reasonable usage of computing resources.

{\em In summary}, it should be clear that the computation of the figure of merit
requires not only the knowledge of the relevant parameters of the ML
procedure, but also a careful assessment of the computing resources
employed in all of the steps of the implementation. Inevitably, then,
our results for such a quantity reflect our own simulations, and are
not meant to be taken as a rigorous indication of what may happen in,
say, a high-energy experiment. With that being said, we believe that
the equations above will help achieve quantitative simulation priorities
within specific computational environments. Furthermore, it turns out
that with an appropriate choice of hyperparameters (which leads to
an accurate learning of $\widetilde{K}$), the difference of costs (first
two terms on the r.h.s.~of eqs.~(\ref{eq:fom:def2}) and~(\ref{eq:fom:def3}))
is a fairly good approximation of the f.o.m..

We conclude by pointing out that eq.~(\ref{eq:fom:def2}) can also
be used to obtain a figure of merit for the case where the simulation 
alternative to the standard MC@NLO one is based on folding. In the case
of a $(n_\xi,n_y,n_\phi)$ folding (see footnote~\ref{ft:fold}), one
simply needs to set:
\beq
\delta_1=n_\xi\,n_y\,n_\phi\,(1+\delta_0)-1\,,\;\;\;\;\;\;\;\;
\ep=0\,,\;\;\;\;\;\;\;\;
\alpha=0\,,\;\;\;\;\;\;\;\;
\beta=0\,,
\eeq
since a folded generation is almost exactly a factor \mbox{$n_\xi n_y n_\phi$}
more expensive\footnote{See ref.~\cite{arXiv:2609.23500}, which focuses on
reducing such a factor. This cost-saving strategy has not been considered
here, in fairness to the ML-based methods for which analogous solutions 
have not been studied yet.} than a non-folded one.
Thus, eq.~(\ref{eq:fom:def2}) becomes:
\beq
 \fom^{({\rm fold})} = c(f_0)-c(f_1)
+r_\tau\!\Big[c(f_0)\delta_0-
c(f_1)(n_\xi\,n_y\,n_\phi\,(1+\delta_0)-1)\Big],
 \label{eq:fom:fold2}
\eeq
or, in the further simplifying assumptions already employed
($\delta_0\to 0$ and \mbox{$n_\tau p_\tau=n_A p_A$})
\beqn
 \fom^{({\rm fold})}\simeq
 c(f_0)-c(f_1)
  -\frac{\tau_+}{\tau_+ +\tau_A}\,
  c(f_1)(n_\xi\,n_y\,n_\phi-1).
 \label{eq:fom:fold3}
\eeqn

\section{Results\label{sec:res}}
In this section we present the results obtained with the DT and NF
approaches we have discussed above. In view of the fact that the sole
focus of this paper is the reduction of negative-weight $\clS$ events,
the fractions of those events reported in the following are w.r.t.~the
total number of $\clS$ events themselves, and {\em not} (as is more common)
w.r.t.~the total number of $\clS+\clH$ events.

Events have been generated with a modified version of
{\tt MadGraph5\_aMC@NLO}~\cite{Alwall:2014hca}; the modifications
have been limited to the dumping\footnote{\label{ft:dump}We have used
a brute-force approach, i.e.~we have not tried either to speed up
the dumping, or to make it compact, and we have not considered
alternative solutions. Since the time spent on dumping is budgeted
as part of the generation time, we expect our results for that
quantity, which we use to compute the figure of merit introduced
in sect.~\ref{sec:fom}, to differ from that relevant to a more
streamlined solution; however, while quantitative results may change,
we expect our qualitative conclusions to remain valid.}
on a file all of the event-by-event information necessary for the ML
setups to learn $\widetilde{K}$ (DT) or $d\sigma^{(\clS)}$ (NF, indirectly).
For each event, this essentially amounts to writing the values
of the Vegas-$x$'s generated by the integrator, the corresponding
transformed variables, including analytically pre-defined importance
samplings and their jacobians, the physical variables $(\meas_n,\meas_r)$,
and the weights associated with all of the contributions to the $\clS$-event
cross section -- more details will be given in
sects.~\ref{par:dtimpl:practical} and~\ref{par:nfimpl:practical}.

We remind the reader that the implementation of MC@NLO within
{\tt MadGraph5\_aMC@NLO} is based on the FKS subtraction
formalism~\cite{Frixione:1995ms,Frixione:1997np}, which partitions
the $(n+1)$-body phase-space so that the cross section is written
as a sum of contributions that have a single soft and a single
collinear singularity. Each of these contributions is referred to
as an FKS sector\footnote{\label{ft:sectors}While the partition of
the phase space into FKS sectors is exact and covers all of it, as
a bookkeeping device it is convenient to identify each sector with
the collinear singularity that dominates it; this is unambiguous,
since such a collinear singularity is unique by construction.};
by default, they are integrated separately\footnote{There are
exceptions. For example, cases in which the difference between FKS
sectors is due to their stemming from different flavour structures,
while being kinematically identical; these sectors are then typically
treated together; we shall see this explicitly later, when considering
hadronic collisions.},
since the underlying phase-space parametrisations generally differ
from one another. In addition to that, {\tt MadGraph5\_aMC@NLO}
rewrites matrix elements as linear combinations of themselves times
weight factors; the latter are defined by means of ratios of the absolute
value squared of individual Feynman diagrams. Each component of such
linear combinations is referred to as a ``channel''. The interested
reader can find more details on these implementation characteristics
in ref.~\cite{Frederix:2009yq}.

The bottom line is that {\tt MadGraph5\_aMC@NLO} outputs sets of events
for a given combination of channel {\em and} FKS sector. By default, ML learns
a $\widetilde{K}$ function for each of these sets.
This is not necessarily the most convenient
thing to do, as far as the reduction of negative weights is concerned,
and in fact it is wise to distinguish the channel$+$FKS-sector labels
from the training partitions; as was said above, generally these
coincide, but we have also studied cases in which the training
partition, and thus the corresponding learned function, was
associated with a channel and all of the FKS sectors relevant to
that channel. Regardless of this, we note that several trees fitted
to distinct partitions do not constitute a boosted ensemble or a
random forest. Furthermore, we point out that the results of a
single channel-plus-FKS-sector(s) may differ if events
are generated only for that quantity, or in the context of a broader
generation (e.g.~one that encompasses many channels, or the whole
processes). Such differences are within the statistical uncertainties
associated with the MC@NLO-plus-ML procedure, and are therefore not significant.
This is consistent with the goal of this paper, which is that of
proving the effectiveness of ML methods in the reduction of negative
weights. For this, what matters most are the relative differences
between two cross sections, rather than their absolute values,
which should not be assumed to be particularly accurate from a
phenomenological viewpoint.

\begin{figure}[thb]
  \begin{center}
  \includegraphics[width=0.47\textwidth]{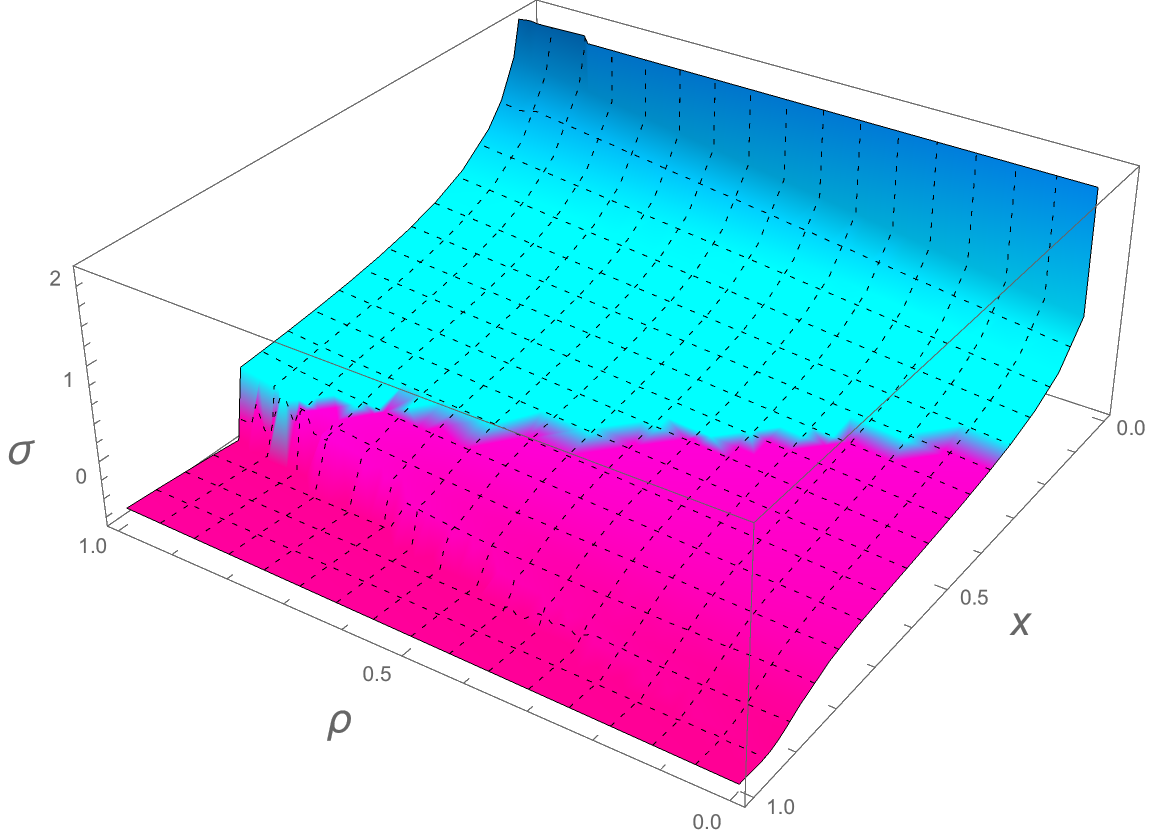}
$\phantom{a}$
  \includegraphics[width=0.47\textwidth]{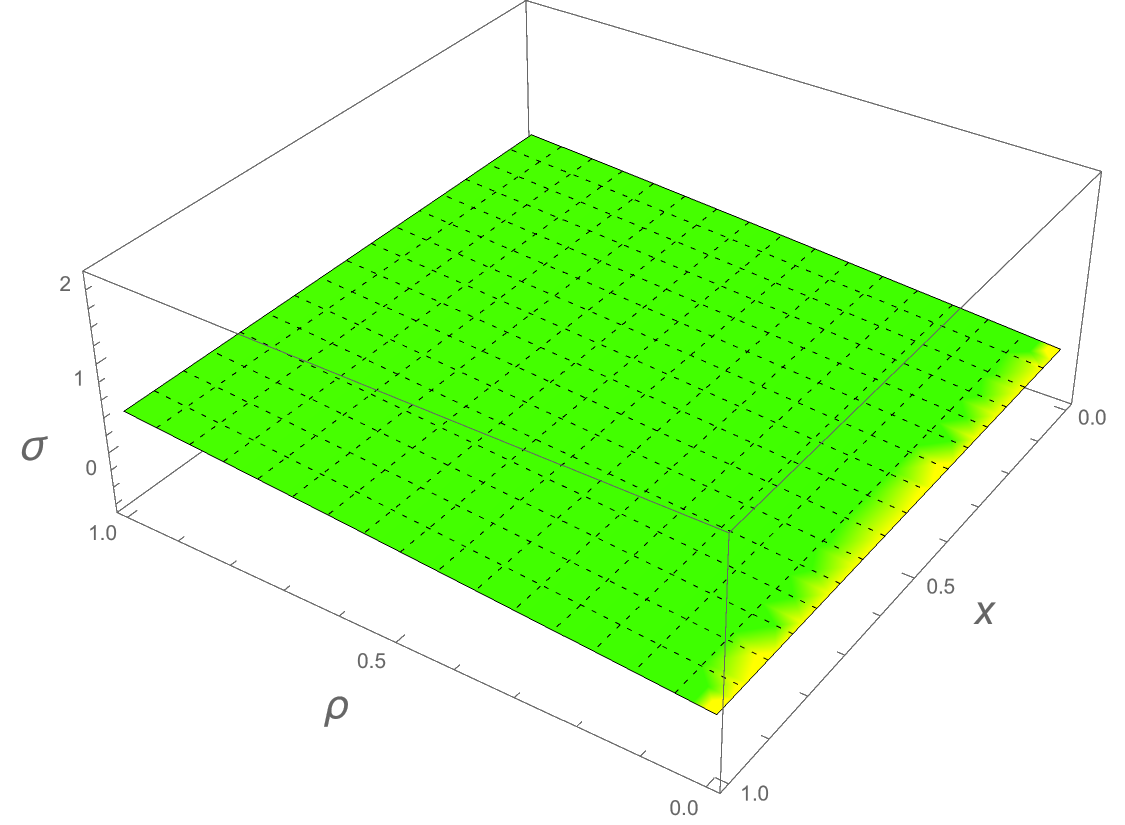}
\caption{\label{fig:toy}
Left panel: original (red is negative, blue is positive);
right panel: after ML (yellow is negative, green is positive).
}
  \end{center}
\end{figure}
A basic understanding of the ML procedure advocated in this work
can be obtained with the help of a two-dimensional toy model (the
extension of the original one-dimensional toy model of
ref.~\cite{Frixione:2002ik}), with the two quantities
\beq
\rho\,,x\;\in\;[0,1]\,,
\eeq
being used as Born and FKS variables, respectively (the former is trivial
in the one-dimensional toy model, being always equal to zero). The assumed
$\clS$-event cross section ($\sigma\equiv K(\rho,x)$ in this simplified case)
is displayed on the left panel of figure~\ref{fig:toy}. The blue (red)
colour shows the region where the cross section is positive
(negative). The model is constructed so that the fraction of
negative weights is extremely large ($\sim 47\%$). Conversely,
the plot on the right panel presents
\mbox{$\sigma\equiv K(\rho,x)-\widetilde{K}(\rho,x)+I_K(\rho)$}
with the green (yellow) colour associated with the region where
$\sigma$ is positive (negative). The yellow region is extremely
small, and this is directly connected with the fact that, post-ML,
the fraction of negative weights of this model is equal to about $10^{-3}$.
In this case, the ML procedure is as good as folding with infinite
statistics\footnote{In the actual DT implementation, there may be
local violations of eq.~(\ref{KminusK}), owing to the piecewise
continuous nature of the DT regressor, as is discussed in
appendix~\ref{app:bdt-granularity}.}, so that
\beq
K(\rho,x)-\widetilde{K}(\rho,x)\simeq 0\,.
\label{KminusK}
\eeq
In other words, the projections onto the $x$ axis at fixed $\rho$
of the ML-modified cross section are essentially constant.
These results are fully unbiased, and therefore
\beq
\int dx K(\rho,x) = I_K(\rho)+
\int dx \left(K(\rho,x)-\widetilde{K}(\rho,x)\right)\,,
\eeq
with extremely high precision.
What is presented above typifies the behaviour in realistic cases. Namely,
the pre-ML MC@NLO $\clS$-event cross section has a non-trivial shape
as a function of $\hat{\xi}$ and $\hat{y}$; post-ML, the same distributions
have a shape which is fairly flat\footnote{Strictly speaking, this
depends on the samplings of the variables in which the regressor is
learnt, and on the impact of the non-complete coverage of the physical
domain by the candidate domain, per eq.~(\ref{eq:dtimpl:L-corner})
in the current applications.} -- in fact, the lack of significant
flatness\footnote{More precisely, this is a DT characteristics,
whereas in the NF approach one mostly finds near-flatness, as we
shall see in more details in sect.~\ref{subsec:results-nf}.}
signals an imperfect understanding of the cross section on the part
of the ML machinery. It is interesting to observe that the post-ML
flatness in $\hat{\xi}$ and $\hat{y}$ is not the result of imposing
that characteristics through the loss function.
The bottom line is that the flatness in the $\hat{\xi}$ and
$\hat{y}$ rescaled FKS variables is a very useful diagnostic tool
for the effectiveness of the ML learning procedure. For example, if
the distributions in such variables are quite flat, but the fraction
of negative weights has not decreased significantly w.r.t.~the
original simulation, then it is likely that in order to eliminate negative
weights a different type of approach would be required.

We shall consider processes in both $e^+e^-$ and $pp$ collisions,
at c.m.~energies equal to 1~TeV and 13~TeV respectively; the former
ignores ISR contributions (i.e.~the lepton PDFs are equal to $\delta(1-z)$
functions), while in the latter we employ NNPDF2.3~\cite{Ball:2012cx} (the
{\tt MadGraph5\_aMC@NLO} default). Other relevant parameters are set as follows:
\beqn
m_b&=&4.7~{\rm GeV},
\\
m_t&=&173.0~{\rm GeV},
\\
m_H&=&125.0~{\rm GeV},
\\
m_Z&=&91.18~{\rm GeV}.
\eeqn
For processes which feature intermediate $Z$ bosons, we use
\mbox{$\Gamma_Z=2.4414$}~GeV. Dynamical factorization and renormalization
scales are used throughout; both are set equal to half the total transverse
energy of the event. Finally, the MC@NLO and MC@NLO-$\Delta$ cross sections
are relevant to matching with
{\tt Pythia8}~\cite{Sjostrand:2014zea,Bierlich:2022pfr}.

Many of the processes we study feature $b$ quarks in the final state.
This is because it is well known that those lead to involved matching
conditions, and are therefore responsible for large fractions of
negative weights; in other words, we stress-test the ML approaches
we propose within worst-case scenarios.

We also point out that the material in sect.~\ref{subsec:results-dt}
is significantly more copious than that in sect.~\ref{subsec:results-nf}.
This does not reflect a bias towards DTs, nor that DTs have
more issues than NFs, but simply the fact that several of the problems
we are going to discuss have nothing to do with the specific ML
strategy used within the architecture of the method we propose, but
are orthogonal to it, and whose solutions are thus outside of the
scope of the present work. Therefore, such problems are common
to DTs and NFs; needless to say, we discuss them only in one of
the two cases, which happens to be that of DTs.

\subsection{Implementation of Decision Trees}
\label{par:dtimpl:practical}
We describe here the implementation at a fixed channel and FKS sector;
the corresponding labels are understood. The construction follows
the general blueprint of sect.~\ref{sssec:bdt-construction}, but gives
additional implementation details, such as the separation of
the variables sampled by the integrator from those used as inputs to
the regressor. This distinction also separates the jacobian entering
the regression target from the weight assigned to its squared error.

\vskip 0.2truecm
\noindent\emph{Sampling and learning coordinates.}
\par\noindent
Let $x=(x_n,x_r)$ be uniformly sampled on the Born-plus-radiation unit
cube -- these are the variables that Vegas employs, which remain flat
even after adaptive-integration iterations; conversely, $v=(v_n,v_r)$
denote the Vegas variables that are rendered non-flat by adaptive
integration. For a given Vegas grid (i.e.~a frozen adaptation status),
the successive mappings that define the learning variables are
\beq
 x=(x_n,x_r)\;\longrightarrow\;v(x)=(v_n(x_n),v_r(x_r))
 \;\longrightarrow\;
 \bar\meas(v(x))=(\bar\meas_n(v_n(x_n)),\bar\meas_r(v_r(x_r))).
 \label{eq:dtimpl:chain}
\eeq
As the notation implies, and at variance with the \mbox{$\bar\meas\to\meas$}
ones, these maps act variable-by-variable. The additional (w.r.t.~the Vegas
ones) prescribed transformations leave the Born and azimuthal coordinates
unchanged\footnote{It is trivial to relax this constraint.} and can apply
importance samplings in the forms of power-laws to the other two FKS
coordinates:
\beq
 \bar\meas_n=v_n,\qquad
 \bar\meas_r=(\bar\meas_\xi,\bar\meas_y,\bar\meas_\varphi)
 \equiv\big(\hat{\xi},\hat{y},\hat{\varphi}\big)
 =\bigl(v_\xi^{\,k_\xi},v_y^{\,k_y},v_\varphi\bigr),\qquad
 k_\xi,k_y\geq1.
 \label{eq:dtimpl:powers}
\eeq
Needless to say, alternative functional forms of importance samplings
are easy to consider. All coordinates have unit-interval ranges (the
possible rescalings of the physical variables ($\meas_n,\meas_r$) -- so
that their ranges are also the unit intervals per eq.~(\ref{eq:fdomain}) --
and their associated jacobians are understood in the corresponding densities).
Thus, eq.~(\ref{eq:dtimpl:powers}) states that the labels on $v_\xi$
and $v_y$ identify the Vegas variables feeding the $\bar\meas_\xi$ and
$\bar\meas_y$ radiation variables, but are not necessarily equal to them.
Indeed, in our practical applications we have set $k_\xi=k_y=2$
(other integer values have been tested, leading to comparable or
slightly worse performances).

Define the positive jacobian factors
\beqn
 J_{\rm V}(x)&=&
 \left|\frac{\partial(v_n,v_r)}{\partial(x_n,x_r)}\right|
 =J_{{\rm V},n}(x_n)J_{{\rm V},r}(x_r),
 \nonumber\\
 J_{\rm P}(v_r)&=&
 \left|\frac{\partial\bar\meas_r}{\partial v_r}\right|
 =k_\xi k_y v_\xi^{\,k_\xi-1}v_y^{\,k_y-1}.
 \label{eq:dtimpl:jacobians}
\eeqn
Here $J_{{\rm V},n}=|\partial v_n/\partial x_n|$ and
$J_{{\rm V},r}=|\partial v_r/\partial x_r|$. Since $\bar\meas_n=v_n$,
there is no additional Born jacobian in the power-law map\footnote{This
map is chosen for simplicity, and it can be easily generalized without
changing the discussion that follows.}. Thus:
\beq
 J_{\bar\meas x}(x)=J_{\rm V}(x)J_{\rm P}(v(x)),\qquad
 d\bar\meas_n\,d\bar\meas_r
 =J_{\rm V}J_{\rm P}\,dx_n\,dx_r.
 \label{eq:dtimpl:full-jacobian}
\eeq
Should samples from several adaptive iterations be combined, the 
Vegas jacobian appropriate to each event must be used, rather than a
jacobian reconstructed from the final grid alone.

\vskip 0.2truecm
\noindent\emph{Regression target and loss.}
\par\noindent
Denote the $i^{th}$ dumped contribution associated with $d\sigma_K$ by
$w_i^K$. By writing \mbox{$\bar\meas_i=\bar\meas(v(x_i))$}, the relation
between this weight and the density in learning coordinates is
\beq
 w_i^K=J_{\rm V}(x_i)J_{\rm P}(v_{r,i})
 \left.\frac{d\sigma_K}{d\bar\meas_n\,d\bar\meas_r}
 \right|_{\bar\meas_i}.
 \label{eq:dtimpl:dumped-weight}
\eeq
The original physical phase-space jacobian is already contained in
$d\sigma_K$. For the Born-rescaled subtract-and-add construction, the target
of the regression ($y_i$ in eq.~(\ref{eq:ml:display-1})) is
\beq
 Y_i=\frac{w_i^K}{J_{\rm V}(x_i)J_{\rm P}(v_{r,i})
                         f(\bar\meas_{n,i})},\qquad
  f(\bar\meas_n(v_n))=\frac{d\sigma^{(B)}}{dv_n}(v_n),
 \label{eq:dtimpl:target}
\eeq
with $f()$ the function introduced in eq.~(\ref{eq:mc:source-44}).
This function depends on $\bar\meas_n$, and assumes a fixed value under
radiation integration. The formula applies to the finite target retained
for a fit. Undefined physical expressions obey the zero-extension convention
(note, in particular, that if $d\sigma^{(B)}$ cannot be evaluated, neither
$w_i^K$ can, so it is unimportant that $f()$ may not be defined everywhere),
and no ratio (i.e.~$Y_i$) is evaluated at a singular mapping point.

With this, we define our DT loss function thus:
\beqn
 \mathcal L_{\rm DT}(\theta)
 &=&\frac{1}{\sum_{j\in\mathcal T}J_{\rm V}(x_j)}
 \sum_{i\in\mathcal T}J_{\rm V}(x_i)
 \left[\widetilde K_\theta(\bar\meas_{n,i},\bar\meas_{r,i})-Y_i\right]^2,
 \label{eq:dtimpl:loss}
\eeqn
where $\mathcal T$ denotes the collection of indices that
identify the selected regression sample. In particular,
the sample weight features $J_{\rm V}$, whereas conversion of the dumped
target requires $J_{\rm V}J_{\rm P}$. For a fixed leaf $R_\ell$, the
minimizing prediction is
\beq
 a_\ell=
 \frac{\sum_{i\in\mathcal T:\,\bar\meas_i\in R_\ell}J_{\rm V}(x_i)Y_i}
      {\sum_{j\in\mathcal T:\,\bar\meas_j\in R_\ell}J_{\rm V}(x_j)}.
 \label{eq:dtimpl:leaf-value}
\eeq
With the squared-error criterion, the tree splits greedily (see
footnote~\ref{ft:greed} for an explanation about greed) to
reduce the weighted within-leaf error; a global minimization over all
possible tree structures is not implied~\cite{SklearnTreeRegressor}.

The sampling measure in the loss is worth making explicit. With
$e_\theta(\bar\meas)=\widetilde K_\theta(\bar\meas)-Y(\bar\meas)$, its
population numerator, with the same selections understood, is
\beq
 \int dx\,J_{\rm V}(x)e_\theta(\bar\meas(v(x)))^2
 =\int dv\,e_\theta(\bar\meas(v))^2
 =\int d\bar\meas\,
   \frac{e_\theta(\bar\meas)^2}{J_{\rm P}(v_r(\bar\meas_r))}.
 \label{eq:dtimpl:loss-measure}
\eeq
Thus the fitting measure is uniform in $v$, not in $\bar\meas$.
Weighting by $J_{\rm V}J_{\rm P}$ instead would define a different loss,
uniform in the learning coordinates. Likewise, minimizing the error on
$f\widetilde K$ rather than on $\widetilde K$ would insert an additional
factor $f^2$ into the loss. Neither modification is implicit here.

\vskip 0.2truecm
\noindent\emph{A constant candidate domain with a corner removed.}
\par\noindent
For a fixed channel and training partition, one can adopt a model for
a Born-independent candidate domain, rather than working it out
starting from eq.~(\ref{eq:Lconstant}). In particular, one observes
that invalid Born-plus-radiation configurations are typically associated
with the FKS variables being far away from the singular regions.
Therefore, a simple candidate domain is the radiation plane
\mbox{$\langle\hat{\xi},\hat{y}\rangle\equiv
\langle\bar\meas_\xi,\bar\meas_y\rangle$}, with its upper-right
rectangular corner removed ($\hat{\varphi}$ does not play any role
in this transformation, and its $[0,1]$ domain is unchanged). More
precisely, let $a_\xi,a_y\in[0,1]$ be the thresholds expressed in
the learning coordinates. Then
\beqn
 \mathcal L
 &=&\left\{\bar\meas_r\in[0,1]^3:
          \bar\meas_\xi\leq a_\xi\ \hbox{or}\
          \bar\meas_y\leq a_y\right\}
 \nonumber\\
 &=&\bigl([0,a_\xi]\otimes[0,1]\otimes[0,1]\bigr)
 \cup\bigl((a_\xi,1]\otimes[0,a_y]\otimes[0,1]\bigr).
 \label{eq:dtimpl:L-corner}
\eeqn
The two boxes have disjoint interiors. One can also impose cuts on
the non-flat Vegas variables, i.e.~$v_\xi\leq b_\xi$ or $v_y\leq b_y$.
Then, with eq.~(\ref{eq:dtimpl:powers}),
$a_\xi=b_\xi^{k_\xi}$ and $a_y=b_y^{k_y}$.

Choosing $\tilde\Theta(\bar\meas_n)=\mathcal L$ gives
\beq
 \bar V=a_\xi+(1-a_\xi)a_y,\qquad
 I_K(\bar\meas_n)=\int_{\mathcal L}d\bar\meas_r\,
       \widetilde K_\theta(\bar\meas_n,\bar\meas_r).
 \label{eq:dtimpl:L-integrals}
\eeq
The integral is computed by intersecting the radiation slices of the DT
leaves with these boxes, in the $d\bar\meas_r$ measure. Now introduce
the shorthand notation
\beq
 h_\theta(\bar\meas_n,\bar\meas_r)
 =-\widetilde K_\theta(\bar\meas_n,\bar\meas_r)
      +\frac{I_K(\bar\meas_n)}{\bar V},
 \label{eq:dtimpl:h}
\eeq
i.e.~essentially what is added by DT to the original $\clS$-event integrand.
At fixed Born coordinates, the average (over radiation variables) of
$h_\theta$ and $f$ are independent of the radiation variables. On physical
points within the candidate domain, the radiation-dependent residual in the
corrected $K$ contribution is therefore
\mbox{$f\times(Y-\widetilde K_\theta)$}. Reducing the regression error
helps remove radiation-dependent structures from the corrected
contribution, and can therefore reduce its negative-weight
fraction.  This explains the squared-error loss function, but note
that it does not constitute a direct minimization of the negative-weight
fraction of the full $\clS$ sample; heuristically, it does work for
the reasons explained in the context of the toy model.

The contribution of the extra term in the ML-modified $\clS$-event
cross section to the physically restricted integration weight is
\beqn
 \delta w_i^K&=&J_{\rm V}(x_i)J_{\rm P}(v_{r,i})
  f(\bar\meas_{n,i})\,
  \mathbf1_{\rm phys}(\bar\meas_{n,i},\bar\meas_{r,i})
 \nonumber\\
 &&\hspace{1em}\times\mathbf1_{\mathcal L}(\bar\meas_{r,i})
 h_\theta(\bar\meas_{n,i},\bar\meas_{r,i}).
 \label{eq:dtimpl:weight-correction}
\eeqn
As was already made clear in the general discussion about this approach,
the same domain must be used in subtraction, integration and add-back. If
$\bar V=0$, the whole correction is zero and no ratio is evaluated.
Physical points outside $\mathcal L$ retain their original weights.

This restriction is an approximation to the conditional physical domain,
not an analytic proof of inclusion. At a valid Born point its bias density
with respect to $d\bar\meas_n$ can be read from eq.~(\ref{eq:rateNB3})
\beq
 \mathcal B_\theta(\bar\meas_n)
 =-f(\bar\meas_n)\int_{\mathcal L\setminus
                  \Theta_{\rm phys}(\bar\meas_n)}
      d\bar\meas_r\,h_\theta(\bar\meas_n,\bar\meas_r).
 \label{eq:dtimpl:corner-bias}
\eeq
It vanishes if $\mathcal L\subseteq\Theta_{\rm phys}(\bar\meas_n)$;
otherwise this integral quantifies the generally nonzero bias. Tuning
cuts (i.e.~those in eq.~(\ref{eq:dtimpl:L-corner}) in our implementation)
to avoid invalid points in a {\em finite} sample does not establish that
\mbox{$\mathcal L\subseteq\Theta_{\rm phys}(\bar\meas_n)$} holds for
any $\bar\meas_n$. Thus, any check on the value of
$\mathcal B_\theta(\bar\meas_n)$ is therefore empirical, with a precision
determined by the validation sample. Therefore, in the forthcoming section
we have assessed the amount of bias in an equally heuristic, but more
physically intuitive, way. 

\vskip 0.2truecm
\noindent\emph{Decision Trees.}
\par\noindent
The approach we have pursued in this paper uses
individual decision-tree regressors implemented with the scikit-learn
class \texttt{DecisionTreeRegressor}~\cite{SklearnTreeRegressor}.
Random Forest and Gradient Boosted Decision Tree approaches (with 
the Huber Loss function\footnote{Huber Loss has been tested in view of
its robustness against outliers; neither it nor the median absolute
deviation have performed significantly better than our default loss
function choice.} in the latter) have been considered and
employed in exploratory tests, but did not lead to any significant
improvements of the results, while requiring longer running times.

The trees are trained on
$\bar\meas_n$ and $\bar\meas_r$; these coincide with the Vegas variables,
except for $\bar\meas_\xi=v_\xi^{\,k_\xi}$ and $\bar\meas_y=v_y^{\,k_y}$,
since we set $k_\xi=2$ and $k_y=2$ there (see eq.~(\ref{eq:dtimpl:powers})).
The candidate domain
$\mathcal L$ is chosen according to eq.~(\ref{eq:dtimpl:L-corner}),
with the $a_\xi$ and $a_y$ parameters determined dynamically,
as is discussed in appendix~\ref{app:canddom}.

\subsection{Results with Decision Trees}\label{subsec:results-dt}
The hyperparameters are set as follows. We scan values of
\texttt{min\_samples\_leaf}, with 20 used as a reference. We also set
\texttt{criterion=}\allowbreak\texttt{"squared\_error"}
(in keeping with eq.~(\ref{eq:dtimpl:loss})),
\texttt{splitter="best"}, \texttt{max\_depth=None}, and
\texttt{random\_state=0}. With \texttt{max\_depth=None} one leaves the
depth unrestricted, subject to the other stopping conditions.
All of the implementation details are given in sect.~\ref{par:dtimpl:practical}.

\begin{table}[ht]
\centering
\renewcommand{\arraystretch}{1.18}
\begin{tabular}{@{}lrrccrrr@{}}
\toprule
Process & $f_0(\%)$ & $f_1(\%)$ & $\epsml(\%)$ &
          $\eta_{\mathrm{lin}}(\%)$ & $N_{\rm train}$ &
	  $\fom^\star~~~$ & $\fom^\dagger~~~$ \\
\midrule
$e^+e^-\to b\bar b$         &
  20.4 & 3.2 & 2.0 & 0.9 & 0.8M & $1.707^{+0.002}_{-0.003}$ &
  $1.601^{+0.010}_{-0.021}$\\
$e^+e^-\to t\bar t$         &
  0.4 & 0.4 & 2.3 & 1.0 & 0.8M & $-0.005^{+0.002}_{-0.003}$ &
  $-0.112^{+0.011}_{-0.022}$ \\
$e^+e^-\to b\bar b t\bar t$ &
  20.8 & 18.0 & 2.7 & 1.0 & 3.2M & $0.479^{+0.005}_{-0.010}$ &
  $0.340^{+0.030}_{-0.060}$ \\
$e^+e^-\to b\bar b b\bar b$ &
  29.9 & 28.3 & 1.6 & 0.7 & 14M & $0.865^{+0.006}_{-0.013}$ &
  $0.712^{+0.039}_{-0.077}$ \\
$e^+e^-\to Hb\bar b$        &
 23.0 & 17.4 & 2.7 & 1.2 & 2M & $1.066^{+0.005}_{-0.010}$ &
 $0.928^{+0.029}_{-0.059}$ \\
$e^+e^-\to Ht\bar t$        &
 0.1 & 0.1 & 4.0 & 0.9 & 3.2M & $-0.007^{+0.003}_{-0.006}$ &
 $-0.127^{+0.018}_{-0.037}$ \\
$pp\to b\bar b$             &
 17.3 & 10.0 & 11.7 & 12.1 & 2.8M & $0.746^{+0.014}_{-0.028}$ &
 $0.510^{+0.088}_{-0.175}$ \\
$pp\to t\bar t$             &
 8.2 & 3.3 & 1.9 & 0.8 & 2.8M & $0.280^{+0.002}_{-0.003}$ &
 $0.174^{+0.010}_{-0.020}$ \\
$pp\to ZZ$                  &
 2.7 & 1.1 & 4.1 & 1.4 & 3.8M & $0.064^{+0.003}_{-0.006}$ &
 $-0.058^{+0.020}_{-0.040}$ \\
$pp\to HH$                  &
 11.3 & 1.2 & 3.1 & 6.6 & 3.8M & $0.613^{+0.002}_{-0.005}$ &
 $0.494^{+0.017}_{-0.035}$ \\
\bottomrule
\end{tabular}
\caption{Processes considered in this work; $f_0$ and $f_1$ denote the
fraction of negative weights in the standard and ML-improved MC@NLO
simulations. The figures of merit are computed with eqs.~(\ref{fomexp})
and~(\ref{fomth}). The fractional overheads $\epsml$ and $\eta_{\mathrm{lin}}$
are defined in eq.~(\ref{eq:fom:dimensionless-overheads}). The process
$pp\to HH$ is computed in the Higgs EFT, with virtual contributions
set equal to zero.
\label{tab:newsummary}
}
\end{table}
A summary of the results obtained with DTs is given in
table~\ref{tab:newsummary}. For each process, the table reports
the values of $f_0$ and $f_1$, i.e.~of the negative-weight fractions
in the original and ML-improved MC@NLO simulations, in the notation
of sect.~\ref{sec:fom}. For simplicity, the results for $f_1$ 
originate from setting \texttt{min\_samples\_leaf}$=20$, since this
typically leads to either the smallest $f_1$ or to a value close
to it (we study the variations of \texttt{min\_samples\_leaf}
between 2 and 500 -- more details will be given later; one must bear
in mind that the absolute minimum of $f_1$ can be obtained by setting
\texttt{min\_samples\_leaf} different from $20$). The third,
fourth, and fifth result columns report the DT-related overheads
$\epsml$ and $\eta_{\mathrm{lin}}$, and the number of training
events. Then, for any given row, the entries of the five leftmost
columns are replaced in eq.~(\ref{eq:fom:def3}), in addition to
setting the remaining input parameters as follows:
\beqn
\fom^\star&=&\fom\left({\rm eq.}~(\protect\ref{eq:fom:def3});
\frac{\tau_+}{\tau_+ +\tau_A}=0.15\,,
N=100N_{\mathrm{train}}\right)\,,
\label{fomexp}
\\
\fom^\dagger&=&\fom\left({\rm eq.}~(\protect\ref{eq:fom:def3});
\frac{\tau_+}{\tau_+ +\tau_A}=0.9\,,
N=10N_{\mathrm{train}}\right)\,.
\label{fomth}
\eeqn
These correspond, respectively, to a setup with large samples with
significant downstream costs, and to one with smaller samples and
moderate downstream costs. These settings lead to the central values
of the f.o.m.~shown in the two rightmost columns of
the table. The uncertainties on those results are obtained by
subtracting the central values from those of the f.o.m.~computed
with the same parameters as the central values, except for the fact
that $\epsml$ and $\eta_{\mathrm{lin}}$ are multiplied and divided
(coherently) by a factor of two\footnote{This renders it clear that
``uncertainties'' here does not have rigorous statistical implications.}.
These are conservative settings which allow one to effectively
account for possible differences among the computing resources used in
different phases of an ML-improved simulation. When the f.o.m.~central
values are not close to zero, these uncertainties are in general
negligible, thus suggesting a significant robustness of the
f.o.m.~against the assumptions or measurements of said resources.
We point out, however, that this statement is true only for
values of \texttt{min\_samples\_leaf} which are not extreme, in the
sense that the training of the regressor is fast and accurate; elsewhere,
especially when \texttt{min\_samples\_leaf} becomes small, the uncertainties
on the f.o.m.~grow large, as we shall document later for individual processes.
This is not a problem in itself, but it hints at the fact that, in particular
for finer granularities, a sound estimate of the f.o.m.~requires a firm
understanding of the relative weights of the computing capabilities relevant
to the various phases of the simulations.

As has to be expected, the values of $\fom^\dagger$ are always smaller
than their $\fom^\star$ counterparts; for f.o.m.~values which are not
too close to zero, the value of $\fom^\star$ is dominated by the
difference of the costs (the first two terms on the r.h.s.~of
eq.~(\ref{eq:fom:def3})), while in $\fom^\dagger$ the training
costs have larger weights. This tendency becomes more evident
if one looks at the lower uncertainties, which are much larger
for $\fom^\dagger$ than for $\fom^\star$, again because increased
costs related to hard-event operations have a larger impact when
downstream costs are not overwhelming. These results confirm
quantitatively what is intuitively obvious, namely that these
ML-based techniques are more useful in complex mass productions
than in limited-statistics studies. Furthermore, the lower
uncertainties are larger than their upper counterparts since they
correspond to increasing, as opposed to decreasing, the impact of
ML-specific running times by a larger fractional amounts, in spite
of treating $\epsml$ and $\eta_{\mathrm{lin}}$ symmetrically, as is
clear from eq.~(\ref{eq:fom:def3}). As we shall see later, this happens
independently of the chosen granularity.

As is apparent from the table, for certain processes
(e.g.~\mbox{$e^+e^-\to b\bar{b}$}, \mbox{$pp\to HH$}),
the reduction we seek is significant, while for others is more limited
but still helpful (e.g.~\mbox{$e^+e^-\to Hb\bar{b}$}, \mbox{$pp\to t\bar{t}$}),
and for others still it is essentially negligible
(e.g.~\mbox{$e^+e^-\to t\bar{t}$}). The latter class is mostly made
of cases where the value of $f_0$ itself is so small that any effort
to further reduce it is academic. Conversely, for the processes of
the first class it is clear that DTs do a very good job, and one
obtains for all practical purposes what is essentially a positive-definite
$\clS$-event sample. The most interesting class is that of processes
for which the reduction of negative weights is more limited, but whose
f.o.m.'s indicate that ML-improvements are still worth the effort. In
some cases, this may appear surprising, but it essentially stems from
the behaviour of the cost function $c(f)$ of eq.~(\ref{eq:fom:source-1}).
The most striking case is probably that of \mbox{$e^+e^-\to b\bar{b}b\bar{b}$},
where the reduction from $f_0=0.299$ to $f_1=0.283$ seems negligible;
however, $c(f_0)\simeq 6.2$ and $c(f_1)\simeq 5.3$ -- which underlines
how the fraction of negative weights is a potentially misleading quantity.
The bottom line is that the f.o.m.~for this class of processes already justifies
 the use of ML-induced improvements, but at the same time points to the need
of a closer inspection. Indeed, it turns out that in each of these cases
there is a reason for the inability of the DT-based approach to reduce
negative-weight events to a larger extent; importantly, such a reason
has generally nothing to do with the DT (or NF, in fact) approach
itself, but stems from other characteristics of the implementation of
the MC@NLO cross sections; these characteristics , upon which we shall
comment in the subsections which follow, may or may not be amenable to
{\em other} reduction strategies, not necessarily ML-driven, whose
discussion is beyond the scope of the present work.

A related observation is the following: it is important to identify
the reasons why the negative weights for certain processes are impossible
to reduce further by means of the procedure we propose here since, when
this is {\em not} the case, the lack of such a further reduction may
simply be due to an insufficient training, which is something that
could be easily addressed -- a specific example will be presented later,
when analyzing $pp\to b\bar{b}$. For situations like these ones, the results
for the f.o.m.~of table~\ref{tab:newsummary} should be seen as a lower
bound, which could be improved after assessing the costs vs benefits
of a more extended training.

The DT approach provides the largest benefits for
processes with substantial negative weight fractions, where significant
ML-induced reductions in negative weights can be achieved with only a
moderate increase in running time. A rough idea about the latter can be
obtained from the third ($\epsml$) and fourth ($\eta_{\mathrm{lin}}$)
result columns of table~\ref{tab:newsummary}, where we report the fractional
values of the DT-induced overhead costs (their definitions are in
eq.~(\ref{eq:fom:dimensionless-overheads})) associated with setting
\texttt{min\_samples\_leaf}$=$20. As one can see, these quantities are 
generally rather small\footnote{In terms of absolute times, obtained
on widely available CERN-based computing facilities, such as Condor
and SWAN, with comparable amount of parallelization during the different
phases: in our simulations, where we consider dumping as part of the
generation, we generate 1M $e^+e^-\to b\bar{b}$ ($pp\to b\bar{b}$) events
in about 1260s(2470s) which, by using table~\ref{tab:newsummary}, result in
additional ML-related generation and training times of about 25s(290s) and
11s(300s), respectively. Generation times depend significantly on process
complexity; at the other end of the spectrum of the cases considered here,
we find that the generation of 1M $e^+e^-\to b\bar{b}b\bar{b}$ events takes
about 4440s\label{ft:times}.}. It should be clear that this is not a statement
valid everywhere in the space of the hyperparameters; specifically, one
of the reasons for smaller f.o.m.'s is the increase in the ML-related
operations during generation and unweighting. This suggests that in
practical applications an exploratory investigation searching the
hyperparameter space for f.o.m.~maximization may be well worth its costs.

Conversely, processes with already flat
weight distributions and very low negative weight fractions, such as
$e^+e^-\to t\bar t$ and $e^+e^-\to Ht\bar t$, 
show little or no improvement, resulting in an almost constant figure
of merit across granularities (or a decrease most likely due to
overtraining). These results thus only serve the
purpose of showing that the DT approach does not worsen those
cases where an improvement is essentially unnecessary.

For most processes, the figure of merit increases as the granularity
is refined, reaching a maximum around \msl{} = 50, 20, or 10. Further
refinement generally leads to a decrease in performance, {\em possibly}
indicating the onset of overtraining. The results also show that comparable
performance can often be achieved by balancing granularity and
training sample size/time, allowing the computational cost to be reduced
without a significant loss in negative-weight suppression.

We now turn to discussing results for selected processes, by
also presenting predictions for individual channels and/or FKS
sectors. This will allow us to understand the reason(s) behind
the inability of the DT-based approach to reduce negative weights.

\subsubsection{\texorpdfstring{$e^+e^-\to b\bar b$}{eebb}}
\begin{figure}[ht]
\centering
\includegraphics[width=0.7\linewidth]{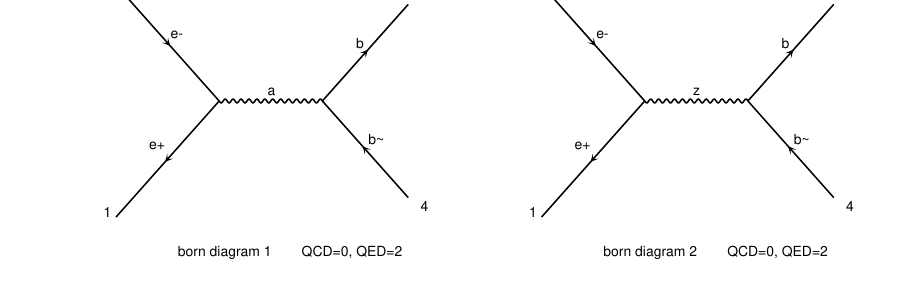}
\caption{Feynman diagrams $e^+e^-\to b\bar b$, underpinning the
two integration channels.}
\label{fig:source1}
\end{figure}

The process $e^+e^-\to b\bar b$ features two channels, denoted by EE1
and EE2, that stem from the Feynman diagrams displayed on the left and
right panel of fig.~\ref{fig:source1}, respectively. Each of them has
two FKS sectors, relevant to the configurations where the emitted gluon is
collinear to the $b$ or $\bar{b}$ quark. The results for each channel, and
for each channel {\em and} FKS sector\footnote{The present public version of
{\tt MadGraph5\_aMC@NLO} adopts an MC-based sum over FKS sectors. Conversely,
we remind the reader that by default we train the regressor given both
the FKS sector and the channel.}, are presented in table~\ref{tab:source1} --
the cross sections (denoted by $\sigma$) and the cross sections
obtained by integrating the absolute value of the integrand (denoted
by $\abs{\sigma}$), both before (``pre-ML'') and after (``post-ML'')
the DT manipulations, and the resulting negative-event fractions.
The agreement (within statistical uncertainties, which we do not show
but which we did assess) between the results for $\sigma$'s pre-
and post-ML implies the absence of ML-induced biases for the rates.
For this process, the reduction of negative weights is largely
independent of the channel and the FKS sector, with $f_1\simeq 3.2\%$
achieved in all cases. Therefore, the best f.o.m.~value of $\sim 1.71$
reported in table~\ref{tab:newsummary} is basically relevant also
to both the individual channels and the individual FKS sectors within
the individual channels.

This gives one the opportunity to compare the performance of our
DT approach with that of folding as is currently implemented in
{\tt MadGraph5\_aMC@NLO}. In order to do that, we consider 
the dominant channel EE2 to be definite.

\begin{table}[ht]
\centering
\setlength{\tabcolsep}{5pt}
\renewcommand{\arraystretch}{1.10}
\small
\begin{tabular}{@{}lcccc@{}}
\toprule
Channel & & $\sigma$ (pb) & $\abs{\sigma}$ (pb) & $f_0,f_1(\%)$\\
\midrule
\multirow{2}{*}{ALL} & pre-ML & $8.752\!\cdot\!10^{-2}$ & $1.478\!\cdot\!10^{-1}$ & 20.4\\
 & post-ML & $8.768\!\cdot\!10^{-2}$ & $9.366\!\cdot\!10^{-2}$ & 3.2\\
\midrule[0.1pt]
\multirow{2}{*}{EE1} & pre-ML & $2.928\!\cdot\!10^{-2}$ & $4.921\!\cdot\!10^{-2}$ & 20.2\\
 & post-ML & $2.926\!\cdot\!10^{-2}$ & $3.124\!\cdot\!10^{-2}$ & 3.2\\
\midrule[0.1pt]
\multirow{2}{*}{-- FKS1} & pre-ML & $1.467\!\cdot\!10^{-2}$ & $2.458\!\cdot\!10^{-2}$ & 20.2\\
 & post-ML & $1.467\!\cdot\!10^{-2}$ & $1.564\!\cdot\!10^{-2}$ & 3.1\\
\midrule[0.1pt]
\multirow{2}{*}{-- FKS2} & pre-ML & $1.462\!\cdot\!10^{-2}$ & $2.464\!\cdot\!10^{-2}$ & 20.3\\
 & post-ML & $1.460\!\cdot\!10^{-2}$ & $1.560\!\cdot\!10^{-2}$ & 3.2\\
\midrule[0.1pt]
\multirow{2}{*}{EE2} & pre-ML & $5.824\!\cdot\!10^{-2}$ & $9.856\!\cdot\!10^{-2}$ & 20.5\\
 & post-ML & $5.842\!\cdot\!10^{-2}$ & $6.241\!\cdot\!10^{-2}$ & 3.2\\
\midrule[0.1pt]
\multirow{2}{*}{-- FKS1} & pre-ML & $2.909\!\cdot\!10^{-2}$ & $4.906\!\cdot\!10^{-2}$ & 20.4\\
 & post-ML & $2.910\!\cdot\!10^{-2}$ & $3.109\!\cdot\!10^{-2}$ & 3.2\\
\midrule[0.1pt]
\multirow{2}{*}{-- FKS2} & pre-ML & $2.915\!\cdot\!10^{-2}$ & $4.949\!\cdot\!10^{-2}$ & 20.6\\
 & post-ML & $2.932\!\cdot\!10^{-2}$ & $3.133\!\cdot\!10^{-2}$ & 3.2\\
\bottomrule
\end{tabular}
\caption{Total rates per channel and FKS sector in $e^+e^-\to b\bar b$.}
\label{tab:source1}
\end{table}
The simple $(4,4,1)$ folding of the weights without any DT manipulations
leads to $f_1=2.7\%$. In turn, by using eq.~(\ref{eq:fom:fold3}), this
fraction of negative weights results in $\fom^{({\rm fold})}=-13.3$ and
$\fom^{({\rm fold})}=-0.78$ for \mbox{$\frac{\tau_+}{\tau_+ +\tau_A}=0.9$}
and $0.15$, respectively. While it is not surprising that the former
figure is large and negative (in the theoretical scenario, most of the
time is spent in generation and unweighting), it is perhaps more
striking that also the latter one, which is directly comparable with
the ML-induced f.o.m.~equal to $1.71$ quoted above, is negative.
The conclusion is that, for the present process, the DT approach is
vastly more convenient w.r.t.~folding.

In view of that, the following observation is borderline academic.
Namely, we have also trained the DT model on events generated with
$(4,4,1)$ folding, and reduced the negative-weight fraction
to 2.8\%, statistically compatible with the 2.7\% folding-only
result obtained before. This represents only a marginal improvement
w.r.t.~the 3.2\% result obtained with training on non-folded events.
We have also applied  folding to ML-generated events, and obtained
$f_1=2.4\%$. In summary , while this shows that any such folding+ML or
ML+folding run is financially unjustified, it also suggests
that, at least in the cases where the reduction of negative weights
with either folding or ML is large, the benefits of these
two methods are not cumulative -- the reduction of negative
weights ``saturates''.

In addition, we have identified the procedure used by {\tt MadGraph5\_aMC@NLO}
for speeding up the computation of the virtual matrix elements
(see sect.~2.4.3 of ref.~\cite{Alwall:2014hca} for more details)
as a source of negative weights. Specifically, an approximate
virtual contribution is evaluated at each phase-space point, while
the exact virtual matrix elements are computed every $n_V$ points,
with $n_V={\cal O}(100)$; the latter serve to guide the former to
achieve a good approximation of the whole differential virtual
contribution. While this procedure works surprisingly well, it
creates event weights which locally can be many times larger than
their immediate neighbours; it is this imbalance which leads to
more negative weights than is strictly necessary. In order to study more
in detail the consequences of this treatment of the virtuals,
$e^+e^-\to b\bar b$ events have also been generated with the exact
virtual contribution computed at each phase-space point, or at every 50
points. No significant effect on the ML-approach performance has been
observed; however, in the former case the fraction of negative weights
has been reduced to 1.9\%, i.e.~the overall best\footnote{This is not
equivalent to obtaining the best f.o.m., owing to the extra costs
stemming from the more frequent evaluation of the virtual matrix elements.}.
This is unimportant per se, but otherwise crucial to uncover a source
of negative weights that cannot be addressed with the present ML-based
method, but which appears to be tractable independently of this work.

\begin{figure}[ht]
\centering
\includegraphics[width=\linewidth]{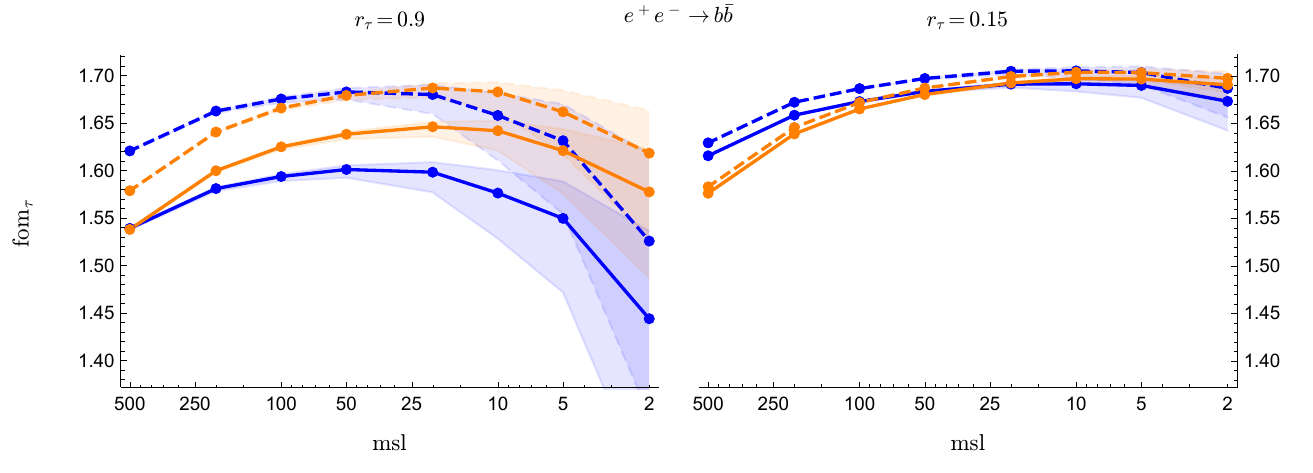}
\caption{Figure of merit for $e^+e^-\to b\bar b$.}
\label{fig:source2}
\end{figure}
Figure~\ref{fig:source2} illustrates the figure of merit for the
process $e^+e^-\to b\bar b$ as a function of \texttt{min\_samples\_leaf},
again for the two scenarios considered before (theoretical on
the left panel, experimental of the right
panel). In each plot there are four curves, with the blue (orange) ones
obtained with the same (half of the) number of training events
reported in table~\ref{tab:newsummary}; furthermore, the solid (dashed)
patterns are associated with assuming the number of generated events
to be 10 (100) times as large as that of the largest training dataset --
in other words, the two solid or two dashed curves correspond to the same
number of generated events, but not to the same number of training events.
The uncertainty bands are obtained as the envelopes of the uncertainties
affecting the individual f.o.m.~results at the current value of
\texttt{min\_samples\_leaf}, by following the same prescription as
in table~\ref{tab:newsummary}.

The results of fig.~\ref{fig:source2} indicate that different
combinations of granularity and training sample size can in fact
achieve comparable performances, and therefore the hyperparameter
set which leads to the ``best'' fom has a significant latitude.
For example, in the theoretical scenario a granularity of \msl{} = 50
trained on 800k events yields a figure of merit similar to that obtained
with \msl{} = 10 trained on 400k events. Conversely, in the experimental
scenario there is the tendency to prefer smaller \msl{} values,
regardless of the training-sample size.

Another characteristic of fig.~\ref{fig:source2} is worth stressing,
since as we shall see it is almost universal across processes. Namely,
the uncertainty bands grow with decreasing \texttt{min\_samples\_leaf}, with the
lower bounds farther away from the central values than the upper bounds,
in keeping with table~\ref{tab:newsummary}. This growth essentially stems
from the increase in training times associated with finer granularities,
in spite of this possibly resulting in smaller negative-weight fractions,
with the largest relative impact seen in the theoretical scenario.

\begin{figure}[thb]
  \begin{center}
  \includegraphics[width=0.47\textwidth]{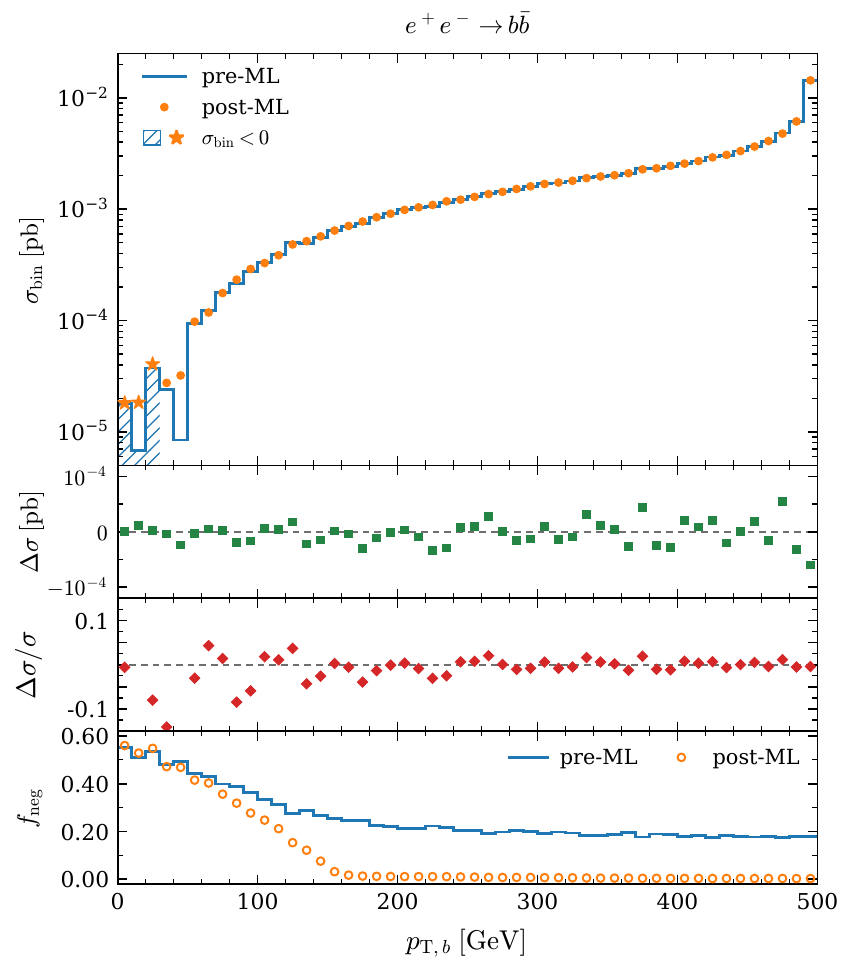}
$\phantom{a}$
  \includegraphics[width=0.47\textwidth]{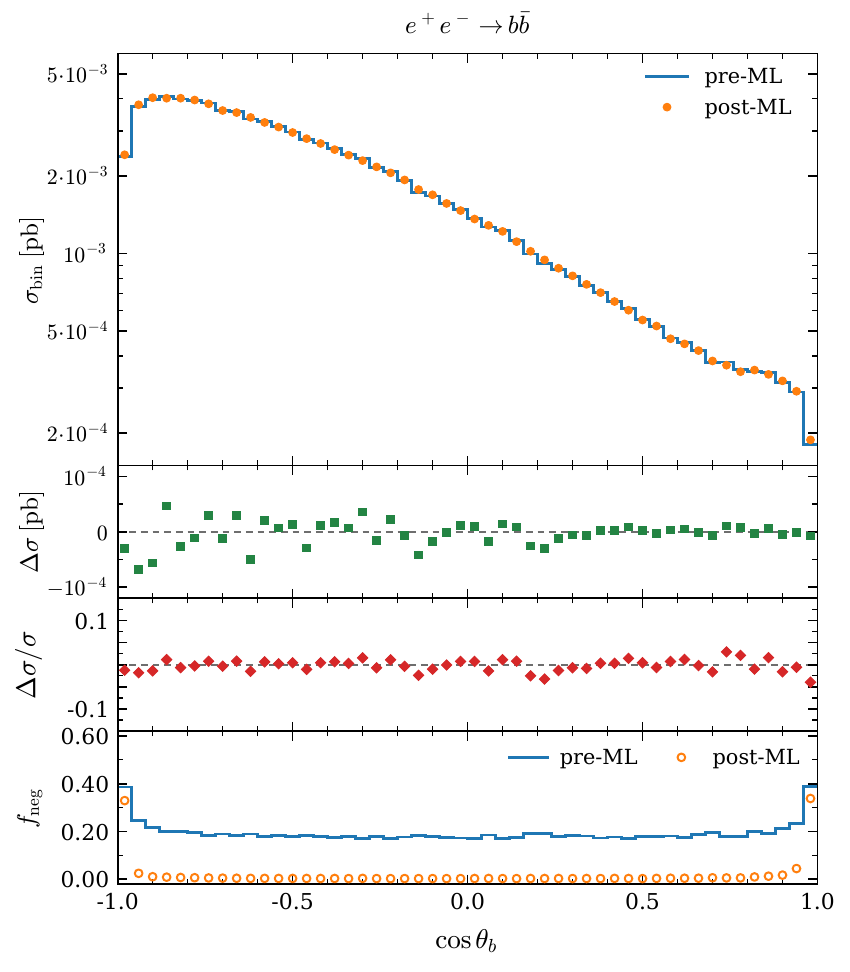}
\caption{\label{fig:eebbdiff}
Left panel: $b$-quark $\pt$ in $e^+e^-\to b\bar{b}$;
right panel: $b$-quark $\cos\theta$.}
  \end{center}
\end{figure}
As far as the differential observables are concerned, in fig.~\ref{fig:eebbdiff}
we present the spectrum of the $b$ quark in transverse momentum (left
panel) and laboratory-frame scattering angle (right panel). These, and all of
the plots of LHE-level ``observables'' henceforth, have the same layout,
as follows: in the main (top) frame, the blue histogram and the full orange
circles are the standard MC@NLO prediction and the ML-modified one,
respectively, for the cross section per bin (in pb). Should the cross
section be negative, its absolute value is what is plotted, with a
hashed pattern in the case of the histogram, and as a full star
for the post-ML prediction.
Below the main frame, the top and middle insets display
\beqn
\Delta\sigma&=&\sigma_{\rm pre-ML}-\sigma_{\rm post-ML}\,,
\label{DSoS1}
\\*
\frac{\Delta\sigma}{\sigma}&=&
2\,\frac{\sigma_{\rm pre-ML}-\sigma_{\rm post-ML}}
{\sigma_{\rm pre-ML}+\sigma_{\rm post-ML}}\,,
\label{DSoS2}
\eeqn
as full green boxes and full red diamonds, respectively. Finally,
in the bottom inset the blue histogram and open orange circles
represent the bin-by-bin fraction of negative weights emerging
from the standard MC@NLO and the ML-modified simulations, respectively.
In essence, we want the ML-modified $\clS$-event cross sections to
give results compatible with the standard ones; for this to happen,
the full circles should be in agreement with the histogram in the
main frame. In order to better appreciate whether this is the case
or not, the first inset presents the bin-by-bin difference of the
two cross sections which appear in the main frame, according to
eq.~(\ref{DSoS1}). In other words, in a given
bin the full green box is equal to the value of the
blue histogram minus the value of the orange full circle in that bin.
Likewise, the second inset shows the relative difference of the same
two cross sections, as is defined in eq.~(\ref{DSoS2}). For the two
cross sections to be compatible, we expect the symbols in the first
and second insets to oscillate, without any discernible pattern,
around zero\footnote{Note that for vanishingly small cross sections the
relative difference of eq.~(\ref{DSoS2}) tends to over-emphasize
discrepancies, and the difference of eq.~(\ref{DSoS1}) gives
one a better idea on the actual level of agreement between the
two simulations; the opposite is true in the vicinity of the
cross section peaks.}.
Finally, the bottom inset allows one to check where the reduction
of negative weights occurs (if any; in that case, the orange circles
must be below the histogram).

The agreement between the standard and ML-improved MC@NLO predictions
shown in fig.~\ref{fig:eebbdiff} is fairly good across the
complete spectra. The reduction of the fraction of negative weights
occurs everywhere in the phase space, but it is less effective at
small $\pt$, where however the cross section becomes quite small.
This may suggest that further improvements are possible with
increased statistics, or a different kind of training. 

\begin{figure}[thb]
  \begin{center}
  \includegraphics[width=0.47\textwidth]{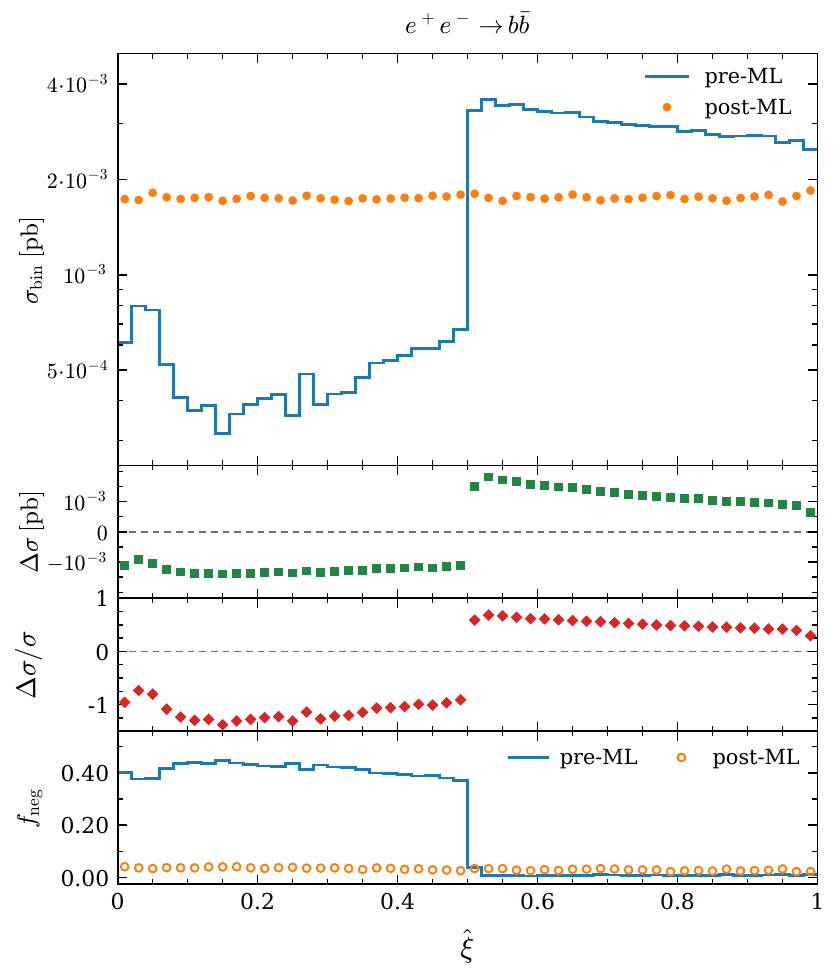}
$\phantom{a}$
  \includegraphics[width=0.47\textwidth]{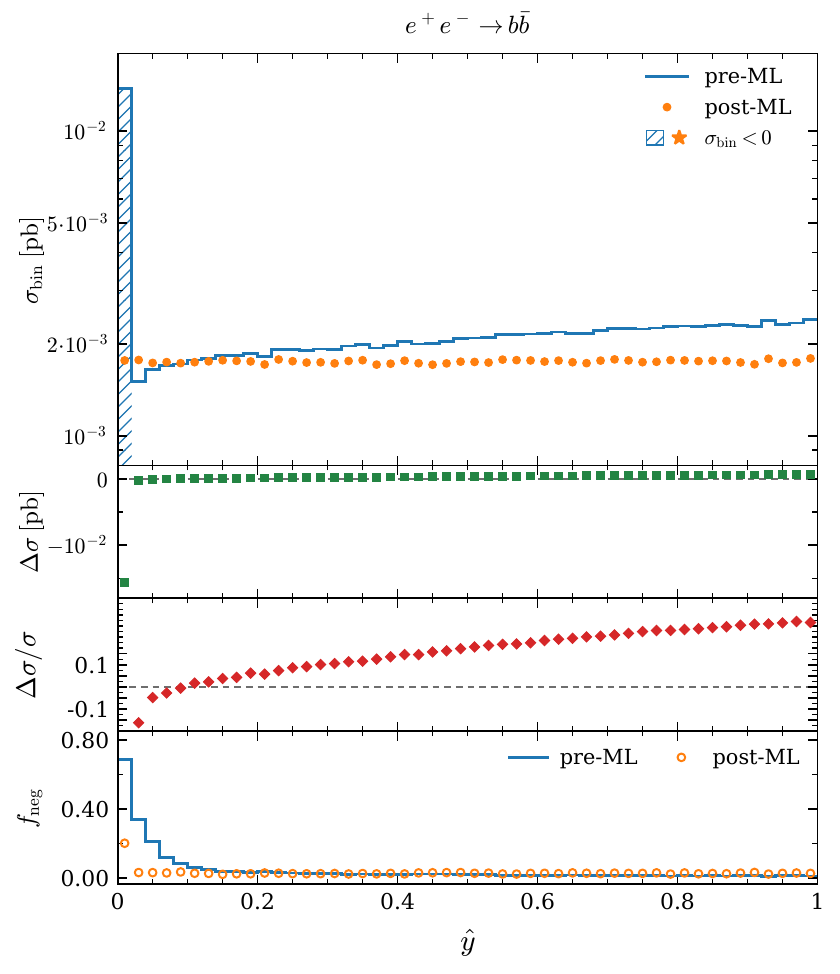}
\caption{\label{fig:eebbxiy}
$\hat{\xi}$ and $\hat{y}$ (rescaled FKS variables) distributions
in $e^+e^-\to b\bar{b}$.}
  \end{center}
\end{figure}
We now show the analogue of the toy-model situation presented in
fig.~\ref{fig:toy}, in the context of the realistic case of
$e^+e^-\to b\bar{b}$, namely that of quantities which {\em are}
biased by the ML procedure, this being the only way in which a
reduction of negative weights can be achieved. Specifically, we
are talking about the rescaled FKS variables $\hat{\xi}$ and $\hat{y}$
defined in eq.~(\ref{FKSvresc}), the analogues of the variable $x$
of the toy model. Their distributions are shown in fig.~\ref{fig:eebbxiy},
whence one sees that the pre- and post-ML results are not in agreement -- 
only their total integral is (first two rows of table~\ref{tab:source1}).
As was anticipated when discussing the toy model, the post-ML distributions
are completely flat, which implies that the regressor has learnt the
shape of the standard-MC@NLO $K$ term in a fairly satisfactory manner,
in spite of the fact that such a shape is non trivial, in that it features
hard edges\footnote{Such hard edges are due to the singularity subtractions
in the short distance cross sections. In the FKS method one limits, by means
of arbitrary parameters, the regions of phase space where the actual
subtractions are carried out; the physical cross section is strictly
independent of such parameters. More details on this point can be
found in refs.~\cite{Frixione:1995ms,Frixione:1997np,Frederix:2009yq}.}
in both $\hat{\xi}$ and $\hat{y}$. This behaviour of the
regressor is ubiquitous across the processes, and since the emphasis
here is to prove the lack of biases on Born-level observables we shall
henceforth refrain from presenting again distributions in $\hat{\xi}$
and $\hat{y}$.

We have also generated $e^+e^-\to b\bar{b}$ with MC@NLO-$\Delta$; the
outcomes of the DT-procedure are fully consistent with those
we have presented before. This suggests that the improvements
introduced in the MC@NLO-$\Delta$ formalism survive the DT-based
manipulations of the cross section; in other words, our ML-driven
approach is agnostic as to whether $\clS$ events are generated
by MC@NLO or MC@NLO-$\Delta$. We see no reason why this conclusion
should not be applied universally, i.e.~to any arbitrary process.

\subsubsection{\texorpdfstring{$e^+e^-\to t\bar t$}{eett}}
\begin{figure}[ht]
\centering
\includegraphics[width=\linewidth]{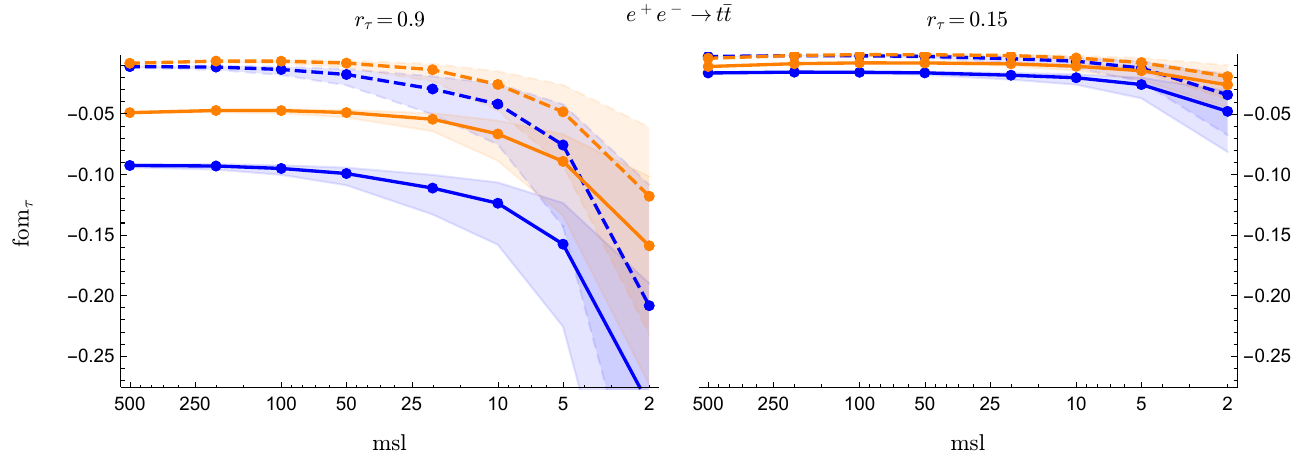}
\caption{Figure of merit for $e^+e^-\to t\bar t$.}
\label{fig:source3}
\end{figure}
The DT-based procedure shows no improvement for this process, with the
negative weight fraction remaining at 0.4\% for both channels. Tests
carried out with different granularities confirmed that no further
reduction of negative weights is possible.

In keeping with that, the figure of merit for the process
\mbox{$e^+e^-\to t\bar t$} exhibits at most a very mild dependence on
the granularity parameter, except when decreasing towards values
smaller than those in the plateau, as is evident from
fig.~\ref{fig:source3}, and which very likely signals overtraining.
As no configuration provides a significant improvement in the
negative-weight fraction, the whole ML-driven machinery is an
overkill, which is fine in view of the already negligible costs for a
standard-MC@NLO sample.

\subsubsection{\texorpdfstring{$e^+e^-\to b\bar b t\bar t$}{eebbtt}}
\begin{table}[ht]
\centering
\setlength{\tabcolsep}{5pt}
\renewcommand{\arraystretch}{1.10}
\fontsize{8.2}{10.2}\selectfont
\setlength{\tabcolsep}{2.3pt}
\begin{tabular}{@{}lcccc@{\hspace{9pt}}lcccc@{}}
\toprule
Channel & & $\sigma$ (pb) & $\abs{\sigma}$ (pb) & $f_0,f_1(\%)$ & Channel & & $\sigma$ (pb) & $\abs{\sigma}$ (pb) & $f_0,f_1(\%)$\\
\midrule
\multirow{2}{*}{ALL} & pre-ML & $1.977\!\cdot\!10^{-3}$ & $3.389\!\cdot\!10^{-3}$ & 20.8 & \multirow{2}{*}{EE5} & pre-ML & $7.515\!\cdot\!10^{-4}$ & $1.299\!\cdot\!10^{-3}$ & 21.1\\
 & post-ML & $1.981\!\cdot\!10^{-3}$ & $3.090\!\cdot\!10^{-3}$ & 18.0 &  & post-ML & $7.524\!\cdot\!10^{-4}$ & $1.182\!\cdot\!10^{-3}$ & 18.2\\
\midrule[0.1pt]
\multirow{2}{*}{EE1} & pre-ML & $5.919\!\cdot\!10^{-7}$ & $8.897\!\cdot\!10^{-7}$ & 16.7 & \multirow{2}{*}{EE6} & pre-ML & $7.594\!\cdot\!10^{-4}$ & $1.292\!\cdot\!10^{-3}$ & 20.6\\
 & post-ML & $5.913\!\cdot\!10^{-7}$ & $6.642\!\cdot\!10^{-7}$ & 5.5 &  & post-ML & $7.613\!\cdot\!10^{-4}$ & $1.180\!\cdot\!10^{-3}$ & 17.7\\
\midrule[0.1pt]
\multirow{2}{*}{EE2} & pre-ML & $5.881\!\cdot\!10^{-7}$ & $8.931\!\cdot\!10^{-7}$ & 17.1 & \multirow{2}{*}{EE7} & pre-ML & $2.303\!\cdot\!10^{-4}$ & $3.946\!\cdot\!10^{-4}$ & 20.8\\
 & post-ML & $5.893\!\cdot\!10^{-7}$ & $6.656\!\cdot\!10^{-7}$ & 5.7 &  & post-ML & $2.310\!\cdot\!10^{-4}$ & $3.601\!\cdot\!10^{-4}$ & 17.9\\
\midrule[0.1pt]
\multirow{2}{*}{EE3} & pre-ML & $1.111\!\cdot\!10^{-6}$ & $1.679\!\cdot\!10^{-6}$ & 16.9 & \multirow{2}{*}{EE8} & pre-ML & $2.327\!\cdot\!10^{-4}$ & $3.975\!\cdot\!10^{-4}$ & 20.7\\
 & post-ML & $1.112\!\cdot\!10^{-6}$ & $1.250\!\cdot\!10^{-6}$ & 5.5 &  & post-ML & $2.326\!\cdot\!10^{-4}$ & $3.646\!\cdot\!10^{-4}$ & 18.1\\
\midrule[0.1pt]
\multirow{2}{*}{EE4} & pre-ML & $1.110\!\cdot\!10^{-6}$ & $1.689\!\cdot\!10^{-6}$ & 17.1 &  &  &  &  & \\
 & post-ML & $1.114\!\cdot\!10^{-6}$ & $1.256\!\cdot\!10^{-6}$ & 5.6 &  &  &  &  & \\
\bottomrule
\end{tabular}
\caption{Total rates per channel in $e^+e^-\to b\bar b t\bar t$.}
\label{tab:source3}
\end{table}
\begin{figure}[ht]
\centering
\includegraphics[width=0.4\linewidth]{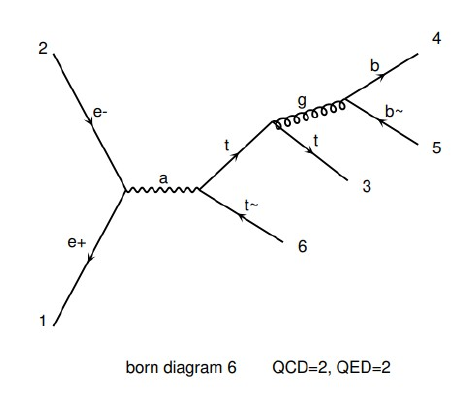}
\caption{Feynman diagram underpinning channel EE6
of $e^+e^-\to b\bar b t\bar t$.}
\label{fig:source5}
\end{figure}
The total pre- and post-ML negative-weight fractions of the
$e^+e^-\to b\bar b t\bar t$ process are equal to 20.8\% and 18.0\%,
respectively. In order to further the analysis of this behaviour, all channels
have been considered separately from one another -- the results are
reported in table~\ref{tab:source3}. From there, one sees that the
dominant contributions to negative weights are due to channels EE5--EE8,
which are underpinned by diagrams that feature a $g\to b\bar{b}$
splitting. This can be seen e.g.~in fig.~\ref{fig:source5}, a diagram
that defines channel EE6, which we have chosen for a closer
inspection. The results of that are reported in
table~\ref{tab:source4}, where the contributions due to the
relevant FKS sectors are shown separately. 
\begin{table}[ht]
\centering
\setlength{\tabcolsep}{5pt}
\renewcommand{\arraystretch}{1.10}
\small
\begin{tabular}{@{}lcccc@{}}
\toprule
FKS sectors & & $\sigma$ (pb) & $\abs{\sigma}$ (pb) & $f_0,f_1(\%)$\\
\midrule
\multirow{2}{*}{TOTAL EE6} & pre-ML & $7.594\!\cdot\!10^{-4}$ & $1.292\!\cdot\!10^{-3}$ & 20.6\\
 & post-ML & $7.613\!\cdot\!10^{-4}$ & $1.180\!\cdot\!10^{-3}$ & 17.7\\
\midrule[0.1pt]
\multirow{2}{*}{-- FKS1} & pre-ML & $3.205\!\cdot\!10^{-4}$ & $3.297\!\cdot\!10^{-4}$ & 1.4\\
 & post-ML & $3.213\!\cdot\!10^{-4}$ & $3.320\!\cdot\!10^{-4}$ & 1.6\\
\midrule[0.1pt]
\multirow{2}{*}{-- FKS2} & pre-ML & $6.380\!\cdot\!10^{-5}$ & $3.150\!\cdot\!10^{-4}$ & 39.9\\
 & post-ML & $6.473\!\cdot\!10^{-5}$ & $2.562\!\cdot\!10^{-4}$ & 37.4\\
\midrule[0.1pt]
\multirow{2}{*}{-- FKS3} & pre-ML & $5.744\!\cdot\!10^{-5}$ & $3.179\!\cdot\!10^{-4}$ & 41.0\\
 & post-ML & $5.737\!\cdot\!10^{-5}$ & $2.595\!\cdot\!10^{-4}$ & 38.9\\
\midrule[0.1pt]
\multirow{2}{*}{-- FKS4} & pre-ML & $3.177\!\cdot\!10^{-4}$ & $3.294\!\cdot\!10^{-4}$ & 1.8\\
 & post-ML & $3.178\!\cdot\!10^{-4}$ & $3.324\!\cdot\!10^{-4}$ & 2.2\\
\bottomrule
\end{tabular}
\caption{Total rates per FKS sector for channel EE6
in $e^+e^-\to b\bar b t\bar t$.}
\label{tab:source4}
\end{table}

\begin{figure}[ht]
\centering
\includegraphics[width=0.45\linewidth,height=0.3\textheight]{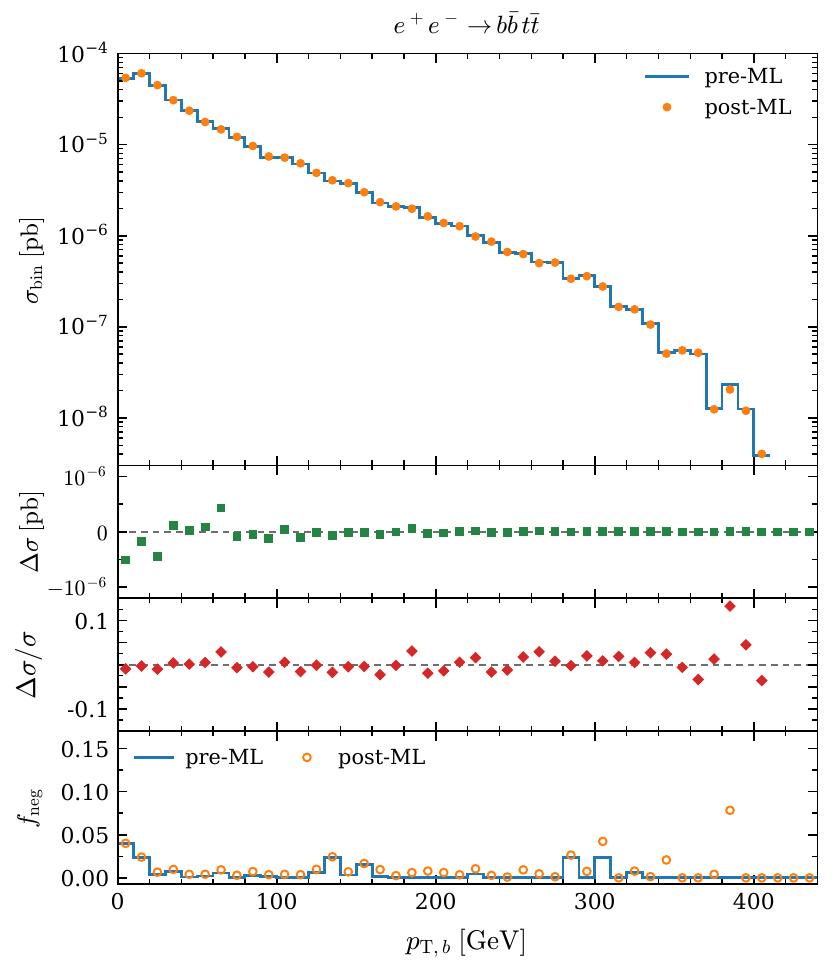}
$\phantom{a}$
\includegraphics[width=0.45\linewidth,height=0.3\textheight]{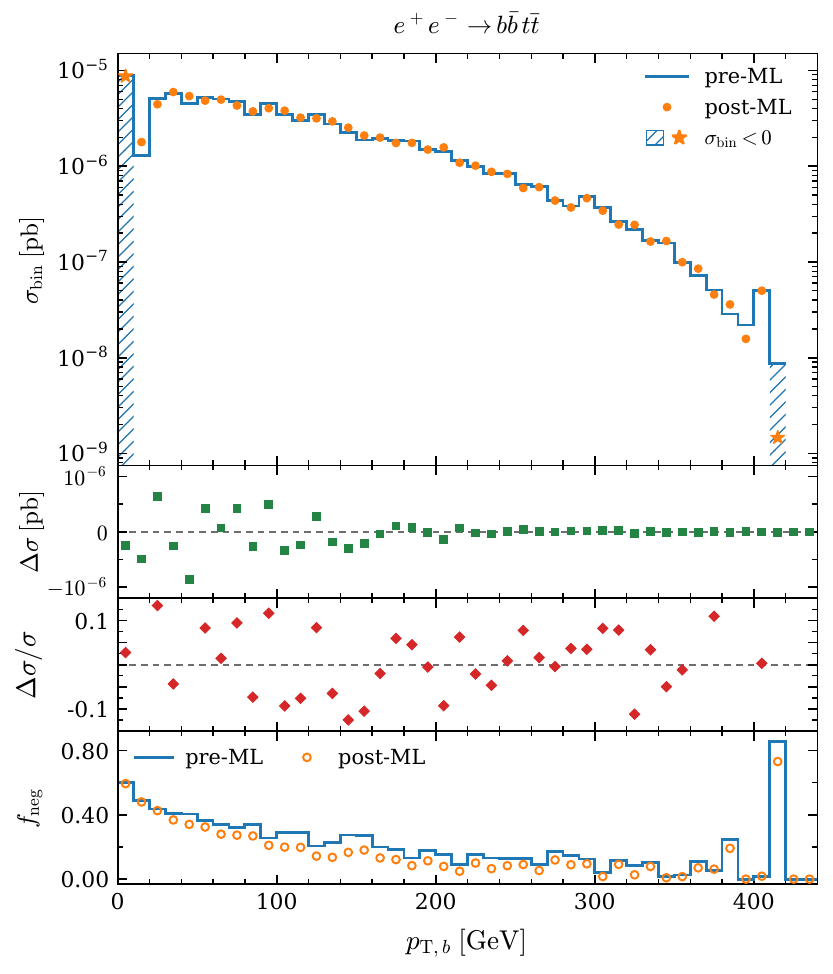}
\includegraphics[width=0.45\linewidth,height=0.3\textheight]{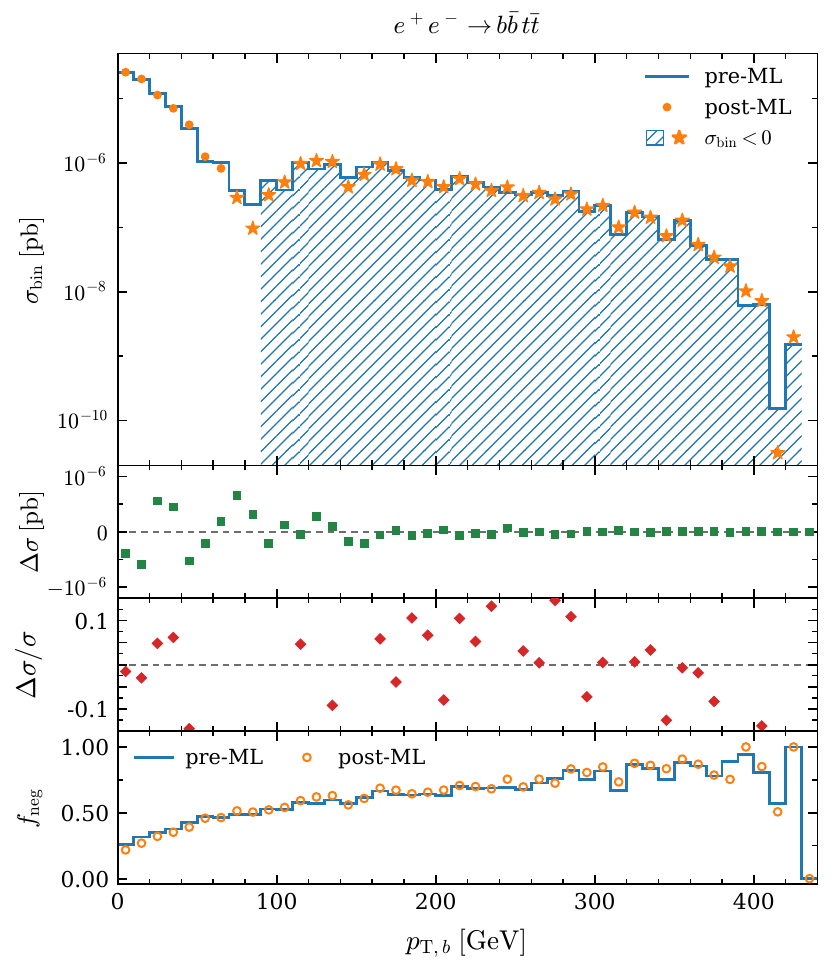}
$\phantom{a}$
\includegraphics[width=0.45\linewidth,height=0.3\textheight]{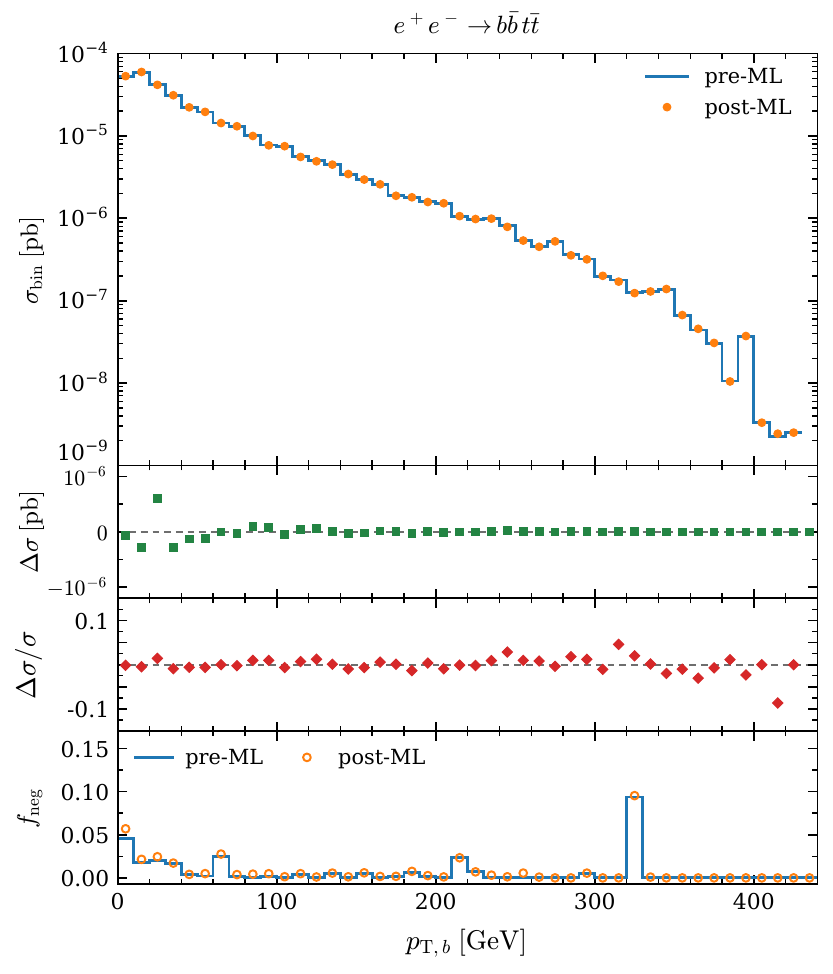}
\caption{$b$-quark $\pt$ distributions due to FKS sectors 1, 2, 3, and 4 
(left to right, top to bottom) of channel EE6, in $e^+e^-\to b\bar b t\bar t$.}
\label{fig:source6}
\end{figure}

\begin{figure}[ht]
\centering
\includegraphics[width=0.60\linewidth]{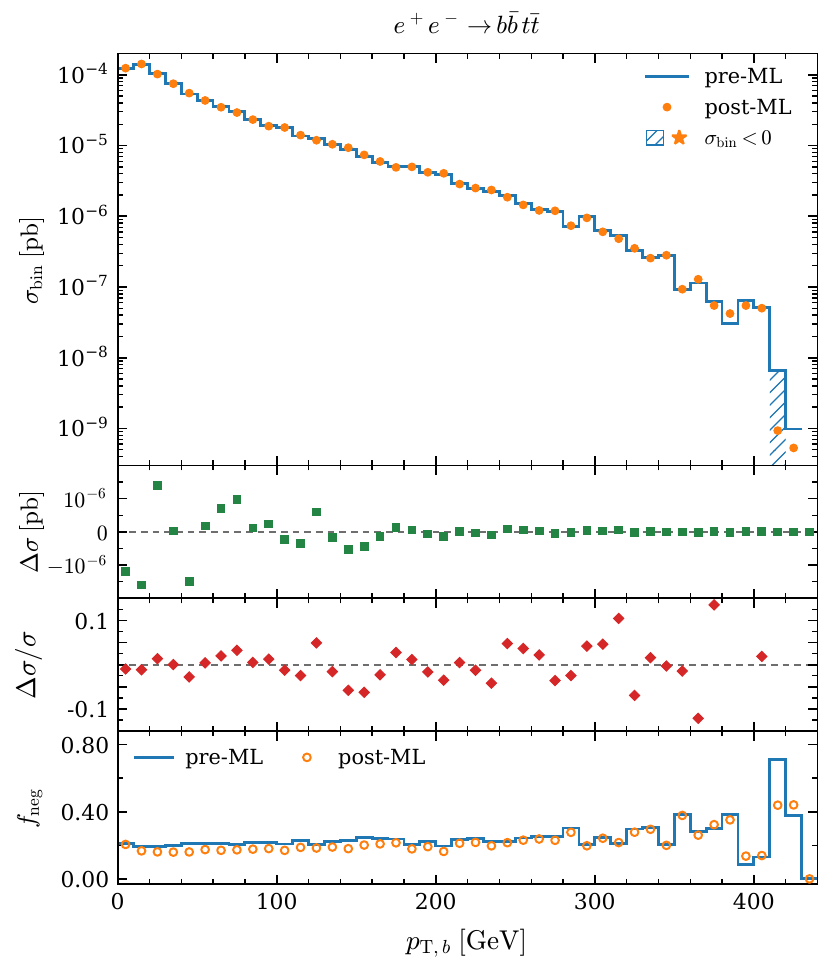}
\caption{$b$-quark $\pt$ distribution due to channel EE6,
in $e^+e^-\to b\bar b t\bar t$.}
\label{fig:source7}
\end{figure}
From the table, we see that negative-weight events are prominent in
FKS sectors 2 and 3 (which correspond to configurations where the
extra gluon is collinear to the $b$ and $\bar b$, respectively),
while they are basically negligible in sectors 1 and 4 (there, the
extra gluon is collinear to the $t$ and $\bar t$, respectively).
In summary, the dominant sources of negative weights in this process
are due to configurations featuring $g\to (b\to bg)\bar{b}$ or
$g\to (\bar{b}\to \bar{b}g)b$ branchings. These findings are
clarified by looking at differential distributions. We have considered
the transverse momentum of the first\footnote{Since we are looking at
LHE-level quantities, these need not be physical, and therefore
the (arbitrary) choice of the first parton in the list of final-state
objects is as good as any other to study the interplay among channels
and FKS sectors, as well as the absence of ML-induced biases.
\label{ft:ordering}} $b$ quark, and plotted in fig.~\ref{fig:source6}
the separate contributions due to FKS sectors
1 (top left), 2 (top right), 3 (bottom left), 4 (bottom right); from
there, it becomes apparent that the sources of negative weights in
sectors 2 and 3 are at the opposite ends of the $\pt$ spectrum. The values
of the differential cross sections suggest that when adding up all of the
four EE6-channel FKS sectors, the negative regions in the $\pt$ distribution
should cancel out; indeed, this can be seen explicitly in fig.~\ref{fig:source7},
and implies that learning a sample which collects the events stemming
simultaneously from all of the FKS sectors should prove beneficial.

Further to this observation, an independent test (with reduced statistics)
has been performed, in which all the EE6-channel FKS sectors have been
summed directly at the {\tt MadGraph5\_aMC@NLO} dump level. By doing so,
the fraction of negative-weight events is reduced to $15.3\%$ (rather than
$20.6\%$) if the FKS-sector contributions are summed before unweighting.
By training the DT on this FKS-summed sample, that fraction is further
reduced to $12.4\%$ (rather than $17.7\%$). These results suggest that it
would be beneficial to study the structure of the Born-level cross sections
before going through the ML procedure, in order to check whether
it would be convenient to first merge different FKS-sector
samples\footnote{While this seems always to be the case,
such a combination may not be always sensible -- a necessary
condition is that the learning variables must have identical
meanings for all of the sectors considered.}.

\begin{figure}[ht]
\centering
\includegraphics[width=0.45\linewidth]{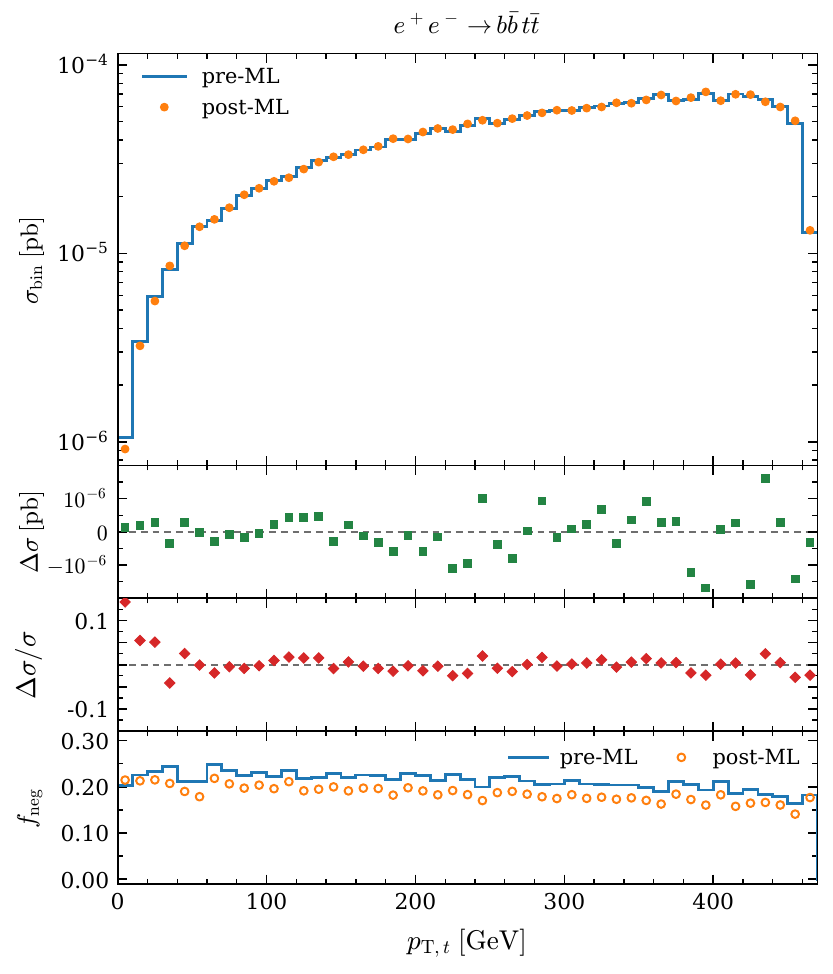}
$\phantom{a}$
\includegraphics[width=0.45\linewidth]{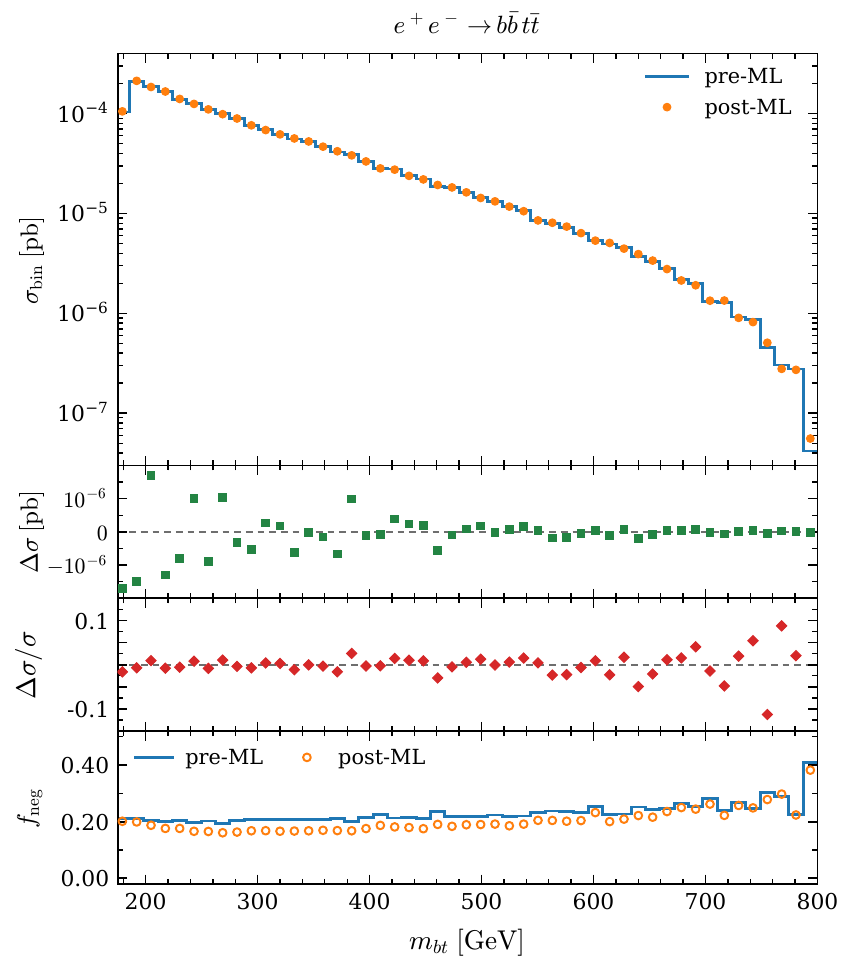}
\caption{Transverse momentum of the top quark (left panel) and invariant
mass of the $(b,t)$ pair in $e^+e^-\to b\bar b t\bar t$.}
\label{fig:eebbtt-ptt}
\end{figure}
Coming back to the all-channel results in the present implementation
with separate training for the various FKS sectors, we employ the
distributions presented in fig.~\ref{fig:eebbtt-ptt} as representatives
of the typical situation we encounter at the differential level.
In that figure, we show the $\pt$ of the top quark, and the invariant
mass of the $(b,t)$ pair, in the left
and right panel, respectively. For both observables, we see an
excellent agreement between the standard MC@NLO simulation and its
ML-improved counterpart. Therefore, the DT-driven machinery does
not lead to any noticeable bias on physical observables.
We also notice that the reduction of negative weights appears to
be rather evenly distributed in the two quantities we have plotted
in fig.~\ref{fig:eebbtt-ptt}.

\begin{figure}[ht]
\centering
\includegraphics[width=\linewidth]{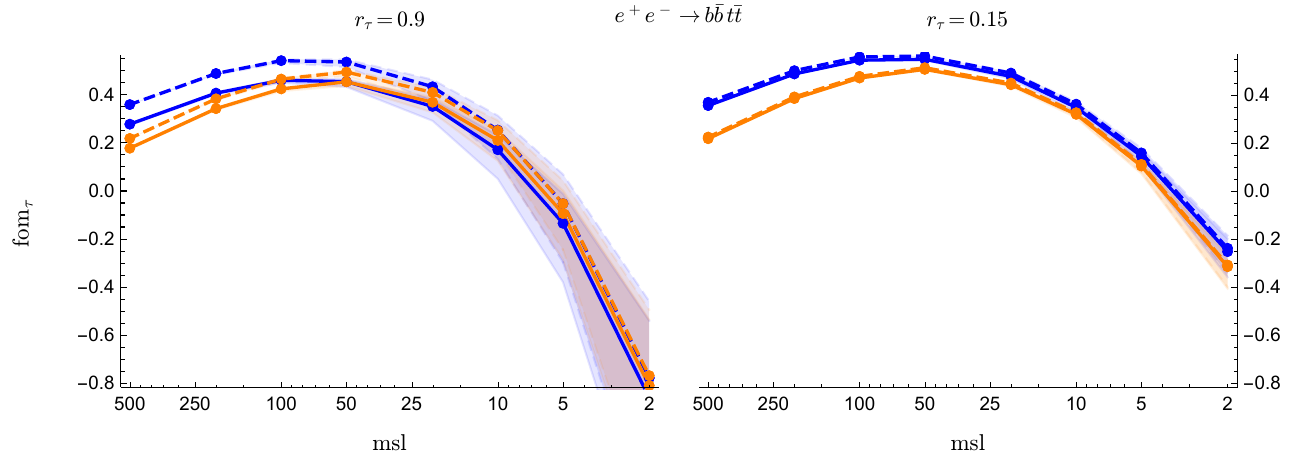}
\caption{Figure of merit for $e^+e^-\to b\bar b t\bar t$.}
\label{fig:source8}
\end{figure}
The figure of merit for this process is shown in
fig.~\ref{fig:source8}. The metric is characterized by a shape which
peaks around \texttt{min\_samples\_leaf}$\simeq$50, i.e.~for values
larger than that used in table~\ref{tab:newsummary}, a trend which is
particularly evident in the theoretical scenario, owing to the larger
relative impact of generation costs w.r.t.~the experimental case.
This preference for a specific granularity could probably be made
milder by means of larger training samples, since the process is
sufficiently complicated to require detailed training without losing
track of its global features. The latter situation is associated with
overtraining, of which we see the possible onset on the right-hand
side of the plots; this is again more pronounced in the theoretical
scenario. 

Lastly, a $(4,4,1)$ folding of the generated events reduces the
negative-weight fraction from 20.8\% to 14.3\%. Interestingly,
by applying a $(4,4,1)$ folding to events already processed by ML
yields a negative-weight fraction of 14.2\%, a result nearly
identical as before. Once again, we therefore find that folding
and Machine Learning do not provide cumulative benefits.
Furthermore, the figures above used in eq.~(\ref{eq:fom:fold3})
result in $\fom^{({\rm fold})}=-25.5$ and $\fom^{({\rm fold})}=-3.44$
for \mbox{$\frac{\tau_+}{\tau_+ +\tau_A}=0.9$} and $0.15$, respectively.
This confirms that folding is not competitive with the DT-driven 
procedure, in keeping with what has been observed for $e^+e^-\to b\bar{b}$.

\subsubsection{\texorpdfstring{$e^+e^-\to b\bar b b\bar b$}{ee4b}}
\begin{figure}[ht]
\centering
\includegraphics[width=0.7\linewidth]{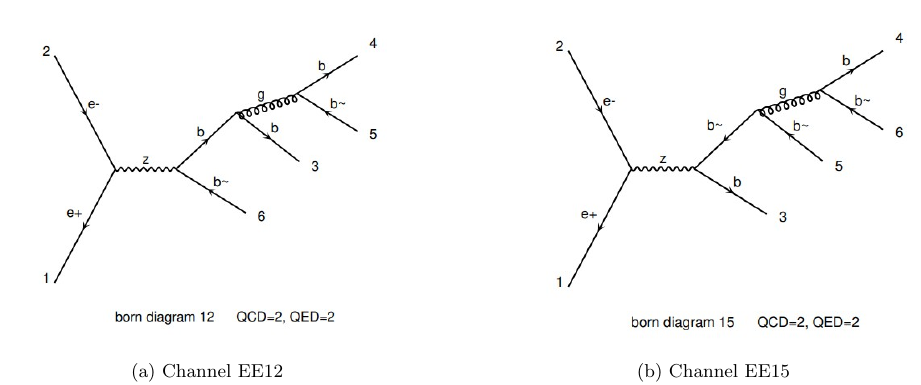}
\caption{Feynman diagrams underpinning the EE12 and EE15 channels
for $e^+e^-\to b\bar b b\bar b$, giving dominant negative contributions.}
\label{fig:source4}
\end{figure}

\begin{figure}[thb]
  \begin{center}
  \includegraphics[width=0.3\textwidth]{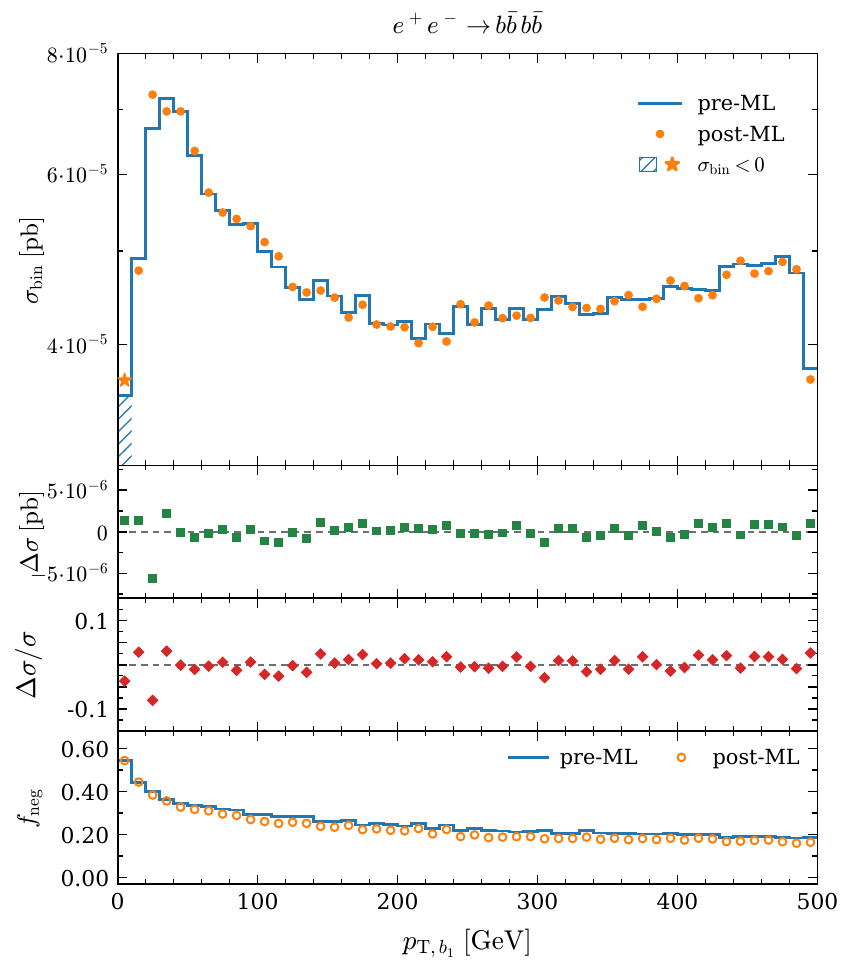}
$\phantom{a}$
  \includegraphics[width=0.3\textwidth]{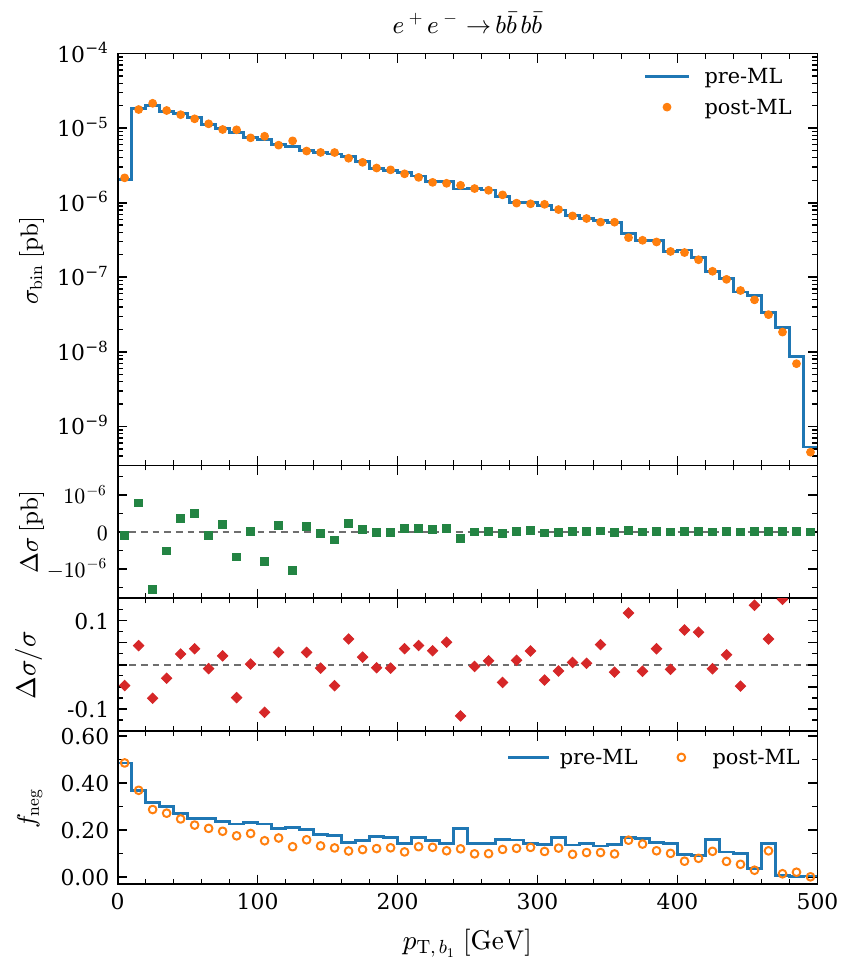}
$\phantom{a}$
  \includegraphics[width=0.3\textwidth]{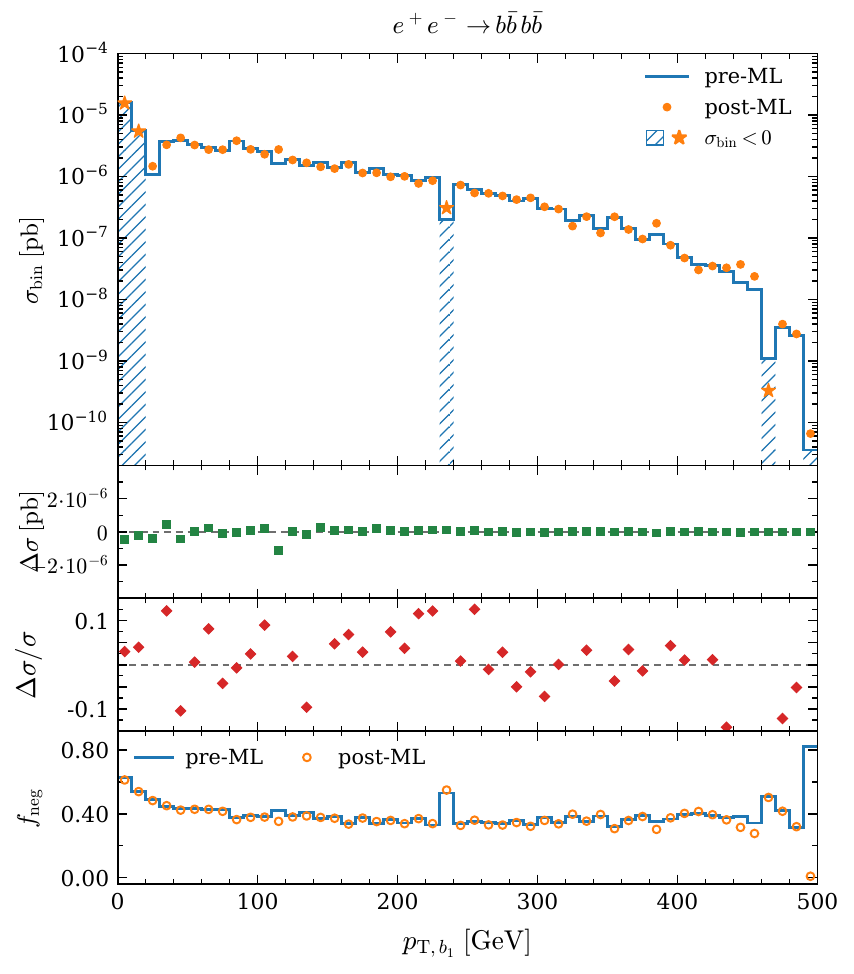}
\caption{\label{fig:eebbbbpt1}
Transverse momentum distribution of the first $b$ quark in
$e^+e^-\to b\bar{b}b\bar{b}$; total (left panel), and contributions
from the EE12 (middle panel) and EE15 (right panel) channels.
}
  \end{center}
\end{figure}

\begin{figure}[ht]
\centering
\includegraphics[width=0.45\linewidth,height=0.3\textheight]{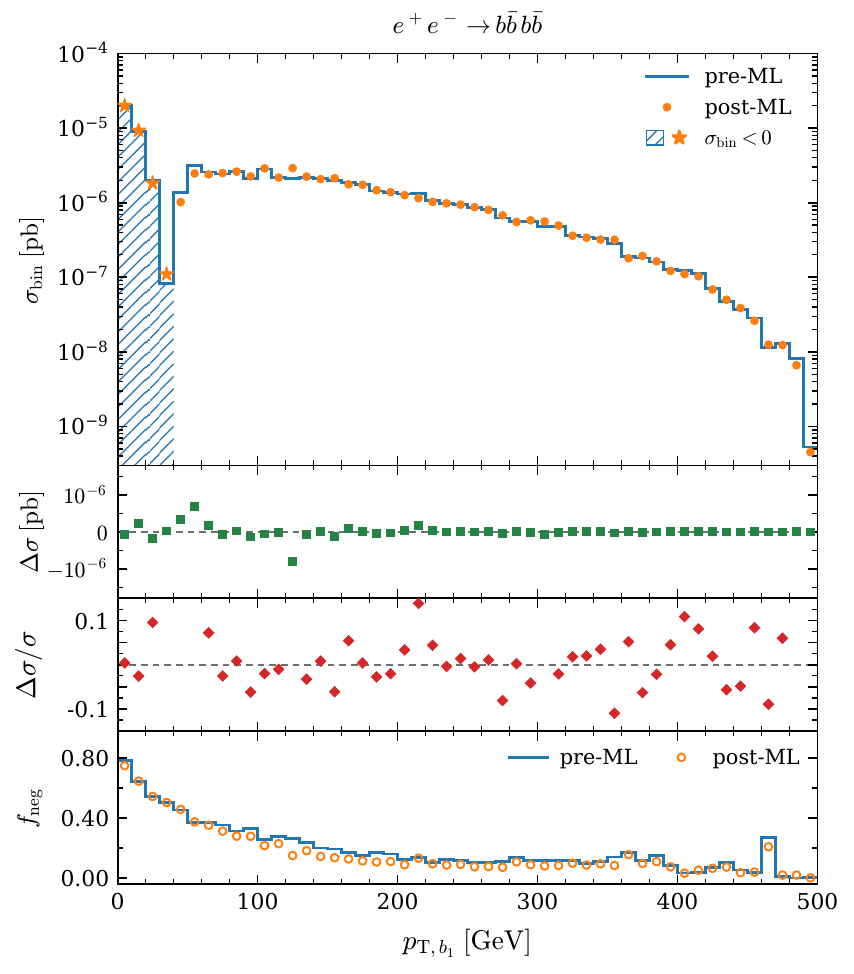}
$\phantom{a}$
\includegraphics[width=0.45\linewidth,height=0.3\textheight]{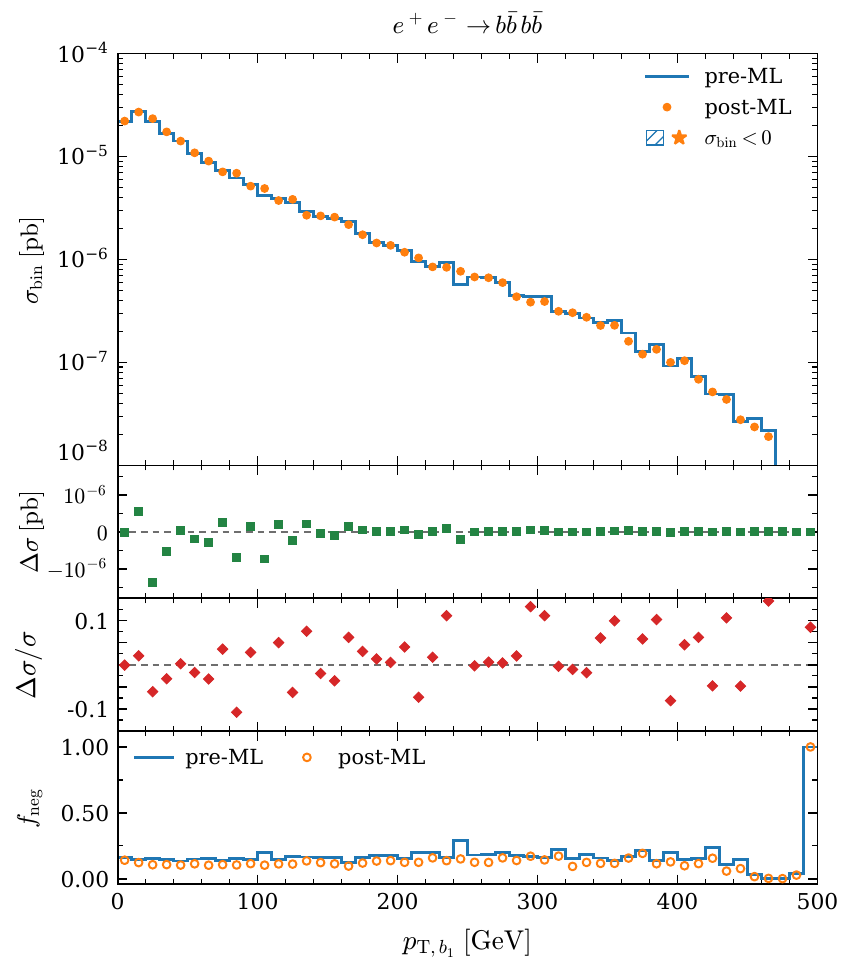}
\includegraphics[width=0.45\linewidth,height=0.3\textheight]{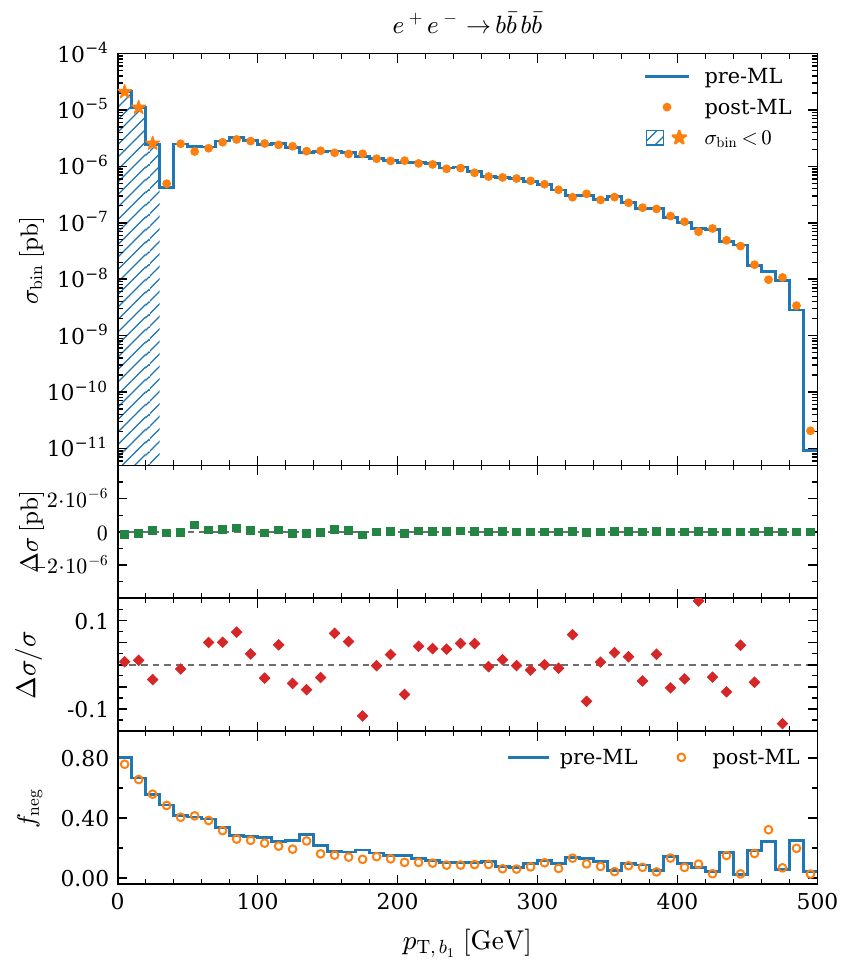}
$\phantom{a}$
\includegraphics[width=0.45\linewidth,height=0.3\textheight]{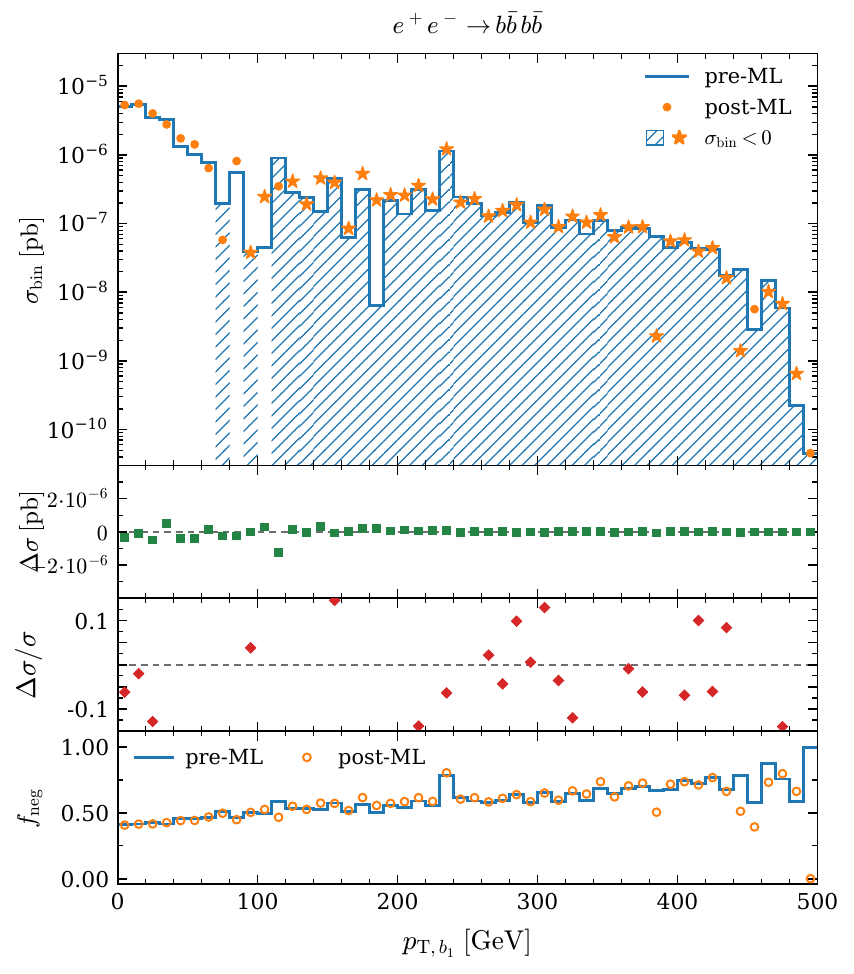}
\caption{\label{fig:eebbbbpt1FKS}
Contributions of FKS sectors 1 (left panels) and 2 (right panels) to
the transverse momentum distribution of the first $b$ quark in
$e^+e^-\to b\bar{b}b\bar{b}$, for the EE12 (top row) and EE15 (bottom
row) channels.}
\end{figure}

The total rate results for $e^+e^-\to b\bar b b\bar b$ are reported
in table~\ref{tab:source2}, and show no substantial reduction of
negative weights following the proposed DT-driven modifications.
The mechanism at play here is the same one as for
\mbox{$e^+e^-\to b\bar b t\bar t$}, namely the separate integration
and training of different FKS sectors; in addition to that, one can
show that the separation of the cross section into integration
channels driven by weights obtained by squaring individual Feynman
diagrams is also a further potential source of negative weights.
In order to document these two facts, one starts by observing that
the rate-per-channel predictions in table~\ref{tab:source2} fall
roughly into three categories, as far as negative weights are
concerned: namely, those for channels EE1--EE4 (relative small $f_0$),
EE5--EE12 (large $f_0$), and EE13-EE16 (very large $f_0$); we shall
henceforth consider only the latter two. While doing that, it turns
out that, qualitatively, it is sufficient to study one channel per
group as representative of the group itself; we have thus chosen
channels EE12 and EE15, whose definitions are underpinned by the
Feynman diagrams\footnote{By comparing these Feynman diagrams with
the results of table~\ref{tab:source2}, one may wonder why the results
they induce are vastly different from one another. The basic reason
stems again from the fact that we are working at the level of LHE,
where the symmetry e.g.~due to the two $b$ quarks being identical does
not apply -- a symmetry-breaking choice is made to parametrize the
phase space. It may be counterintuitive, then, that symmetries can
still be exploited e.g.~to reduce the number of independent FKS
sectors from 4 to 2; more details on this point can be found in
ref.~\cite{Frederix:2009yq}.} depicted in fig.~\ref{fig:source4}.
\begin{table}[ht]
\centering
\setlength{\tabcolsep}{5pt}
\renewcommand{\arraystretch}{1.10}
\fontsize{8.2}{10.2}\selectfont
\setlength{\tabcolsep}{2.3pt}
\begin{tabular}{@{}lcccc@{\hspace{9pt}}lcccc@{}}
\toprule
Channel & & $\sigma$ (pb) & $\abs{\sigma}$ (pb) & $f_0,f_1(\%)$ & Channel & & $\sigma$ (pb) & $\abs{\sigma}$ (pb) & $f_0,f_1(\%)$\\
\midrule
\multirow{2}{*}{ALL} & pre-ML & $2.287\!\cdot\!10^{-3}$ & $5.684\!\cdot\!10^{-3}$ & 29.9 & \multirow{2}{*}{EE9} & pre-ML & $9.346\!\cdot\!10^{-5}$ & $2.341\!\cdot\!10^{-4}$ & 30.0\\
 & post-ML & $2.284\!\cdot\!10^{-3}$ & $5.275\!\cdot\!10^{-3}$ & 28.3 &  & post-ML & $9.369\!\cdot\!10^{-5}$ & $2.204\!\cdot\!10^{-4}$ & 28.7\\
\midrule[0.1pt]
\multirow{2}{*}{EE1} & pre-ML & $1.767\!\cdot\!10^{-4}$ & $2.599\!\cdot\!10^{-4}$ & 16.0 & \multirow{2}{*}{EE10} & pre-ML & $9.764\!\cdot\!10^{-5}$ & $2.340\!\cdot\!10^{-4}$ & 29.1\\
 & post-ML & $1.756\!\cdot\!10^{-4}$ & $2.364\!\cdot\!10^{-4}$ & 12.9 &  & post-ML & $9.745\!\cdot\!10^{-5}$ & $2.160\!\cdot\!10^{-4}$ & 27.4\\
\midrule[0.1pt]
\multirow{2}{*}{EE2} & pre-ML & $1.735\!\cdot\!10^{-4}$ & $2.552\!\cdot\!10^{-4}$ & 16.0 & \multirow{2}{*}{EE11} & pre-ML & $1.842\!\cdot\!10^{-4}$ & $4.777\!\cdot\!10^{-4}$ & 30.7\\
 & post-ML & $1.721\!\cdot\!10^{-4}$ & $2.293\!\cdot\!10^{-4}$ & 12.5 &  & post-ML & $1.830\!\cdot\!10^{-4}$ & $4.468\!\cdot\!10^{-4}$ & 29.5\\
\midrule[0.1pt]
\multirow{2}{*}{EE3} & pre-ML & $3.541\!\cdot\!10^{-4}$ & $5.168\!\cdot\!10^{-4}$ & 15.7 & \multirow{2}{*}{EE12} & pre-ML & $1.928\!\cdot\!10^{-4}$ & $4.725\!\cdot\!10^{-4}$ & 29.6\\
 & post-ML & $3.546\!\cdot\!10^{-4}$ & $4.684\!\cdot\!10^{-4}$ & 12.1 &  & post-ML & $1.953\!\cdot\!10^{-4}$ & $4.362\!\cdot\!10^{-4}$ & 27.6\\
\midrule[0.1pt]
\multirow{2}{*}{EE4} & pre-ML & $3.520\!\cdot\!10^{-4}$ & $5.122\!\cdot\!10^{-4}$ & 15.6 & \multirow{2}{*}{EE13} & pre-ML & $1.531\!\cdot\!10^{-5}$ & $2.153\!\cdot\!10^{-4}$ & 46.4\\
 & post-ML & $3.516\!\cdot\!10^{-4}$ & $4.605\!\cdot\!10^{-4}$ & 11.8 &  & post-ML & $1.472\!\cdot\!10^{-5}$ & $2.040\!\cdot\!10^{-4}$ & 46.4\\
\midrule[0.1pt]
\multirow{2}{*}{EE5} & pre-ML & $9.819\!\cdot\!10^{-5}$ & $2.378\!\cdot\!10^{-4}$ & 29.4 & \multirow{2}{*}{EE14} & pre-ML & $1.875\!\cdot\!10^{-5}$ & $2.188\!\cdot\!10^{-4}$ & 45.7\\
 & post-ML & $9.686\!\cdot\!10^{-5}$ & $2.218\!\cdot\!10^{-4}$ & 28.2 &  & post-ML & $1.848\!\cdot\!10^{-5}$ & $2.053\!\cdot\!10^{-4}$ & 45.5\\
\midrule[0.1pt]
\multirow{2}{*}{EE6} & pre-ML & $9.223\!\cdot\!10^{-5}$ & $2.387\!\cdot\!10^{-4}$ & 30.7 & \multirow{2}{*}{EE15} & pre-ML & $2.616\!\cdot\!10^{-5}$ & $4.335\!\cdot\!10^{-4}$ & 47.0\\
 & post-ML & $9.232\!\cdot\!10^{-5}$ & $2.230\!\cdot\!10^{-4}$ & 29.3 &  & post-ML & $2.700\!\cdot\!10^{-5}$ & $4.182\!\cdot\!10^{-4}$ & 46.8\\
\midrule[0.1pt]
\multirow{2}{*}{EE7} & pre-ML & $1.938\!\cdot\!10^{-4}$ & $4.732\!\cdot\!10^{-4}$ & 29.5 & \multirow{2}{*}{EE16} & pre-ML & $2.974\!\cdot\!10^{-5}$ & $4.337\!\cdot\!10^{-4}$ & 46.6\\
 & post-ML & $1.922\!\cdot\!10^{-4}$ & $4.385\!\cdot\!10^{-4}$ & 28.1 &  & post-ML & $3.011\!\cdot\!10^{-5}$ & $4.097\!\cdot\!10^{-4}$ & 46.3\\
\midrule[0.1pt]
\multirow{2}{*}{EE8} & pre-ML & $1.888\!\cdot\!10^{-4}$ & $4.702\!\cdot\!10^{-4}$ & 29.9 &  &  &  &  & \\
 & post-ML & $1.895\!\cdot\!10^{-4}$ & $4.407\!\cdot\!10^{-4}$ & 28.5 &  &  &  &  & \\
\bottomrule
\end{tabular}
\caption{Total rates per channel in $e^+e^-\to b\bar b b\bar b$.}
\label{tab:source2}
\end{table}
In fig.~\ref{fig:eebbbbpt1} we show the momentum of the
first (see footnote~\ref{ft:ordering} for an explanation of ``first'')
$b$ quark of the event. The left panel presents the total distribution,
obtained by summing all channels; there, we see the irreducible
(but moderate in size) presence of negative contributions (in the
leftmost bin). Having said that, the first two insets confirm the
absence of biases due to DT manipulations. In the middle and right
panels of the figure we present the contribution to the $b$ $\pt$
due to the EE12 (middle panel) and EE15 (right panel) channels. We see that
there are negative-weight regions due to the EE15 channel that will eventually
be compensated by positive weights stemming from the EE12 channel in the total
distribution, which implies that an event generation which treats
channels separately is responsible for a reducible (but not reduced)
fraction of negative weights. Reassuringly, also for the individual-channel
results there are no DT-induced biases.
The comparison of the two rightmost plots in fig.~\ref{fig:eebbbbpt1}
with those in fig.~\ref{fig:eebbbbpt1FKS} gives a message similar
to that emerging from our previous study of $e^+e^-\to b\bar{b}t\bar{t}$;
namely, that the separation of a given channel into FKS sectors
increases the number of negative weights, which could be avoided
by summing the FKS-sector contributions before unweighting.
More in detail, the two top-row plots of fig.~\ref{fig:eebbbbpt1FKS}
show the contributions of FKS sectors 1 and 2 to channel EE12:
the negative weights at small $\pt$ in the left-panel figure are
compensated by the other FKS-sector contributions, since the total
(middle panel of fig.~\ref{fig:eebbbbpt1}) is positive there.
Conversely, the two bottom-row plots of fig.~\ref{fig:eebbbbpt1FKS}
are the contributions to channel EE15 due to FKS sectors 1 and 2.
These exhibit negative weights at the opposite end of the spectrum,
which are eventually compensated, both mutually and by the other two
FKS sectors (not shown), so that the total (right panel of
fig.~\ref{fig:eebbbbpt1}) is positive almost everywhere.
Once again, while the statistics is poor, we can be confident that
none of this individual-sector results is biased by our DT approach.
As a final comment on this matter, we point out that while summing
$\clS$-event contributions over FKS sectors is entirely feasible,
summing over channels appears to be much more complicated, in that
it could create numerical instabilities owing to the potential inability
of the integrator to figure out the peak structure of the cross section.

\begin{figure}[ht]
\centering
\includegraphics[width=\linewidth]{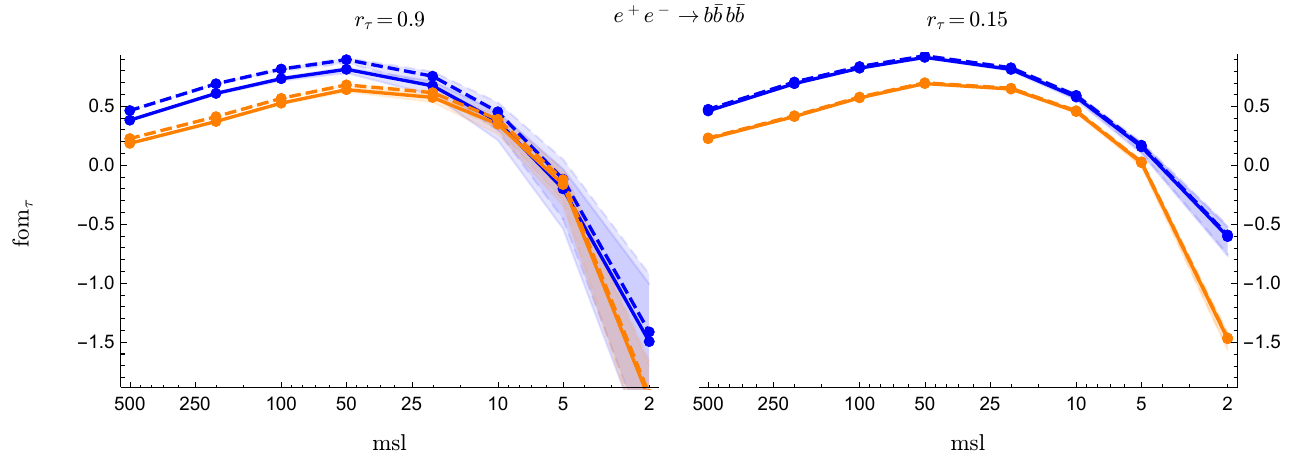}
\caption{Figure of merit for $e^+e^-\to b\bar b b\bar b$.}
\label{fig:eebbbbscan}
\end{figure}
Finally, the figure of merit for $e^+e^-\to b\bar b b\bar b$ is
presented in fig.~\ref{fig:eebbbbscan}. The shapes of the curves have
visible similarities with those relevant to \mbox{$e^+e^-\to b\bar b t\bar t$},
but they are flatter than those for large \texttt{min\_samples\_leaf}
values (left-hand side of the plots). While the final state here is
kinematically simpler w.r.t.~\mbox{$b\bar b t\bar t$} (since all
masses are identical), this process has been trained with many more
events w.r.t.~\mbox{$e^+e^-\to b\bar b t\bar t$}; as we argued in the
discussion of the latter process, this could be helpful to give one
a larger leeway in the choice of the granularity.

\subsubsection{\texorpdfstring{$e^+e^-\to Hb\bar b$}{eeHbb}\label{sec:eeHbb}}
\begin{figure}[ht]
\centering
\includegraphics[width=0.4\linewidth]{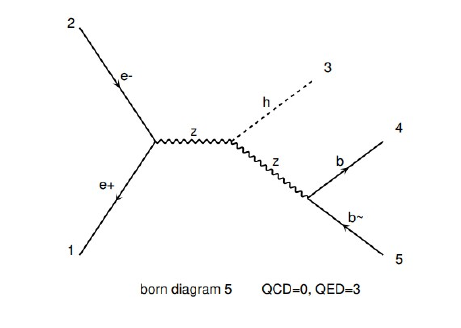}
\caption{Feynman diagram for channel EE5 in $e^+e^-\to Hb\bar b$.}
\label{fig:source13}
\end{figure}

\begin{figure}[ht]
\centering
\includegraphics[width=0.45\linewidth]{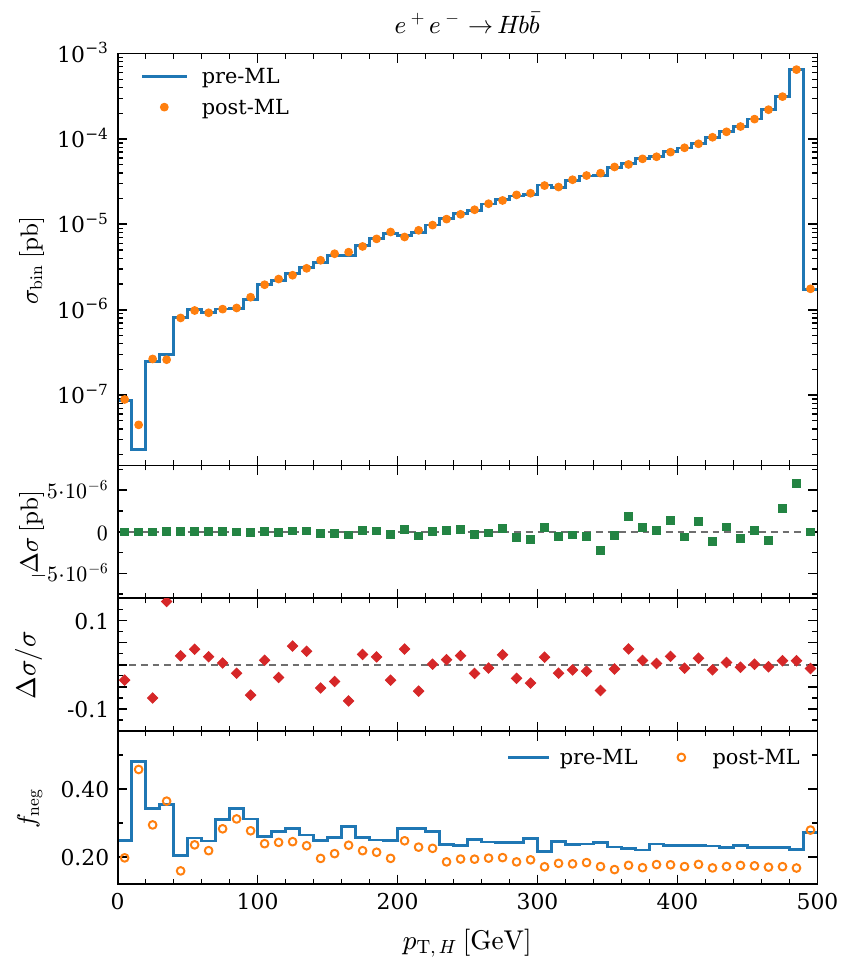}
$\phantom{a}$
\includegraphics[width=0.45\linewidth]{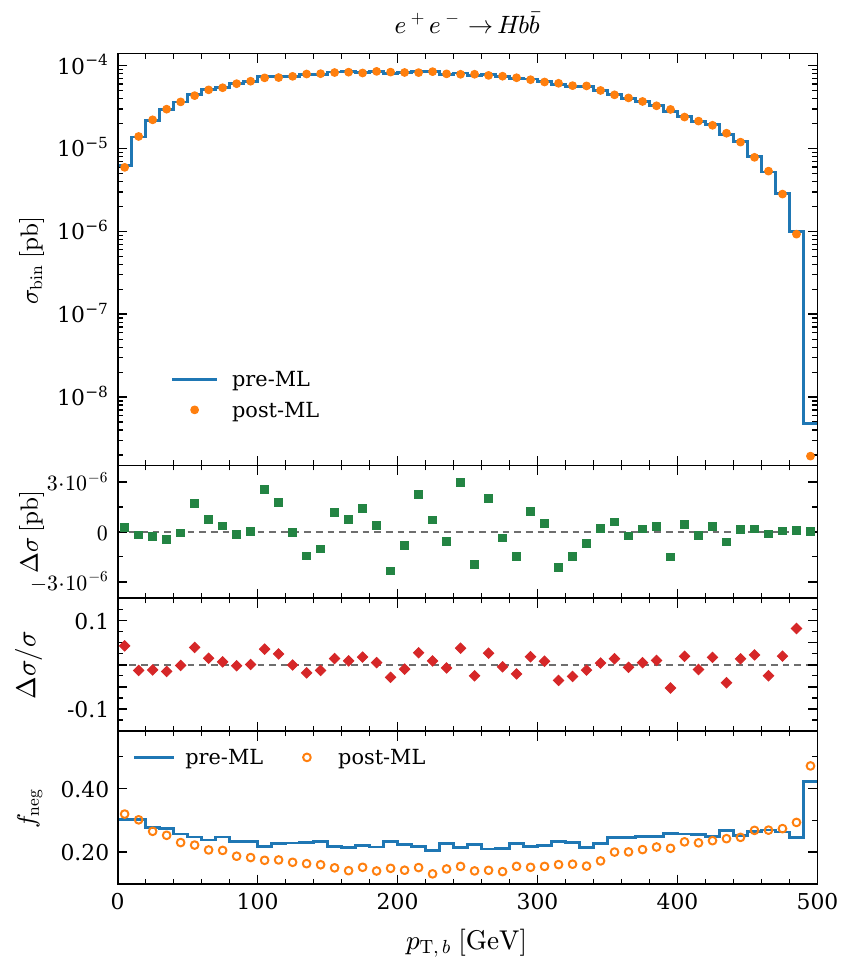}
\caption{Transverse momentum of the Higgs (left panel) and $b$ quark
(right panel) in $e^+e^-\to Hb\bar b$.}
\label{fig:eeHbbpts}
\end{figure}

Table~\ref{tab:source6} shows an overall reduction of the fraction
of negative weights from 23\% to 17.4\% for the $e^+e^-\to Hb\bar b$
process. The vastly dominant EE5 channel, where a $Z$ boson splits into a
$H$ boson and another $Z$, which further branches into a $b\bar{b}$
pair (see fig.~\ref{fig:source13}), has been analysed in detail with
increased statistics, without leading to any further visible improvements.
Interestingly, at variance with previous cases, no negative-cross-section
regions have been found in single-inclusive Born variables. This can
be seen e.g.~in fig.~\ref{fig:eeHbbpts}, where we plot the transverse
momenta of the Higgs (left panel) and of the $b$ quark, which are
positive-definite everywhere. In addition to that, we stress once
again the absence of biases due to the DT-induced modifications to
the cross section.
\begin{table}[ht]
\centering
\setlength{\tabcolsep}{5pt}
\renewcommand{\arraystretch}{1.10}
\small
\begin{tabular}{@{}lcccc@{}}
\toprule
Channel & & $\sigma$ (pb) & $\abs{\sigma}$ (pb) & $f_0,f_1(\%)$\\
\midrule
\multirow{2}{*}{ALL} & pre-ML & $2.540\!\cdot\!10^{-3}$ & $4.709\!\cdot\!10^{-3}$ & 23.0\\
 & post-ML & $2.535\!\cdot\!10^{-3}$ & $3.889\!\cdot\!10^{-3}$ & 17.4\\
\midrule[0.1pt]
\multirow{2}{*}{EE1} & pre-ML & $5.825\!\cdot\!10^{-8}$ & $1.157\!\cdot\!10^{-7}$ & 24.8\\
 & post-ML & $5.846\!\cdot\!10^{-8}$ & $8.193\!\cdot\!10^{-8}$ & 14.3\\
\midrule[0.1pt]
\multirow{2}{*}{EE2} & pre-ML & $1.059\!\cdot\!10^{-7}$ & $2.111\!\cdot\!10^{-7}$ & 24.9\\
 & post-ML & $1.060\!\cdot\!10^{-7}$ & $1.489\!\cdot\!10^{-7}$ & 14.4\\
\midrule[0.1pt]
\multirow{2}{*}{EE3} & pre-ML & $5.833\!\cdot\!10^{-8}$ & $1.153\!\cdot\!10^{-7}$ & 24.7\\
 & post-ML & $5.853\!\cdot\!10^{-8}$ & $8.146\!\cdot\!10^{-8}$ & 14.1\\
\midrule[0.1pt]
\multirow{2}{*}{EE4} & pre-ML & $1.052\!\cdot\!10^{-7}$ & $2.090\!\cdot\!10^{-7}$ & 24.8\\
 & post-ML & $1.058\!\cdot\!10^{-7}$ & $1.474\!\cdot\!10^{-7}$ & 14.1\\
\midrule[0.1pt]
\multirow{2}{*}{EE5} & pre-ML & $2.540\!\cdot\!10^{-3}$ & $4.708\!\cdot\!10^{-3}$ & 23.0\\
 & post-ML & $2.535\!\cdot\!10^{-3}$ & $3.888\!\cdot\!10^{-3}$ & 17.4\\
\bottomrule
\end{tabular}
\caption{Total rates per channel in $e^+e^-\to Hb\bar b$.}
\label{tab:source6}
\end{table}

We have also observed that, while by increasing the granularity
of the DTs employed one generally lowers the fraction of negative weights
in a significant way, the same happens with the test set, but to a
much lower extent. This may point to an insufficient understanding
of the cross section structure owing to lack of statistics in tails
(i.e.~few-event regions) of distributions. It is interesting to
observe that the NF-driven results for this process, to be presented
later, are always self-consistent, and show a more marked reduction
of negative weights -- see sect.~\ref{sec:epemNFres}.

\begin{figure}[ht]
\centering
\includegraphics[width=\linewidth]{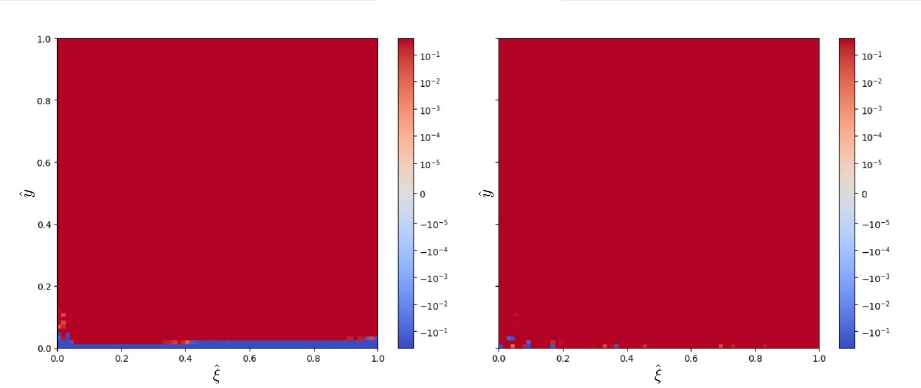}
\caption{Distribution of event weights in the $\langle\hat{\xi},\hat{y}\rangle$
plane for channel EE5 in $e^+e^-\to Hb\bar b$, pre- (left panel) and
post-DT (right panel).}
\label{fig:source14}
\end{figure}

\begin{figure}[ht]
\centering
\includegraphics[width=\linewidth]{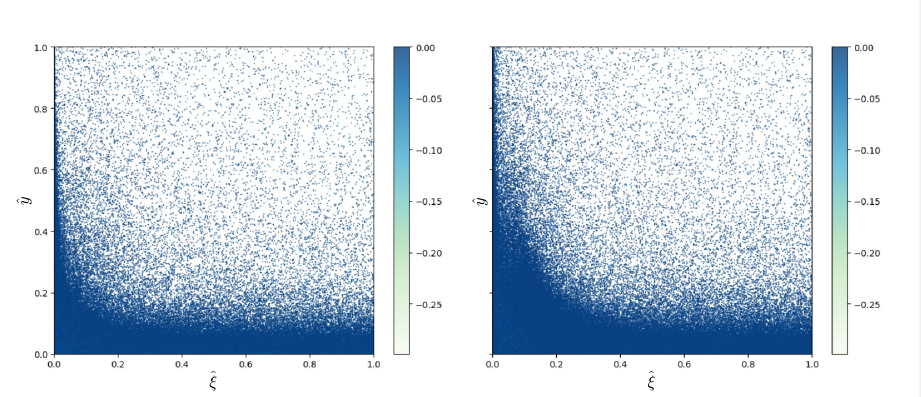}
\caption{As in fig.~\ref{fig:source14}, by considering only negative weights.}
\label{fig:source15}
\end{figure}
Leaving NF results aside for the moment, the persistent presence of
negative weights in spite of the positivity of single-inclusive
Born-level distributions (i.e.~a situation
not encountered in the processes analysed so far), has led us to
check that the $\hat{\xi}$ and $\hat{y}$ distributions are
flattened by the DT approach, which indeed is the case. Then, a deeper
understanding can be obtained by looking at correlations; in
particular, we consider the distribution of weights in the
\mbox{$\langle\hat{\xi},\hat{y}\rangle$} plane, before and
after DT-induced modifications (left and right panel, respectively,
of fig.~\ref{fig:source14}). 
The post-ML distribution shows only a marginal residual presence
of the negative weights found in the pre-ML case, which tend to
aggregate towards $\hat{y}=0$; it is clear why in the projections
onto the two axes those residual negative weights are not visible.
When the results of fig.~\ref{fig:source14} are restricted to
including only negative-event weights, one sees why those events
are barely visible overall (fig.~\ref{fig:source15}) -- they are
still numerous, but with fairly small absolute weights\footnote{Note
the colour code in the figure: larger blue regions are associated with
negative weights with {\em smaller absolute} values.}. Their
distribution suggests that an alternative possibility would be
that of learning $\log\hat{y}$ rather than $\hat{y}$. Independently
of this, these results may be the evidence of how the piecewise-continuity of
the regressor found by means of DT (see appendix~\ref{app:bdt-granularity})
prevents one from maximizing the negative-weight reduction potential
of the physical case. This interpretation is consistent with the NF
approach (which is mostly continuous and differentiable; a more precise
qualification can be found in appendix~\ref{app:nf-regularity}) resulting
in a smaller number of negative weights.

\begin{figure}[ht]
\centering
\includegraphics[width=0.45\linewidth]{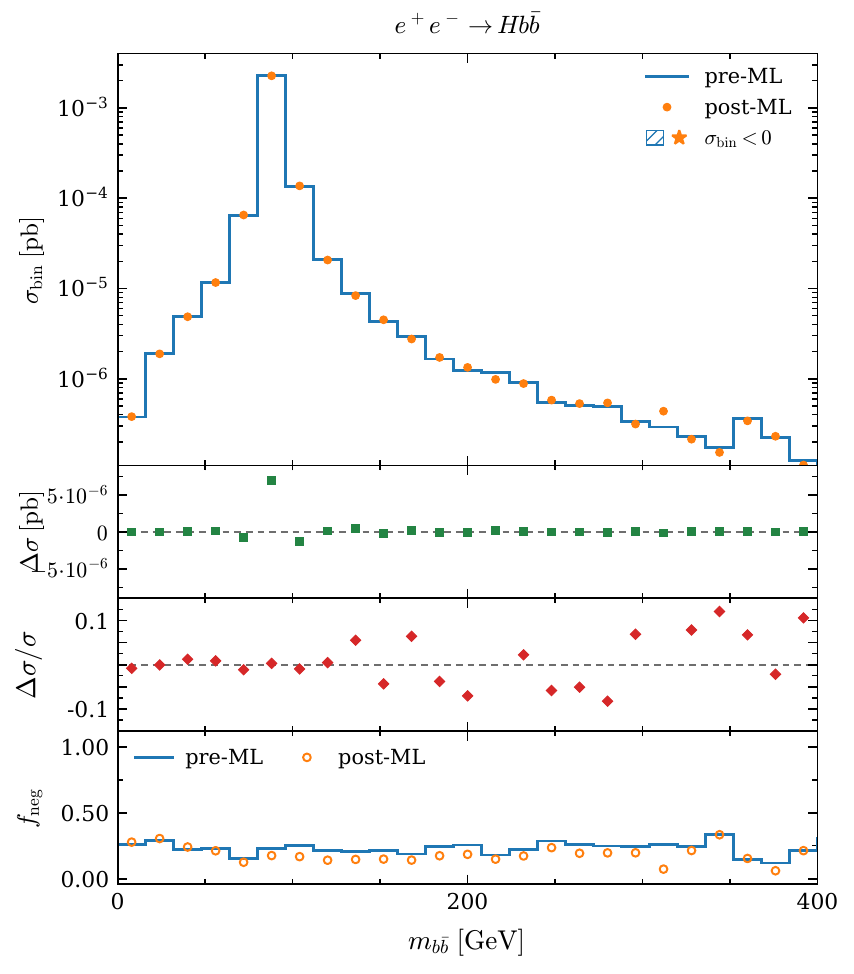}
$\phantom{a}$
\includegraphics[width=0.45\linewidth]{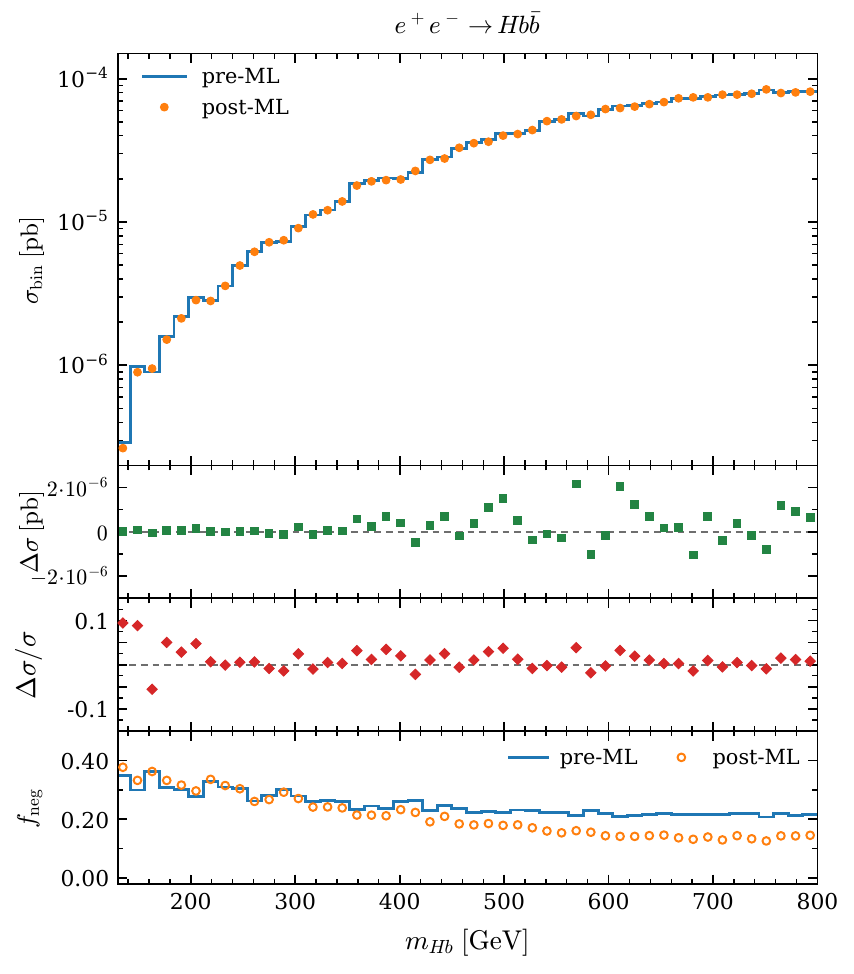}
\caption{Invariant mass of the $b\bar{b}$ pair (left panel) and
of the $Hb$ pair (right panel) in $e^+e^-\to Hb\bar b$.}
\label{fig:eeHbbcorr}
\end{figure}
The analysis of other kinematical variables, including correlations,
leads to the same conclusions as before: namely, generally positive-definite
distributions, unbiased by ML-driven manipulations. An example of
correlations is given in fig.~\ref{fig:eeHbbcorr}, where we show the
invariant mass of the $b\bar{b}$ pair and of the $Hb$ pair.
The absence of biases is manifest. In the tail of the $b\bar{b}$-pair
invariant mass, there are negative bins (not shown), which however appear as
being due to insufficient statistics rather than to a genuine region
of non-null measure where the cross section is negative.

\begin{figure}[ht]
\centering
\includegraphics[width=\linewidth]{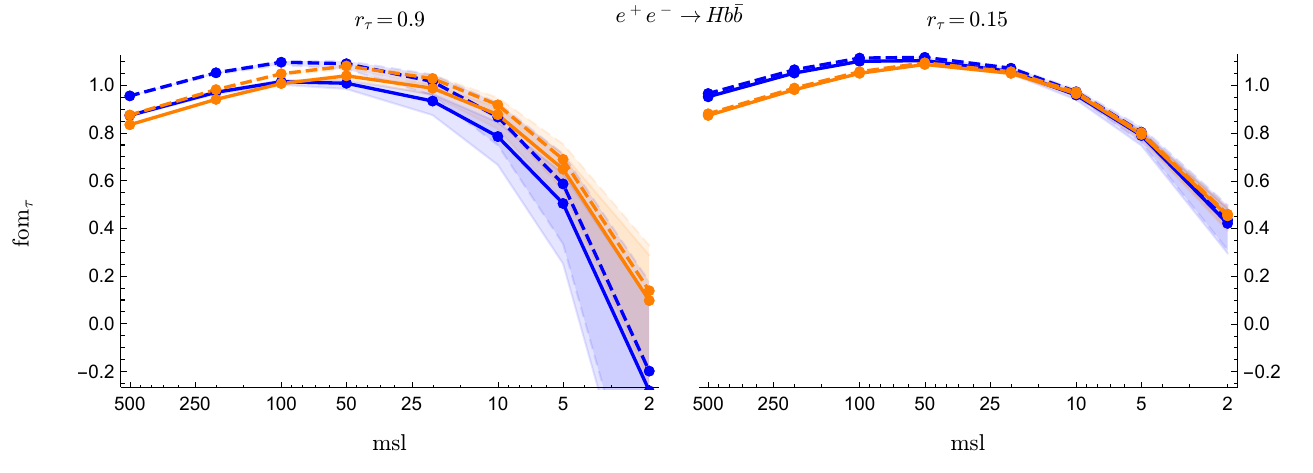}
\caption{Figure of merit for $e^+e^-\to Hb\bar b$.}
\label{fig:source16}
\end{figure}

Finally, the figure of merit for $e^+e^-\to Hb\bar b$ is shown in
fig.~\ref{fig:source16}. In general, one observes that the best
results are obtained on average with coarser granularities w.r.t.~that
relevant to table~\ref{tab:newsummary}. Also, the results obtained
with the larger dataset (2M events) are not clearly better than
those emerging from training on the smaller dataset (1M events).
Thus, in this case the number of training events is likely sufficient
to capture the complexity of the process; in that, it certainly helps
that the Higgs is colour-neutral, which simplifies radiation patterns.
Regardless of the sizes of the datasets, one sees a marked decrease
of the f.o.m.~values moving towards finer granularities, again
possibly signalling the effect of overtraining, as well as the increased
costs of finer training on a complicated process.

\subsubsection{\texorpdfstring{$e^+e^-\to Ht\bar t$}{eeHtt}}
The negative weight fraction for the $e^+e^-\to Ht\bar t$ process is
approximately equal to 0.1\%, with values ranging between 0.1\% and 0.3\%
across all channels. Given this already minimal baseline, no further
reduction was achieved by employing the DT-driven approach.

\subsubsection{\texorpdfstring{$pp\to b\bar b$}{ppbb}}
Hadronic-collisions processes differ from $e^+e^-$ ones owing to,
among other things, the integration over Bjorken  $x$'s which
enter the PDFs. Related to this is the fact that the flavour
structure of real-emission processes is richer than its Born-level
counterpart. Adopting the picture underpinning the backward evolution
in parton shower Monte Carlos, a Born-level initial-state gluon may
branch backwards into either a gluon-gluon pair or a $q\bar{q}$ pair;
FKS-wise, this corresponds to having three FKS sectors (FKS is flavour
neutral) associated with the same kinematics. Conversely, in the
forward-looking FKS approach, these stem from the three initial-state
branchings $g\to gg$, $q\to qg$, and $\bar{q}\to \bar{q}g$, which
crucially are kinematically identical to one another. From what we
have learned, such three FKS sectors will conveniently be learned
simultaneously in our ML-based approaches. In addition to that,
\begin{figure}[ht]
\centering
\includegraphics[width=0.96\linewidth]{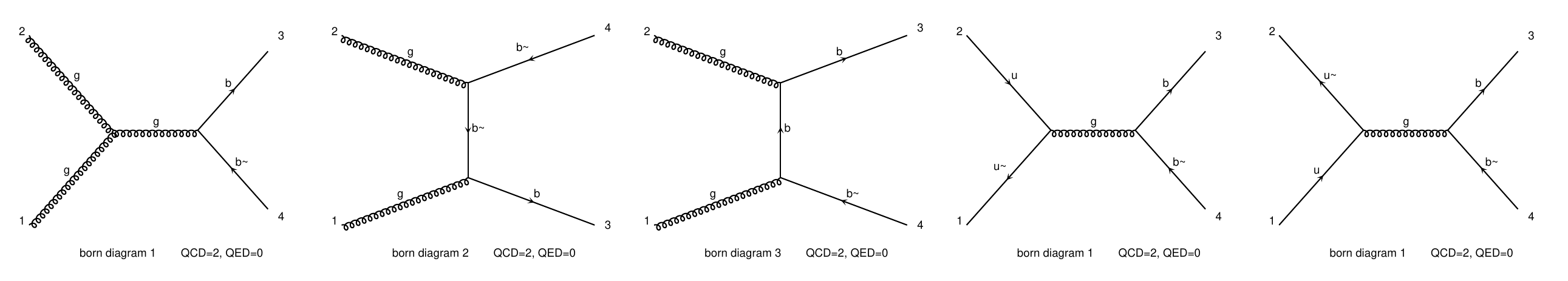}
\caption{Feynman diagrams underpinning the five channels of $pp\to b\bar b$.}
\label{fig:source17}
\end{figure}
for the specific case of $pp\to b\bar{b}$, there are five channels, three of
which stem from the $gg$ initial state, while the remaining two are relevant
to $q\bar{q}$ and $\bar{q}q$ initial states (see fig.~\ref{fig:source17}).
As has been done before, we shall consider both individual channels
and individual channel+FKS sector(s) for training the regressor.
The latter are defined according to the following conventions: FKS sectors
1, 5, and 6 (2, 7, and 8) correspond to configurations in which the extra
final-state parton becomes collinear to the parton incoming from the
left (from the right). Finally, FKS sectors 3 and 4 stems from configurations
in which the emitted gluon becomes collinear to the $b$ and $\bar{b}$,
respectively. 

\begin{table}[ht]
\centering
\setlength{\tabcolsep}{5pt}
\renewcommand{\arraystretch}{1.10}
\small
\begin{tabular}{@{}lcccc@{}}
\toprule
Channel & & $\sigma$ (pb) & $\abs{\sigma}$ (pb) & $f_0,f_1(\%)$\\
\midrule
\multirow{3}{*}{ALL} & pre-ML & $4.379\!\cdot\!10^{8}$ & $6.706\!\cdot\!10^{8}$ & 17.4\\
 & post-ML & $4.344\!\cdot\!10^{8}$ & $5.429\!\cdot\!10^{8}$ & 10.0\\
 & post-ML$^\star$ & $4.344\!\cdot\!10^{8}$ & $5.967\!\cdot\!10^{8}$ & 13.6\\
\midrule[0.1pt]
\multirow{3}{*}{GG1} & pre-ML & $3.290\!\cdot\!10^{7}$ & $5.058\!\cdot\!10^{7}$ & 17.5\\
 & post-ML & $3.257\!\cdot\!10^{7}$ & $3.927\!\cdot\!10^{7}$ & 8.5\\
 & post-ML$^\star$ & $3.303\!\cdot\!10^{7}$ & $4.229\!\cdot\!10^{7}$ & 10.9\\
\midrule[0.1pt]
\multirow{3}{*}{GG2} & pre-ML & $1.979\!\cdot\!10^{8}$ & $3.040\!\cdot\!10^{8}$ & 17.5\\
 & post-ML & $1.966\!\cdot\!10^{8}$ & $2.466\!\cdot\!10^{8}$ & 10.1\\
 & post-ML$^\star$ & $1.957\!\cdot\!10^{8}$ & $2.722\!\cdot\!10^{8}$ & 14.1\\
\midrule[0.1pt]
\multirow{3}{*}{GG3} & pre-ML & $1.992\!\cdot\!10^{8}$ & $3.069\!\cdot\!10^{8}$ & 17.5\\
 & post-ML & $1.972\!\cdot\!10^{8}$ & $2.485\!\cdot\!10^{8}$ & 10.3\\
 & post-ML$^\star$ & $1.977\!\cdot\!10^{8}$ & $2.736\!\cdot\!10^{8}$ & 13.9\\
\midrule[0.1pt]
\multirow{3}{*}{UXU} & pre-ML & $3.983\!\cdot\!10^{6}$ & $4.560\!\cdot\!10^{6}$ & 6.3\\
 & post-ML & $3.981\!\cdot\!10^{6}$ & $4.260\!\cdot\!10^{6}$ & 3.3\\
 & post-ML$^\star$ & $3.984\!\cdot\!10^{6}$ & $4.299\!\cdot\!10^{6}$ & 3.7\\
\midrule[0.1pt]
\multirow{3}{*}{UUX} & pre-ML & $3.981\!\cdot\!10^{6}$ & $4.542\!\cdot\!10^{6}$ & 6.2\\
 & post-ML & $3.977\!\cdot\!10^{6}$ & $4.243\!\cdot\!10^{6}$ & 3.1\\
 & post-ML$^\star$ & $3.981\!\cdot\!10^{6}$ & $4.284\!\cdot\!10^{6}$ & 3.5\\
\bottomrule
\end{tabular}
\caption{Total rates per channel in $pp\to b\bar b$. A star denotes
results obtained with low-statistics training.}
\label{tab:source7}
\end{table}

In table~\ref{tab:source7} we present the pre- and post-DT results
for individual channels and total cross sections. The
former are underpinned by training on FKS sectors some of which are
taken together -- specifically the $(156)$ and $(278)$ sets are considered,
as well as $3$ and $4$ individually. In this setup, the total fraction of
negative weights goes from 17.4\% to 10.0\%. We stress that this happens
by training over ${\cal O}(1{\rm M})$ events. That is a significant
increase w.r.t.~the size of the samples employed in $e^+e^-\to b\bar{b}$
(by a factor $3.5$ for the largest samples in either process), which
is however seemingly important. Indeed, training on samples of sizes
similar to those used in $e^+e^-$ leads to the post-ML results in
table~\ref{tab:source7} marked with an asterisk, i.e.~an overall
$f_1=13.6\%$, which is noticeably worse than the 10\% quoted above.
This is the first evidence of the rather general observation that
hadron-collision processes need more training w.r.t.~their $e^+e^-$
counterparts with the same final states; at the most trivial
level, this is easy to interpret as the consequence of the two
extra integration variables present in the hadronic case (the
Bjorken $x$'s), which render the learning task much more involved.

\begin{figure}[ht]
\centering
\includegraphics[width=\linewidth]{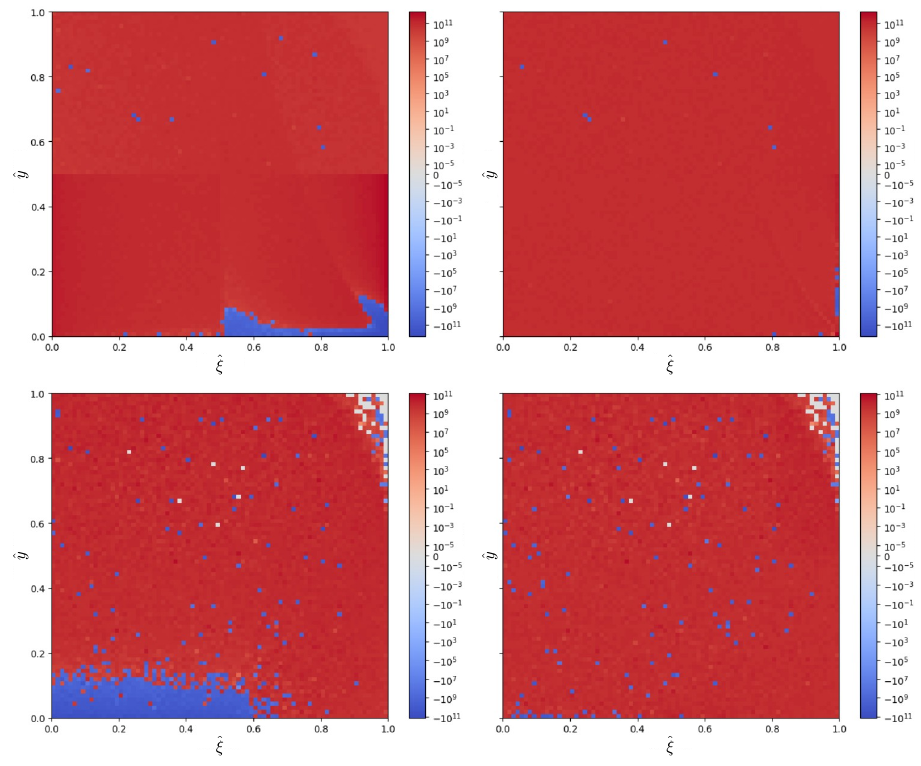}
\caption{Distribution of event weights in the $\langle\hat{\xi},\hat{y}\rangle$
plane for channel GG3 in $pp\to b\bar b$, pre- (left panels) and
post-DT (right panels). The plots in the top and bottom row are obtained
by combining all FKS sectors and by considering only FKS sector 4,
respectively.}
\label{fig:source18}
\end{figure}

Given the analyses of the processes considered thus far, we can assume
that $f_1=10.0\%$ may stem from sources both further reducible with
the DT approach proposed here and irreducible ones (although these
may potentially be addressed with other methods). In order to
investigate the former, we have started directly from plotting the
weight distributions in the $\langle\hat{\xi},\hat{y}\rangle$ plane
in fig.~\ref{fig:source18}; we have chosen channel GG3 just to be
definite. The two panels in the upper row are obtained by training
with all of the FKS sectors taken together, while in the lower row
the individually-learned FKS sector 4 (as a representative for the
its group) has been chosen. The panels on the left (right) side
present the pre-ML (post-ML) distributions.
As has been already observed in several cases prior, while the majority
of negative weights (clustered around the $\hat{y}=0$ axis) disappear with
either training set, the individually-trained FKS sector 4 is responsible
for a sparse negative-weight distribution, which may have different
explanations (compatible with one another): learning anomalies,
insufficient training statistics, or a consequence of the coarse
nature of the final DT regressor.

Within the separate-FKS-sector computations, we have re-generated
the events by requiring the actual virtual matrix elements (as opposed
to their approximation) be generated for every single phase-space point.
By doing so we have observed a general reduction of negative weights,
in keeping with our previous findings, and in plots such as that
in the bottom left corner of fig.~\ref{fig:source18} the negative-weight
distribution becomes less coarse. As a consequence of this, by applying
the DT-driven approach on a training set generated (with high statistics)
in this way, we have obtained a fraction of negative weights equal to $7.6\%$,
i.e.~a much better result w.r.t.~that obtained by training over events
where the actual virtuals are computed every now and then.

\begin{figure}[ht]
\centering
\includegraphics[width=0.45\linewidth]{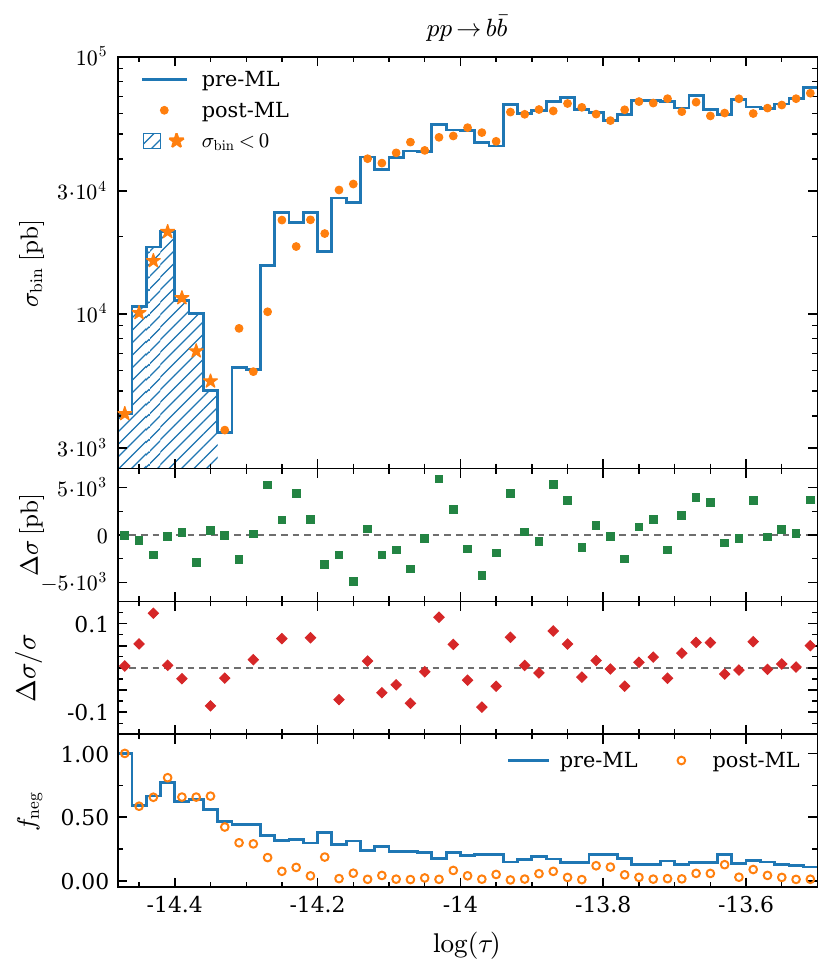}
$\phantom{a}$
\includegraphics[width=0.45\linewidth]{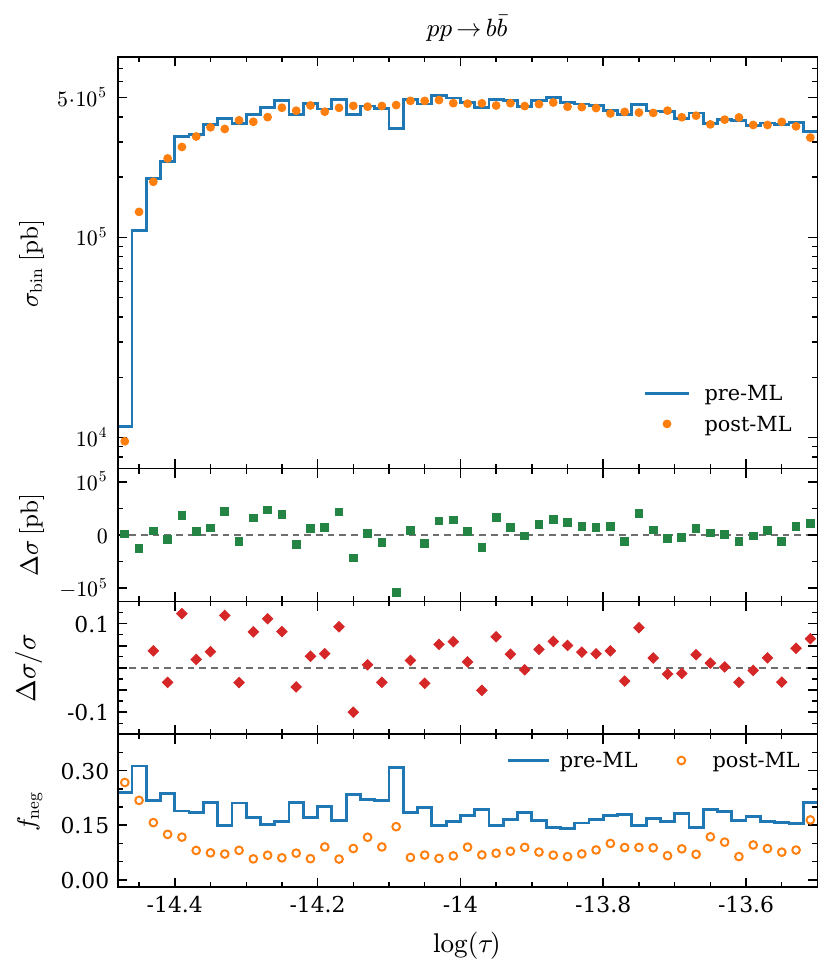}
\caption{$\log\tau$ distribution in channel GG1 of $pp\to b\bar{b}$,
stemming from the sum of FKS sectors 3 and 4 (left panel), and of all
FKS sectors (right panel).}
\label{fig:source21}
\end{figure}

The residual, or most of them anyhow, negative weights are likely
to be irreducible with our current approach, since they are associated
with non-zero measure regions in the Born phase-space. We can show that,
once again, that irreducibility stems from the separate treatment
of different FKS sectors, and we do so by employing a quantity
defined as follows:
\beq
\tau=x_1x_2=\frac{m_{b\bar{b}}^2}{S}\,,
\eeq
with $x_i$ the Bjorken variables, $m_{b\bar{b}}^2$ the $b\bar{b}$-pair
invariant mass, and $S$ the c.m.~energy squared. We present in
fig.~\ref{fig:source21} the distribution in $\log\tau$, by considering
(left panel) the sum of the contributions due to FKS sectors 3 and
4\footnote{These sectors are symmetric, thus by considering their sum we
obtain a result statistically compatible with either of them multiplied by
two, but numerically more stable.}, and the sum of all FKS sectors (right
panel); both are for the GG1 channel. In a by-now familiar pattern, the
negative-weight region present in the former is compensated in the sum
that determines physical distributions, so that a probably significant
reduction of negative weights would occur if the DT were trained onto the sum
of all sectors. The second message of fig.~\ref{fig:source21} is the absence of
DT-induced biases; thus, the very preliminary indication is that also in hadronic
collisions our ML approach to $\clS$ events is physically sound.

\begin{figure}[thb]
  \begin{center}
  \includegraphics[width=0.3\textwidth]{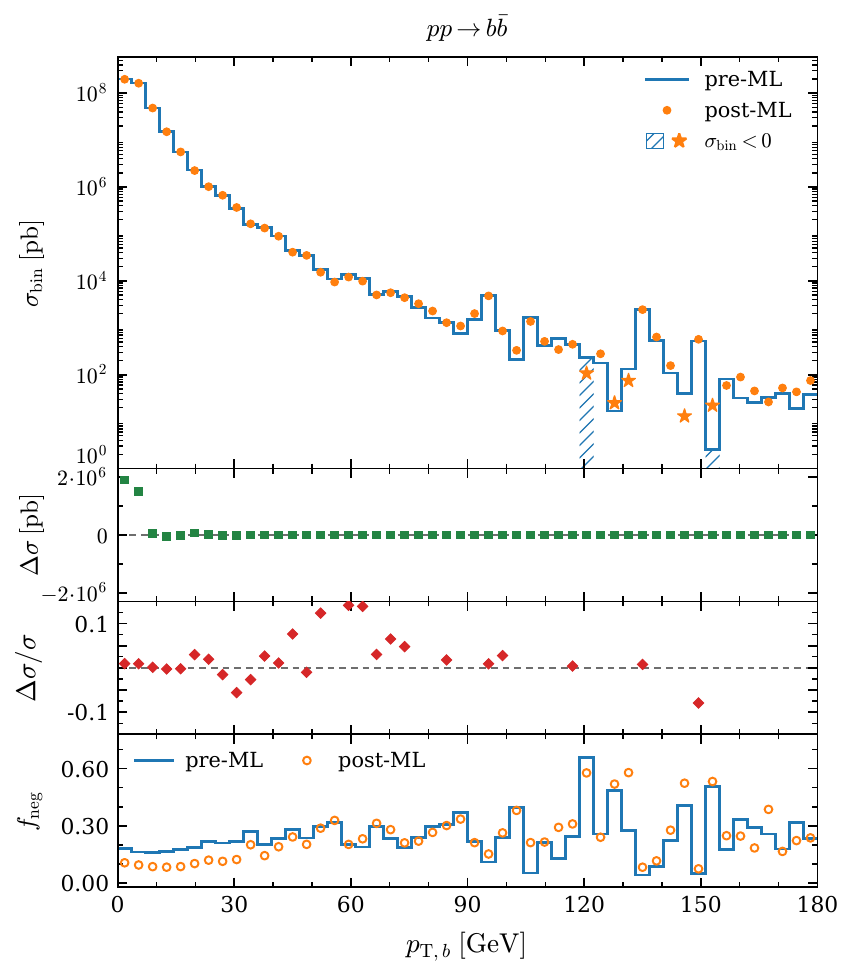}
$\phantom{a}$
  \includegraphics[width=0.3\textwidth]{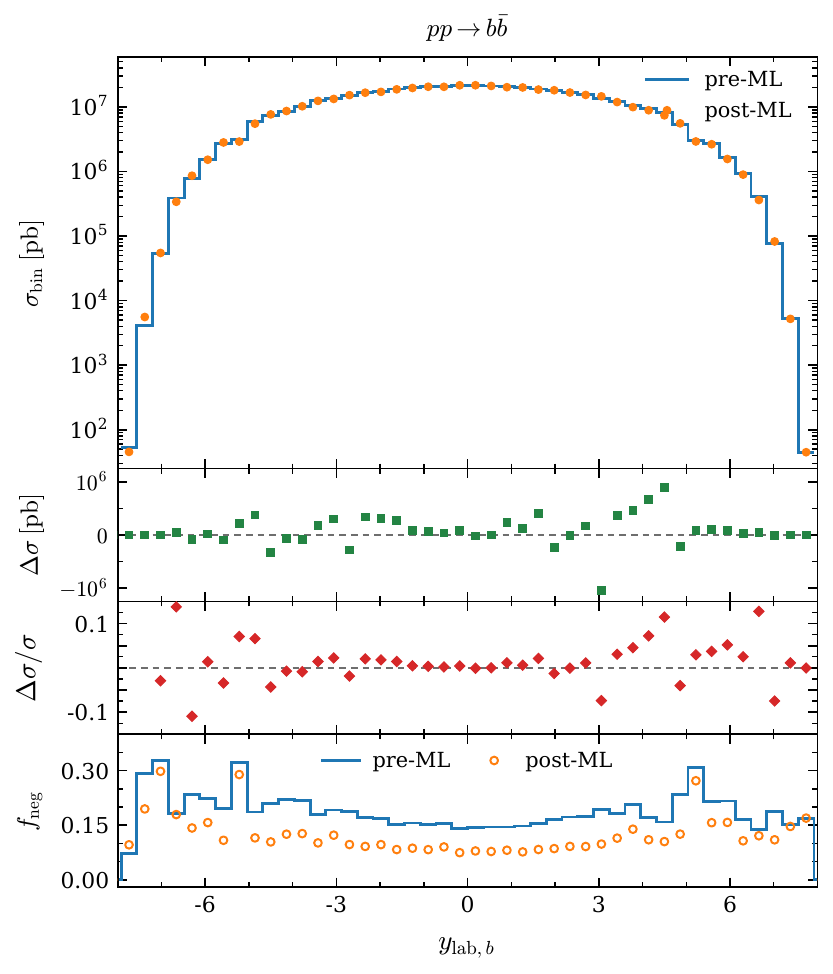}
$\phantom{a}$
  \includegraphics[width=0.3\textwidth]{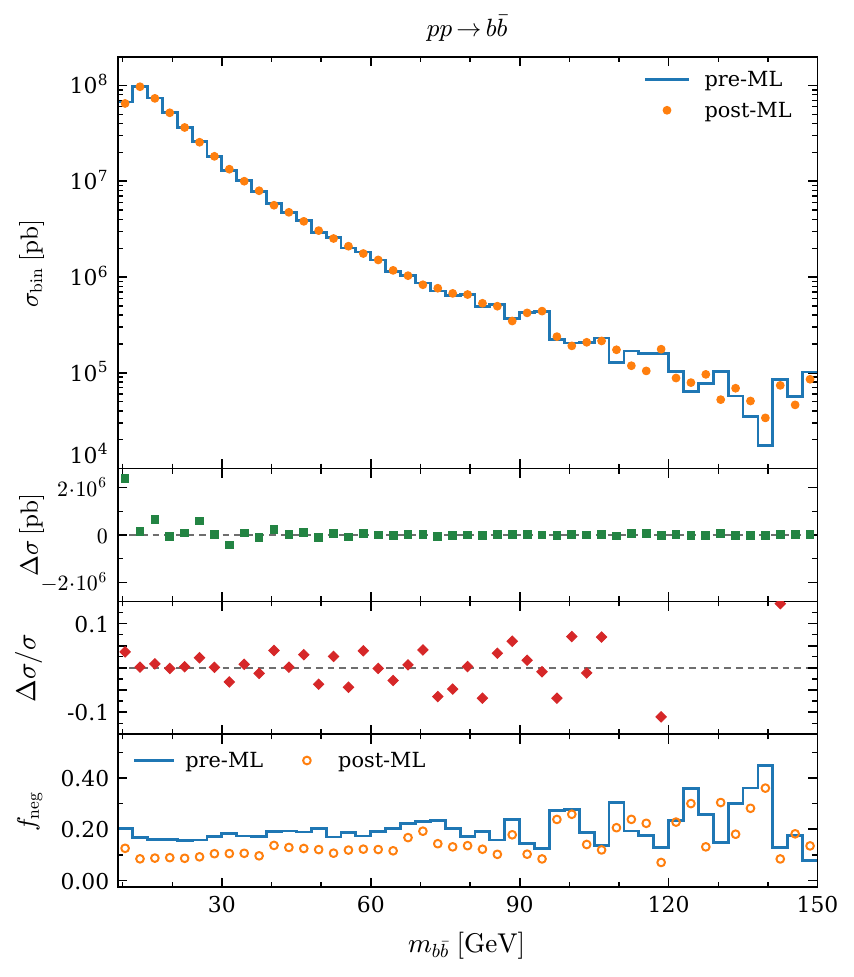}
\caption{\label{fig:ppbbdistr}
$b$-quark $\pt$ (left panel) and lab-frame rapidity (middle panel),
and $b\bar{b}$-pair invariant mass in $pp\to b\bar{b}$.
}
  \end{center}
\end{figure}
In order to further this point, we consider in fig.~\ref{fig:ppbbdistr}
the single-inclusive $b$-quark distributions in the transverse
momentum (left panel) and rapidity in the laboratory frame (middle
panel), as well as the $b\bar{b}$-pair invariant mass (right panel).
While the agreement appears to reasonable, in particular given the
absence of clear patterns in the relative differences, and therefore
any biases can be reasonably excluded in this case, it is evident
that the bin-by-bin level of agreement is worse here than in any
of the $e^+e^-$-physics results. This is due to both the generally
coarser predictions pre-ML (indeed, hadronic cross sections are
more difficult to integrate than $e+e^-$ ones), and the possibly
still-insufficient statistics used in training, in keeping with
the observations made for the total rates of table~\ref{tab:source7}.

\begin{figure}[ht]
\centering
\includegraphics[width=\linewidth]{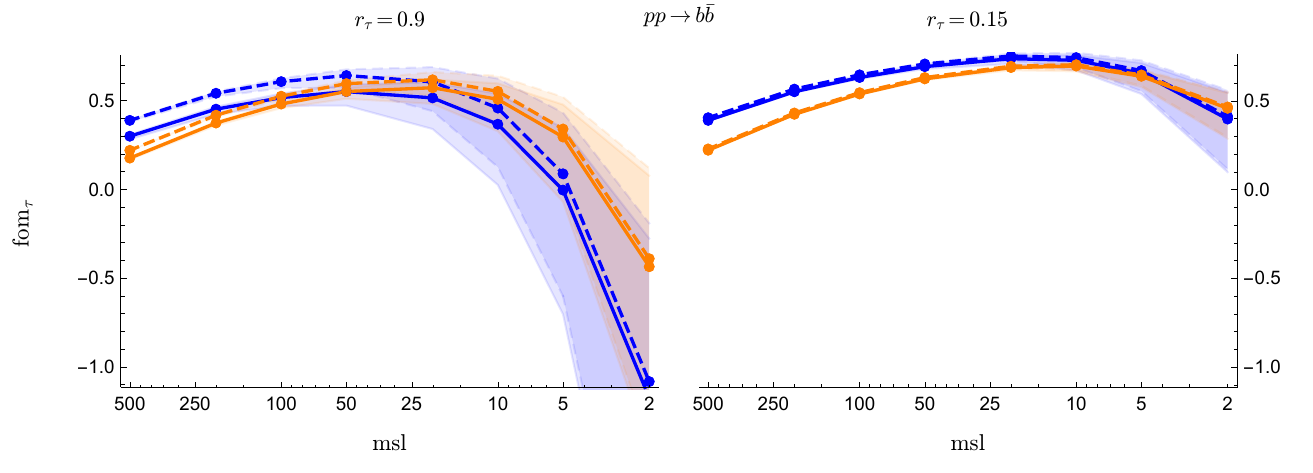}
\caption{Figure of merit for $pp\to b\bar b$.}
\label{fig:source22}
\end{figure}

With a $(4,4,1)$ folding, the negative-weight fraction goes 
down to 4.7\%, i.e. a better result than those obtained by DTs so far.
In addition to that, by applying the DT-induced approach to the
folded sample, we have obtained a further reduction, with the
final result being $f_1=3.5\%$. However, as usual the folding
step is extremely costly: with eq.~(\ref{eq:fom:fold3}) and $f_1=4.7\%$
we obtain $\fom^{({\rm fold})}=-15.3$ and $\fom^{({\rm fold})}=-1.61$
for \mbox{$\frac{\tau_+}{\tau_+ +\tau_A}=0.9$} and $0.15$, respectively.
In order to put these numbers in perspective, in fig.~\ref{fig:source22}
we present the f.o.m.~of the DT approach (in the usual layout) which,
even in the worst of cases, is much larger than that obtained with
folding -- again, the conclusion is that the latter is financially
not quite justifiable.

Figure~\ref{fig:source22} also shows a somewhat narrow range of ``ideal''
granularities w.r.t.~the $e^+e^-$ examples studied so far, with final states of
{\em comparable complexity}, and this in spite of a larger training dataset used
here.

\subsubsection{\texorpdfstring{$pp\to t\bar t$}{pptt}}
\begin{table}[ht]
\centering
\setlength{\tabcolsep}{5pt}
\renewcommand{\arraystretch}{1.10}
\small
\begin{tabular}{@{}lcccc@{}}
\toprule
Channel & & $\sigma$ (pb) & $\abs{\sigma}$ (pb) & $f_0,f_1(\%)$\\
\midrule
\multirow{2}{*}{ALL} & pre-ML & $7.742\!\cdot\!10^{2}$ & $9.252\!\cdot\!10^{2}$ & 8.2\\
 & post-ML & $7.746\!\cdot\!10^{2}$ & $8.296\!\cdot\!10^{2}$ & 3.3\\
\midrule[0.1pt]
\multirow{2}{*}{GG1} & pre-ML & $6.885\!\cdot\!10^{1}$ & $8.387\!\cdot\!10^{1}$ & 9.0\\
 & post-ML & $6.892\!\cdot\!10^{1}$ & $7.295\!\cdot\!10^{1}$ & 2.8\\
\midrule[0.1pt]
\multirow{2}{*}{GG2} & pre-ML & $3.099\!\cdot\!10^{2}$ & $3.756\!\cdot\!10^{2}$ & 8.7\\
 & post-ML & $3.097\!\cdot\!10^{2}$ & $3.335\!\cdot\!10^{2}$ & 3.6\\
\midrule[0.1pt]
\multirow{2}{*}{GG3} & pre-ML & $3.091\!\cdot\!10^{2}$ & $3.751\!\cdot\!10^{2}$ & 8.8\\
 & post-ML & $3.098\!\cdot\!10^{2}$ & $3.332\!\cdot\!10^{2}$ & 3.5\\
\midrule[0.1pt]
\multirow{2}{*}{UXU} & pre-ML & $4.310\!\cdot\!10^{1}$ & $4.526\!\cdot\!10^{1}$ & 2.4\\
 & post-ML & $4.309\!\cdot\!10^{1}$ & $4.498\!\cdot\!10^{1}$ & 2.1\\
\midrule[0.1pt]
\multirow{2}{*}{UUX} & pre-ML & $4.318\!\cdot\!10^{1}$ & $4.535\!\cdot\!10^{1}$ & 2.4\\
 & post-ML & $4.310\!\cdot\!10^{1}$ & $4.500\!\cdot\!10^{1}$ & 2.1\\
\bottomrule
\end{tabular}
\caption{Total rates per channel in $pp\to t\bar t$.}
\label{tab:source8}
\end{table}

From the diagrammatic viewpoint, this process is identical to
$pp\to b\bar{b}$; however, MC@NLO cross sections behave quite
differently, owing essentially to the fact that matching a hard
$b$ quark with a shower is kinematically more complicated than
for a $t$ quark. Because of this, the $t\bar{t}$ $\clS$-event
cross section has a fraction of negative weights equal to a more
modest $8.2\%$ -- the DT approach allows one to reduce it down
to $3.3\%$. As is shown in table~\ref{tab:source8}, this is
driven by the reduction in the $gg$ channels, with the $q\bar{q}$
ones largely unchanged by ML.

\begin{figure}[thb]
  \begin{center}
  \includegraphics[width=0.3\textwidth]{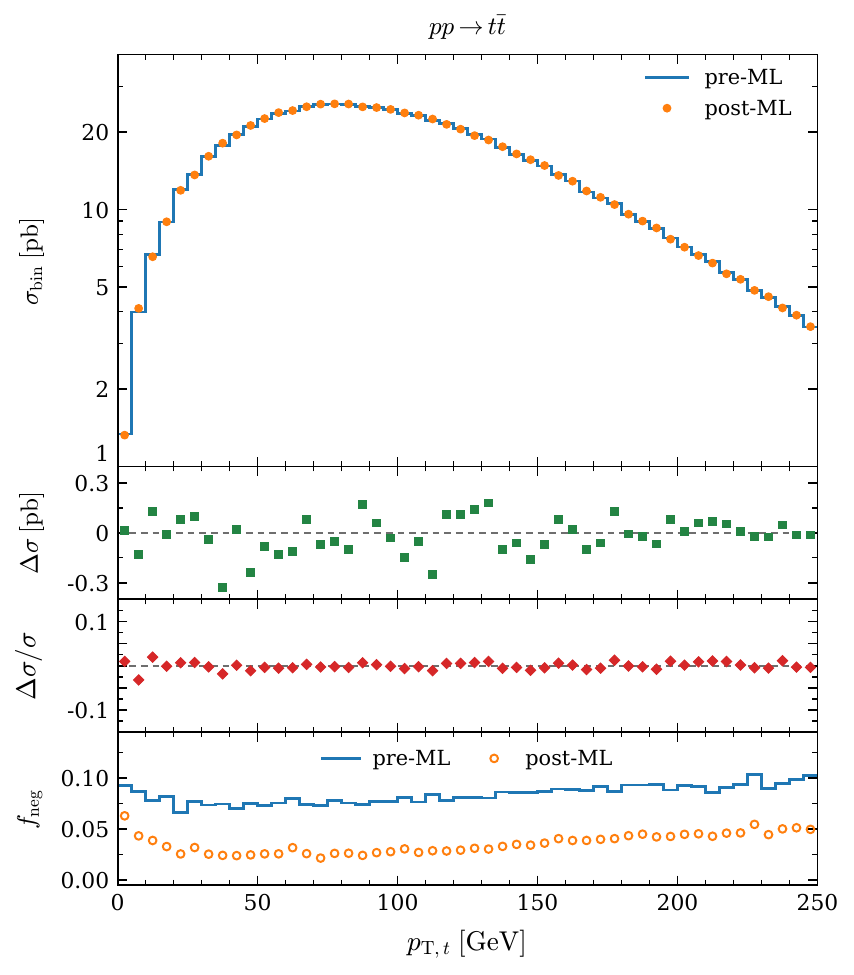}
$\phantom{a}$
  \includegraphics[width=0.3\textwidth]{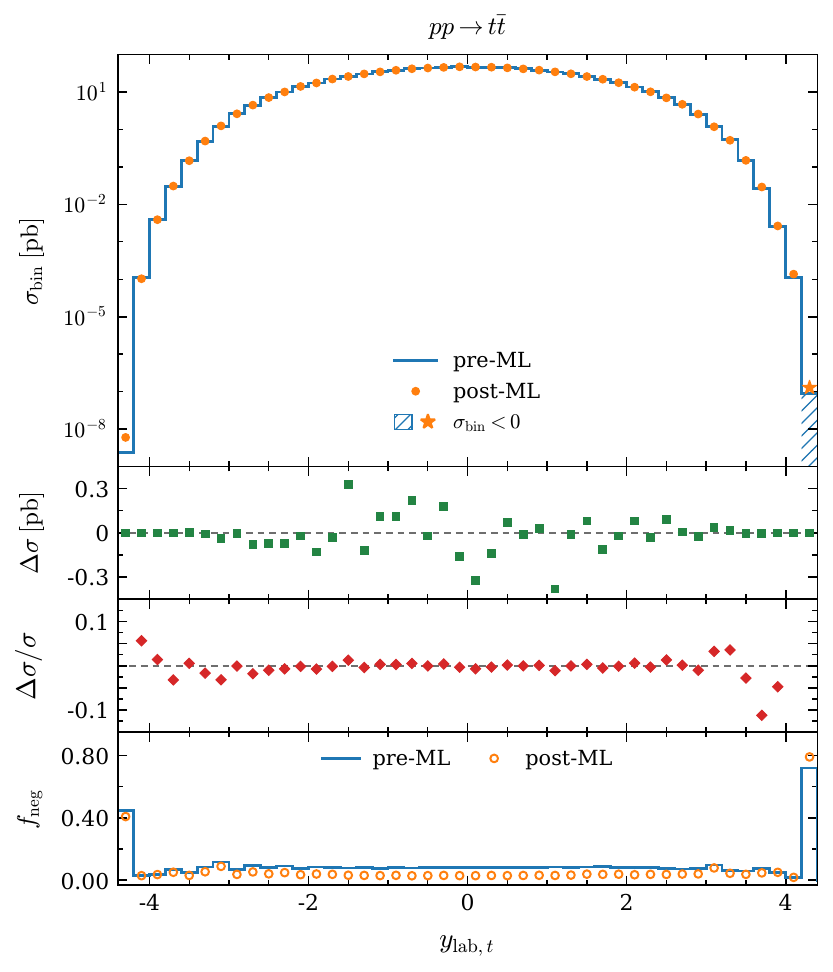}
$\phantom{a}$
  \includegraphics[width=0.3\textwidth]{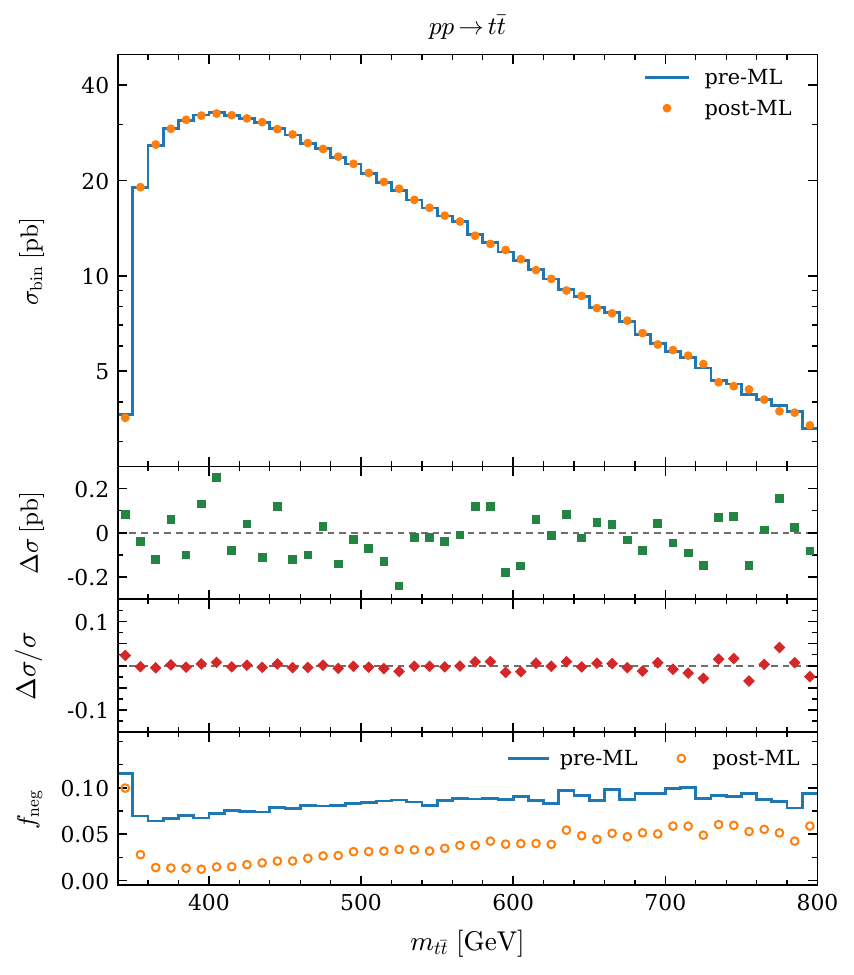}
\caption{\label{fig:ppttdistr}
Top-quark $\pt$ (left panel) and lab-frame rapidity (middle panel),
and $t\bar{t}$-pair invariant mass in $pp\to t\bar{t}$.
}
  \end{center}
\end{figure}
The differential counterpart of table~\ref{tab:source8} is presented
in fig.~\ref{fig:ppttdistr}; from left to right, the three panels
show the $\pt$ of the top quark, its laboratory-frame rapidity,
and the $t\bar{t}$ invariant mass. The absence of biases is evident,
thanks also to the fact that the fraction of negative weights, both
pre- and post-ML, is significantly smaller than the one relevant to
$pp\to b\bar{b}$. Thus, the regressor for the current process
is easier to learn than the regressor for $b\bar{b}$ production,
and the results do not fluctuate very much.

\begin{figure}[ht]
\centering
\includegraphics[width=\linewidth]{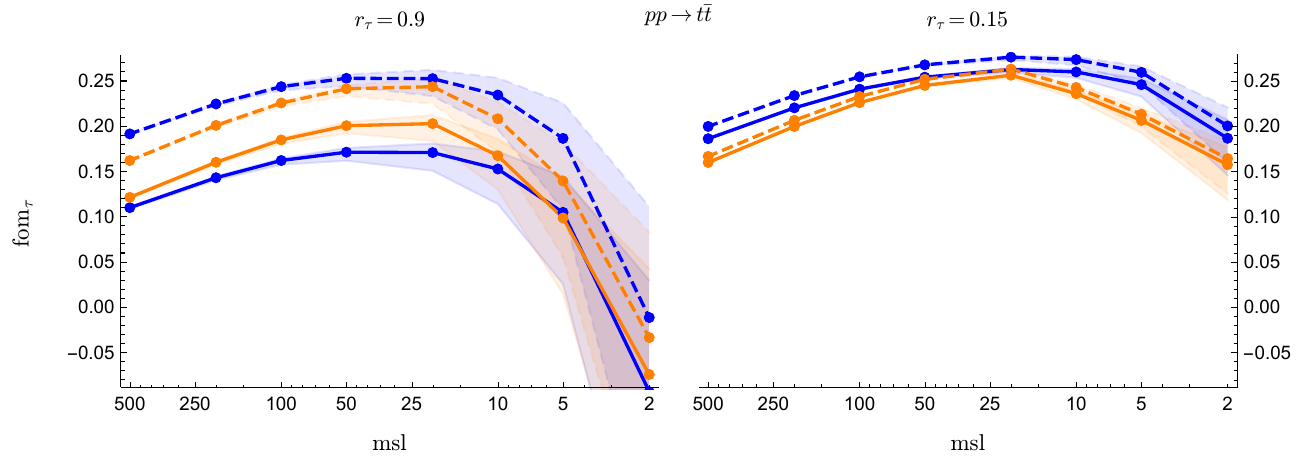}
\caption{Figure of merit for $pp\to t\bar t$.}
\label{fig:source23}
\end{figure}

As in the case of $pp\to b\bar{b}$, the f.o.m.~has a rather narrow shape
(see fig.~\ref{fig:source23}), where the improvements w.r.t.~choosing
a coarse granularity (left sides of the plots) and the degradation
associated with possible overtraining (right sides of the plots) are rather
clear. 

The same process has also been generated with MC@NLO-$\Delta$, and the
results are largely consistent with what has been found previously,
namely: the weight reduction primarily affects $\clH$-events, with
some further reduction for $\clS$ events. Specifically, for the
MC@NLO-$\Delta$ $pp\to t\bar t$ process, the negative-weight
$\clS$-event fraction has been reduced by DTs from 7.9\% to 2.9\%.
These values of $f_0$ and $f_1$  are nearly identical to those induced
by the standard MC@NLO cross sections, if marginally better; this confirms
that any improvements on the $\clH$-event side will not be lost due
to ML acting on the $\clS$-event side.

\subsubsection{\texorpdfstring{$pp\to ZZ$}{ppZZ}}

The process $pp\to ZZ$ receives contributions from four channels:
two initiated by down quarks (DDX and DXD) and two by up quarks
(UUX and UXU). All channels exhibit relatively small negative-weight
fractions (see table~\ref{tab:source9}). Overall, the negative-weight
fraction decreases from $2.7\%$ to $1.1\%$ thanks to the DT-driven
procedure. We have verified that the latter number is fairly robust
w.r.t.~increases in the training-sample size, and there are no
sizable regions at the Born level where the cross section is
negative.
\begin{table}[ht]
\centering
\setlength{\tabcolsep}{5pt}
\renewcommand{\arraystretch}{1.10}
\small
\begin{tabular}{@{}lcccc@{}}
\toprule
Channel & & $\sigma$ (pb) & $\abs{\sigma}$ (pb) & $f_0,f_1(\%)$\\
\midrule
\multirow{2}{*}{ALL} & pre-ML & $11.46$ & $12.11$ & 2.7\\
 & post-ML & $11.46$ & $11.71$ & 1.1\\
\midrule[0.1pt]
\multirow{2}{*}{DDX} & pre-ML & $3.511$ & $3.683$ & 2.3\\
 & post-ML & $3.512$ & $3.573$ & 0.9\\
\midrule[0.1pt]
\multirow{2}{*}{DXD} & pre-ML & $3.516$ & $3.712$ & 2.6\\
 & post-ML & $3.514$ & $3.597$ & 1.2\\
\midrule[0.1pt]
\multirow{2}{*}{UUX} & pre-ML & $2.202$ & $2.325$ & 2.6\\
 & post-ML & $2.202$ & $2.238$ & 0.8\\
\midrule[0.1pt]
\multirow{2}{*}{UXU} & pre-ML & $2.233$ & $2.388$ & 3.2\\
 & post-ML & $2.233$ & $2.303$ & 1.5\\
\bottomrule
\end{tabular}
\caption{Total rates per channel in $pp\to ZZ$.}
\label{tab:source9}
\end{table}

\begin{figure}[ht]
\centering
\includegraphics[width=\linewidth]{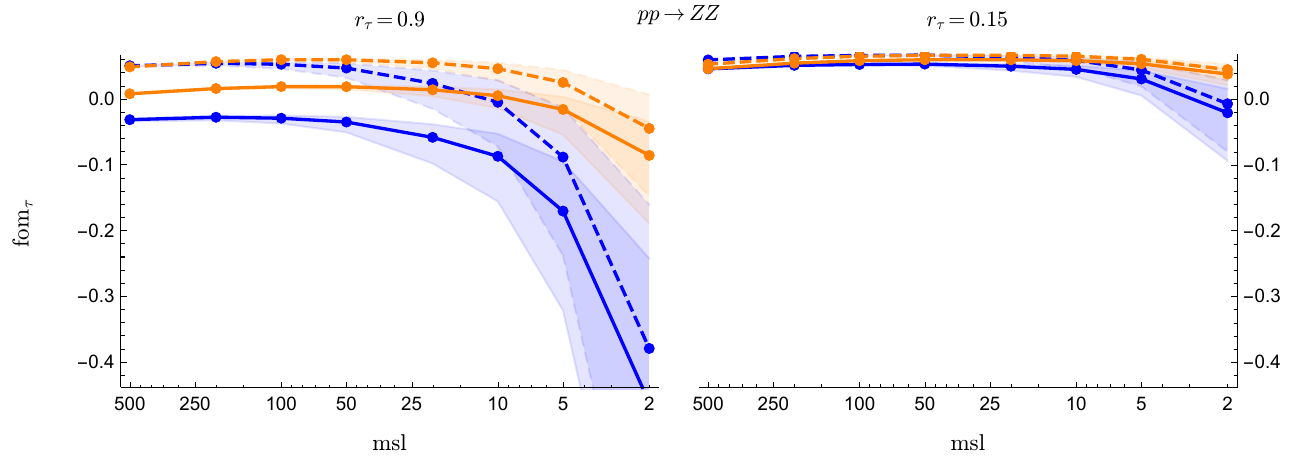}
\caption{Figure of merit for $pp\to ZZ$.}
\label{fig:source24}
\end{figure}

Both observations are in keeping with the fact that the reduction
of negative weights is close to being saturated, i.e.~the resulting
$f_1$ is quite small. Indeed, the figure of merit for
this process, shown in fig.~\ref{fig:source24},  exhibits a behaviour
on the l.h.s.~of the plots which is flatter w.r.t.~the hadronic
processes discussed so far, with the results of the smaller-size
training dataset equivalent or better than those obtained with
the larger-size ones; thus, in the present case one never risks
undertraining.

\subsubsection{\texorpdfstring{$pp\to HH$}{ppHH}}
\begin{figure}[ht]
\centering
\includegraphics[width=0.4\linewidth]{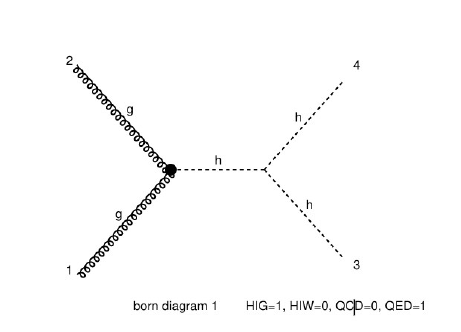}
\caption{Feynman diagram underpinning $pp\to HH$ in EFT.}
\label{fig:source25}
\end{figure}
\begin{table}[ht]
\centering
\setlength{\tabcolsep}{5pt}
\renewcommand{\arraystretch}{1.10}
\small
\begin{tabular}{@{}lcccc@{}}
\toprule
FKS sectors & & $\sigma$ (pb) & $\abs{\sigma}$ (pb) & $f_0,f_1(\%)$\\
\midrule
\multirow{2}{*}{ALL} & pre-ML & $2.383\!\cdot\!10^{-3}$ & $3.078\!\cdot\!10^{-3}$ & 11.3\\
 & post-ML & $2.382\!\cdot\!10^{-3}$ & $2.441\!\cdot\!10^{-3}$ & 1.2\\
\midrule[0.1pt]
\multirow{2}{*}{-- FKS134} & pre-ML & $1.191\!\cdot\!10^{-3}$ & $1.539\!\cdot\!10^{-3}$ & 11.3\\
 & post-ML & $1.191\!\cdot\!10^{-3}$ & $1.220\!\cdot\!10^{-3}$ & 1.2\\
\midrule[0.1pt]
\multirow{2}{*}{-- FKS256} & pre-ML & $1.192\!\cdot\!10^{-3}$ & $1.539\!\cdot\!10^{-3}$ & 11.3\\
 & post-ML & $1.191\!\cdot\!10^{-3}$ & $1.221\!\cdot\!10^{-3}$ & 1.2\\
\bottomrule
\end{tabular}
\caption{Total rates per group of FKS sectors in $pp\to HH$. This
process receives contributions from a single channel (GG1).}
\label{tab:source10}
\end{table}
This process represents a test case of the ML machinery in a simplified
hadronic environment, rather than a fully-fledged physics
simulation, in that it has been generated in the effective theory
where one writes the top loop as a vertex, and by ignoring the
virtual contribution. The process has a single channel, stemming
from the diagram depicted in fig.~\ref{fig:source25}. In turn, that
channel has six FKS sectors, the complete analogues of the $(156)$
and $(278)$ sectors relevant to $pp\to b\bar{b}$ production. Here,
these are re-labelled as $(134)$ and $(256)$ respectively, owing
to the absence of FKS sectors stemming from final-state emissions.
The total rate results are presented in table~\ref{tab:source10}; we see
that DTs do an excellent job, with the fraction of negative weights
reduced from 11.3\% to 1.2\% (i.e.~at a physiologically negligible
level).
\begin{figure}[thb]
  \begin{center}
  \includegraphics[width=0.3\textwidth]{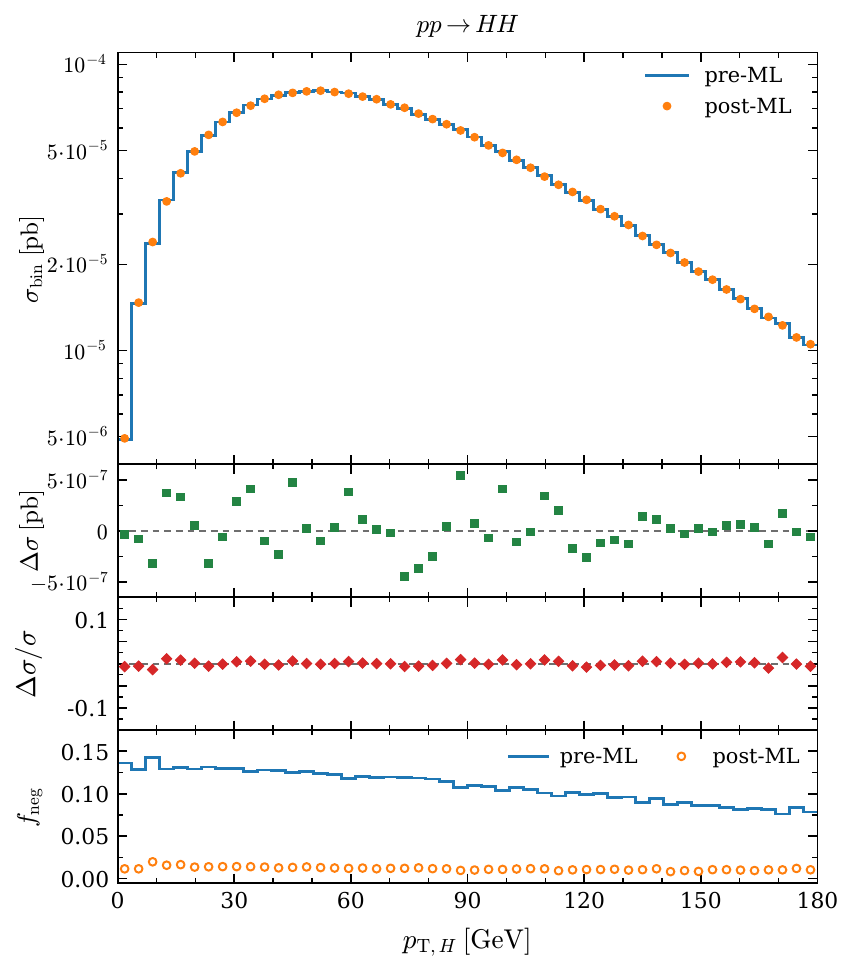}
$\phantom{a}$
  \includegraphics[width=0.3\textwidth]{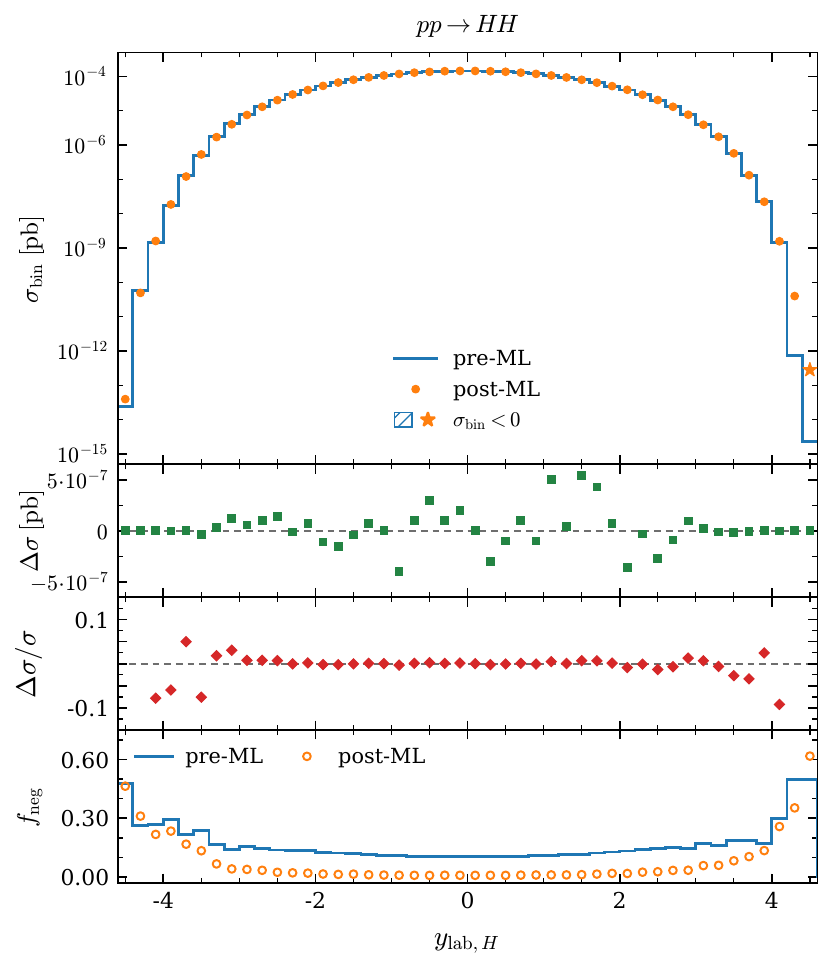}
$\phantom{a}$
  \includegraphics[width=0.3\textwidth]{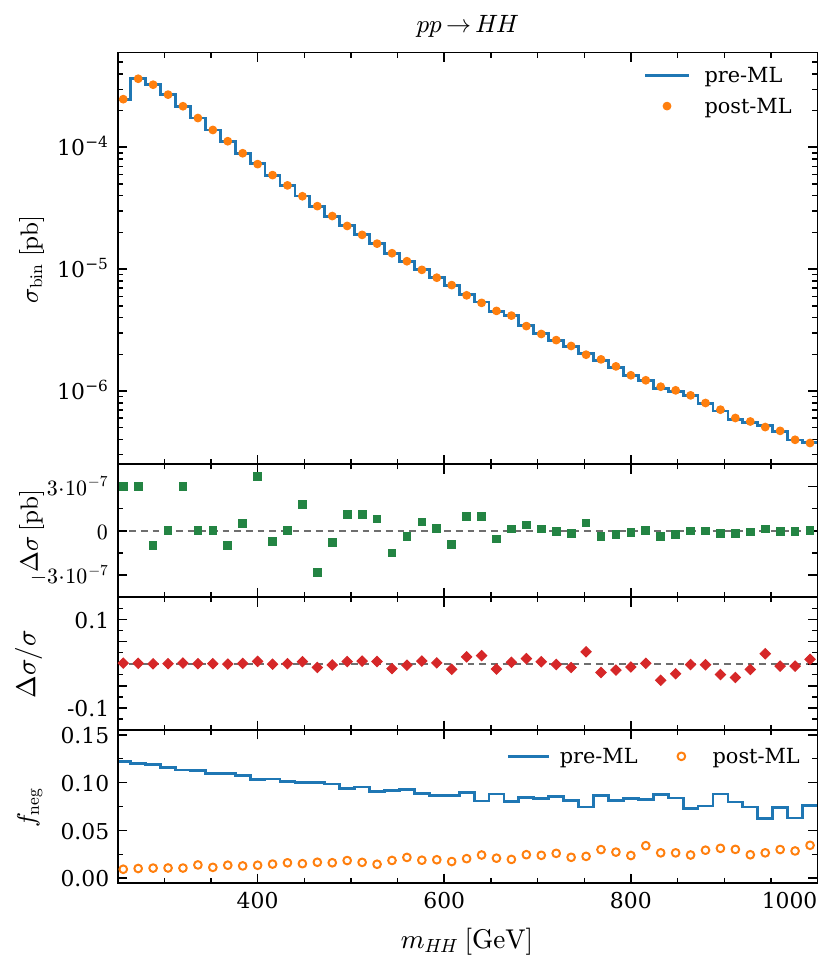}
\caption{\label{fig:ppHHdistr}
Higgs $\pt$ (left panel) and lab-frame rapidity (middle panel),
and $HH$-pair invariant mass in $pp\to HH$.
}
  \end{center}
\end{figure}

At the differential level, we obtain the results of
fig.~\ref{fig:ppHHdistr}, for the Higgs transverse momentum (left plot),
the Higgs laboratory-frame rapidity, and the Higgs-pair invariant
mass; the agreement between the pre- and post-ML predictions is
extremely good, and biases can be firmly excluded
\begin{figure}[ht]
\centering
\includegraphics[width=\linewidth]{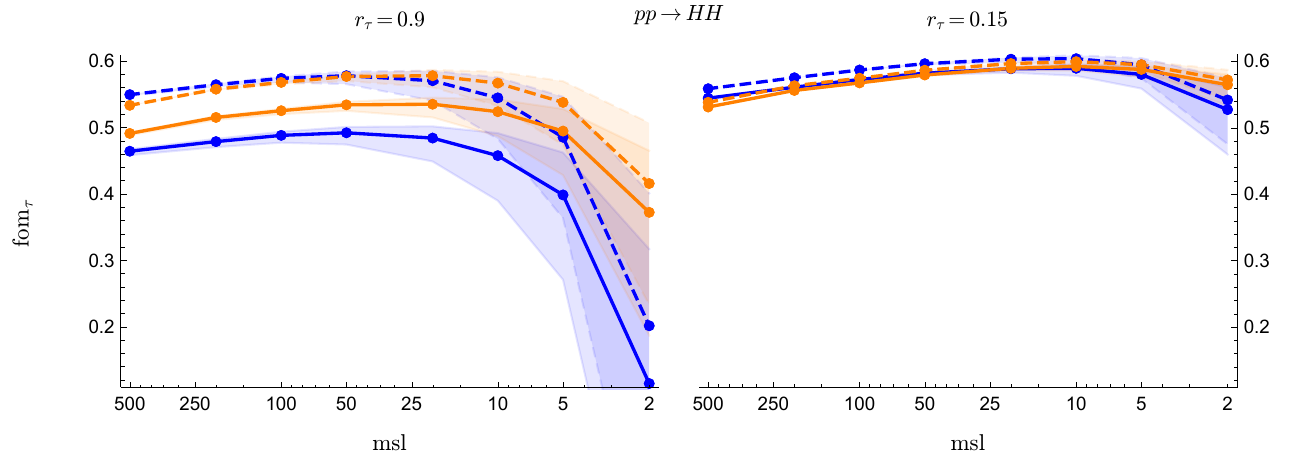}
\caption{Figure of merit for $pp\to HH$.}
\label{fig:source26}
\end{figure}

Finally, the figure of merit is presented in fig.~\ref{fig:source26},
which has several traits of that relevant to $pp\to ZZ$ -- namely, a
moderate growth to reach the absolute maximum before overtraining kicks
in, and the evidence that training with a yet larger number of events
could even be counterproductive.

\subsubsection{Conclusions on Decision Trees\label{sec:DTconcl}}
There are several distinct reasons why negative weights persist after
the DT-driven procedure. It is important to distinguish between limitations
intrinsic to the weight-redistribution approach, and limitations of its
specific DT implementation. The former cannot be overcome
simply by improving the ML model; the latter might be.

\noindent
The main issues are as follows:

1. {\em Negative differential Born cross sections}. The ML-modified
cross section is unbiased; therefore, it preserves the integral
over radiation variables at fixed Born kinematics.
Consequently, where the original radiation-integrated $\clS$-event cross
section is negative, negative weights must remain, and therefore this
problem is not within the scope of the ML-based procedure suggested here
(while it could still be addressed by ML methods). For example, the
results for \mbox{$e^+e^-\to b\bar b t\bar t$} and \mbox{$pp\to b\bar b$}
exhibit significant negative-weight regions in Born-level distributions
consistent with this scenario.

2. {\em Cancellations among different channels and FKS sectors}. When the
ML-driven cross-section contribution is constructed independently in
each FKS sector, it cannot exploit cancellations between contributions
belonging to different sectors. The occurrence of such cancellations
is particularly evident in $e^+e^-\to b\bar b t\bar t$ and
$e^+e^-\to b\bar b b\bar b$  among the cases we have studied (see
e.g.~figs.~\ref{fig:source6} and~\ref{fig:eebbbbpt1FKS}). Summing over
FKS sectors before applying ML gives substantially better results\footnote{For
a very recent approach to this and to related issues, see
ref.~\cite{Frederix:2026qsf}.}. Addressing this problem requires
modifying the way in which the contributions are combined, rather than
merely improving the regression within each sector. Still,
even after summing different FKS sectors a single regressor makes
sense only if such a regressor is learned in terms of variables
whose ranges and physical meanings are common to all of the FKS
sectors involved (e.g.~this approach cannot be attempted if the
learning variables are physical). Similar considerations apply
when cancellations among different channels are sizable; addressing
this issue appears to be more complicated than that of the FKS-sector
sums, since its solution may hamper unweighting efficiency.

3. {\em Negative weights associated with the treatment of virtual
corrections}. Some negative weights originate from the infrequent evaluation
of the exact virtual contribution (see ref.~\cite{Alwall:2014hca}),
rather than from fluctuations in the function which the DT approximates.
Such fluctuations cannot possibly be modelled by the present $K$-regression
procedure (or by any other regular-approximant method, in fact), and their
reduction calls for changes to the numerical treatment of the virtual
contribution. In particular, the default computation evaluates virtual
corrections only once every $\ord(100)$ phase-space points, and uses
a surrogate otherwise. We have verified that by increasing the frequency
of the former evaluation one significantly reduces certain negative-weight
fractions, notably in \mbox{$e^+e^-\to b\bar b$} and \mbox{$pp\to b\bar b$}.
Clearly, being too time-consuming this strategy is not viable for
high-precision computations.

4. {\em Unphysical phase-space points}. The fact that phase-space generation
does not have efficiency equal to one (i.e.~the fact that
\mbox{$\Theta_{\rm phys}(\bar\meas_n)\ne [0,1]^3$} in the language
of sect.~\ref{sec:mcnlo}) presents another difficulty. The
restricted-domain solution used to obtain most of the results
presented here is problematic, because  it leads to a decrease
in efficiency, and intrinsically limits the
amount of reduction of negative weights. This demonstrates the need
for an explicit and exact treatment of the candidate integration domain
rather than relying exclusively on a fixed FKS-space mask.

Among the above problems, which one is genuinely unavoidable,
e.g.~in the sense that its solution would cost an amount of money
comparable to that one wishes to save?

The first problem is a mathematical obstruction: if the
Born-differential cross section is negative, no redistribution that
preserves it can make the event weights everywhere positive.
However, a solution in principle exists, which is that of
modifying the channel definitions, and/or that of the ${\cal S}$
functions in the underlying FKS subtraction; both modifications are
liable to be tackled by means of ML techniques.

The second problem is an obstruction within the present FKS sector-by-sector
construction, but not necessarily within a more general approach
that treats several FKS contributions jointly. We have shown how to
address this problem, but we did not give an automatization
prescription in order to do so without human intervention.

The remaining problems are associated with the chosen approximation
target, the learning variables, the numerical treatment of virtual
corrections, and finite training statistics. These are implementation
issues, the solutions of some of which have been tackled in the text;
generally-speaking, it appears to be relatively easy to fix such issues,
since they do not involve complicated conceptual problems.

\subsection{Implementation of Normalizing Flows}
\label{par:nfimpl:practical}
\label{subsec:nfimpl:practical}
Similarly to the case of DT-modified cross sections, the practical
implementation of the NF-based strategy also employs Vegas variables;
conceptually, it strictly follows sect.~\ref{sssec:nf-construction}.
We shall therefore use the results in the latter section, and express
them by means of the sampling variables introduced in
sect.~\ref{par:dtimpl:practical}. This change of variables does not alter
the two-flow ansatz, its candidate domain, or its physical-domain
restriction.  

We start by giving operational definitions of the loss functions
introduced in sect.~\ref{sssec:nf-construction}.
Let $\mathcal T$ be the collection of indices of the configurations used
in a loss evaluation, and write
\beq
 N_{\mathcal T}=|\mathcal T|,\qquad
 z_a=(\bar\meas_{n,a},\bar\meas_{r,a}),\qquad a\in\mathcal T.
 \label{eq:nf:training-sample}
\eeq
With this, we mean that $N_{\mathcal T}$ is the number of those configurations.
Also, here and elsewhere we make use of the shorthand notation:
\beq
\bar\varsigma_a\equiv
 \bar\varsigma^{(\clS)}_{\mathrm{NF}}(z_a),\qquad
 F_{i,a}\equiv F_i(\bar\meas_{r,a}\mid\bar\meas_{n,a}),\qquad i=1,2.
 \label{eq:nf:sampled-loss-quantities}
\eeq
For any quantity $A$ evaluated on these configurations, angle brackets
denote the arithmetic sample average, namely:
\beq
 \langle A\rangle_{\mathcal T}
 \equiv\frac{1}{N_{\mathcal T}}
       \sum_{a\in\mathcal T}A(z_a),\qquad N_{\mathcal T}>0.
 \label{eq:nf:sample-average}
\eeq
For the losses defined in eq.~(\ref{eq:nf:loss-total}), $\mathcal T$
includes all physical and unphysical configurations. In the current
context, their inputs are the barred variables -- this condition will
be relaxed later. The physical weight is zero on invalid configurations,
but these entries still count in $N_{\mathcal T}$; removing them would
change the normalisation of the average. The subset relevant to the
unphysical-event penalty (i.e.~the quantity responsible for minimizing
the $q_i$ integrals defined before) is
\beq
 \begin{aligned}
 \mathcal T_{\rm unph}
 &=\bigl\{a\in\mathcal T:\
       \bar\meas_{r,a}\in
{\tilde{\Theta}(\bar\meas_{n,a})
\bigcap\Theta_{\rm phys}^c(\bar\meas_{n,a})}\bigr\},
 \\
 N_{\rm unph}&=|\mathcal T_{\rm unph}|.
 \end{aligned}
 \label{eq:nf:unphysical-sample}
\eeq
The restricted average is normalised by the number of entries in this
subset, rather than by the size of the complete sample:
\beq
 \langle A\rangle_{\mathcal T_{\rm unph}}
 \equiv\frac{1}{N_{\rm unph}}
       \sum_{a\in\mathcal T_{\rm unph}}A(z_a),\qquad N_{\rm unph}>0.
 \label{eq:nf:restricted-sample-average}
\eeq
In particular,
\beq
 \left\langle
\mathbf{1}_{\tilde{\Theta}(\bar\meas_n)
\bigcap\Theta_{\rm phys}^c(\bar\meas_n)}(\bar\meas_r)A
 \right\rangle_{\mathcal T}
 =\frac{N_{\rm unph}}{N_{\mathcal T}}
   \langle A\rangle_{\mathcal T_{\rm unph}},
 \qquad N_{\rm unph}>0.
 \label{eq:nf:average-restriction}
\eeq
Thus inserting the indicator in an average over the full sample is not
the same as averaging over the unphysical subset. If $N_{\rm unph}=0$,
the restricted average is not evaluated and the unphysical penalty is
set to zero for that loss evaluation.

With the definitions given above, the loss in eq.~(\ref{eq:nf:loss-negative})
is computed as follows:
\beq
 L_{\rm neg}
=\frac{1}{M_\varsigma N_{\mathcal T}^{\,2}}
 \left[
   \left(\sum_{a\in\mathcal T}|\bar\varsigma_a|\right)^2
  -\left(\sum_{a\in\mathcal T}\bar\varsigma_a\right)^2
 \right].
 \label{eq:nf:loss-negative2}
\eeq
with $M_\varsigma$ a normalization factor equal to the maximum of
the absolute values of the weights (or of their average if the
weights are partitioned into batches) -- see eq.~(\ref{avgNormepch}).
Conversely, by using the
restricted average of eq.~(\ref{eq:nf:restricted-sample-average}),
the loss of eq.~(\ref{eq:nf:loss-unphysical}) is evaluated thus:
\beq
 L_{\rm unph}
=\frac{1}{N_{\rm unph}}
   \sum_{a\in\mathcal T_{\rm unph}}\sum_{i=1}^{2}
       \log(F_{i,a}+\epsilon),
 \qquad N_{\rm unph}>0,\quad\epsilon>0.
 \label{eq:nf:loss-unphysical2}
\eeq
In our applications, we set $\ep=10^{-10}$.
Finally, the regularising loss of eq.~(\ref{eq:nf:loss-regularization})
can be computed as follows:
\beq
 L_{\rm reg}
=\frac{1}{N_{\mathcal T}}
   \sum_{a\in\mathcal T}\sum_{i=1}^{2}F_{i,a}^{\,2}.
 \label{eq:nf:loss-regularization2}
\eeq

\vskip 0.2truecm
\noindent\emph{Epochs and batches.}
\par\noindent
If we want to be more precise as far as the usage of epochs,
and batches within them, are concerned, we proceed as follows.
Let $\mathcal T_{\mathrm{train}}$ denote the indices of the complete training
sample, and let $\mathcal T_{e,b}$ denote those in batch $b$ of epoch $e$.
For a loss evaluation used in a mini-batch update, the notation introduced
above is interpreted as
\beq
 \mathcal T=\mathcal T_{e,b},\qquad
 N_{\mathcal T}=n_{e,b}:=|\mathcal T_{e,b}|.
 \label{eq:nftrain:batch-index-set}
\eeq
In particular, $N_{\mathcal T}$ is neither the complete training-sample
size nor the number of configurations processed over all epochs.
Writing $\theta_{e,b}$ for the parameters before this update, one has
\beq
 \langle A\rangle_{\mathcal T_{e,b}}
 =\frac{1}{n_{e,b}}\sum_{a\in\mathcal T_{e,b}}
       A(z_a;\theta_{e,b}).
 \label{eq:nftrain:batch-average}
\eeq
All entries in a batch are evaluated at the same parameter values; the
next batch is evaluated after the update. The denominator is the actual
batch size, including when the final batch is incomplete. With this,
the parameter $M_\varsigma$ of eq.~(\ref{eq:nf:loss-negative2}) reads
as follows:
\beq
M_\varsigma =
\langle\max\Big\{\abs{\bar\varsigma_a}\Big\}_{\mathcal
T_{e,b}}\rangle_e\,,
\label{avgNormepch}
\eeq
which is implemented as follows:
\beq
M_\varsigma \equiv
M_{\varsigma,e}
=
\frac{1}{B_e}
\sum_{b=1}^{B_e}
\max_{a\in\mathcal T_{e,b}}
\big\{\abs{\bar\varsigma(z_a;\theta_{e,b})}\big\},
\eeq
where $B_e$ is the number of batches in epoch $e$; as was said before,
the trainable parameters $\theta_{e,b}$ are updated after each batch
is processed. Full-epoch quantities are taken at their previous-epoch values.

The set $\mathcal T_{\rm unph}$ correspondingly identifies the subset of the
current batch lying in the candidate domain but outside the physical domain.
The unphysical-event loss is normalised by the number of entries in this
subset, not by $n_{e,b}$; it is set to zero when the subset is empty.
Invalid configurations still count in $n_{e,b}$ for the full-batch averages,
even though their corrected cross-section contribution vanishes. Batching
does not change the sampling measure or the jacobian factors, and need not
group configurations with identical Born variables.

Batch-wise training losses must be distinguished from a full-sample loss
reported for a fixed model. If all batches partition the training sample
and $A$ is evaluated at the same fixed parameters throughout, then
\beq
 \langle A\rangle_{\mathcal T_{\mathrm{train}}}
 =\frac{1}{N_{\mathrm{train}}}
   \sum_b n_{e,b}\langle A\rangle_{\mathcal T_{e,b}}.
 \label{eq:nftrain:pooled-average}
\eeq
The analogous restricted average uses the numbers of unphysical entries
as weights. For the negative-weight loss, the signed and absolute first
moments must first be pooled and only then squared: averaging the losses
computed separately on the batches does not in general give the full-sample
loss, even with the same fixed normalization. For example, a one-event
batch makes the difference of squared first moments identically zero.
This is a direct consequence of the loss definition, not a convention for
reporting it.

A loss of the model obtained at the end of an epoch should therefore be
evaluated in a separate pass with its parameters fixed, on the training
sample or an independent validation sample. An average of the losses
recorded during the training pass instead mixes successive model states.

\vskip 0.2truecm
\noindent\emph{Change of coordinates.}
\par\noindent
We give here an example analogous to that of sect.~\ref{par:dtimpl:practical},
i.e.~we employ the previous results for NF-based manipulations and express
them in terms of the un-flat Vegas variables ($v$), as we do in practical
applications. In order to do that, we use exactly the same
relationships as in eq.~(\ref{eq:dtimpl:chain})
\beqn
 &&x=(x_n,x_r)\longrightarrow v(x)=(v_n(x_n),v_r(x_r))
 \longrightarrow\bar\meas(v(x)),
 \nonumber\\
 &&\bar\meas_n=v_n,\qquad
 \bar\meas_r=(v_\xi^{k_\xi},v_y^{k_y},v_\varphi).
 \label{eq:nfimpl:chain}
\eeqn
The corresponding jacobians are given in eqs.~(\ref{eq:dtimpl:jacobians})
and~(\ref{eq:dtimpl:full-jacobian}). We remind the reader that all
variables have unit-interval ranges.  

The candidate and physical radiation domains must now be expressed in the
$v$ coordinates, as follows:
\beqn
 \tilde\Theta_v(v_n)
 &=&\bigl\{v_r\in[0,1]^3:
       \bar\meas_r(v_r)\in\tilde\Theta(v_n)\bigr\},
 \nonumber\\
 \Theta_{{\rm phys},v}(v_n)
 &=&\bigl\{v_r\in[0,1]^3:
       \bar\meas_r(v_r)\in\Theta_{\rm phys}(v_n)\bigr\}.
 \label{eq:nfvimpl:domains}
\eeqn
Their indicators are the original indicators evaluated at the mapped
coordinates, with the same zero-extension convention as in
sect.~\ref{sssec:nf-construction}. Next, one must define the
conditional densities in the $dv_r$ measure, as follows:
\beq
 Q_i^{(v)}(v_r\mid v_n)
 =J_{\rm P}(v_r)F_i\bigl(\bar\meas_r(v_r)\mid v_n\bigr),
 \qquad i=1,2.
 \label{eq:nfvimpl:densities}
\eeq
The normalizations follow directly:
\beq
 \int_{\tilde\Theta_v(v_n)}dv_r\,Q_i^{(v)}(v_r\mid v_n)
 =\int_{\tilde\Theta(v_n)}d\bar\meas_r\,
       F_i(\bar\meas_r\mid v_n)=1.
 \label{eq:nfvimpl:normalization}
\eeq
The densities are extended by zero outside their candidate domains, not
outside the physical domain. For $k_\xi=k_y=2$,
eq.~(\ref{eq:dtimpl:jacobians}) gives
$J_{\rm P}=4v_\xi v_y$ and $Q_i^{(v)}=4v_\xi v_y F_i$.

At any valid Born configuration, the common prefactor becomes:
\beq
 C^{(v)}(v_n)
 :=C\bigl(\meas_n(v_n),v_n\bigr)
 =\exp[\operatorname{MLP}(v_n)]\,
      \frac{d\sigma^{(B)}}{dv_n}(v_n),
 \label{eq:nfvimpl:prefactor}
\eeq
which is independent of $v_r$. A physical-domain indicator multiplies the
complete added contribution, not this Born-only function. The Born
quantity in eq.~(\ref{eq:nfvimpl:prefactor}) is a Born-differential
density, not a dumped contribution containing radiation-dependent
sampling factors.

We now write the original and NF-modified $\clS$-event cross
section as follows:
\beqn
 w^{(\clS,v)}(v)&:=&\frac{d\sigma^{(\clS)}}{dv_n\,dv_r},
 \\
 w_{\rm NF}^{(\clS,v)}(v)&:=&
 \frac{d\sigma_{\rm NF}^{(\clS)}}{dv_n\,dv_r}
 =w^{(\clS,v)}(v)+\delta w_{\rm NF}^{K,(v)}(v)
 =J_{\rm P}(v_r)\,
   \bar\varsigma_{\rm NF}^{(\clS)}\bigl(v_n,\bar\meas_r(v_r)\bigr).
 \label{eq:nfvimpl:weights}
\eeqn
By employing the two-flow ansatz, eq.~(\ref{eq:nf:two-flow}),
we then obtain:
\beqn
 w_{\rm NF}^{(\clS,v)}(v)
 &=&w^{(\clS,v)}(v)+\delta w_{\rm NF}^{K,(v)}(v),
 \nonumber\\
 \delta w_{\rm NF}^{K,(v)}(v)
 &=&\mathbf1_{\Theta_{{\rm phys},v}(v_n)}(v_r)
    \mathbf1_{\tilde\Theta_v(v_n)}(v_r)\,C^{(v)}(v_n)J_{\rm P}(v_r)
 \nonumber\\
 &&\quad\times\Bigl[
 F_1\bigl(\bar\meas_r(v_r)\mid v_n\bigr)
 -F_2\bigl(\bar\meas_r(v_r)\mid v_n\bigr)\Bigr]
 \nonumber\\
 &=&\mathbf1_{\Theta_{{\rm phys},v}(v_n)}(v_r)
    \mathbf1_{\tilde\Theta_v(v_n)}(v_r)\,C^{(v)}(v_n)
 \nonumber\\
 &&\quad\times\bigl[Q_1^{(v)}(v_r\mid v_n)-Q_2^{(v)}(v_r\mid v_n)\bigr].
 \label{eq:nfvimpl:correction}
\eeqn
Equivalently,
$d\sigma_{K,\rm NF}=d\sigma_K+dv_n\,dv_r\,\delta w_{\rm NF}^{K,(v)}$.
Thus the only new factor relative to the barred-coordinate expression is
$J_{\rm P}$, exactly as required by the change of measure. It multiplies
the difference of the two flows and is not inserted again after that
factor has been absorbed into $Q_i^{(v)}$. The physical phase-space
jacobian remains in the original densities. The original Born and virtual
terms are unchanged, and the same construction applies to MC@NLO-$\Delta$
as before. We note that the dumped weights associated with uniformly
sampled $x$ are related to the weights above by the following
relationships
\beq
w_{\rm dump}^\prime(x)=
J_{\rm V}(x)\,
w_{\rm NF}^{(\clS,v)}(v(x))\,,\;\;\;\;\;\;\;\;
\delta w_{\rm dump}^{K}(x)
=
J_{\rm V}(x)\,
\delta w_{\rm NF}^{K,(v)}(v(x))\,.
 \eeq
At fixed valid $v_n$, the candidate-domain integral of
$C^{(v)}(Q_1^{(v)}-Q_2^{(v)})$ vanishes. Its physically restricted
integral need not vanish. The probabilities integrated over the
complement of the physical region are
\beqn
 q_i(v_n)
 &=&\int_{\tilde\Theta_v(v_n)\setminus\Theta_{{\rm phys},v}(v_n)}
       dv_r\,Q_i^{(v)}(v_r\mid v_n)
 \nonumber\\
 &=&\int_{\tilde\Theta(v_n)\setminus\Theta_{\rm phys}(v_n)}
       d\bar\meas_r\,F_i(\bar\meas_r\mid v_n).
 \label{eq:nfvimpl:excluded-probabilities}
\eeqn
These are the same probabilities as in sect.~\ref{sssec:nf-construction}.
Therefore, the quantity
\beq
 \mathcal B_{\rm NF}^{(v)}(v_n)
 :=\int_{[0,1]^3}dv_r\,\delta w_{\rm NF}^{K,(v)}(v_n,v_r)
 =C^{(v)}(v_n)\,[q_2(v_n)-q_1(v_n)],
 \label{eq:nfvimpl:bias}
\eeq
which measures the (possible) bias, is essentially unchanged
w.r.t.~that which implicitly appears in eq.~(\ref{eq:rateNF3}).
There is no Born jacobian in this formula because $dv_n=d\bar\meas_n$.
For a Born observable, the corresponding bias-inducing integral is
\beqn
 \frac{d\sigma_{K,\rm NF}}{d\Omega}(\Omega_0)
 -\frac{d\sigma_K}{d\Omega}(\Omega_0)
 &=&\int_{[0,1]^m}dv_n\,
 \delta\!\left(\Omega_0-\Omega(\meas_n(v_n),0)\right)
 \mathcal B_{\rm NF}^{(v)}(v_n).
 \label{eq:nfvimpl:observable-bias}
\eeqn
which is the analogue of eq.~(\ref{eq:NFbiasdiff}).
The integral is understood to vanish at invalid Born points. In particular,
\beq
 |\mathcal B_{\rm NF}^{(v)}(v_n)|
 \leq |C^{(v)}(v_n)|\,[q_1(v_n)+q_2(v_n)].
 \label{eq:nfvimpl:bias-bound}
\eeq
The domain and normalization conditions are therefore identical to those
already established. Suppression of densities at a finite set of invalid
points does not by itself bound the excluded probabilities.

The loss functions must now be computed by means of the Vegas variables.
In order to do that, we start again with eq.~(\ref{eq:nf:training-sample}),
with $\mathcal T$ the same collection of evaluated configurations as
before, but now denoted by $v_a$. We now set
\beqn
 w_a^{\prime(v)}&=&w_{\rm NF}^{(\clS,v)}(v_a),\qquad
 Q_{i,a}^{(v)}=Q_i^{(v)}(v_{r,a}\mid v_{n,a}),\qquad
 J_{{\rm P},a}=J_{\rm P}(v_{r,a}),
 \nonumber\\
 \bar\varsigma_a&=&\frac{w_a^{\prime(v)}}{J_{{\rm P},a}},\qquad
 F_{i,a}=\frac{Q_{i,a}^{(v)}}{J_{{\rm P},a}}.
 \label{eq:nfvimpl:loss-quantities}
\eeqn
For practical training, we now choose objectives acting directly on
the $v$-differential quantities. These are not the literal
reparametrization of the preceding barred-density losses, but
alternative loss definitions compatible with the same cross-section
construction.
\beqn
 L_{\rm neg}^{(v)}
 &=&\frac{\langle|w^{\prime(v)}|\rangle_{\mathcal T}^{\,2}
            -\langle w^{\prime(v)}\rangle_{\mathcal T}^{\,2}}{M_\varsigma},
 \label{eq:nfvimpl:direct-loss-negative}
 \\
 L_{\rm unph}^{(v)}
 &=&\sum_{i=1}^{2}
     \left\langle\log\bigl(Q_i^{(v)}+\epsilon\bigr)
     \right\rangle_{\mathcal T_{\rm unph}},
 \label{eq:nfvimpl:direct-loss-unphysical}
 \\
 L_{\rm reg}^{(v)}
 &=&\sum_{i=1}^{2}
     \left\langle\bigl(Q_i^{(v)}\bigr)^2\right\rangle_{\mathcal T},
 \label{eq:nfvimpl:direct-loss-regularization}
 \\
 L_{\rm tot}^{(v)}
 &=&L_{\rm neg}^{(v)}+\lambda_{\rm unph}^{(v)}L_{\rm unph}^{(v)}
                    +\lambda_{\rm reg}^{(v)}L_{\rm reg}^{(v)}.
 \label{eq:nfvimpl:direct-loss-total}
\eeqn
Here the quantity inside the auxiliary losses is
$Q_i^{(v)}=J_{\rm P}F_i$, with no adaptive Vegas factor. Likewise, the
first loss uses the complete corrected density in $dv_n\,dv_r$.
These definitions specify a different objective from the preceding
barred-density losses on the same sample; the cross-section and bias
formulae are unaffected. The coefficients must be determined for the
objectives actually trained.

\vskip 0.2truecm
\noindent\emph{Normalizing Flows.}
\par\noindent
We use the MadNIS~\cite{Heimel:2023ngj} normalizing flow, a coupling-block
flow with monotonic rational-quadratic spline
transformations~\cite{Durkan:2019nsq}. The transformations are invertible
and spline-based, rather than affine.
Each flow is built as a sequence of four coupling blocks, acting on the
three radiation variables. In each block, the variables are split into two
groups: one group is left unchanged, while each variable in the other group
is reshaped by a smooth, monotonic one-dimensional function (a
rational-quadratic spline). The shape of each spline is not fixed, but is
predicted by a neural network from the unchanged variables and the Born
inputs. The roles of the two groups alternate from block to block, so that,
after the four blocks, each radiation variable has been reshaped twice, and
its transformation has been informed by both of the other radiation
variables. In total, each flow therefore applies six one-dimensional spline
transformations.

\subsection{Results with Normalizing Flows\label{subsec:results-nf}}
As far as the hyperparameters are concerned, the most relevant
for our applications are set as follows.
\begin{itemize}
\item[{\em a)}]
Network size ($L$, $n_{\mathrm{units}}$): The NF spline parameters are predicted
by a ReLU MLP with $L=3$ hidden layers of $n_{\mathrm{units}}=64$ units each,
conditioned on the untransformed radiation variables ($\bar\meas_r$),
and the Born-level inputs. The separate Born-dependent prefactor
($C(\meas_n)$) is parameterized by the exponential of a
ReLU MLP with $L=3$ layers of $n_{\mathrm{units}}=32$ units each.
\item[{\em b)}]
Number of spline bins: Each rational-quadratic spline uses $10$ bins.
The spline parameterizes the invertible transformation; the number of
spline bins is kept fixed, i.e.~it is not fitted to data.
\item[{\em c)}]
Training settings: The learning rate is $\eta_{\mathrm{NF}}=10^{-3}$, the
batch size is $n_{\mathrm{batch}}=2048$. For the epochs during training
we used both $N_{\mathrm{epoch}}=10$ and $N_{\mathrm{epoch}}=20$.
\end{itemize}

\begin{table}[ht]
\centering
\renewcommand{\arraystretch}{1.18}
\begin{tabular}{@{}lrrccrr|r@{}}
\toprule
Process & $f_0(\%)$ & $f_1(\%)$ &
          $\epsml(\%)$ & $\eta_{\mathrm{lin}}(\%)$ & $N_{\rm train}$ &
	  $\fom^\star~~~$ & $\fom^\star[{\rm DT}]~$\\
\midrule
$e^+e^-\to b\bar b$         &
  20.4 & 3.4 & 1.6 & 59.5 & 0.4M & $1.698^{+0.002}_{-0.003}$ &
               $1.707^{+0.002}_{-0.003}$ \\
$e^+e^-\to t\bar t$         &
  0.4 & 0.4 & 1.9 & 72.1 & 0.4M & $-0.004^{+0.002}_{-0.003}$ & 
              $-0.005^{+0.002}_{-0.003}$ \\
$e^+e^-\to b\bar b t\bar t$ &
  21.2 & 16.4 & 0.9 & 33.8 & 1.6M & $0.796^{+0.002}_{-0.003}$ &
                $0.479^{+0.005}_{-0.010}$ \\
$e^+e^-\to b\bar b b\bar b$ &
  29.8 & 26.2 & 0.5 & 17.0 & 3.2M & $1.710^{+0.002}_{-0.003}$ &
                $0.865^{+0.006}_{-0.013}$\\
$e^+e^-\to Hb\bar b$        &
 23.0 &  6.8 & 1.3 & 49.3 & 1M & $2.086^{+0.001}_{-0.003}$ &
               $1.066^{+0.005}_{-0.010}$ \\
$e^+e^-\to Ht\bar t$        &
 0.1 & 0.1 & 1.2 & 45.2 &   1M & $-0.002^{+0.001}_{-0.002}$ &
             $-0.007^{+0.003}_{-0.006}$  \\
$pp\to b\bar b$             &
 17.5 &  6.0 &  0.8 &  30.4 &   1M & $1.073^{+0.001}_{-0.002}$ &
               $0.746^{+0.014}_{-0.028}$ \\
$pp\to t\bar t$             &
 8.2 & 1.0 & 1.2 & 43.9 &   1M & $0.387^{+0.001}_{-0.002}$ &
             $0.280^{+0.002}_{-0.003}$  \\
$pp\to ZZ$                  &
 2.5 & 0.7 & 1.2 & 46.0 & 0.4M & $0.077^{+0.001}_{-0.002}$ &
             $0.064^{+0.003}_{-0.006}$  \\
$pp\to HH$                  &
 11.3 & 0.2 & 1.9 & 71.4 & 0.2M & $0.658^{+0.001}_{-0.003}$ &
              $0.613^{+0.002}_{-0.005}$ \\
\bottomrule
\end{tabular}
\caption{As in table~\ref{tab:newsummary}, for the results obtained
with NFs. The column denoted by $\fom^\star$ is obtained
according to eq.~(\ref{fomexp}).
\label{tab:newsummaryNF}
}
\end{table}
A summary of the results for, chiefly among others, the fractions of
negative weights and the figures of merit, is presented in
table~\ref{tab:newsummaryNF}; these have been obtained by setting the
parameters as specified above, with $N_{\mathrm{epoch}}=10$.
As far as the input parameters are
concerned (leftmost five columns), this table has the same format
as table~\ref{tab:newsummary}. Conversely, the results are
presented only for the f.o.m.~dubbed $\fom^\star$, which is computed
according to eq.~(\ref{fomexp}) (i.e.~in the experimental scenario),
including the fact that the number of training events employed
{\em there}\footnote{I.e.~solely in eq.~(\ref{fomexp}); the training
of NFs is carried out using the number of events which appear
in table~\ref{tab:newsummaryNF}.}
to determine the number of generated events $N$ is that of the DT-driven
simulations, rather than that proper to the NF-driven ones; this
guarantees an equal-footing comparison between the DT- and NF-induced
f.o.m.'s in terms of event numbers, irrespective of the sizes of the
training samples. Having said that, as we have observed
before $\fom^\star$ is largely insensitive to the terms in 
square brackets of eq.~(\ref{eq:fom:def3}), and the choice made here
for the number of events, as opposed to that which defines their number
relative to the NF-training samples, has an impact at the level of the third
significant digits in the quoted values of $\fom^\star$. Finally, the column
of $\fom^\dagger$ results is omitted from table~\ref{tab:newsummaryNF}, and
replaced by that of the $\fom^\star$ values taken from table~\ref{tab:newsummary}
in order to facilitate a more direct comparison between DT- and NF-driven
predictions. In any case, the NF-induced $\fom^\dagger$ values relative
to the NF-induced $\fom^\star$ values follow closely the pattern relevant
to their DT-induced counterparts.

In fact, the general pattern of the f.o.m.~values, and their hierarchy
w.r.t.~process type, is the same in DT- and NF-based simulations.
Quantitatively, NFs appear to usually outperform DTs,
leading to smaller fractions of negative weights and larger
f.o.m.'s. While an analytical understanding of any ML result
is extremely difficult to obtain, it is plausible that one of
the reasons for that is the fact that NFs are smooth and continuous,
and it is the lack of these characteristics in DTs which has been
advocated in sect.~\ref{subsec:results-dt} as a reducible source
of negative weights. We point out that, in spite of the training samples
being smaller for NFs than for DTs, the training effort is not
necessarily less demanding in the former case w.r.t.~the latter
one\footnote{One must bear in mind that the $\epsml$ and
$\eta_{\mathrm{lin}}$ parameters are normalized by the same quantity
in DT- and NF-driven approaches. Thus for example if the absolute
times for ML-induced generation and training operations are 25s and
11s, respectively, for 1M $e^+e^-\to b\bar{b}$ events with DTs
(see footnote~\ref{ft:times}), their NF counterparts are equal 
to 20s and 750s, respectively, where we have used table~\ref{tab:newsummaryNF}
as well as the absolute generation time of 1260s quoted before.},
as is shown by the values of the $\eta_{\mathrm{lin}}$ parameter
in the two cases\footnote{In particular, with NFs the same event is
used in training once {\em per epoch}, and this implies that
$\eta_{\mathrm{lin}}$ has essentially a linear dependence on the
number of epochs.}. 

We also stress that since our DT and NF simulations make use of
statistically compatible but not identical datasets, which may
differ in their sizes and/or how the events that enter them
are randomly selected, some figures of nominally identical
quantities (e.g.~$f_0$) may actually be slightly different in
the two cases.

We now turn to presenting predictions at the differential level.
As we have shown in sect.~\ref{subsec:results-dt}, there exist
classes of negative weights which cannot be eliminated by the
procedure we have proposed in this paper; their nature is such
that this statement applies to any ML implementation, thus
including NFs. Therefore, here we shall limit ourselves to
presenting results based on NF-driven short-distance cross section
modifications for ``physical'' observables, namely for quantities
summed over all channels and FKS sectors. We shall do so for those
processes for which analogous quantities have been obtained by
means of the DT-based procedure, distinguishing between $e^+e^-$
initiated processes ($b\bar{b}$, $b\bar{b}t\bar{t}$, and $Hb\bar{b}$
final states, sect.~\ref{sec:epemNFres}), and $pp$ initiated processes
($b\bar{b}$, $t\bar{t}$, and $HH$ final states, sect.~\ref{sec:ppNFres}).
The figures have the same layouts as those of sect.~\ref{subsec:results-dt},
except for the fact that there is a dotted histogram in each of the
three lower insets. Such a dotted histogram shows the quantity of
relevance for the specific inset (namely, $\Delta\sigma$,
\mbox{$\Delta\sigma/\sigma$}, and $f_{neg}$ from top to bottom)
as is computed by means of the DT-based approach. In other words,
the values in the bins of the dotted histograms in, say, the middle
inset coincide with the values of the red diamonds in the corresponding
plot of sect.~\ref{subsec:results-dt}, which allows one to do an immediate
direct comparison between the DT and NF predictions. Having said that,
one must bear in mind that the training and test phases of the DT-
and NF-induced regressors are different, also because possibly
different training and test samples have been used; therefore, the
blue histogram in the main frame of a DT-based figure generally does
not coincide, but is statistically compatible, with the same quantity
in an NF-based figure.

\subsubsection{$e^+e^-$ collisions\label{sec:epemNFres}}

\noindent
$\bullet~e^+e^-\longrightarrow b\bar{b}$\\\noindent
\begin{figure}[thb]
  \begin{center}
  \includegraphics[width=0.47\textwidth]{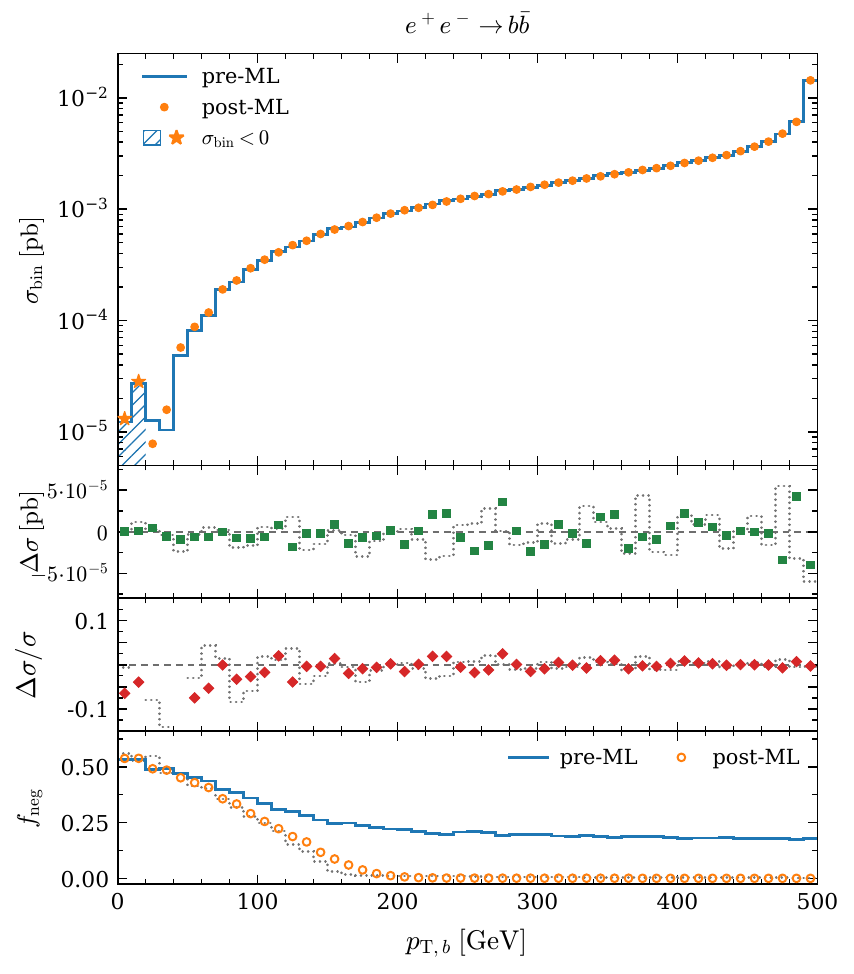}
$\phantom{a}$
  \includegraphics[width=0.47\textwidth]{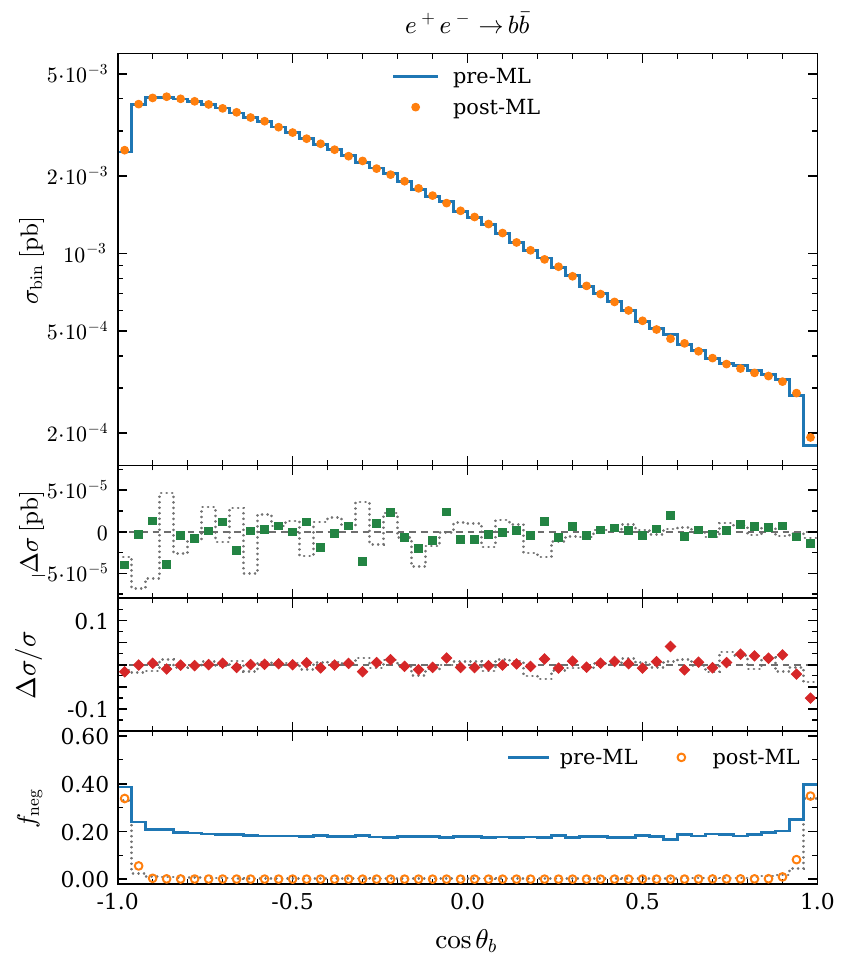}
\caption{\label{fig:NFeebbdiff}
Left panel: $b$-quark $\pt$ in $e^+e^-\to b\bar{b}$;
right panel: $b$-quark $\cos\theta$.}
  \end{center}
\end{figure}
\begin{figure}[thb]
  \begin{center}
  \includegraphics[width=0.47\textwidth]{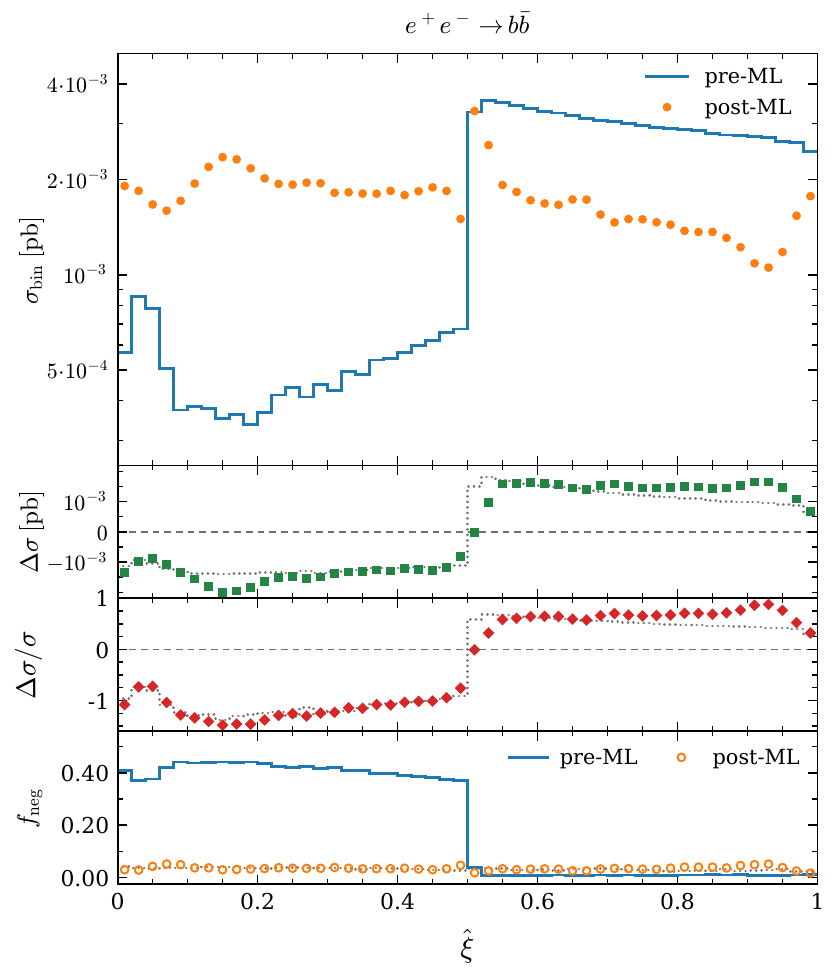}
$\phantom{a}$
  \includegraphics[width=0.47\textwidth]{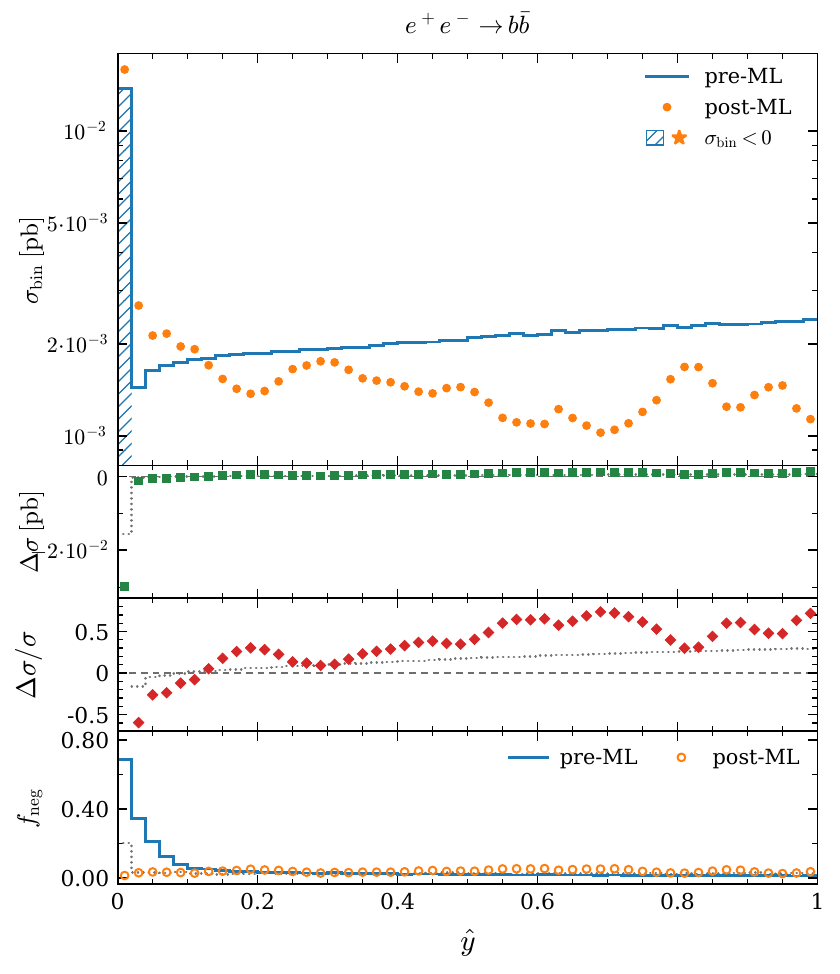}
\caption{\label{fig:NFeebbxiy}
$\hat{\xi}$ and $\hat{y}$ (rescaled FKS variables) distributions
in $e^+e^-\to b\bar{b}$.}
  \end{center}
\end{figure}
We begin by presenting, in fig.~\ref{fig:NFeebbdiff}, the $\pt$ of the
$b$ quark and the cosine of its scattering angle in $e^+e^-\to b\bar{b}$
production; this is the analogue of the DT-based fig.~\ref{fig:eebbdiff}.
For both observables the agreement between pre- and post-NF predictions is 
fairly good, with no signs of biases. In that respect, NFs and DTs are
seen to perform equally well. What is perhaps more remarkable is that
both ML methods have the essentially same behaviour in the small-$\pt$
region as far as the fraction of negative weights is concerned, in that
either fails to reduce that fraction significantly. While this is not
a concern, since in that region the cross section is very small, it
may have some technical relevance, pointing to the necessity of a more
granular learning in Born phase-space regions characterized by a small number
of events, e.g.~through the choice of different learning variables at
the Born level, or by means of a different reweighting function 
$f(\bar\meas_n)$ (see e.g.~eq.~(\ref{eq:mc:source-44})). Having said
that, it is worth bearing in mind that we have already discussed how
this process features a larger negative-weight fraction than physics
would imply owing to the technique used to generate virtual matrix
elements, which may also have an oversized impact on this phase-space
region.

In fig.~\ref{fig:NFeebbxiy} we present the analogue of fig.~\ref{fig:eebbxiy},
namely the distributions in the reduced FKS variables $\hat{\xi}$ and $\hat{y}$.
We remind the reader that these variables are expected to be biased by
the ML learning procedure, which is indeed what we see in the plots.
The different characteristics of the DT regressors and the NFs can be clearly
seen in these figures -- while the DT-induced distributions are extremely
flat, this is not the case for the NF-induced ones, which seemingly
bear memory of their underlying spline structures; this is particularly
evident around the hard edge of the $\hat{\xi}$ distribution. These
observations are, at this stage, largely academic, since as was discussed
before in terms of negative-weight reduction on physical observables the
two ML methods perform comparably well.

\vskip 0.3truecm
\noindent
$\bullet~e^+e^-\longrightarrow b\bar{b}t\bar{t}$\\\noindent
\begin{figure}[ht]
\centering
\includegraphics[width=0.3\linewidth]{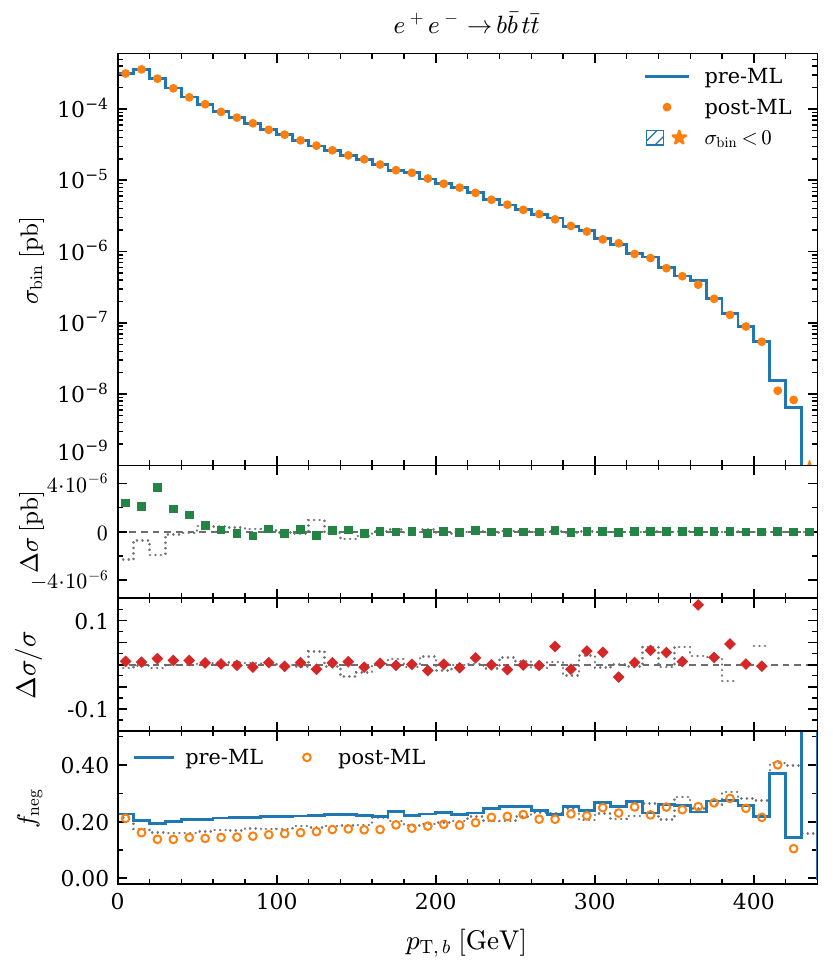}
$\phantom{a}$
\includegraphics[width=0.3\linewidth]{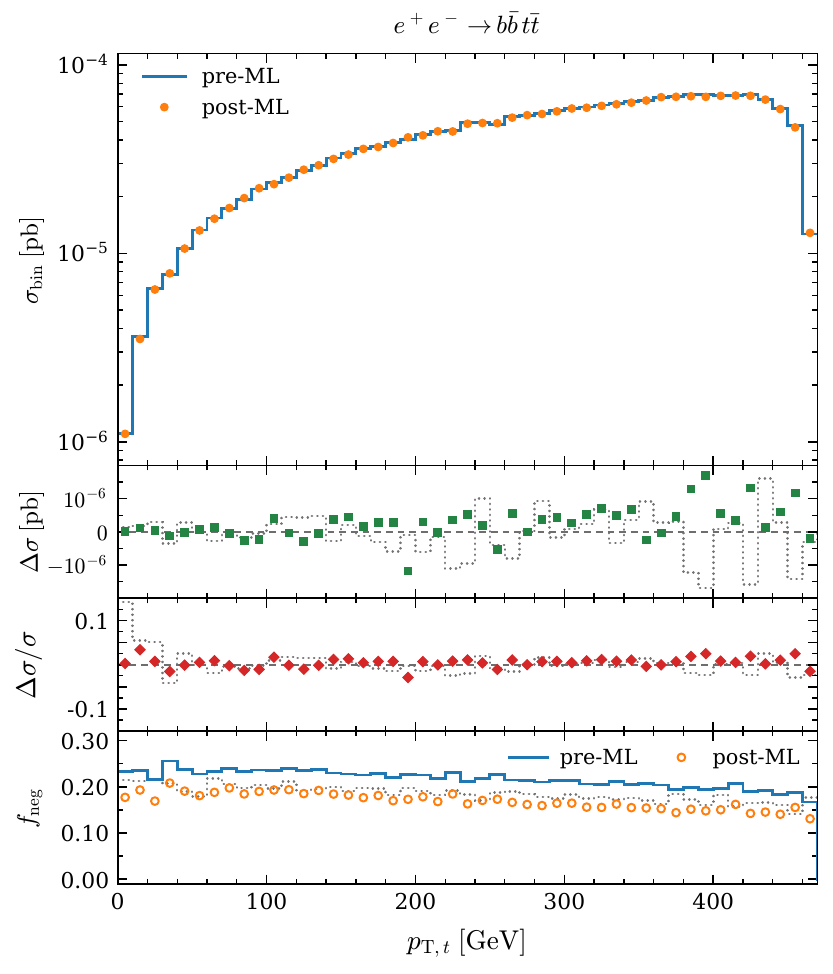}
$\phantom{a}$
\includegraphics[width=0.3\linewidth]{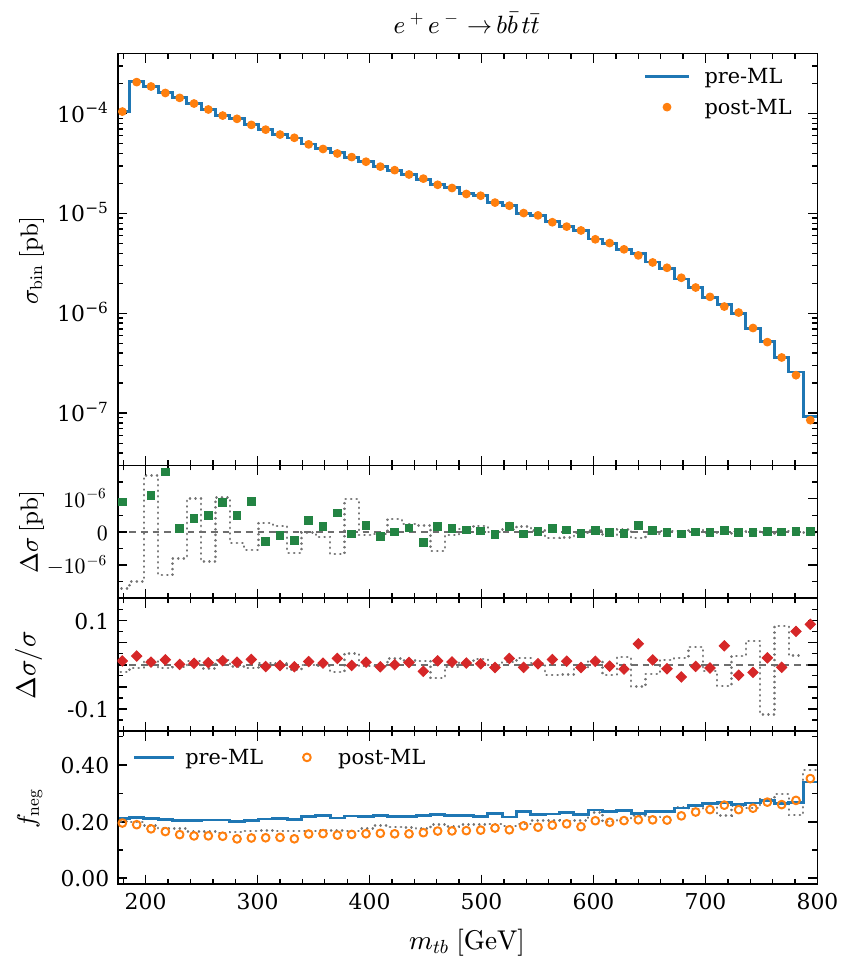}
\caption{Transverse momentum of the bottom and top quarks (left and
middle panels) and invariant mass of the $(b,t)$ pair in
$e^+e^-\to b\bar b t\bar t$.}
\label{fig:NFeebbtt-pttM}
\end{figure}
Sample results for $e^+e^-\longrightarrow b\bar{b}t\bar{t}$ are shown
in fig.~\ref{fig:NFeebbtt-pttM}, with distributions in the $\pt$ of
the bottom (left panel) and top (middle panel) quarks, as well as the
$bt$-pair invariant mass (right panel). The DT analogue of this figure
is fig.~\ref{fig:eebbtt-ptt}, where however the transverse momentum of
the $b$ quark is not presented (this observable is indeed plotted in
fig.~\ref{fig:source7}, but that is limited to the EE6-channel contribution).

The overarching message is that the pre- and post-NF distributions are
in very good agreement, as was the case for the DT-based approach. In fact,
NFs and DTs perform on a similar level; while the former has a smaller
overall fraction of negative weights, for both methods that fraction is
close to a physiological minimum, and post-ML cross sections are essentially
positive everywhere.

\vskip 0.3truecm
\noindent
$\bullet~e^+e^-\longrightarrow Hb\bar{b}$\\\noindent
\begin{figure}[ht]
\centering
\includegraphics[width=0.45\linewidth]{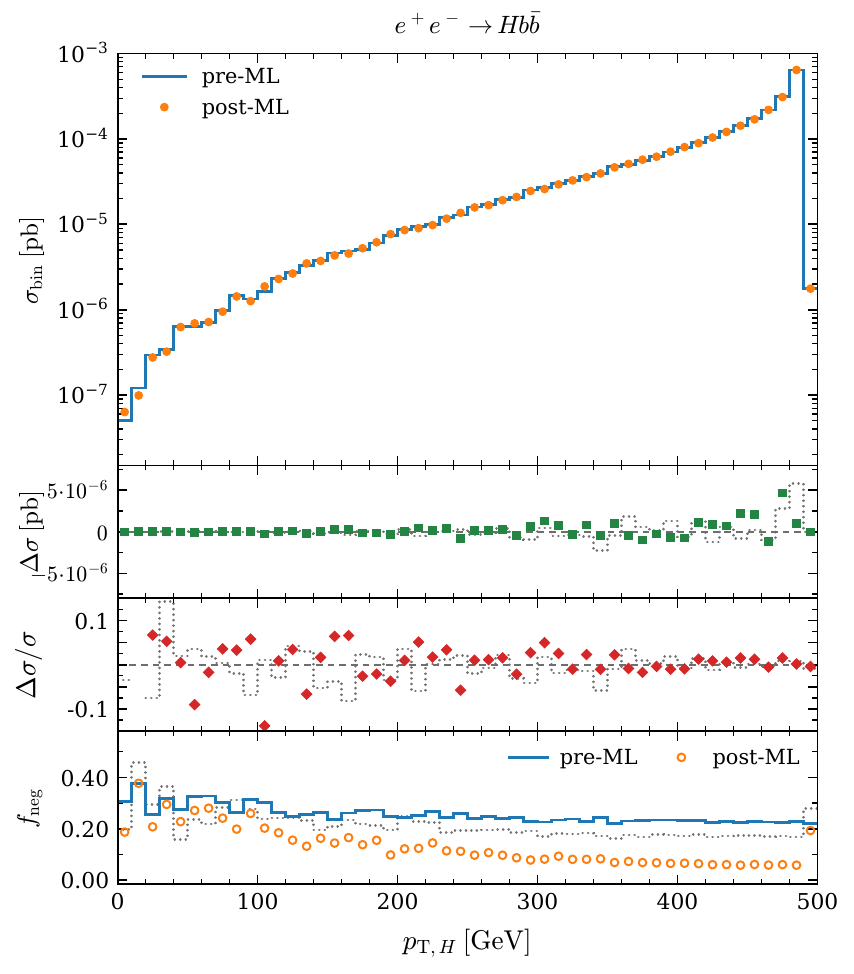}
$\phantom{a}$
\includegraphics[width=0.45\linewidth]{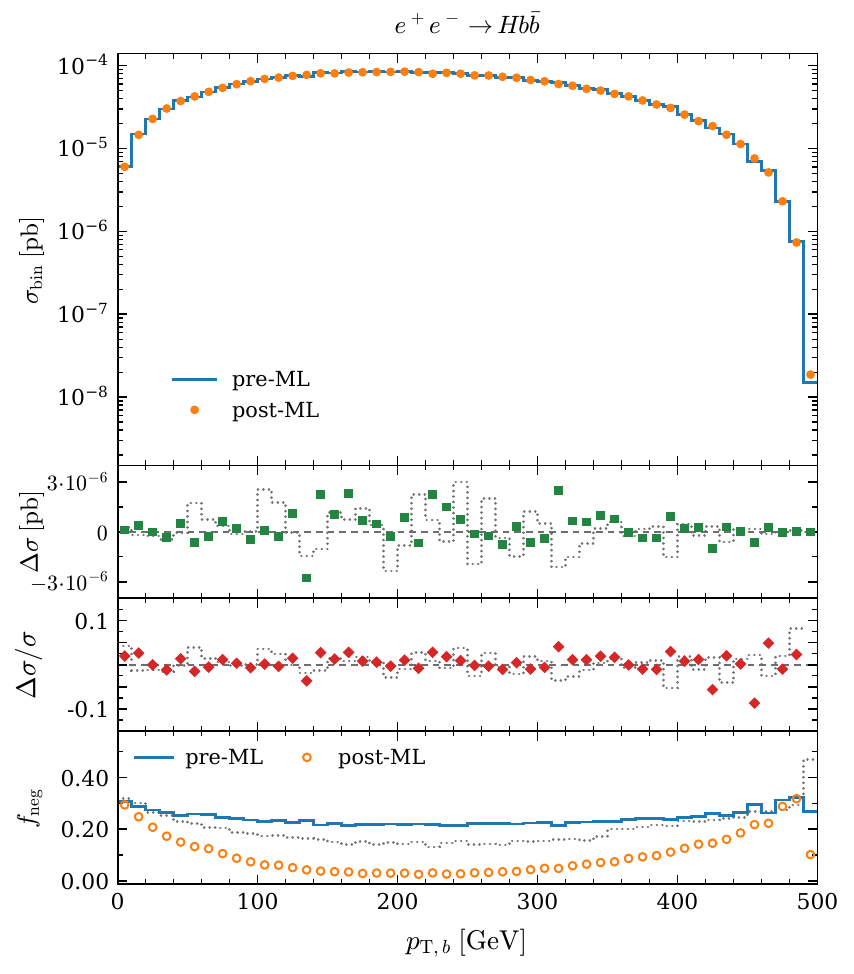}
\caption{Transverse momentum of the Higgs (left panel) and $b$ quark
(right panel) in $e^+e^-\to Hb\bar b$.}
\label{fig:NFeeHbbpts}
\end{figure}
\begin{figure}[ht]
\centering
\includegraphics[width=0.45\linewidth]{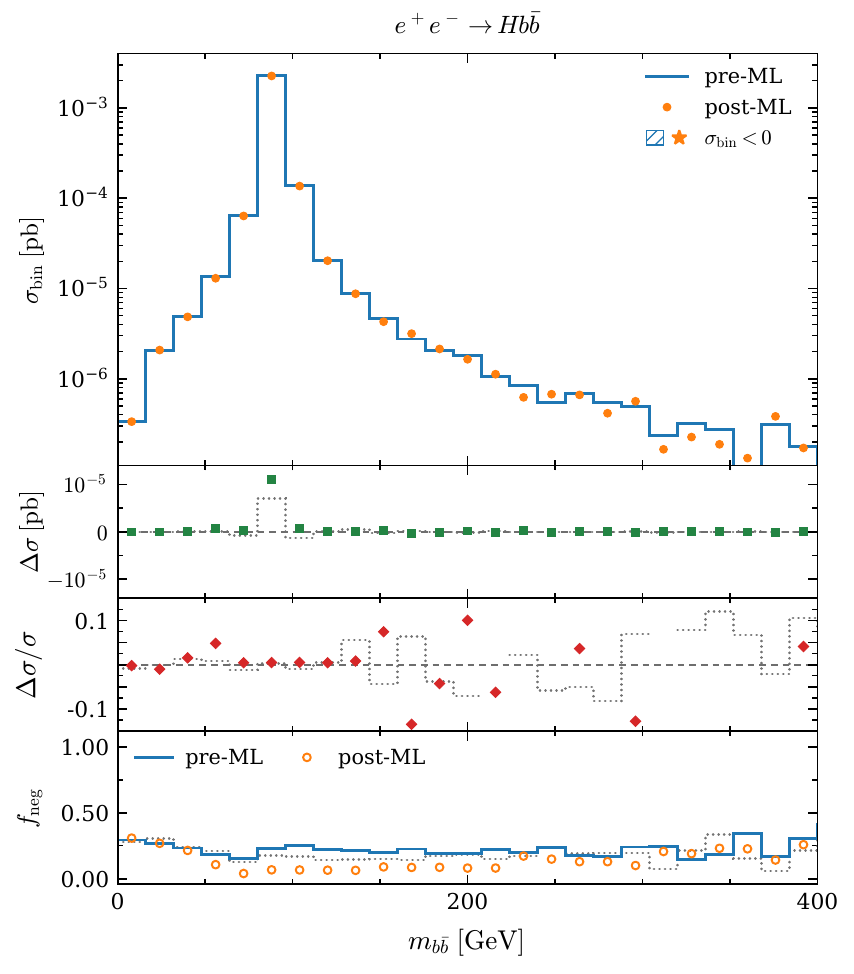}
$\phantom{a}$
\includegraphics[width=0.45\linewidth]{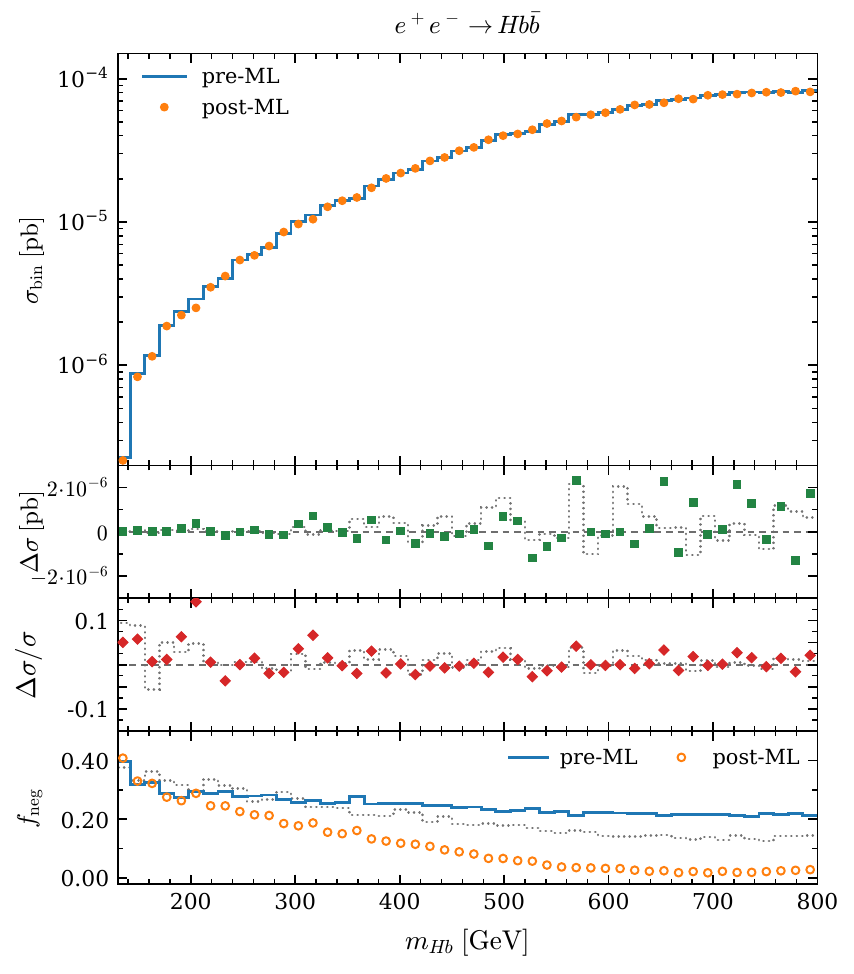}
\caption{Invariant mass of the $b\bar{b}$ pair (left panel) and
of the $Hb$ pair (right panel) in $e^+e^-\to Hb\bar b$.}
\label{fig:NFeeHbbcorr}
\end{figure}
Predictions for $Hb\bar{b}$ production are presented in
fig.~\ref{fig:NFeeHbbpts} ($\pt$ of the Higgs and of the $b$ quark;
its DT analogue is fig.~\ref{fig:eeHbbpts}), and fig.~\ref{fig:NFeeHbbcorr}
(invariant mass of the $b\bar{b}$ and the $Hb$ pair; its DT analogue
is fig.~\ref{fig:eeHbbcorr}). For this process, there is a non-negligible
difference between the DT- and NF-driven performance as far as the
reduction of negative weights is concerned. As can be seen from the
plots, that difference tends to be larger in the regions where the
cross section is large, thus possibly implying that poor training and
inability to learn the regressor on the part of DTs is not an overwhelming
factor here. This also lends support to the alternative explanation put
forward in sect.~\ref{sec:eeHbb}, namely that NF perform better in
view of their smoothness and continuity across the whole phase space.
 
With that being said, the relative comparisons between pre- and post-ML
in the DT and NF cases are quite similar to one another, with the latter
only marginally flatter w.r.t.~the former. In either case, the cross
sections appear not to be biased by the ML procedures.

\subsubsection{$pp$ collisions\label{sec:ppNFres}}
In keeping with what has been done in the DT-based approach,
in hadronic collisions we consider $2\to 2$ processes, and plot
the transverse momentum and laboratory-frame rapidity of the
first particle (in the particle list; this is arbitrary but it
does not matter for our purposes -- see footnote~\ref{ft:ordering}),
and the invariant mass of the two particles.

\vskip 0.3truecm
\noindent
$\bullet~pp\longrightarrow b\bar{b}$\\\noindent
\begin{figure}[thb]
  \begin{center}
  \includegraphics[width=0.3\textwidth]{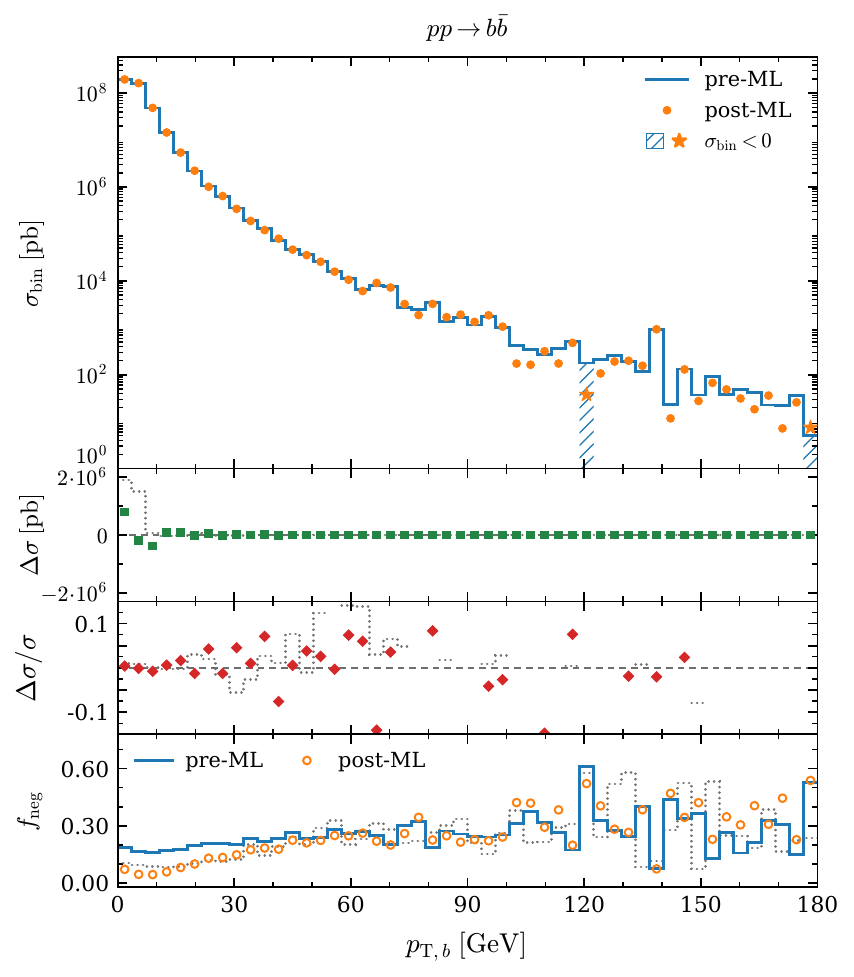}
$\phantom{a}$
  \includegraphics[width=0.3\textwidth]{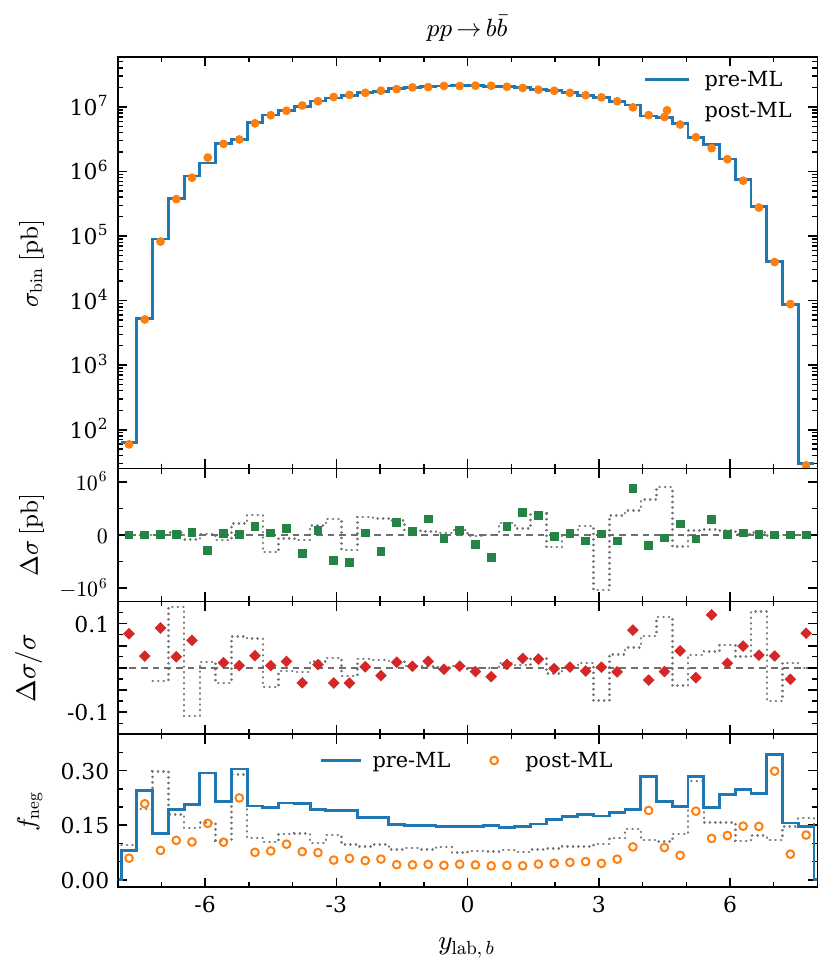}
$\phantom{a}$
  \includegraphics[width=0.3\textwidth]{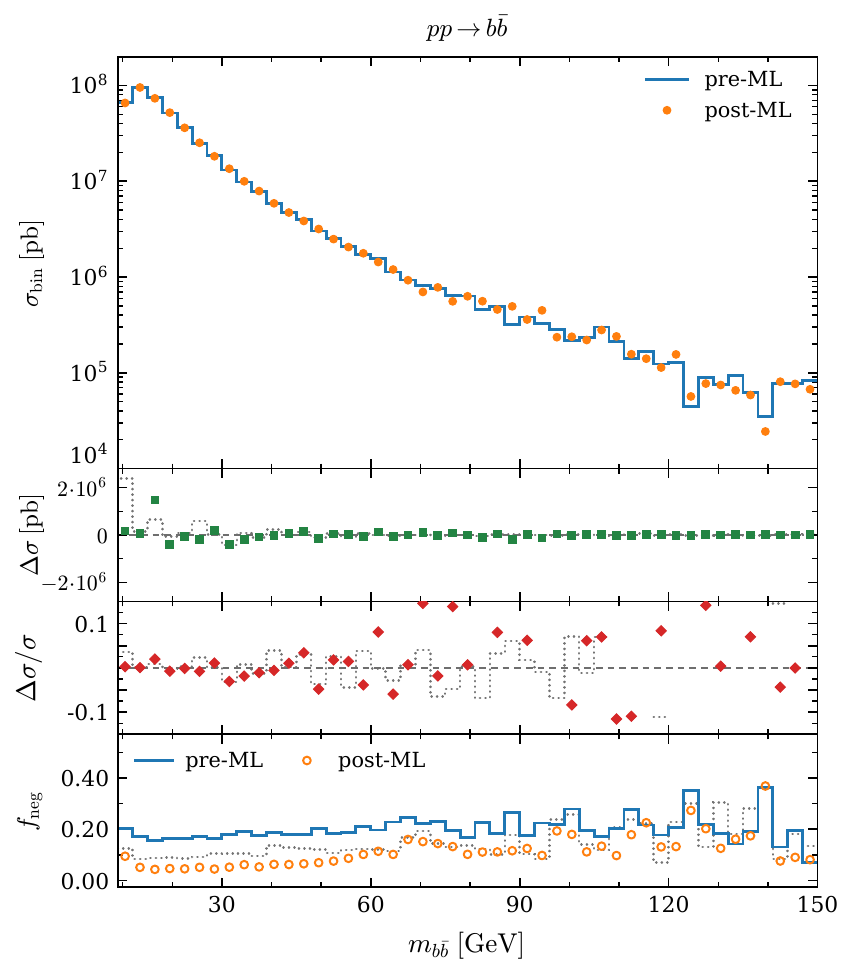}
\caption{\label{fig:NFppbbdistr}
$b$-quark $\pt$ (left panel) and lab-frame rapidity (middle panel),
and $b\bar{b}$-pair invariant mass in $pp\to b\bar{b}$.
}
  \end{center}
\end{figure}
We begin with $b\bar{b}$ production, and show the relevant kinematical
distributions in fig.~\ref{fig:NFppbbdistr}; its DT analogue is
fig.~\ref{fig:ppbbdistr}. This process has a notoriously difficult
convergence, and indeed the results are not as smooth as elsewhere.
Yet, there is no clear pattern emerging from the (relative) difference
between the pre- and post-ML distributions, whence one can confidently
exclude the presence of a non-negligible bias. This is perhaps remarkable,
since we remind the reader that neither of our present treatments of
unphysical configurations {\em establishes the exact} absence of biases
(but it renders it plausible that they are minimized, if they exist).
For DTs, the candidate domain is constructed heuristically, as is described
in appendix~\ref{app:canddom}, without proving its inclusion in the physical
domain for every Born configuration. For NFs, the two densities are exactly
normalized on the candidate domain, but their suppression in its unphysical
part is achieved through training rather than imposed as an exact
constraint. Consequently, the vanishing of the residual proportional
to $q_2-q_1$ (see eq.~(\ref{eq:NFbiasdiffint})) is not guaranteed.
In both approaches, the size of any remaining bias must therefore
be assessed numerically. This is especially true in the case of hadronic
collisions, where e.g.~the DT candidate-domain procedure is seen to return values
of $a_\xi$ and $a_y$ larger than for analogous $e^+e^-$ processes. In the DT-based
approach, we have some evidence that the construction of the candidate
domain would lead to better results if it were sliced in regions of
$\tau$ or $\log\tau$; we have not explored this option any further,
since a better and comprehensive solution will come from an actual
learning of the candidate domain. The analogue of that in the case
of NFs is to condition the probability densities to $\log\tau$.

\vskip 0.3truecm
\noindent
$\bullet~pp\longrightarrow t\bar{t}$\\\noindent
\begin{figure}[thb]
  \begin{center}
  \includegraphics[width=0.3\textwidth]{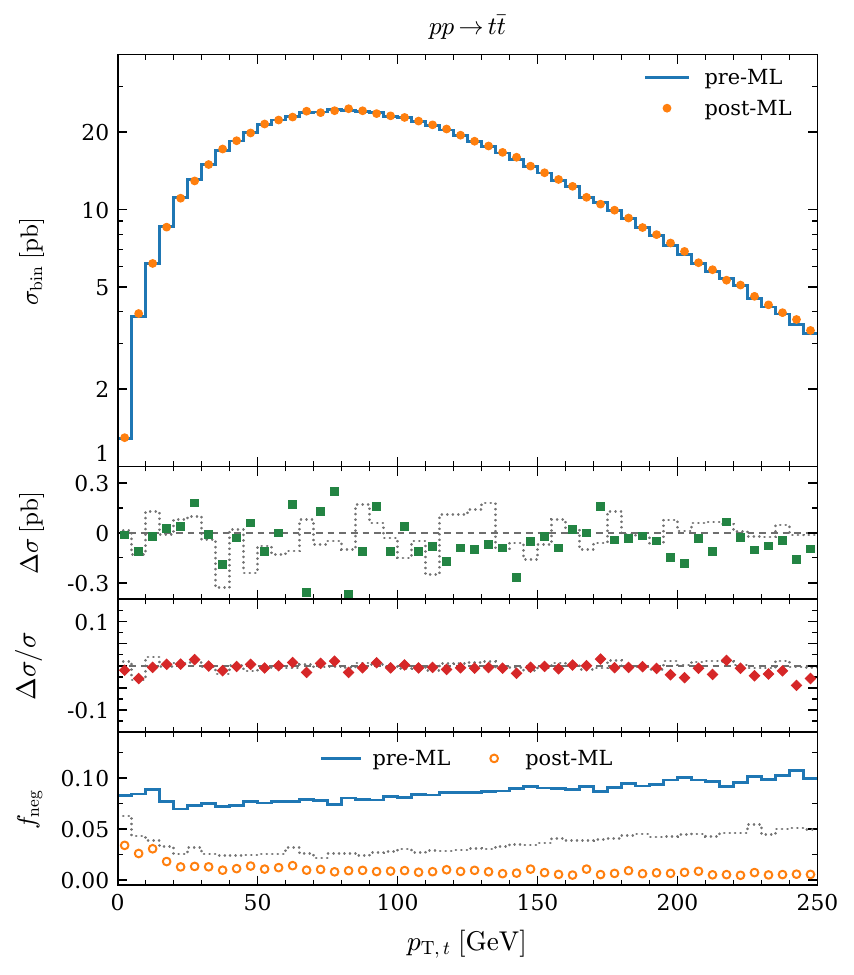}
$\phantom{a}$
  \includegraphics[width=0.3\textwidth]{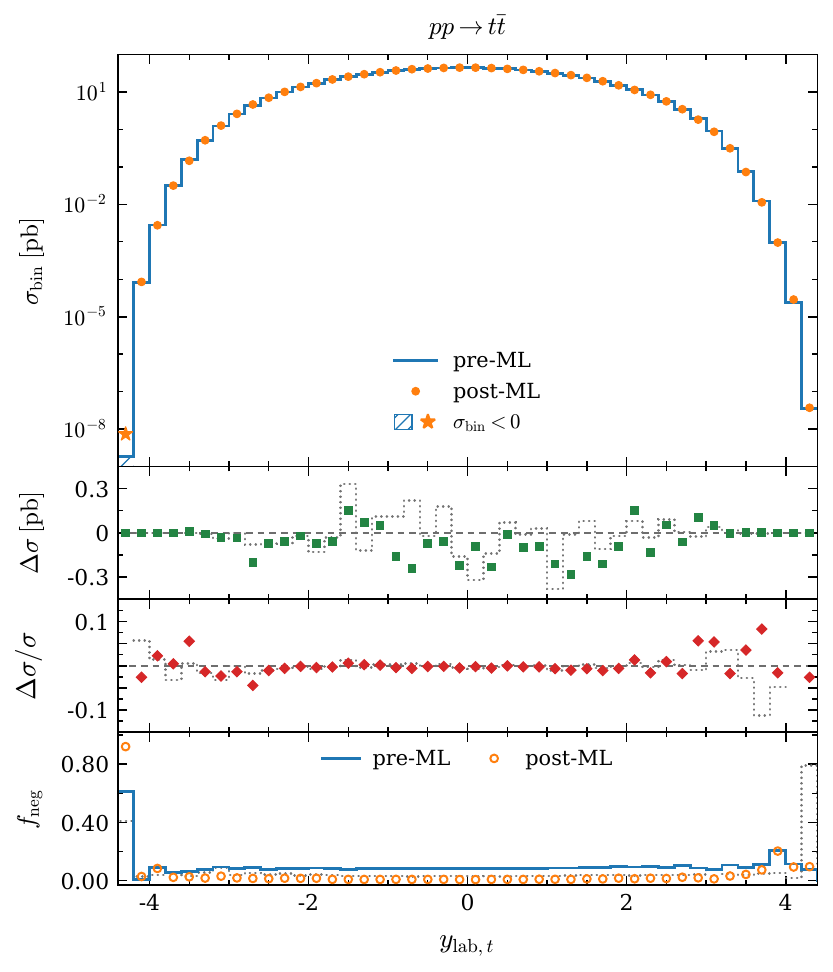}
$\phantom{a}$
  \includegraphics[width=0.3\textwidth]{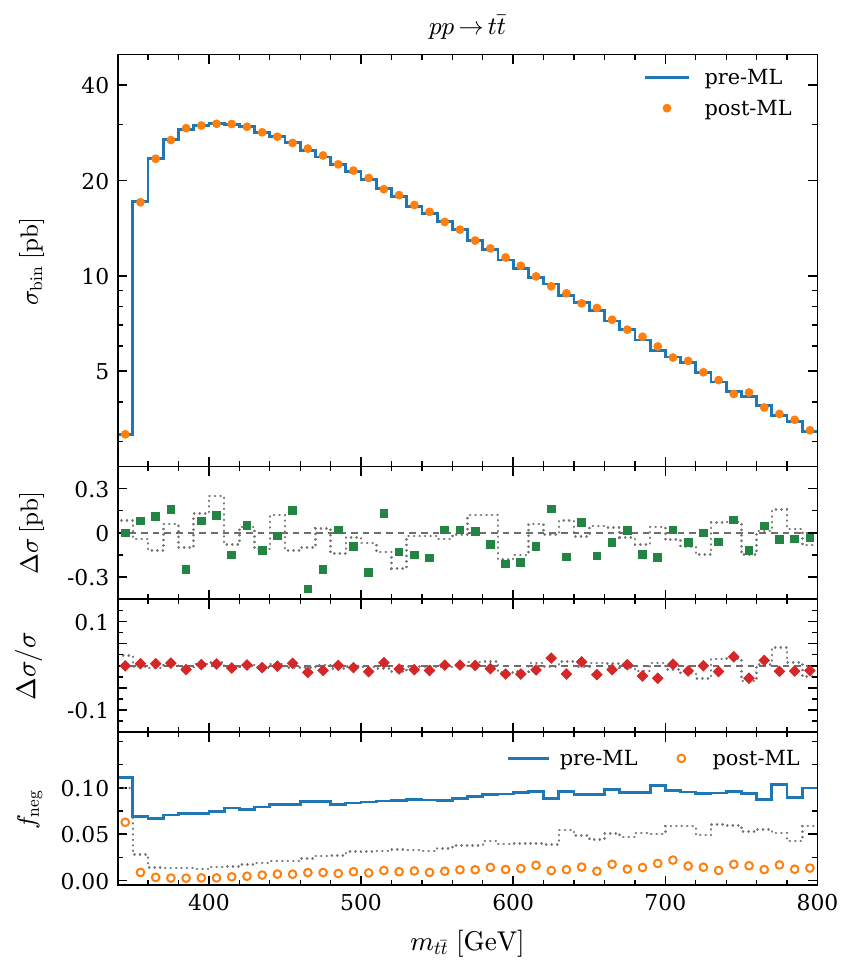}
\caption{\label{fig:NFppttdistr}
Top-quark $\pt$ (left panel) and lab-frame rapidity (middle panel),
and $t\bar{t}$-pair invariant mass in $pp\to t\bar{t}$.
}
  \end{center}
\end{figure}
While from a diagrammatic standpoint the hadroproduction of a $t\bar{t}$
pair is identical to that of $b\bar{b}$ discussed in the previous
section, as far as MC-matching is concerned the two are actually quite
different, since in $t\bar{t}$ production FKS sectors related to final-state
emissions induce much smaller cross-section contributions than their
initial-state counterparts, which is not the case for $b\bar{b}$
production. In addition
to that, since the natural scale for both of these processes is the
heavy quark mass, the perturbative convergence is much faster for
the top than for the bottom. This is evident by looking at our
NF-based predictions in fig.~\ref{fig:NFppttdistr} (whose DT analogue
is fig.~\ref{fig:ppttdistr}), which are quite smooth across the phase
space, both before and after the ML-induced modifications. The pre-
and post-NF distributions agree well with each other, as was the
case in the DT-based approach, and therefore no significant biases
are present.

\vskip 0.3truecm
\noindent
$\bullet~pp\longrightarrow HH$\\\noindent
\begin{figure}[thb]
  \begin{center}
  \includegraphics[width=0.3\textwidth]{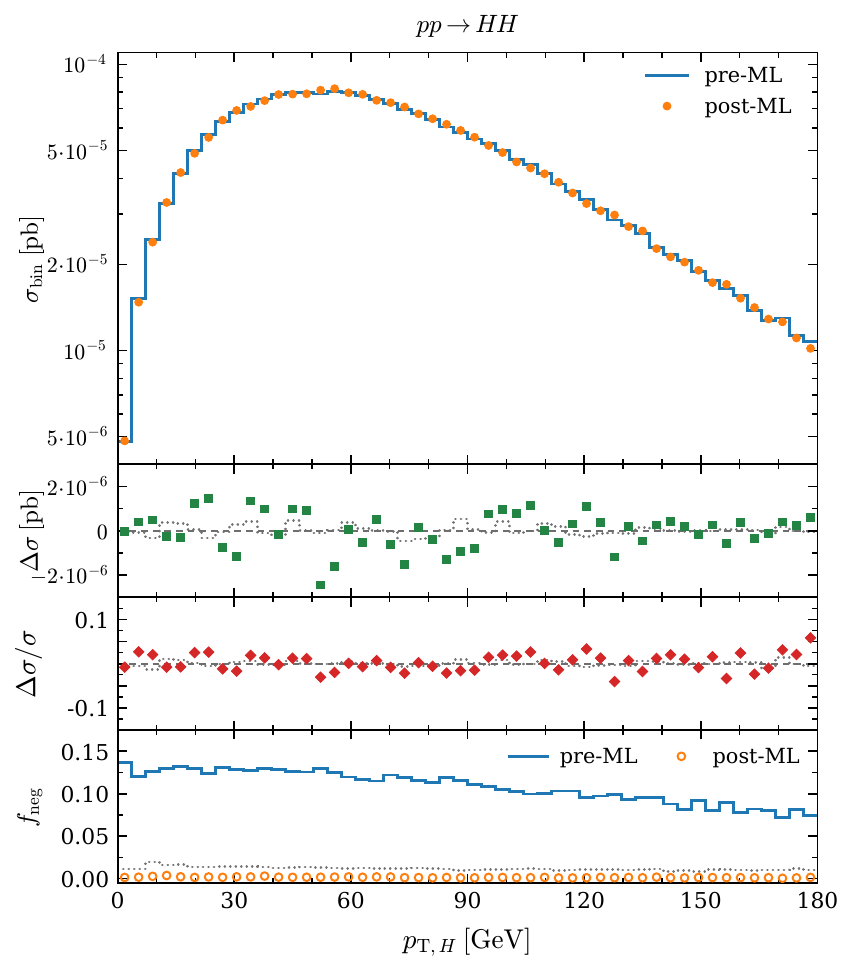}
$\phantom{a}$
  \includegraphics[width=0.3\textwidth]{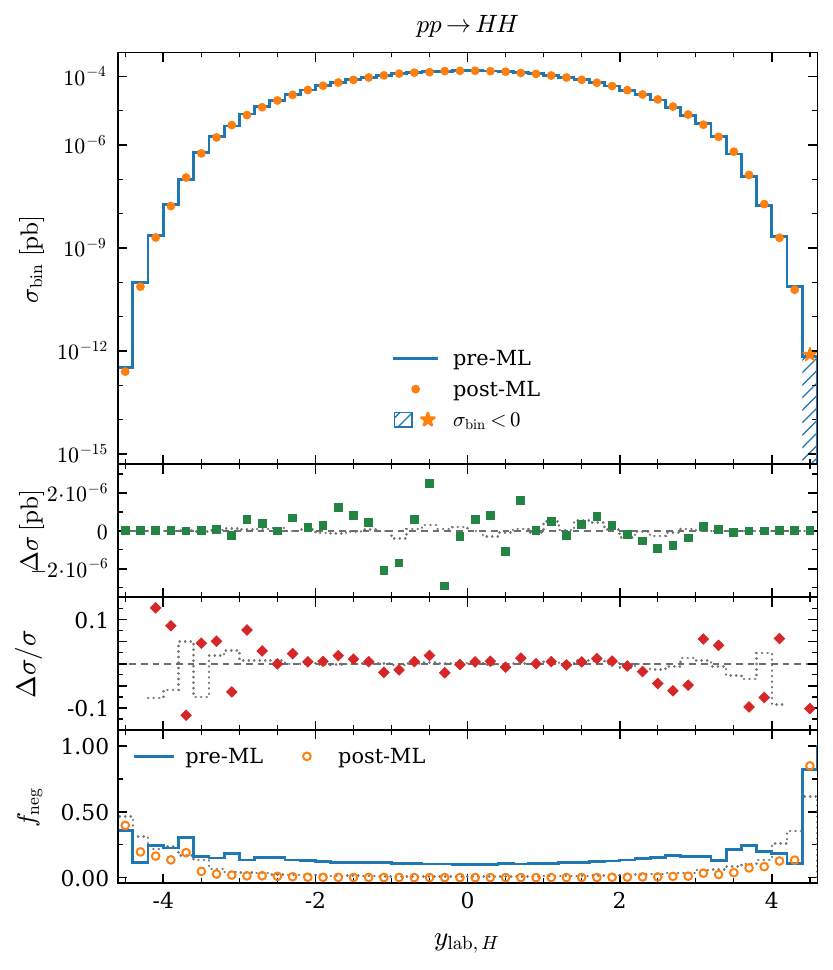}
$\phantom{a}$
  \includegraphics[width=0.3\textwidth]{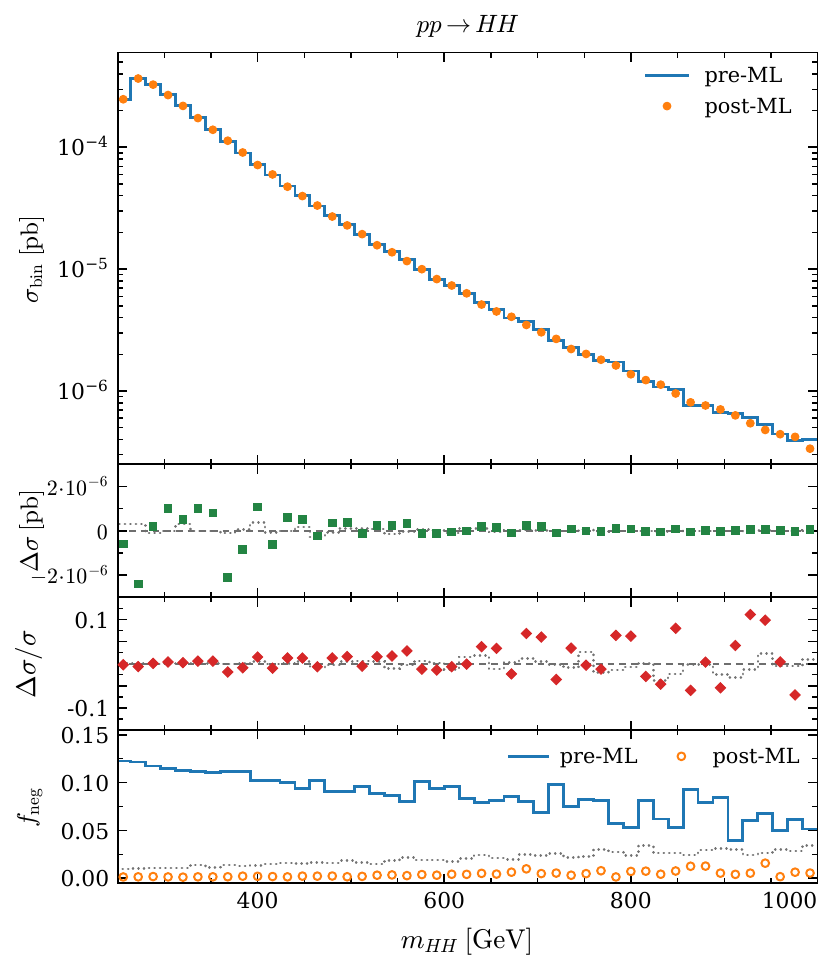}
\caption{\label{fig:NFppHHdistr}
Higgs $\pt$ (left panel) and lab-frame rapidity (middle panel),
and $HH$-pair invariant mass in $pp\to HH$.
}
  \end{center}
\end{figure}
We finally consider Higgs-pair production in fig.~\ref{fig:NFppHHdistr}
(its DT analogue is fig.~\ref{fig:ppHHdistr}). The agreement of pre-
and post-ML predictions is apparently very good, as was already the
case for the DT approach, with the two ML methods performing on a
similar level, despite the NFs having an edge over DTs in the
attained fraction of negative weights. Interestingly,
that advantage appears to grow with the hardness, which may be
beneficial for the stability of large-$\pt$ tails.

\subsubsection{Conclusions on Normalizing Flows\label{sec:NFconcl}}
We note that the underlying issues identified in items~1--4 of
sect.~\ref{sec:DTconcl} also arise in the NF-based procedure whose
results we have presented above, for the simple reason that they are
neither DT- nor NF-specific (while clearly their specific treatment can
differ, as is discussed above e.g.~for unphysical configurations).

Inspection of table~\ref{tab:newsummary} and~\ref{tab:newsummaryNF}
shows how the training phase of NFs is more involved than that of DTs,
which is not surprising in view of the fact that the former are more
complicated mathematical objects than the latter. In general, however,
this is not a limiting factor for the use of NFs, since the f.o.m.'s
we obtain from them are generally larger (i.e.~better) than those obtained 
by means of DTs. The simple reason behind this is that, in either case,
the training time is only a fraction of the generation time, which
dominates the budget (and increasingly so for more complicated
processes).

At the differential level, we see a fair agreement between NF-based
predictions with standard MC@NLO ones, as well as with their DT-based
counterparts. The local reduction of negative weights due to employing
NFs reflects its slightly superior performance w.r.t.~of DTs, and
generally does not follow a constant pattern. This is interesting,
because it suggests that further refining strategies should be
specifically tailored to each method, i.e.~involving different
choices for alternative learning variables.

In the radiation space, the NF-modified cross sections are less flat
than those obtained by DTs. In itself, this is not an unambiguous
signal that something can still be improved. One may speculate that
tests where the number of spline bins is varied could elucidate
this matter. The possibility remains that these shapes are dictated
by the two-flow structure of the model, and in particular its
non-uniform distribution of the subtraction term, per
eq.~(\ref{eq:nf:subtract-add-identity}). In any case, this does not
seem to have a major impact on the model performances, if any.

\section{Conclusions\label{sec:concl}}
In this paper we have introduced a strategy for the reduction of
negative-weight $\clS$ events in MC@NLO-type simulations based on local
modifications of the corresponding short-distance cross section,
dictated by Machine Learning techniques. In particular, we have
explored two different approaches, based on Decision Trees and
Normalizing Flows. The basic idea is more general than either of
these mechanisms, but a pre-requisite for any alternative
is that the chosen method must allow one to compute integrals
over the radiation variables almost instantaneously, in order
for it to be viable.

Our Decision Trees and Normalizing Flows approaches have been tested
on fully-realistic processes in both $e^+e^-$ and $pp$ collisions,
with an emphasis on final states featuring $b$ quarks, which pose
notoriously difficult MC-matching conditions and large fractions
of negative weights. Heuristically, we observe how Normalizing Flows
perform slightly better than Decision Trees, which we tentatively
interpret as the ability of a more sophisticated mathematical model
to describe complicated objects such as MC@NLO cross sections.
In both approaches, the reduction of negative weights is achieved
while maintaining agreement with the original predictions for total
rates and Born-level distributions.
In any case, either method generally constitutes a much cheaper
option w.r.t.~folding, which ideally is the best solution for the
reduction of negative-weight $\clS$ events, but which in practice
is extremely expensive and slow, although progress is being made
to speed it up.

The practical usefulness of these Machine Learning methods must be assessed by
balancing the reduction of negative weights against the costs of
training and of evaluating the learned correction. The figure of merit
introduced in this work provides one such criterion; without being unique,
it can help guide the choice of the training configuration. In the scenarios
considered, the gains are particularly significant for large event productions
with substantial downstream processing costs; the more involved
training of Normalizing Flows is then generally compensated by their
improved negative-weight suppression.

Importantly, while the procedures we have devised are seen to be
quite effective (the reader can have a quick idea of that by inspecting
tables~\ref{tab:newsummary} and~\ref{tab:newsummaryNF}),
the existing results indicate that improving the accuracy or
complexity of the Machine Learning model is not, by itself, sufficient to
eliminate all residual negative weights. The most significant
limitations arise either from negative contributions that survive
integration over radiation variables, or from cancellations between
separately treated contributions that the present Machine Learning procedures
cannot exploit. Additional limitations arise from the numerical treatment of
virtual corrections, the choice of learning variables, and the handling of
unphysical phase-space points. In particular, a more complete
understanding of the physical radiation domain at fixed Born kinematics
would provide further scope for improving the negative-weight reduction.
The distinction between these mechanisms is essential: only the first
represents a fundamental obstruction to radiation-only redistribution
at fixed Born kinematics, while the others suggest directions for
extensions of the current implementation.

The strategy we have pursued is agnostic w.r.t.~the specific approach
which generates negatively-weighted $\clS$ events; in particular, we
have verified that it performs equally well on MC@NLO and MC@NLO-$\Delta$
samples, and is compatible with both of them.

\section*{Acknowledgements}
The authors are grateful to F.~Giuli, T.~Heimel, Z.~Marshall, O.~Mattelaer,
R.~Winterhalder, and E.~Yazgan for having provided us with useful information.
SF thanks the CERN TH Division for hospitality during the course
of this work, M.~Mangano, G.~Vita, and U.~Wiedemann for support,
technical and otherwise, and J.~Bendavid for discussions.
VF and SF have been partially funded by the Eric \& Wendy Schmidt Fund for
Strategic Innovation through the CERN Next Generation Triggers project under
grant agreement number SIF-2023-004.
VF has been the recipient of a Short-Term Scientific Mission Grant,
funded by the COST Action CA24146 ``Machine Learning and Quantum
Computing for Future Colliders''.
MZ~acknowledges the financial support from the MUR (Italy), with
funds of the European Union (NextGenerationEU), through the
PRIN2022 grant 2022EZ3S3F.
SF~warmly thanks Uno the Cat for being there, in the good and
in the bad times. Godspeed, Uno.

We acknowledge the usage of ChatGPT for assisting us in technical
aspects of this work: general cross-checks, and the creation of figures
out of data. Claude has been used for coding.

\appendix

\section{Additional material on Decision Trees\label{sec:appDT}}

\subsection{Construction of candidate domain\label{app:canddom}}
The candidate domain used in the DT-driven simulations of this work
has the form given in eq.~(\ref{eq:dtimpl:L-corner}). In keeping with the
general strategy we follow, a domain $\mathcal L$ is determined for
\begin{figure}[thb]
  \begin{center}
  \includegraphics[width=0.47\textwidth]{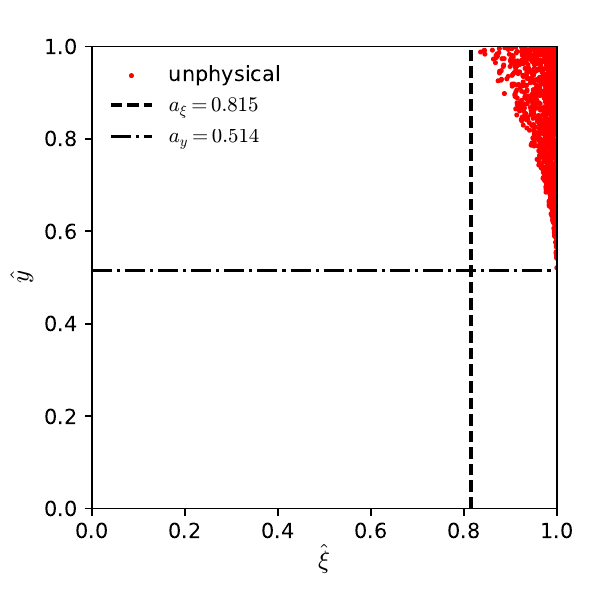}
$\phantom{a}$
  \includegraphics[width=0.47\textwidth]{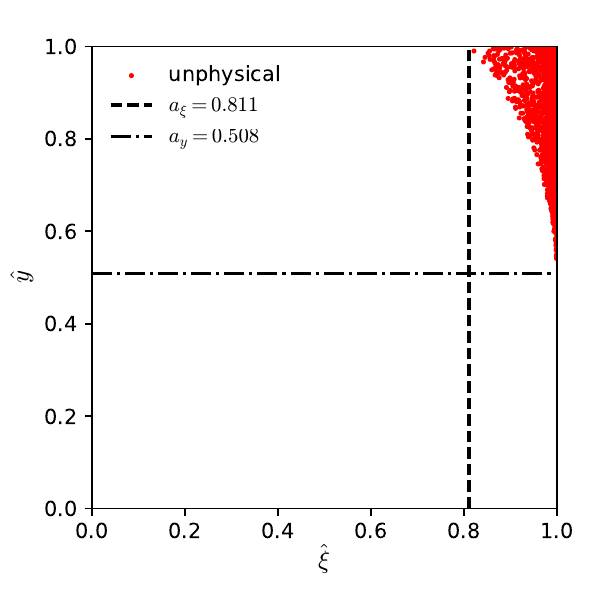}
\caption{\label{fig:canddom}
$pp\to b\bar{b}$ production, channel GG1. Left panel: candidate domain
for FKS sector 3;  right panel: candidate domain for FKS sector 4.}
  \end{center}
\end{figure}
each relevant training set; most often, this means that $\mathcal L$ is
associated with a given combination of a channel and a FKS sector.
The procedure determines $a_\xi$ and $a_y$ separately, by scanning
the (possibly enlarged) training set {\em prior to training}, separately
in $\hat{\xi}$ and $\hat{y}$. Let us denote by $\hat{q}$ either of
these variables, and by \mbox{$q_M=(q_{\min}+q_{\max})/2$}, the
midpoint between two parameters \mbox{$0\le q_{\min}<q_{\max}\le 1$}.
One starts by setting $q_{\min}=0$, $q_{\max}=1$, and checks whether,
in the training sets, there are any unphysical points stemming
from \mbox{$q\in(q_{\min},q_M]$}. If the answer is yes, one sets
\mbox{$q_{\max}=q_M$}, otherwise one sets \mbox{$q_{\min}=q_M$}.
The procedure is then iterated till $q_{\min}$ and $q_{max}$ differ
by less than a pre-established threshold (e.g.~$0.001$), whereby
one sets \mbox{$a_q=q_{\min}$}. By construction, in this way there are
no unphysical configurations associated with values of $q$ smaller
than $a_q$.

An example relevant to an actual physical process, $pp\to b\bar{b}$,
is shown in fig.~\ref{fig:canddom}. There, the red dots correspond
to unphysical configurations. The left and right panels depict the
results of the procedure above for channel GG1 and FKS sectors
3 and 4, respectively. As one can see from the figures,
the values of $a_\xi$ and $a_y$ are similar, but not identical in
the two cases. Since FKS sectors 3 and 4 in this process are symmetric,
one expects identical values only in the infinite statistics limit,
so the non-zero differences between them are a consequence of the
finite size of the training samples.

The plots also give an idea about the loss of efficiency stemming
from determining the candidate domain in this way: essentially,
the white region in the upper-right rectangle is included
in the physical domain, but not in the candidate domain.
According to eq.~(\ref{eq:Thetanobias}), this is a sufficient
condition to avoid any biases; at the same time, it leaves open
the possibility of further improving the reduction of negative
weights.

Finally, one must bear in mind that this construction does not
{\em prove} that the candidate domain so obtained is a subset of the
physical domain -- in a much larger sample, an unphysical (i.e.~red)
point could be generated with $\hat{\xi}<a_\xi$ and/or $\hat{y}<a_y$,
thus inducing a bias\footnote{Strictly speaking, a single such point
would not bias the cross section, since it would constitute a
zero-measure set in the \mbox{$\langle\hat{\xi},\hat{y}\rangle$}
plane. The statement actually assumes that that point would belong
to a part of the unphysical region of non-zero measure which was not
found by the procedure above.}.
However, our results suggest that this situation is quite unlikely.

\subsection{Piecewise-constant approximants and adaptive cell sizes}
\label{app:bdt-granularity}

Consider the ordinary constant-leaf regression tree defined in
eq.~\eqref{eq:app:single-tree}, with learning variables
$z=(\bar\meas_n,\bar\meas_r)$.
The prediction $a_\ell$ is constant within each leaf region $R_\ell$.
For axis-aligned splits, each region is a hyperrectangle in the
learning-variable space:
\beq
 R_\ell=\prod_{j=1}^{d}[x_{\ell j},y_{\ell j}).
 \label{eq:app:leaf-region}
\eeq
The side lengths can differ both between coordinates and between leaves. 

There is normally no hyperparameter that directly specifies a geometrical
side length. Split variables and thresholds are learned from the data;
hyperparameters constrain how finely the tree may partition the space.
Using standard scikit-learn terminology, the principal individual-tree
controls are listed below~\cite{SklearnTreeRegressor}, together with
their main consequences; their settings are given in
section~\ref{subsec:results-dt}.
\begin{center}
\small
\renewcommand{\arraystretch}{1.22}
\begin{tabular}{@{}>{\raggedright\arraybackslash}p{0.34\linewidth}>{\raggedright\arraybackslash}p{\dimexpr0.66\linewidth-2\tabcolsep\relax}@{}}
\toprule
\textbf{Parameter} & \textbf{Effect on the partition}\\
\midrule
\texttt{min\_samples\_leaf} ($n_{\min}$)&
Minimum training occupancy per leaf; smaller values permit finer partitions.\\
\texttt{max\_depth} & Maximum number of successive splits along a root-to-leaf path.\\
\texttt{max\_leaf\_nodes} & Maximum number of terminal regions in one tree.\\
\texttt{min\_samples\_split} &
Minimum node occupancy required for a split to be considered.\\
\texttt{min\_impurity\_decrease} & Minimum improvement required to accept a split.\\
\texttt{ccp\_alpha} & Cost-complexity pruning strength, which limits unnecessary subdivision.\\
\bottomrule
\end{tabular}
\end{center}
In the text, by ``granularity'' we typically refer to
\texttt{min\_samples\_leaf}. There is an interplay of that parameter
with the training-sample size, and we observe what is most likely
overtraining when the partition becomes too fine
(see sect.~\ref{subsec:results-dt}).

The distinction between \emph{sample occupancy} and \emph{geometrical size}
is important. For example, \texttt{min\_samples\_leaf}$=20$ does not
directly prescribe an interval in $\hat{\xi}$ or $\hat{y}$. Heuristically,
a small leaf close to its minimum occupancy obeys
\beq
 N_{\mathrm{train}}\,p_{\mathrm{train}}(z)\,
 \operatorname{Vol}(R_\ell)\sim n_{\min},
 \label{eq:app:leaf-occupancy}
\eeq
provided that the training-point density $p_{\mathrm{train}}$ is
approximately constant across the leaf and the tree is trained on the
sample of size $N_{\mathrm{train}}$ under consideration. Hence
\beq
 \operatorname{Vol}(R_\ell)
 \sim\frac{n_{\min}}
 {N_{\mathrm{train}}\,p_{\mathrm{train}}(z)}.
 \label{eq:app:leaf-volume}
\eeq
Stemming from (approximate) occupancy estimates, this is not a geometrical
bound: the same occupancy can correspond to different volumes and aspect
ratios. Here $p_{\mathrm{train}}$ is the density of training points in the
chosen learning variables, i.e.~generally not a physical cross-section density.

By using DTs, we are bound to obtain piecewise-constant quantities. Note,
however, that the piecewise-constant property refers to the \emph{learned
regressor in its learning coordinates}. Multiplying it by a nonconstant
known scaling function or a change-of-measure jacobian can produce a
function that varies within those cells. Thus the complete correction
supplied to the integrator need not itself be piecewise constant in
the original integration variables, as is explicit in the construction
with alternative learning variables and a Born-dependent scaling
function in subsection~\ref{sssec:bdt-construction}, but it will bear
memory of such an underlying structure.

\subsection{Extensions to tree ensembles}\label{app:tree-ensembles}
Combining finitely many constant-leaf trees does not, by itself, make the
prediction continuous. For example, an additive boosted ensemble has the
form~\cite{SklearnEnsembles}
\beq
 \widetilde K_{\mathrm{BDT}}(z)
 =b+\sum_{t=1}^{N_{\mathrm{trees}}}\nu_t\,T_t(z),
 \label{eq:app:boosted-ensemble}
\eeq
where $b$ is a constant initial prediction and $\nu_t$ specifies the
contribution of tree $t$. This expression illustrates a possible
gradient-boosted extension; it does not describe the individual-tree
regressors used for the main numerical results. Since every $T_t$ is constant
within its own leaves, the sum is constant on the \emph{common refinement} of
all their partitions: regions
on which the leaf assignment remains unchanged in every tree.
Consequently, $\widetilde K_{\mathrm{BDT}}$ is constant, and hence
differentiable with zero derivative, in the interiors of these regions,
but generally jumps across their boundaries. The regions of constancy of
the ensemble can therefore be (and generally are) smaller than the
leaves of any individual tree.

The number of trees affects the ensemble's effective partition because
additional trees can introduce additional boundaries. However, a
minimum-occupancy requirement on each tree's leaves need not hold for the
smaller intersections defining their common refinement. Therefore,
the emerging structure may be more suitable to describing a complex
short-distance cross section than an individual tree with finer
granularity, in a way possibly mimicking what happens with NFs;
clearly, this implies additional computational costs.

For the additive ensemble in eq.~\eqref{eq:app:boosted-ensemble},
linearity gives its conditional radiation integral on the prescribed
support:
\beqn
 I_{K,\mathrm{BDT}}(\bar\meas_n)
 &=&b\,\bar V(\bar\meas_n)
 +\sum_{t=1}^{N_{\mathrm{trees}}}\nu_t I_t(\bar\meas_n),
 \nonumber\\
 I_t(\bar\meas_n)
 &=&\int_{\widetilde{\Theta}(\bar\meas_n)}d\bar\meas_r\,
 T_t(\bar\meas_n,\bar\meas_r).
 \label{eq:app:ensemble-integral}
\eeqn
Here $b$ and the coefficients $\nu_t$ are fixed after training, and
$\bar V$ is the volume of the same radiation domain. The constant $b$
cancels between subtraction and normalized add-back (i.e.~the integrand
in eq.~(\ref{eq:Ktildenormbar}), and its analogues). Each tree can be
integrated separately, so the common refinement need not be constructed
explicitly to evaluate the integral. 

The same reasoning applies to a random-forest regressor whose prediction
is the arithmetic mean of its constituent trees~\cite{SklearnEnsembles}.
Either way, the effectiveness of the methods, their training times, and
recurring evaluation and integration costs must be assessed through the
figure of merit of section~\ref{sec:fom}.

\section{Additional material on Normalizing Flows\label{sec:appNF}}

\subsection{The constructed function and its regularity with
Normalizing Flows}
\label{app:nf-regularity}
We use the two conditional densities and the common Born-dependent
coefficient introduced in sect.~\ref{sssec:nf-construction}. In particular,
$C$ abbreviates $C(\meas_n(\bar\meas_n),\bar\meas_n)$ and is independent
of the radiation variables. Keeping this coefficient outside the learned
shape, the identification in eq.~(\ref{eq:nf:bdt-identification}) is
\beqn
 \widetilde K_{\mathrm{NF}}(\bar\meas_n,\bar\meas_r)
 &=&-F_1(\bar\meas_r\mid\bar\meas_n),
 \nonumber\\
 I_{K,\mathrm{NF}}(\bar\meas_n)
 &=&\int_{\tilde\Theta(\bar\meas_n)}d\bar\meas_r\,
            \widetilde K_{\mathrm{NF}}(\bar\meas_n,\bar\meas_r)=-1.
 \label{eq:app:nf-approximant}
\eeqn
The integral is added back with the normalized distribution $F_2$, rather
than with a uniform distribution. Thus
$-\widetilde K_{\mathrm{NF}}+I_{K,\mathrm{NF}}F_2=F_1-F_2$, and the
actual density added to the original remainder is
\beqn
 \frac{d\delta\sigma_{K,\mathrm{NF}}}
      {d\bar\meas_n\,d\bar\meas_r}
 &=&\mathbf1_{\rm phys}(\bar\meas_n,\bar\meas_r)
    \mathbf1_{\tilde\Theta(\bar\meas_n)}(\bar\meas_r)\,
    C\bigl[F_1(\bar\meas_r\mid\bar\meas_n)
             -F_2(\bar\meas_r\mid\bar\meas_n)\bigr].
 \label{eq:app:nf-physical-density}
\eeqn
We remind the reader that the algebraic identification above
does not imply that the NFs are trained by direct regression of $K$.

For each flow, its conditional invertible transformation can be
written as follows:
\beq
 \bar\meas_r=T_{\theta_i}(u\mid\bar\meas_n),\qquad
 u\sim p_0(u),\qquad i=1,2.
 \label{eq:app:flow-map}
\eeq
The Born coordinates are fixed when taking the radiation jacobian. The
change-of-variables formula gives~\cite{Papamakarios2021}
\beq
 F_i(\bar\meas_r\mid\bar\meas_n)
 =p_0(u_i)
  \left|\frac{\partial T_{\theta_i}^{-1}
                   (\bar\meas_r\mid\bar\meas_n)}{\partial\bar\meas_r}\right|,
 \qquad
 u_i=T_{\theta_i}^{-1}(\bar\meas_r\mid\bar\meas_n).
 \label{eq:app:flow-density}
\eeq
Note that the learned density already contains a derivative of the
transformation, so the order of its differentiability is not the same
as that of the transformation map.

For a simplified one-dimensional illustration at fixed Born coordinates, let
$s=T_i(u)$, with a uniform base on the unit interval and $T_i'(u)>0$.
Within the transformed interval,
\beq
 F_i(s)=\frac{1}{T_i'(u)},\qquad u=T_i^{-1}(s).
 \label{eq:app:one-dimensional-density}
\eeq
Wherever $T_i''$ exists, differentiation gives
\beq
 \frac{dF_i}{ds}=-\frac{T_i''(u)}{[T_i'(u)]^3}.
 \label{eq:app:density-derivative}
\eeq
Hence a continuous first derivative of the map gives a continuous density,
whereas differentiability of the latter involves a second derivative of
the map. Rational-quadratic spline transformations match their values
and first derivatives at the internal knots; they are $C^1$
(i.e.~continuous with a continuous first derivative), but their
second derivatives need not match there~\cite{Durkan:2019nsq}. The induced
density can therefore be continuous with kinks at the knots.
The distinction between map regularity and density regularity is also
relevant to the smooth-flow constructions of ref.~\cite{Kohler2021}.

For conditional coupling transformations, the networks that supply the
spline parameters matter as well. ReLU conditioners are continuous and
piecewise smooth, but need not be differentiable on their switching
surfaces. The scalar spline derivatives entering the triangular
jacobian are continuous functions of the coordinates and spline
parameters. Subject to nonsingular spline maps, the resulting densities
are therefore continuous and piecewise smooth in the interior, but they
need not be differentiable on the knot or conditioner switching surfaces.
Therefore, in general, one should not assume that the complete multidimensional
map is globally $C^1$ merely because each scalar spline has that property for fixed
conditioning inputs. It should be clear that, by construction, the difference
of flows relevant to our strategy, $F_1-F_2$, retains continuity and
differentiability wherever both are continuous and differentiable, but
it does not in general cancel their individual kinks.

\phantomsection
\addcontentsline{toc}{section}{References}
\bibliographystyle{JHEP}
\bibliography{negwgtML}
\end{document}